\documentclass{article}
\usepackage{graphicx}

\usepackage[margin=1in]{geometry}

\usepackage[T1]{fontenc}
\usepackage{comment}

\usepackage{float}
\usepackage[labelfont=bf,labelsep=period]{caption}
\usepackage{authblk}

\usepackage{amsmath}
\usepackage{amssymb}
\usepackage{siunitx}
\usepackage[super]{nth}

\usepackage{xr-hyper}
\usepackage{titlesec}
\usepackage{scalerel}
\usepackage{csquotes}
\usepackage{setspace}
\usepackage{orcidlink}
\usepackage{graphicx}
\usepackage{xcolor, hyperref}

\usepackage{booktabs}

\hypersetup{
    colorlinks,
    citecolor=electricultramarine,
    filecolor=.,
    linkcolor=mediumtealblue,
    urlcolor=smokyblack,
    linktoc=all
}
\definecolor{smokyblack}{rgb}{0.06, 0.05, 0.03}
\definecolor{mediumtealblue}{rgb}{0.0, 0.33, 0.71}
\definecolor{electricultramarine}{rgb}{0.25, 0.0, 1.0}

\usepackage[noabbrev,nameinlink]{cleveref}
\crefdefaultlabelformat{#2\textbf{#1}#3}
\Crefname{figure}{\textbf{Figure}}{\textbf{Figures}}

\newcommand\scht[1]{\stretchrel*{$\textsc{#1}$}{\textsc{x}}}

\usepackage{graphicx}
\usepackage[margin=1in]{geometry}
\usepackage[T1]{fontenc}
\usepackage{comment}
\usepackage{float}
\usepackage{amsmath}
\usepackage{siunitx}
\usepackage[labelfont=bf,labelsep=period]{caption}
\usepackage{subcaption}
\usepackage{xr-hyper}
\usepackage{xcolor, hyperref}

\hypersetup{
    colorlinks,
    citecolor=electricultramarine,
    filecolor=.,
    linkcolor=mediumtealblue,
    urlcolor=smokyblack,
    linktoc=all
}
\definecolor{smokyblack}{rgb}{0.06, 0.05, 0.03}
\definecolor{mediumtealblue}{rgb}{0.0, 0.33, 0.71}
\definecolor{electricultramarine}{rgb}{0.25, 0.0, 1.0}
\usepackage[noabbrev,nameinlink]{cleveref}
\crefdefaultlabelformat{#2\textbf{#1}#3}
\Crefname{figure}{\textbf{Figure}}{\textbf{Figures}}

\titleformat*{\section}{\bfseries\LARGE}
\titleformat*{\subsection}{\sloppy\bfseries\large\hyphenchar\font=-1}

\usepackage[main=english]{babel}
\usepackage[sorting=ynt,backend=biber,giveninits,citestyle=numeric-comp,autocite=superscript]{biblatex}
\DeclareNameAlias{author}{family-given}

\DeclareFieldFormat{doi}{%
  \hfil\penalty90\hfilneg\space DOI\addcolon\addnbspace
  \ifhyperref
    {\href{https://doi.org/#1}{\nolinkurl{#1}}}
    {\nolinkurl{#1}}}
\DeclareFieldFormat{labelnumberwidth}{{#1\adddot}}

\newcommand*{\mkbibbracketsuperscript}[1]{%
  \mkbibsuperscript{%
  {#1}}}

\DeclareCiteCommand{\supercite}[\mkbibbracketsuperscript]
  {\usebibmacro{cite:init}%

   \iffieldundef{prenote}
     {}
     {\BibliographyWarning{Ignoring prenote argument}}%
   \iffieldundef{postnote}
     {}
     {\BibliographyWarning{Ignoring postnote argument}}}
  {\usebibmacro{citeindex}%
   \usebibmacro{cite:comp}}
  {}
  {\usebibmacro{cite:dump}}

\title{A mathematical framework for centromere-aware evaluation of human genome assemblies}
\author[c,*]{Luca Franco\,\orcidlink{0000-0003-0107-6755}\,}
\author[b,c,*]{Matteo Migliarini\,\orcidlink{0009-0004-2333-1010}\,}
\author[a,*]{Matteo Tommaso Ungaro\,\orcidlink{0000-0002-1107-3307}\,}
\author[b,*]{\authorcr Egnald \c{C}ela\, \orcidlink{0009-0002-1602-7817}}
\author[a]{Luca Corda\,\orcidlink{0009-0003-5294-9285}}
\author[a]{Andreas Giannis\,}
\author[a]{Ester Mondelli\,}
\author[b,c,$\mathsection$]{\authorcr Fabio Galasso\,\orcidlink{0000-0003-1875-7813}\,}
\author[a,$\mathsection$]{Simona Giunta\,\orcidlink{0000-0002-6666-9271}}

\affil[a]{Department of Biology and Biotechnologies “Charles Darwin”, University of Rome “La Sapienza”, 00185 Rome, Italy}
\affil[b]{Department of Computer Science, University of Rome “La Sapienza”, 00185 Rome, Italy}
\affil[c]{ItalAI, 00186 Rome, Italy \url{https://italailabs.com/}}
\affil[*]{{\footnotesize These authors contributed equally to the work}}
\affil[$\mathsection$]{{\footnotesize For correspondence, email \url{galasso@di.uniroma1.it} and \url{simona.giunta@uniroma1.it}}}
\date{}

\begin{document}

\maketitle

\begin{abstract}
    \noindent Accurate evaluation of genome assemblies within highly repetitive regions, such as centromeres, remains a major open challenge in genomics. Conventional benchmarking relies on sequence alignment, which becomes problematic in regions of high homogeneity and divergence. Here, we framed centromere assembly evaluation as a comparative distribution problem in a compact centeny representation by computing genomic distances between functional motifs, rather than relying on nucleotide sequence. Our distribution-based metric assesses agreement between a query and a target chromosome by comparing their centromeric inter-motif distances rendered by \textsc{kl} divergence. When applied genome-wide to currently available human telomere-to-telomere (\textsc{t\scht{2}t}) genomes, this approach yields an accuracy ranking for the entire assembly and for each individual chromosome. Altogether, we present a rapid and robust scoring system based on genomes’ numerical rendering of inter-motif distances, that provides a quantitative standard of assembly integrity in repetitive \textsc{dna} regions and establishes a \textit{bona fide} framework for chromosome-level genome-to-genome comparison.
\end{abstract}

\section*{Introduction}
\label{sec:introduction}

Assembly of complete human genomes, defined as telomere-to-telomere (\textsc{t\scht{2}t}) and spanning every chromosome from one end to the other, has recently become possible due to long-read sequencing technologies and computational advances. \textsc{t\scht{2}t} assemblies are set to be the new standard for genome biology. \supercite{nurk2022complete} High-fidelity (\textsc{h}i-\textsc{f}i) long reads and ultra-long Oxford Nanopore (\textsc{ont}) sequencing integrated with graph-based assemblers, such as \texttt{Verkko}\supercite{rautiainen2023telomere} and \texttt{hifiasm},\supercite{rautiainen2023telomere} routinely resolve previously intractable megabase-scale repeats. The \textsc{t\scht{2}t} era has thus transformed our capacity to determine segmental duplications, satellite arrays, and other complex repeats that were previously collapsed or largely absent from the first and following iterations of the \textsc{grc}h\textsc{\scht{38}} human reference.\supercite{international2001initial} These advances have catalysed comparative and functional genomics at the population scale, including the construction of human pangenomes aiming at representing genomic diversity \supercite{wang2022} beyond a single linear reference.\supercite{international2001initial} Routine generation of haplotype-resolved assemblies has also addressed the long-staining reference bias by providing new genome references for a cell-line model system.\supercite{volpe2025reference,pavcar2026,ranallo2026} In turn, the generation of such assemblies improves reads alignment and downstream results, especially within centromeres and highly divergent loci (\textsc{hdl}), when reads and reference are matched.\supercite{volpe2025reference, corda2025cell, corda2025chromosome} 
Yet, despite the proliferation of nominally complete or near-complete human genomes, systematic validation of assembly quality, including highly repetitive regions – most notably centromeres – remains an unmet challenge.\supercite{miga2021variation} Centromeres are essential chromosomal domains that serve as a template for the kinetochore formation and ensure faithful chromosome segregation during mitosis and meiosis, due to the presence of the centromere-specific \textsc{h}3 histone variant \textsc{cenp-a}.\supercite{giunta202540, earnshaw1985identification} In humans, the \textsc{cenp-a} domain and adjacent regions are composed predominantly of $\alpha$-satellite arrays \textsc{dna}, further organized into higher-order repeat (\textsc{hor}) units.\supercite{willard1987} \textsc{hor}s are characterized by an exceptionally high density of \textsc{cenp-b} box motifs, the 17 bps sequence elements that bind centromere protein \textsc{b} (\textsc{cenp-b}), with an enrichment approximately 130-fold greater within centromeres than the genome-wide background.\supercite{corda2025chromosome} This rich landscape of inter-motif distance values provides numerical rendering of centromeric \textsc{dna} capturing chromosome-specific architecture, expressed in a specific barcode of distances conserved in humans and certain non-human primates;\supercite{corda2025chromosome} conversely, \textsc{cenp-b} boxes outside the centromere are sparse and were named ectocentromeric sites (\textsc{ecs}).\supercite{corda2025chromosome} Genome-wide, \textsc{cenp-b} boxes position, organization and orientation define the human centeny map, a visual representation of the aforementioned inter-motif distances which provides banding patterns specific to each chromosome.\supercite{corda2025chromosome} Within centromeres, the inter-motif distances are confined to just ~42 values in all genomes currently analysed, with chromosome-specific set of distances that builds a chromosome-specific barcode of recognition for each chromosome’s centromere.\supercite{corda2025chromosome} While these distances are preserved, human centromere sequence, composition, \textsc{hor} structure and array length vary substantially among haplotypes\supercite{volpe2025reference} and individuals.\supercite{logsdon2024variation, gao2025global} Across species, the centromere paradox showcases rapid sequence divergence, even when centromere function is conserved,\supercite{henikoff2001,balzano2020} implying that the structural organization of \textsc{cenp-b} boxes is being preserved, and it is likely ancestral to the underlying sequence as well as less prone to tolerate variation. Such rapid evolutionary turnover, coupled with pervasive polymorphism in repeat organization, renders single-reference benchmarking of centromeres fundamentally insufficient.\supercite{logsdon2024variation} Reference-guided evaluation, or polishing, risk to “force” structural conformity to an arbitrary template, particularly in regions where alignment is ambiguous, paralogous or biologically variable. This includes current use of machine learning (\textsc{ml}) driven error correction, as it risks to overpolish subtle variation within the centromeres’ consensus sequences just because they are reiterated within each near-identical \textsc{hor} ($>$98\% sequence identity). Thus, as \textsc{t\scht{2}t} assemblies become widespread, there is an urgent need for validation frameworks that can assess centromere correctedness, correct chromosomal assignment and structural fidelity without relying exclusively on nucleotide-level alignment to a single reference genome.\supercite{miga2020centromere}
Here, we leverage our discovery that centromeres harbour a conserved architectural principle dictated by the position, orientation, and spacing of functional \textsc{cenp-b} boxes.\supercite{corda2025chromosome} We define centeny maps as genome-wide matrices of inter-motif distances that numerically encode this higher-order organization. Owing to evolutionary constraints on these distance distributions, centeny provides a structural coordinate system that is comparable across different human assemblies without requiring nucleotide-level alignment. Accordingly, we introduce a distance-based scoring framework for centromeres assembly quality which compares centeny-derived inter-motif distance statistics using similarity measures such as Jaccard and Kullback-Leibler (\textsc{kl}), rather than relying on sequence-identity criteria. Applied genome-wide to all currently available human \textsc{t\scht{2}t} assemblies
, this approach establishes a ranking of assembly accuracy at chromosome and centromere resolution. Collectively, we provide a quantitative, distribution-based metric for benchmarking genome reconstruction fidelity in repetitive \textsc{dna}. This enables rapid, automated, interpretable assessment of assembly integrity and advances centromere-aware comparison across existing and newly generated genomes.
\section*{Results}

\subsection*{Human centeny maps enable genome-to-genome comparison of T2T assemblies}
\label{sec:1_chapter}

\begin{figure}[htbp]
    \centering
    \includegraphics[width=5.5in]{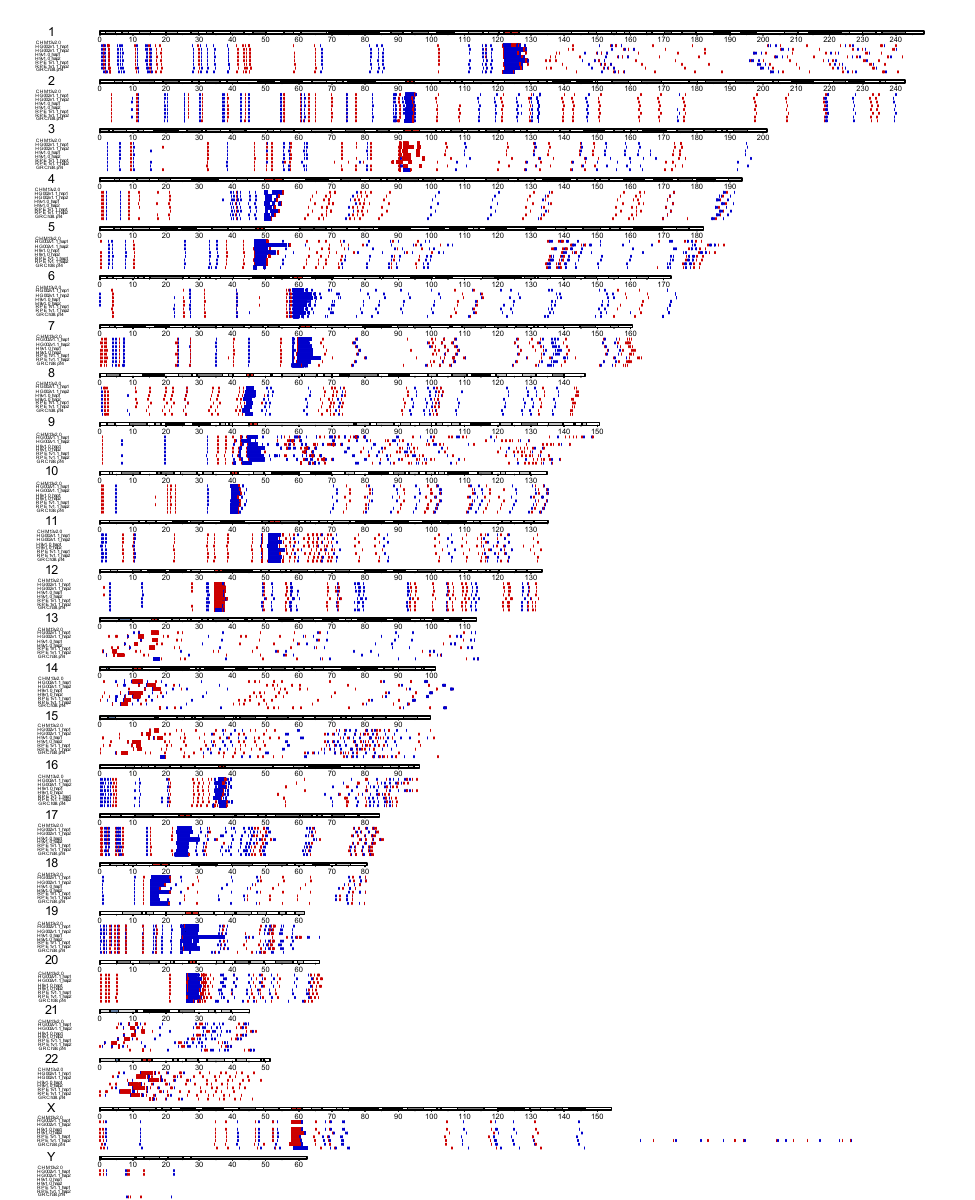}
    \caption{\textbf{Overlay visualization of \textit{centeny} maps for three diploid human genomes and the two human references.} Centeny maps depict chromosome-specific banding patterns using a single sequence motif (\textsc{cenp-b} boxes), in its forward (blue) and reverse complement (red) orientation, found within the same \textsc{dna} strand. These banding patterns are largely conserved in distance and position from one another across human genome assemblies.}
    \label{fig:1_figure}
\end{figure}

The completion of multiple telomere-to-telomere (\textsc{t\scht{2}t}) human genome assemblies provides an unprecedented opportunity to compare chromosomes at full length across individuals and cell types. However, direct nucleotide alignment breaks down in highly repetitive regions, particularly centromeres. To overcome this limitation, we implemented centeny maps, a structural representation of genome organization defined by the distances between functional sequence motifs,\supercite{corda2025chromosome} to enable alignment-independent comparison across assemblies. We generated centeny maps for multiple human references, such as \textsc{chm\scht{13}}, \textsc{grc}h\textsc{\scht{38}}, \textsc{hg\scht{002}}, \textsc{rpe\scht{1}} and \textsc{h\scht{9}}.\supercite{international2001initial,nurk2022complete,jarvis2022,rautiainen2023telomere,volpe2025reference,hansen2025,pavcar2026} Using a single biologically meaningful motif, the \textsc{cenp-b} box, we rendered each chromosome as a banded pattern reflecting motif orientation and spacing along the \textsc{dna} strand. Forward-oriented \textsc{cenp-b} boxes were plotted in blue and reverse-complement orientations in red, generating chromosome-specific visual signatures (\Cref{fig:1_figure}). Because \textsc{cenp-b} boxes occur at exceptionally high density within $\alpha$-satellite higher-order repeat (\textsc{hor}) arrays — approximately 130-fold enriched relative to the genome-wide background,\supercite{corda2025chromosome} centromeres produce distinctive, information-rich centeny patterns. Strikingly, despite substantial nucleotide divergence across assemblies, centeny maps revealed conserved chromosome-specific banding patterns that were reproducible across all \textsc{t\scht{2}t} genomes examined. Each chromosome exhibited a characteristic distribution of inter-motif placement and orientation switches, generating a unique architectural fingerprint (\Cref{fig:1_figure}). These fingerprints were stable across haplotypes and individuals, even when underlying $\alpha$-satellite sequences differed in copy number or local composition, as we have shown between the two haplotypes of \textsc{rpe\scht{1}}v1.1.\supercite{volpe2025reference}

To quantify these observations, we implemented two complementary analytical frameworks, as previously conceptualized in Corda \& Giunta (2025).\supercite{corda2025chromosome} In Model 1, centeny captures pairwise inter-motif spacing along each chromosome, rendering genomes from letters into a series of numerical values. Model 2 incorporates positional encoding within $\alpha$-satellite monomers, integrating motif directionality and local distance variance to detect structural rearrangements that preserve motif density but alter higher-order organization. Across \textsc{chm\scht{13}} and \textsc{grc}h\textsc{\scht{38}}, centeny maps demonstrated near-complete conservation of chromosome-level architecture outside centromeres, while revealing previously unappreciated discrepancies within centromeric arrays (\hyperlink{supp-fig:S1}{\textcolor{mediumtealblue}{\Cref{supp-fig:S1}}}). In \textsc{hg\scht{002}}, \textsc{rpe\scht{1}} and \textsc{h\scht{9}} assemblies, centeny visualization uncovered haplotype-specific centromere inversions, orientation flips and localized expansions (\Cref{fig:1_figure}, for details see \hyperlink{supp-fig:S1}{\textcolor{mediumtealblue}{\Cref{supp-fig:S1}}}) that were not readily apparent from sequence alignment alone. These inversions manifested as mirror-image banding patterns in centeny space, accompanied by shifts in orientation-specific distance distributions captured by Model 2. Importantly, such structural variants were distinguishable from assembly artefacts by their preservation of conserved inter-motif spacing hierarchies, a feature captured numerically by low divergence scores in Model 1 despite orientation reversal. These advantages are especially evident in the short arms of acrocentric chromosomes, for which distances are anomalous (\Cref{fig:1_figure}).

Together, these results demonstrate that conserved centeny architecture constitutes a robust structural coordinate system for whole-genome comparison. By transforming highly repetitive centromeric \textsc{dna} into interpretable banding patterns and distance distributions, centeny maps enable \textit{bona fide} genome-to-genome evaluation across \textsc{t\scht{2}t} assemblies, establishing a foundation for structural benchmarking independent of sequence identity.
\subsection*{Numerical rendering of human genomes' distance values between CENP-B boxes}
\label{sec:2_chapter}

\begin{figure}[htbp]
    \centering
    \includegraphics[width=6in]{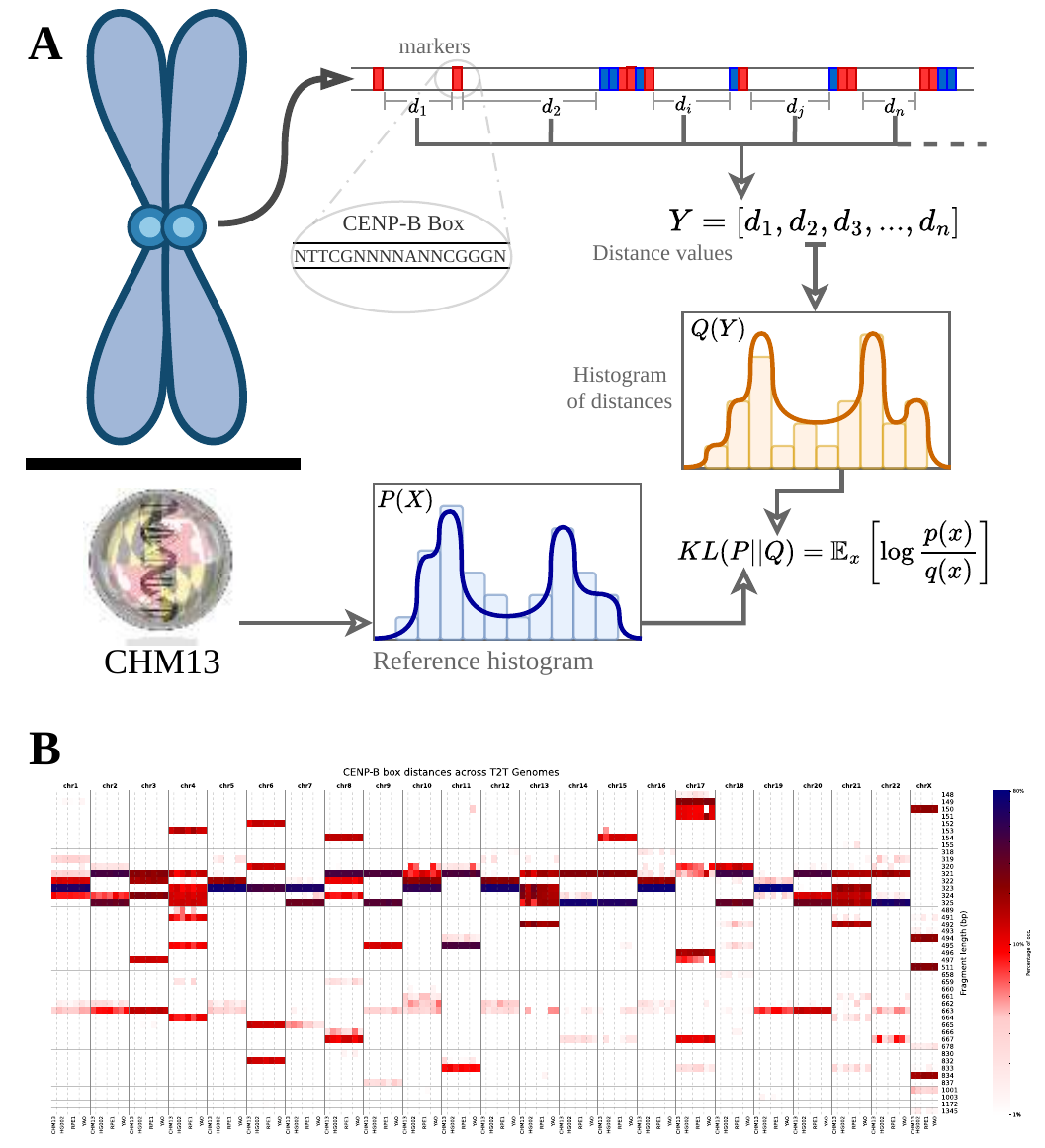}
    \caption{\textbf{Overview of the approach and motif boxes conservation.} \textbf{(A)} To evaluate centromere reconstruction fidelity without linear sequence alignment, we extracted the relative distances between functional \textsc{cenp-b} box motifs (\enquote{\textsc{nttcgnnnnanncgggn}}), creating a 1-dimensional representation of the target chromosome of consecutive inter-motif distances ($Y = [d_1, d_2, d_3, \dots, d_n]$). This vector is converted into a discrete probability density distribution ($Q(Y)$) that captures the holistic structural topology and natural polymorphic variance of the satellite array. This query distribution is evaluated against a high-quality standardized reference histogram, $P(X)$ (e.g., the haploid \textsc{chm}\oldstylenums{13} assembly), using Kullback-Leibler divergence (\textsc{kl}) based metric to measure the conservation, or lack thereof, of centromere structure. \textbf{(B)} Heatmap showing the frequency of discrete inter-\textsc{cenpb} box distances (\emph{y}-axis, bp) across all chromosomes (chr1-22, chrX; top labels) in four \textsc{t}\oldstylenums{2}\textsc{t} genome assemblies: \textsc{chm}\oldstylenums{13}, \textsc{hg}\oldstylenums{002}, \textsc{rpe}\oldstylenums{1}, and \textsc{yao} (bottom labels). Color intensity indicates the percentage of a given distance among all inter-box spacings observed within the centromeric array of each chromosome-haplotype combination (log scale: 1-80\%). Only distances exhibiting \textgreater$\,1\%$ recurrence across the dataset are shown.}
    \label{fig:2_figure}
\end{figure}

Our novel contribution lies in interpreting the centeny map as a density of inter-motif distance occurrences, establishing a rigorously quantitative framework to evaluate the structural rhythm of chromosomes (\Cref{fig:2_figure}). For a given chromosome, we first identify genomic coordinates of all \textsc{cenp-b} boxes, defined by the highly conserved 17 bps sequence motif \enquote{\textsc{nttcgnnnnanncgggn}}, rather than analyzing the intervening sequence. Then we compute the linear array of consecutive genomic distances $X=[d_1,d_2,d_3,\dots,d_n]$ between adjacent motifs. This transformation effectively distills megabase-scale, complex $\alpha$-satellite arrays into compact, 1-dimensional vectors of inter-motif distances, rendering the chromosomes’ higher-order architectural signature directly computable. As such, we are able to effectively target centromeric sequences and assess their assembly correctness across \textsc{t\scht{2}t} human genomes.

To enable robust comparisons across highly divergent regions in different human genomes, these distance vectors are binned into normalized density histograms, denoted as $P(X)$ (\Cref{fig:2_figure}A). By representing centromeric architecture through these density histograms, we capture the overarching structural topology while accommodating the natural, biologically permissible variance inherent to satellite \textsc{dna} arrays (e.g., minor local expansions or contractions). To benchmark newly assembled genomic contigs against a standardized baseline, our pipeline directly compares a query histogram $P(X)$ with a reference histogram $Q(Y)$ derived from the high-quality \textsc{chm\scht{13}} haploid assembly. This comparison forms the mathematical foundation for evaluating assembly fidelity via a distribution-based metric, directly quantifying the structural divergence between the two profiles.

Applying this numerical rendering globally across all 23 chromosomes (chr1-chrX) in four fully resolved \textsc{t\scht{2}t} human genome assemblies (\textsc{chm\scht{13}}v\textsc{\scht{2.0}}, \textsc{hg\scht{002}}v\textsc{\scht{1.1}}, \textsc{rpe\scht{1}}v\textsc{\scht{1.1}}, and \textsc{yao}v\textsc{\scht{2.0}}) validates the biological authenticity of the distance-based metric (\Cref{fig:2_figure}B). Heatmap visualization of the genome-wide distance matrices reveals that inter-\textsc{cenpb} box distances are not distributed continuously. Instead, they are quantized to integer multiples of approximately 171 base pairs ($n \times $171 bps). This manifests as discrete horizontal bands of high occupancy at predictable intervals (e.g., $\sim$150, $\sim$320, $\sim$490, $\sim$660, $\sim$830, $\sim$1001, $\sim$1172, and $\sim$1345 bps). This rigid quantization corresponds precisely to the length of the fundamental $\alpha$-satellite monomer and its associated nucleosomal repeat unit, indicating that \textsc{cenp-b} box positioning is not stochastic; rather, it is strictly constrained to defined nodal positions within the underlying $\alpha$-satellite chromatin lattice.

Furthermore, this quantization exposes three critical features of human centromeric architecture. First, the dominant spacing cluster differs markedly by chromosome, reflecting the distinct, chromosome-specific higher-order repeat (\textsc{hor}) structures (e.g., the prevalence of shorter spacings on acrocentric chromosomes versus longer spacing intervals on larger metacentric arms). Second, the distance distribution pattern exhibits striking allelic symmetry; the quantized signatures are nearly identical between haplotypes of the same individual (hap1 $\approx$ hap2 for all diploid genomes), demonstrating highly conserved \textsc{cenp-b} organization across homologous alleles. Third, the near-homozygous \textsc{chm\scht{13}} assembly perfectly recapitulates the periodic signatures observed in the diverse diploid cohorts, confirming that this quantized architecture is a universal, diploid-independent property of human centromeres.

These results support a model in which our centeny distance vectors capture fundamental biological constraints. The strict nucleosomal phasing defined by these arrays likely provides the structural basis required for the precise spatial coordination of \textsc{cenp-b} interactions at functional centromeres. Because this structural signature is universally dictated by nucleosomal mechanics, yet it has its specific chromosomal identity, it provides a perfect alignment-independent numerical foundation for cross-genome comparisons.
\subsection*{Comparative evaluation of metrics for inter-chromosomal distance assessment}
\label{sec:3_chapter}

\begin{figure}[htbp]
    \centering
\includegraphics[width=6.5in]{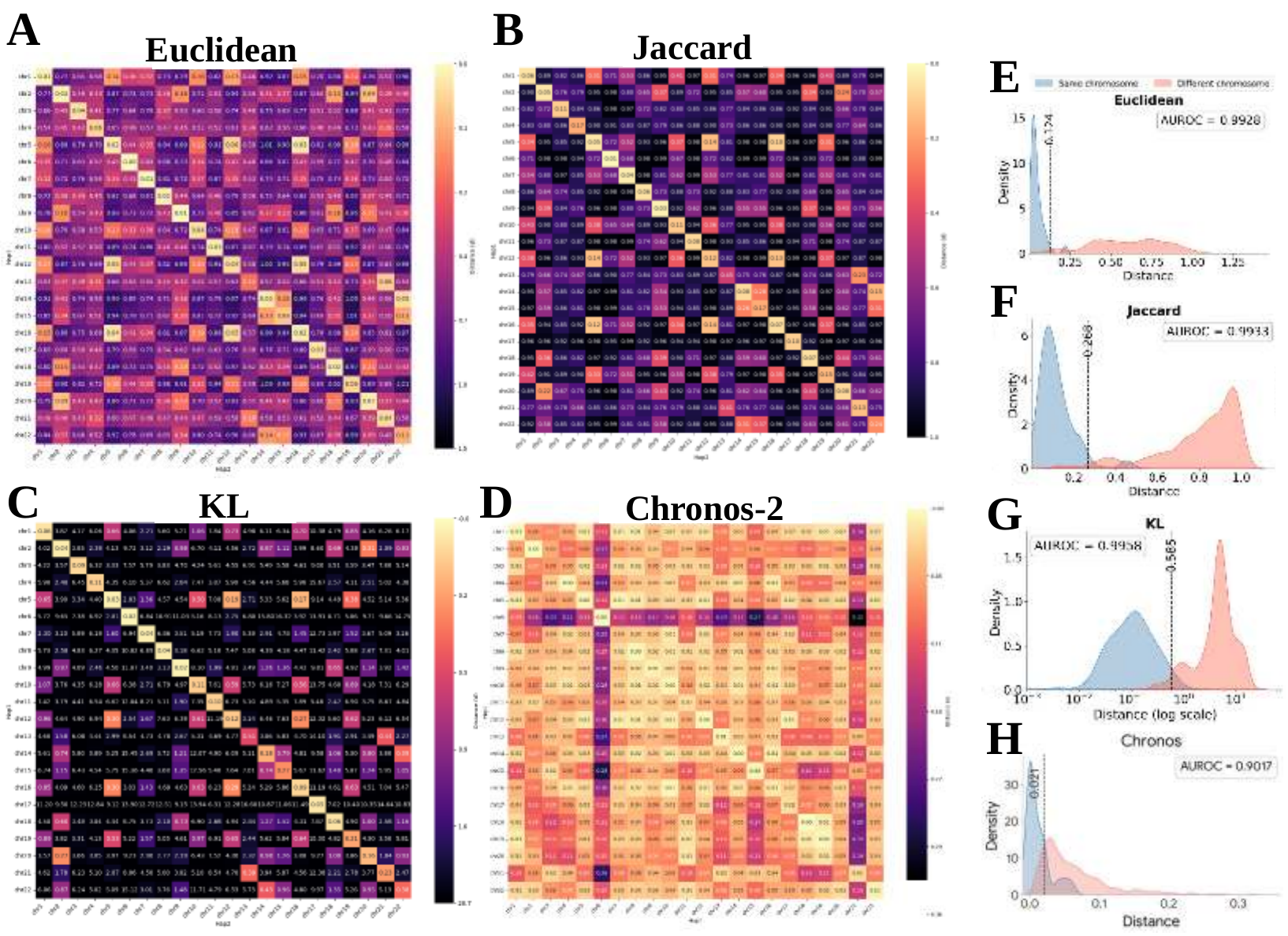}
    \caption{\textbf{Comparison of four quantitative metrics to evaluate inter-chromosomal structural similarity.} To identify a robust mathematical framework that best captures the conserved biological identity of centromeres, we compared a standard metric (Euclidean distance), two distribution-based metrics (Jaccard distance and Symmetric Kullback-Leibler divergence), and a machine learning-based sequence encoder (Chronos-2). \textbf{(A-D)} Comparative genomic distance heatmaps for the diploid \textsc{hg}\oldstylenums{002} genome. Pairwise distance matrices between \textsc{hg}\oldstylenums{002} haplotype 1 (\emph{y}-axis) and haplotype 2 (\emph{x}-axis) were evaluated using \textbf{(A)} Euclidean distance, \textbf{(B)} Jaccard distance, \textbf{(C)} Symmetric \textsc{kl} divergence, and \textbf{(D)} Chronos-2 Sequence Encoder (cosine distance). Each cell $(i,j)$ represents the normalized distance between specific chromosomes of the two haplotypes. The \textsc{kl} \textbf{(C)} demonstrates the sharpest contrast between the diagonal and the background, confirming that centromeric identity is driven by overarching probabilistic spacing distributions rather than strict sequential ordering, providing superior sensitivity for identifying chromosome-specific structural fingerprints. \textbf{(E-H)} Kernel density estimates (\textsc{kde}) of pairwise inter-haplotype distances pooled across 22 chromosomes (chr1-chr22) from multiple \textsc{t}\oldstylenums{2}\textsc{t} genome assemblies. Distributions are stratified by diagonal entries (hap1 chromosome $i$ vs. hap2 chromosome $i$; homologous pairs, blue) and off-diagonal entries (hap1 chromosome $i$ vs. hap2 chromosome $j$, $i \neq j$; non-homologous pairs, red). Performance is shown for the four metrics: \textbf{(E)} Euclidean and \textbf{(F)} Jaccard distance ($1 - \text{Jaccard index}$) as well as \textbf{(G)} \textsc{kl} divergence (log scaled) and \textbf{(H)} Chronos-2. Dashed vertical lines mark optimal separation thresholds. All metrics exhibited near-perfect separability in distinguishing homologous from non-homologous pairs, with \textsc{kl} divergence achieving the highest \textsc{auroc} score (0.9958) compared to Jaccard (0.9933) and Euclidean (0.9928) distances.}
    \label{fig:3_figure}
\end{figure}

To quantitatively measure the differences in centromeres' centeny maps with respect to a robust mathematical framework for inter-chromosomal comparison, we systematically adopted four distinct quantitative metrics. By representing the inter-motif distances of each chromosome as high-dimensional frequency distributions, we sought to identify a metric that best captures the conserved biological identity of centromeric architecture while remaining resilient to background noise.

Let $X = [d_1, d_2, d_3, \dots, d_n]$ be the vector of inter-motif distances, we first define the (unnormalized) histogram of occurrences

\begin{equation}
h_d(X) \;=\; \sum_{i=1}^{n} \mathbf{1}\!\left[d_i=d\right].
\end{equation}

Then, we obtain the empirical probability density $P(X)$ over discrete distances by normalization

\begin{equation}
P_d(X) \;=\; \frac{h_d(X)}{\sum_{d'} h_{d'}(X)}
\end{equation}

We first adopted a standard quantitative baseline using Euclidean distance, operating in a high-dimensional space where each dimension represents a distance of the centeny map (\Cref{fig:3_figure}A). Considering $P(X)$ and $Q(Y)$ as the empirical discrete distributions of two centromeres $X$ an $Y$, we define Euclidean distance as

\begin{equation}
d_{\mathrm{Euc}}\!\big(P(X),Q(Y)\big)
\;=\;
\sqrt{\sum_{d} \left(P_d(X) - Q_d(Y)\right)^2 }
\end{equation}

This geometric value calculates the absolute, \enquote{straight-line} magnitude of divergence between two chromosomal profiles. While mathematically intuitive, Euclidean distance assigns uniform weight to all structural bins. Consequently, in the context of highly repetitive genomic data, minor, biologically irrelevant fluctuations in common inter-motif distances can disproportionately obscure the signal from rare, highly specific structural markers. This susceptibility to high-dimensional noise fundamentally limits its discriminatory power in our clustering benchmarks.
We next evaluated Jaccard distance (\Cref{fig:3_figure}B), reformulating this classical set theory metric to quantify the intersection over union of our distance histograms 

\begin{equation}
d_{Jacc}\!\big(P(X),Q(Y)\big)
\;=\;
1 - \frac{\sum_{d} \min\!\big(P_d(X),\,Q_d(Y)\big)}
     {\sum_{d} \max\!\big(P_d(X),\,Q_d(Y)\big)}
\end{equation}

From a genomic perspective, Jaccard effectively measures the strict proportion of perfectly overlapping structural intervals. Because it requires exact bin-for-bin overlap, Jaccard distance frequently penalizes slight, biologically permissible shifts in motif spacing, such as minor local array expansions or contractions, as absolute mismatches, thereby failing to recognize broader underlying structural homology. 

To overcome the limitations of geometric and set-based comparisons, we ultimately implemented a symmetric Kullback-Leibler (\textsc{kl}) divergence model (\Cref{fig:3_figure}C). Rooted in information theory, \textsc{kl} divergence treats centeny maps not as rigid physical shapes, but as dynamic, probabilistic signatures. In biological terms, rather than looking for exact sequence matches, this metric measures how much the overall centeny blueprint of one centromere deviates from another, capturing the overarching structural rhythm of the repetitive array. The \textsc{kl} divergence between the empirical distributions $P(X)$ and $Q(Y)$ is defined as 

\begin{equation}
d_{\mathrm{KL}}\!\big(P(X)\,\|\,Q(Y)\big)
\;=\;
\sum_{d} P_d(X)\,\log\!\left(\frac{P_d(X)}{Q_d(Y)}\right)
\end{equation}

We used a symmetrized iteration of this formula to ensure bidirectional distance parity

\begin{equation}
d_{\mathrm{sKL}}\!\big(P(X),Q(Y)\big)
\;=\;
\frac{1}{2}\Big(
d_{\mathrm{KL}}\!\big(P(X)\,\|\,Q(Y)\big)
+
d_{\mathrm{KL}}\!\big(Q(Y)\,\|\,P(X)\big)
\Big).
\end{equation}

Crucially, symmetric \textsc{kl} divergence captures the unique \enquote{fingerprint} of each chromosome by intelligently penalizing structurally disruptive anomalies that stray away from, or compromise, the underlying distribution while accommodating the natural, probabilistic variance inherent to satellite \textsc{dna} arrays. Because it is exquisitely sensitive to the overarching distribution of probability rather than strict pointwise overlap, symmetric \textsc{kl} divergence provides the highest resolution of homologous versus non-homologous chromosomes.

Given the recent success of deep learning foundation models in deciphering genomic and structural patterns,\supercite{ji2021dnabert, brixi2025genome, avsec2025alphagenome} we also tested whether a state-of-the-art sequence encoder model could provide zero-shot separable latent representations for chromosomal identity. We treated the 1-dimensional vector of inter-\textsc{cenp-b} box distances as a sequence and encoded it using Chronos-2.\supercite{ansari2025chronos} By mapping each chromosome’s centeny vector into a high-dimensional latent space, we computed pairwise cosine distances between the resulting latent embeddings to distinguish chromosomes. 
Given two centromeres $X$ and $Y$, we obtain latent embeddings by encoding them with the Chronos-2 encoder

\begin{equation}
z_X = f_{\mathrm{Chr}}(X), 
\qquad
z_Y = f_{\mathrm{Chr}}(Y),
\end{equation}

where $f_{\mathrm{Chr}}(\cdot)$ denotes the Chronos-2 pretrained encoder.
We then define the distance between $X$ and $Y$ as the cosine distance in latent space:

\begin{equation}
d_{\cos}(X,Y)
\;=\;
1-\frac{\langle z_X, z_Y\rangle}{\|z_X\|_2\,\|z_Y\|_2}
\end{equation}

Surprisingly, despite its architectural complexity, the sequence encoder underperformed compared to standard distribution-based metrics. The comparative genomic distance heatmap for Chronos-2 (\Cref{fig:3_figure}D) showed a noisier diagonal and higher off-diagonal background signal, indicating frequent confusion between chromosomes. Rather than an inherent limitation of deep learning, this failure mode is characteristic of an out-of-distribution (\textsc{ood}) generalization problem. Foundation models rely heavily on the statistical priors established during their training. Centeny maps constitute a fundamentally novel and highly specialized data modality. This highlights a critical limitation of adopting \enquote{black-box} foundation models for genomic representations: without massive, specialized datasets for domain adaptation, they fail to recognize underlying biological homology.

\paragraph{Metric Separability and Chromosome Identification}
A high-performing genomic distance metric is characterized by a strong, prominent diagonal, indicating low distance (high structural homology) between homologous chromosomes (e.g., \textsc{hg\scht{002}} hap1 chr1 vs. \textsc{hg\scht{002}} hap2 chr1). Conversely, off-diagonal values should remain high, reflecting clear differentiation between non-homologous chromosomes. The \textsc{kl} metric (\Cref{fig:3_figure}C) demonstrates the sharpest contrast between the diagonal and the background, suggesting superior sensitivity for chromosome-specific identification compared to Jaccard and Euclidean metrics.

To rigorously quantify this discriminative power, we computed the Area Under the Receiver Operating Characteristic curve (\textsc{auroc}). The \textsc{auroc} serves as a comprehensive statistical measure of classification performance; in this context, it evaluates a metric's ability to correctly distinguish same-chromosome pairs (diagonal) from different-chromosome pairs (off-diagonal) based on structural distance. An \textsc{auroc} of 1.0 indicates perfect separation between the two distributions, whereas 0.5 would indicate random assignment.

All three distance metrics achieved high separability. Euclidean distance attained an \textsc{auroc} of 0.9928, Jaccard distance 0.9933, and symmetric \textsc{kl} divergence 0.9958. The near-ceiling performance of all three metrics ($\text{AUROC} > 0.99$) demonstrates that the inter-\textsc{cenpb} box spacing distribution is highly chromosome-specific and robustly identifies homologous chromosomes across haplotypes without any sequence alignment. Symmetric \textsc{kl} divergence marginally outperforms the other metrics ($\Delta \text{AUROC} \approx$ 0.003 over Jaccard), consistent with its sensitivity to the full shape of the spacing distribution rather than only the overlap support. The well-separated \textsc{kde} distributions (\Cref{fig:3_figure}E-G) confirm that the optimal threshold cleanly partitions same- from different-chromosome pairs, establishing centeny maps as a viable alignment-free fingerprint for chromosome identity.

\paragraph{Statistical Benchmark of Different Quantitative Metrics}

Making use of our generated similarity and distance matrices, we have deployed a group of metrics to establish a statistical benchmark.

To quantify pairwise chromosome distance between any two given individuals, let $M_{G_1, G_2} \in \mathbb{R}^{22 \times 22}$ be the similarity matrix for the 22 autosomes of two distinct genomes ($G_1$ and $G_2$), given a specific similarity metric. The rows correspond to the chromosomes of $G_2$, while the columns correspond to the chromosomes of $G_1$. Each entry $M_{i,j}$ encodes the similarity score $s$ between the chromosome $i$ of $G_2$ and the chromosome $j$ of $G_1$ given by:

\[s = \exp(-d_{\mathrm{sKL}}(G_2 \parallel G_1)) \]
ensuring a bounded score \( s \in [0,1]\).

In a perfect setting $M_{G_1, G_2} \approx I$, where $I$ is the identity matrix. In this ideal mapping, the similarity approaches $1$ for pairs of the same chromosome along the diagonal ($i = j$), whereas it approaches $0$ for all non-homologous pairs ($i \neq j$) which would indicate zero cross-chromosomal interference.

To rigorously determine how closely our generated matrices approach this ideal identity mapping, we have applied three mathematical evaluations: Accuracy, Entropy, and the InfoNCE loss. 

The true matches lie along the diagonal ($i = j$), thus we define Accuracy as the number of times the highest score of each row falls in the entries of the diagonal.
For each row $i$, we define the predicted match as the column with the highest similarity score:
\begin{equation}
    \hat{j}_i = \underset{j \in \{1,\dots,22\}}{\arg\max} \ M_{i,j}
\end{equation}

We then compute the matching accuracy as the fraction of rows for which the predicted match corresponds to the ground-truth diagonal entry:
\begin{equation}
    \mathrm{Acc} =
    \frac{1}{22}
    \sum_{i=1}^{22}
    \mathbf{1}\left[\hat{j}_i = i\right],
\end{equation}

where $\mathbf{1}[\cdot]$ is the indicator function. This metric measures how often each element in ($G_1$) is more similar to the corresponding one in ($G_2$). A perfect value of $1.0$ confirms that every chromosome uniquely maps to its correct counterpart without being confused by inter-chromosomal noise. We want to maximize accuracy.

Further, we have evaluated column-wise Shannon entropy, which in our case serves as a way to ascertain how confidently the model preserves the distinct biological identity of a specific chromosome. A good model will push each column toward a one-hot encoding representation, meaning it would record 1 in the entry $M_{i,j}$ where chromosome $i$ is the same as chromosome $j$ and 0 everywhere else, yielding a near-zero entropy. We want to minimize entropy. We first normalized each column $j$ into a valid discrete probability distribution:
\begin{equation}
    P_{i,j} = \frac{M_{i,j}}{\sum_{k=1}^{22} M_{k,j}}
\end{equation}

The total system entropy was then defined as the mean Shannon entropy across all $n=22$ columns:
\begin{equation}
    H(M) = \frac{1}{n} \sum_{j=1}^{n} \left( - \sum_{i=1}^{n} P_{i,j} \log P_{i,j} \right)
\end{equation}


Finally, to quantify the contrastive separation between the true structural matches and the non-homologous pairs, we have exploited the Information Noise-Contrastive Estimation (InfoNCE) loss. InfoNCE explicitly and heavily penalizes \enquote{hard negatives}, non-homologous chromosomes that yield high similarity scores. This considers the model's ability to filter out \enquote{genomic noise}, where different chromosomes appear similar due to shared repetitive sequences, ensuring the true biological match stands out distinctly. We want to minimize the InfoNCE loss.
 Fixing a temperature scaling hyperparameter $\tau=0.1$, we extract the diagonal elements $M_{i,i}$ as the positive matches, yielding the row-wise loss:
\begin{equation}
    \mathcal{L}_{\text{InfoNCE}}(M) = \frac{1}{n} \sum_{i=1}^{n} \left( - \log \frac{\exp(M_{i,i} / \tau)}{\sum_{j=1}^{n} \exp(M_{i,j} / \tau)} \right)
\end{equation}

Similarity matrices were computed for the genomes \textsc{bj}, \textsc{imr\scht{90}}, \textsc{chm\scht{13}}v2.0, \textsc{grc}h\textsc{\scht{38}}, \textsc{hg\scht{002}}v1.1, \textsc{i\scht{002}c}v0.7, \textsc{yao}v2.0, \textsc{h\scht{9}}v0.1, and \textsc{rpe\scht{1}}v1.1 using \textsc{kl} similarity, Jaccard similarity, and Euclidean similarity, with the latter normalized to the interval $[0,1]$ to ensure comparability across metrics. For each resulting similarity matrix, we output accuracy, entropy, and InfoNCE. The results we report correspond to the average values of these measures across all genome pairs for each similarity metric, as per \textbf{\Cref{tab:metric_comparison}} below.

\begin{table}[htbp]
\centering
\caption{Comparison of average evaluation metrics across different similarity measures.}
\label{tab:metric_comparison}
\begin{tabular}{lccc}
\toprule
\textbf{Similarity Metric} & \textbf{Avg Accuracy} & \textbf{Avg InfoNCE} & \textbf{Avg Entropy} \\
\midrule
KL Similarity & 0.9053 & 0.6173 & 1.8719 \\
Jaccard Similarity & 0.8712 & 0.6517 & 2.6623 \\
Euclidean Similarity & 0.8194 & 1.0987 & 3.0617 \\
\bottomrule
\end{tabular}
\end{table}

Overall, \textsc{kl} similarity achieved the highest average accuracy with lowest entropy and InfoNCE, indicating more consistent and discriminative similarity scores across genome pairs. In contrast, Euclidean similarity exhibited the lowest accuracy with highest entropy and InfoNCE, suggesting reduced separability and greater uncertainty, while Jaccard similarity showed intermediate performance across the evaluated metrics.
\subsection*{High conservation of haplotypic distance values shows chromosome specificity}
\label{sec:5_chapter}

\begin{figure}[htbp]
    \centering
    \includegraphics[width=5in]{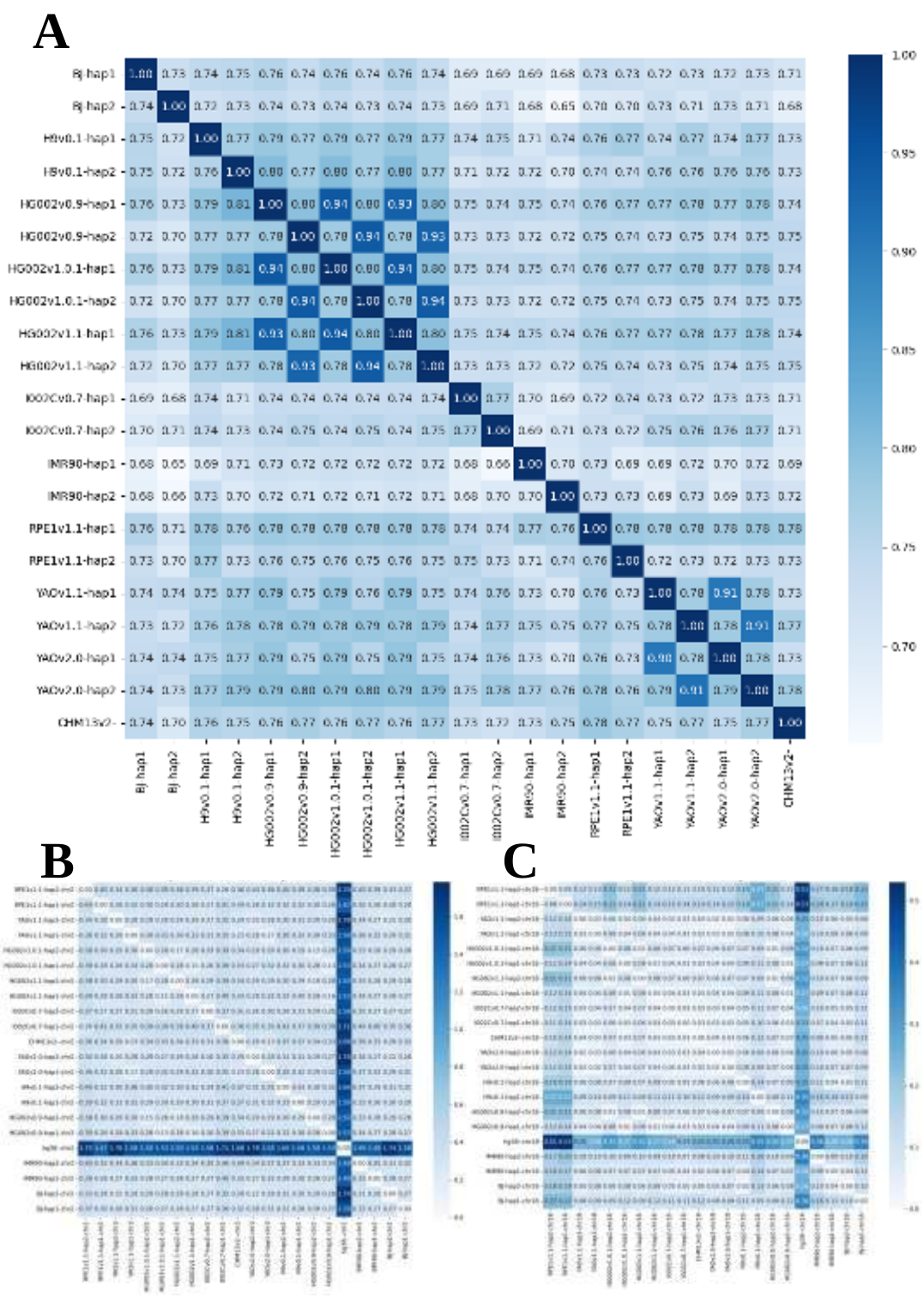}
    \caption{\textbf{Scoring chromosome similarity among individuals.} The matrices are evaluated considering seven \textsc{t}\oldstylenums{2}\textsc{t} genomes in \textsc{hg}\oldstylenums{002}, \textsc{i}\oldstylenums{002}\textsc{c}, \textsc{yao}, \textsc{rpe}\oldstylenums{1}, \textsc{h}\oldstylenums{9}, \textsc{imr}\oldstylenums{90}, and \textsc{bj} along with the two human references (\textsc{grc}h\oldstylenums{38} and \textsc{chm}\oldstylenums{13}), including several assembly versions of the same genome when applicable. \textbf{(A)} Mean \textsc{kl} divergence similarity score for all genomes, computed as the average score across all chromosomes per pair of comparison, providing a general panorama of the ability to recognize the same chromosome among different individuals. The similarity score is the \textsc{kl} divergence normalized to $[0,1]$, where 1 refers to “perfect similarity”. \textbf{(B)} Scores of the symmetric \textsc{kl} divergence for chromosome 2, highlighting higher divergence with a deeper blue color, reflecting that the distribution of the distances of chromosome 2 in \textsc{grc}h\oldstylenums{38} is different from the distribution of the same chromosome in other genomes. \textbf{(C)} Scores of the symmetric \textsc{kl} divergence for chromosome 18, highlighting higher divergence with a deeper blue color. This matrix shows a low \textsc{kl} divergence score for each entry, with the maximum value reported at 0.53.}
    \label{fig:5_figure}
\end{figure}

To evaluate whether our metric reliably identifies the same chromosomes across individuals, we analyzed the constructed chromosome-level probability distributions of \textsc{cenp-b} box distances for multiple human assemblies. The genomes considered include \textsc{hg\scht{002}}, \textsc{i\scht{002}c}, \textsc{yao}, \textsc{rpe\scht{1}}, \textsc{h\scht{9}}, \textsc{imr\scht{90}}, and \textsc{bj}, plus the two human references \textsc{grc}h\textsc{\scht{38}} and \textsc{chm\scht{13}}. Whenever available, haplotypes were analyzed separately, and multiple assembly versions of the same genome were included when applicable.

The objective of this analysis is to demonstrate that the proposed metric captures chromosome-specific probability distributions of the \textsc{cenp-b} boxes’ distances that are conserved across individuals. Using the symmetric \textsc{kl} divergence we reveal a high degree of consistency among modern \textsc{t\scht{2}t} assemblies. We computed symmetric \textsc{kl} divergence matrices for pairwise chromosome comparisons, with divergence values normalized to the interval $[0,1]$, where 1 denotes perfect similarity (\Cref{fig:5_figure}A). The considered metric effectively captures the low divergence corresponding to high similarity score observed when comparing the same chromosome across different individuals. (\Cref{fig:5_figure}A).

The results indicate that each chromosome maintains its own characteristic probability distribution of \textsc{cenp-b} box distances, largely preserved across genomes, as reflected by consistently low divergence scores.

In contrast, the \textsc{grc}h\textsc{\scht{38}} genome stands out as a clear outlier in the heatmaps by chromosomes. We report chromosome 2 and chromosome 18 (\Cref{fig:5_figure}B-C), where it is noticeable how \textsc{grc}h\textsc{\scht{38}} consistently exhibits substantially higher divergence scores. This observed “blue stripe” (\Cref{fig:5_figure}B) effect highlights the gap between legacy reference assemblies and modern \textsc{t\scht{2}t} genomes. The symmetric \textsc{kl} divergence thus provides a straightforward and quantitative measure of assembly completeness, illustrating the major structural improvements introduced by \textsc{t\scht{2}t}-level references.

Notably, chromosome 18 displays exceptionally low divergence scores among all individuals. This can indicate that centromeric distance distributions and surrounding structural features are remarkably consistent across individuals, potentially reflecting fewer expansions, divergent satellite regions, or other structural variations that could alter the underlying numerical distributions.
\subsection*{Testing the robustness of this metric and tolerance for synthetic noise}
\label{sec:6_chapter}

\begin{figure}[htbp]
    \centering
    \includegraphics[width=\textwidth]{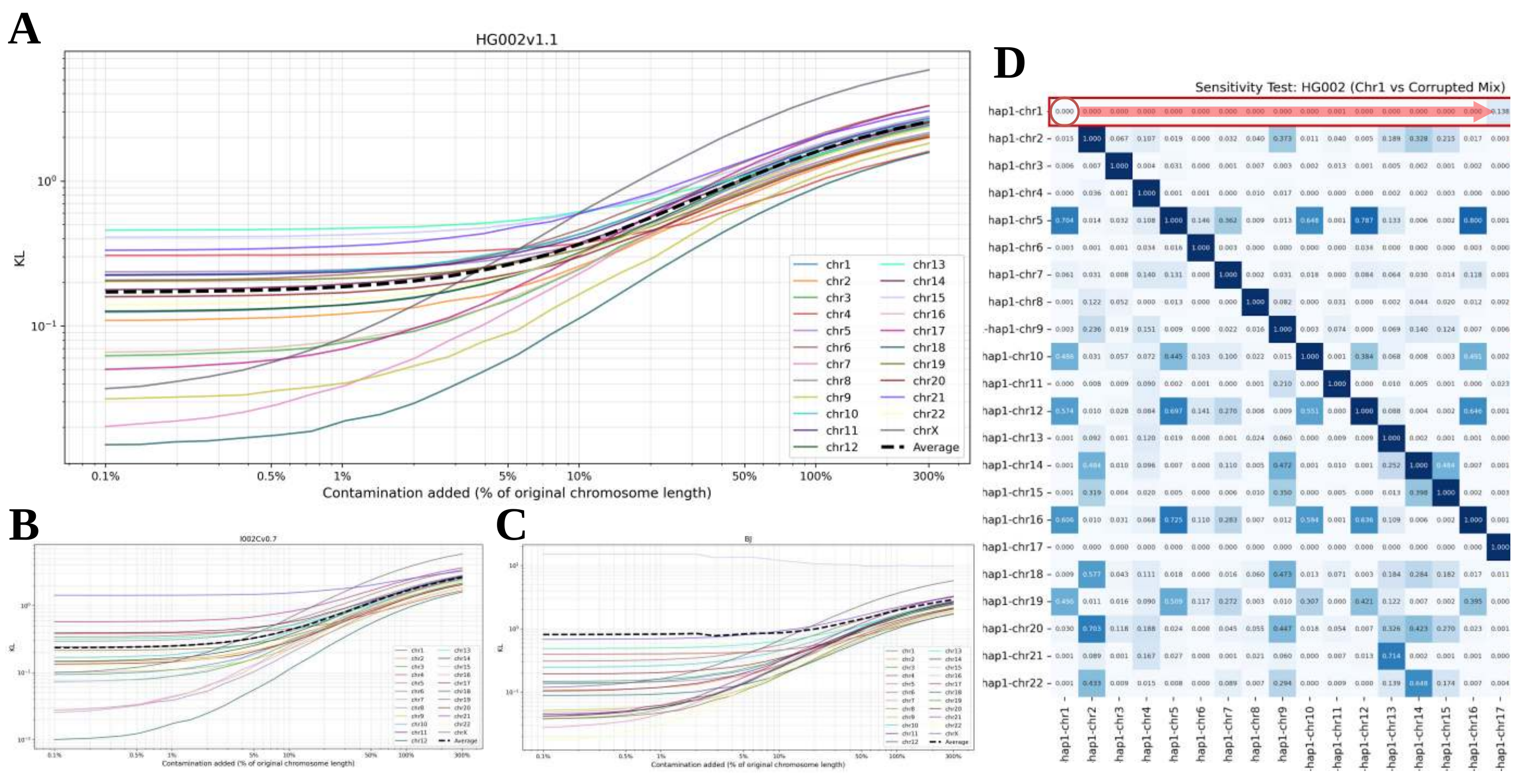}
    \caption{\textbf{Synthetic perturbation profiles demonstrate metric robustness and diagnose catastrophic assembly errors.} Quantitative stress-test profiles for three \textsc{t}\oldstylenums{2}\textsc{t} genomes: \textsc{hg}\oldstylenums{002}v\oldstylenums{1.1} \textbf{(A)}, \textsc{i}\oldstylenums{002}\textsc{c}v\oldstylenums{0.7} \textbf{(B)}, and \textsc{bj} \textbf{(C)}. Each target chromosome was systematically corrupted by appending proportionally scaled sequences ($0.1\%$ to $300\%$ on the \emph{x}-axis, log scaled) from non-homologous chromosomes. The structural integrity of the resulting chimeras was evaluated by measuring their symmetric \textsc{kl} divergence (\emph{y}-axis, log-scaled) against the pristine \textsc{chm}\oldstylenums{13} reference genome. At low contamination levels, trajectories remain stable, reflecting the metric's resilience and its capacity to capture true biological divergence (baseline \textsc{kl}) despite trace assembly noise. As contamination increases, \textsc{kl} divergence escalates, marking the threshold where biological identity is lost. The composite average trajectory is denoted by the dashed black line. Notably, the \textsc{bj} assembly \textbf{(C)} reveals a severe diagnostic anomaly: \textsc{bj} chromosome 15 begins with an extreme baseline divergence that \textit{decreases} as random contamination is added. This paradoxical downward trajectory indicates that the underlying native assembly is so structurally flawed that the addition of random genomic noise mathematically rescues its distribution toward a physiological baseline, providing definitive proof of a collapsed or deeply erroneous assembly. To evaluate the sensitivity and interpretability of our chromosome-level similarity metric, we performed a controlled corruption experiment on \textsc{hg}\oldstylenums{002}v\oldstylenums{1.1} \textbf{(D)}. Specifically, chromosome 1 (chr1) was synthetically modified by replacing 70\% of its sequence with material from chromosome 17 (chr17) of the same genome, while retaining 20\% of the original chr1 sequence and introducing 10\% from chromosome 3 (chr3). The resulting corrupted chromosome was then compared against the same uncorrupted genome using our chromosome-to-chromosome similarity scoring framework. As shown in the heatmap, the modified chr1 no longer exhibits strong similarity to the original chr1, confirming that the method correctly detects the corruption. Importantly, the highest similarity signal shifts toward chr17. Thus, the score not only quantifies the deviation from the expected chromosomal identity but also accurately identifies the predominant source of the inserted material, demonstrating that the approach can both detect structural corruption and trace its genomic origin (chr17).}
    \label{fig:6_figure}
\end{figure}

\paragraph{Robustness of the centeny metric to structural contamination and assembly noise}

To rigorously evaluate the resilience of our information-theoretic framework against potential assembly artifacts, we engineered a synthetic perturbation model. Real-world \textit{de novo} genome assemblies are frequently compromised by critical failure modes, including erroneous chromosomal assignments, chimeric contig formation, and pervasive structural noise within highly repetitive satellite arrays. Our objective was to quantify the exact threshold at which the symmetric Kullback-Leibler (\textsc{kl}) divergence of a chromosome's \textsc{cenp-b} box distance distribution degrades when systematically corrupted by artificially introduced genomic material. 

For each evaluated genome (including \textsc{hg\scht{002}}, \textsc{i\scht{002}c}, and \textsc{bj}), we generated synthetic chimeras across all 22 autosomes. Contamination was introduced by digitally appending randomly sampled subsequences from every other non-homologous chromosome within the same genome. To ensure that the magnitude of structural perturbation scaled proportionally with the target chromosome’s inherent architecture, the volume of appended material was strictly defined relative to the original chromosome’s length ($p \times L_{orig}$). We modeled contamination levels across an extreme logarithmic spectrum, ranging from trace contamination (0.1\% of original length) to massive structural corruption (300\%). Following the generation of these synthetic chimeras, the corrupted assemblies were converted into normalized centeny maps. Crucially, rather than comparing the corrupted sequence against its original, uncontaminated counterpart, we evaluated the symmetric \textsc{kl} divergence strictly against the standardized \textsc{chm\scht{13}} reference genome. This approach directly tests a vital, real-world application: the capacity to correctly identify and validate a newly, potentially flawed assembled chromosome against a gold-standard reference.

The resulting quantitative stress-test profiles reveal the precise dynamics of structural signal degradation (\Cref{fig:6_figure}A-C). A highly robust genomic metric is characterized by a stable horizontal trajectory at low contamination levels, maintaining its baseline \textsc{kl} value, which represents the natural, physiological structural polymorphism between the test individual and \textsc{chm\scht{13}}. The metric only escalates significantly when the volume of foreign material becomes mathematically overwhelming. Furthermore, these profiles expose the inherent structural uniqueness of individual chromosomes. When possessing a highly distinct, idiosyncratic \textsc{cenp-b} box spacing distribution, chromosomes demonstrate profound resilience, as trace contamination fails to meaningfully shift their overarching probability histogram. The composite average trajectory (dashed black line) synthesizes this behavior genome-wide, establishing a standardized baseline that enables immediate, quantitative comparison of assembly robustness.

Crucially, this stress test acts as a highly sensitive diagnostic tool for pre-existing assembly collapse. A striking example is visible in the \textsc{bj} genome profile. Under normal circumstances, adding non-homologous contamination raises a chromosome's \textsc{kl} score, as the chromosome diverges further from the \textsc{chm\scht{13}} baseline. However, the \textsc{bj} chromosome 15 (light blue/purple trajectory) exhibits an anomalous, inverted behavior: it begins with an extraordinarily high baseline \textsc{kl} divergence, and as random structural contamination is added, its \textsc{kl} score actually decreases. This paradox perfectly illustrates a catastrophic assembly error. The native \textsc{bj} chr15 assembly is so structurally aberrant and non-physiological that diluting it with random human genomic noise actually pushes its probability distribution closer to a physiological baseline.

\paragraph{Applications and limitations of the perturbation model}

While this synthetic perturbation framework effectively simulates chimeric contig formation (analogous to massive unassigned insertions or deletions), we note specific boundaries to its current scope. The applied mathematical augmentation primarily tests the robustness of the overarching distance histogram against additive sequence noise. Because centeny distance vectors fundamentally measure spacing \textit{between} motifs, the current synthetic pipeline does not explicitly augment pure complex inversions, balanced translocations, or a mixture of structural rearrangements. For instance, an inversion of an entire intact higher-order repeat array might preserve internal inter-motif distances while flipping orientation. To further validate the diagnostic behavior of our score (\Cref{fig:6_figure}D), we constructed a synthetic case where chr1 was replaced by: 70\% chr17 sequence, 10\% chr3 sequence, and retained only 20\% of its original one; we observed that the similarity signal shifted from chr1 to chr17, confirming that the metric both detects corruption and identifies the dominant genomic source. This feature is captured beautifully by our visual centeny maps (\hyperref[sec:1_chapter]{\textcolor{mediumtealblue}{\textbf{Chapter 1}}}), but could be drowned in the noise when considering very long purely distance-based \textsc{\scht{1}d} vectors. Expanding this stress-test to encompass synthetic modeling of all four major \textsc{sv} classes remains a critical next step for comprehensive genomic benchmarking.
\subsection*{KL-score as a metric of centromere assembly quality}
\label{sec:7_chapter}

Our framework, which can be extended to an arbitrary number of genomes, allows one to establish a ranking of the \textsc{t\scht{2}t} genomes at hand, based solely on the divergence scores obtained from a reference distribution. The nature of a \textsc{kl} divergence requires a reference probability distribution, by its definition. We initially performed our experiments considering \textsc{chm\scht{13}} as the reference genome.

To rank genomes beyond the limitations of a single reference genome, we also constructed a population average baseline. We introduce the concept of a consensus genome, which we define as the distribution composed of \textsc{yao}v\textsc{\scht{2.0}} and \textsc{hg\scht{002}v\textsc{\scht{1.1}}} genomes.

To construct the consensus probability distribution for a given autosome $c$, we aggregate the discrete distance values from the chromosomes of the genomes we selected for the consensus.

Let $\mathcal{G}$ be the set of reference genomes (e.g., $\mathcal{G} = \{\textsc{hg\scht{002}}\text{v\textsc{\scht{1.1}}}, \textsc{yao}\text{v\textsc{\scht{2.0}}}\}$). For each genome $g \in \mathcal{G}$, haplotype $h \in \{1, 2\}$ and chromosome $c \in \{1,2, ... , 22\}$, let $X_{g,h}^{(c)}$ denote the set of raw distance values for that specific chromosome, haplotype, and genome.

We pool these observations to form a combined set $X^{(c)}$ representing the set of the raw distances for that chromosome formed by gathering these distances from different haplotypes of different genomes of that same chromosome. 

$$X^{(c)} = \bigcup_{g \in \mathcal{G}} \bigcup_{h \in \{1,2\}} X_{g,h}^{(c)}$$

For any discrete distance value $x$, we compute its absolute frequency $C^{(c)}(x)$ by summing the occurrences across the pooled data:

$$C^{(c)}(x) = \sum_{v \in X^{(c)}} \mathbf{1}_{\{v = x\}}$$

where $\mathbf{1}_{\{v = x\}}$ is the indicator function, which equals $1$ if $v = x$ and $0$ otherwise.

Finally, we normalize these counts to define the consensus probability distribution $P_{\text consensus}^{(c)}$. Let $M$ be the maximum observed distance in $X^{(c)}$. The probability mass for each distance $x \in [0, M]$ is given by dividing its count by the total number of observations:

$$P_{\text consensus}^{(c)}(x) = \frac{C^{(c)}(x)}{\sum_{k=0}^{M} C^{(c)}(k)}$$

This normalization guarantees that $\sum_{x=0}^{M} P_{\text consensus}^{(c)}(x) = 1$. The resulting distribution $P_{\text consensus}^{(c)}$ acts as the generalized population baseline for evaluating the symmetric \textsc{kl} divergence of other assemblies.

\begin{figure}[htbp]
    \centering
    \includegraphics[width=6.5in]{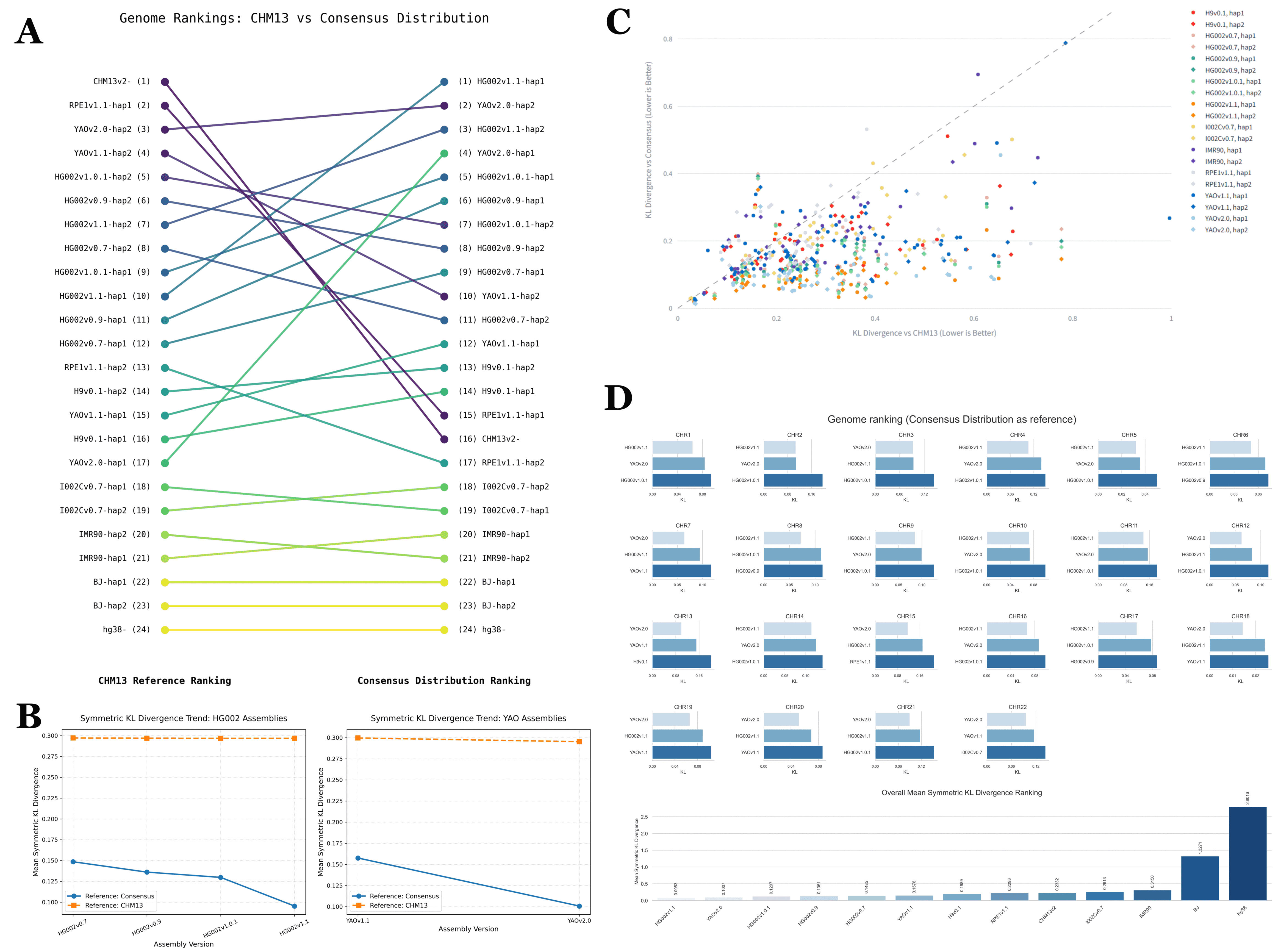}
    \caption{\textbf{Ranking of the telomere-to-telomere genomes with respect to a reference distribution.} \textbf{(A)} Side-by-side ranking of the \textsc{t}\oldstylenums{2}\textsc{t} genomes based on different reference baselines. On the left, \textsc{chm}\oldstylenums{13} as the reference, on the right, the constructed consensus distribution as the reference. \textbf{(B)} Newer genome assemblies yield progressively lower \textsc{kl} divergence scores, proving that the metric reliably reflects assembly improvement \textbf{(C)} Scatterplot in which each data point consists of the pair of the two \textsc{kl} scores, one per reference, displaying that most of those place under the dashed line — where the score in both genomes is calculated against the same \textsc{chm}\oldstylenums{13} reference. \textbf{(D)} Comprehensive genome ranking using the constructed consensus genome as the reference. The upper grid shows the top 3 best-ranked genomes per autosome, and the bottom panel presents the overall mean symmetric \textsc{kl} divergence ranking.}
    \label{fig:7_figure}
\end{figure}

We provide two rankings of the \textsc{t\scht{2}t} genomes, indicating the change in rank for each genome.
To assess the effect of reference selection, we compared the rankings obtained from the \textsc{chm\scht{13}} reference scoring and the consensus reference scoring using a slopegraph (\Cref{fig:7_figure}A). Several genomes change position when the consensus reference distribution is used. 

The \textsc{chm\scht{13}} assembly, which ranks first when compared to its own reference distribution, drops significantly to \nth{16} position when evaluated relative to the population average baseline. In contrast, assemblies from the \textsc{hg\scht{002}} and \textsc{yao} lines occupy the highest positions in the population-based ranking. For example, \textsc{hg\scht{002}v\textsc{\scht{1.1}}}-hap1 moves from \nth{10} to first position, while \textsc{yao}v\textsc{\scht{2.0}}-hap2 moves from third to second position. Other notable improvements include \textsc{yao}v\textsc{\scht{2.0}}-hap1, which climbs from \nth{17} to fourth, and \textsc{hg\scht{002}v\textsc{\scht{1.1}}}-hap2, which moves from \nth{7} to third. In contrast, the lowest-ranking assemblies, including \textsc{imr\scht{90}}, \textsc{bj}, and \textsc{grc}h\scht{38}, remain in the lowest positions under both scoring approaches, showing that no matter the reference these genomes perform poorly .

The open question remains whether we can make statements on genome assembly quality based solely on the scores our framework assigns to these genomes. To answer this question we tracked the symmetric Kullback-Leibler (\textsc{kl}) divergence across newer assembly releases for the \textsc{hg\scht{002}} and \textsc{yao} samples (\Cref{fig:7_figure}B). When evaluated against the population consensus distribution, the mean symmetric \textsc{kl} divergence exhibits a consistent decline, reflecting the progressive refinement of the assemblies. For instance, the \textsc{hg\scht{002}} shows a reduction from approximately 0.15 in v0.7 to below 0.10 in v1.1.

Conversely, when \textsc{chm\scht{13}} is employed as the sole reference, the divergence remains nearly constant at a significantly higher baseline ($\approx 0.30$). This disparity suggests that the consensus reference provides a more sensitive and dynamic metric for capturing assembly quality improvements.

This illustrates that the population-based consensus distribution offers a more effective benchmark for evaluating the quality of new assemblies, since it is able to reflect better and newer assemblies by means of lower divergence.

We plot the symmetric \textsc{kl} divergence of various genomes and their assemblies (\Cref{fig:7_figure}C) calculated against the \textsc{chm\scht{13}} reference ($x$-axis) versus the consensus distribution ($y$-axis). The distribution of data points predominantly below the dashed identity line ($y < x$) demonstrates that the \textsc{kl} divergence is consistently lower when evaluated against the consensus distribution. This indicates that the consensus based distribution serves as a more representative benchmark for diverse human assemblies by mitigating the reference bias associated with a single-haploid reference like \textsc{chm\scht{13}}.

\Cref{fig:7_figure}D provides a high-resolution breakdown of this performance metric across individual autosomes, identifying the top three assemblies per chromosome. The results indicate that \textsc{hg\scht{002}v\textsc{\scht{1.1}}} and \textsc{yao}v\textsc{\scht{2.0}} largely dominate the local rankings. While \textsc{hg\scht{002}v\textsc{\scht{1.1}}} frequently secures the top position, \textsc{yao}v\textsc{\scht{2.0}} demonstrates superior accuracy on specific chromosomes, such as chr3, chr7, and chr12, highlighting its ability to capture regional variations in assembly quality.

Ultimately, these results demonstrate that our framework effectively overcomes the limitations of single-reference bias. By transitioning to a consensus distribution, we provide a more sensitive and representative benchmark that accurately reflects the progressive technical refinements of \textsc{t\scht{2}t} assemblies. This implies that the consensus-based \textsc{kl} divergence may be a robust indicator of assembly quality, capable of distinguishing the most accurate genomic representations within a diverse population.

\newpage
\section*{Discussion}
\label{sec:discussion}

In this work we have explored an innovative approach to assess human genome assemblies by implementing a probabilistic model encapsulating the highly ordered and conserved organization of chromosomes functional motifs, as in centeny maps, adopting a Kullback-Leibler (\textsc{kl}) divergence measure. We applied it to \textsc{t\scht{2}t} human genomes and sought to determine the nature and boundaries to which we are able to detect assembly errors, with a particular focus on centromeric regions. Notably, this approach leverages the discovery of a chromosome-specific structural architecture identified by \textsc{cenp-b} box motifs genome-wide\supercite{corda2025chromosome} and, most importantly, it defines for the first time a relationship between numeric distances and their implication for detecting potential misassemblies in repetitive regions. Because this matter has been difficult to address even with modern-day sequencing technologies coupled with state-of-the-art genome assemblers, our solution provides a rapid and robust evaluation of virtually any genomes in a matter of minutes. This was possible by introducing a distribution-based scoring system that operates in centeny space rather than at nucleotide level, where genomic \textsc{dna} is translated into numerical values (distances) for each chromosome. Quantifying how well a pre-determined set of input assemblies preserves centromeric architecture can be obtained by measuring the divergence between a query and a target distribution. Summarized in a \textsc{kl} divergence score, this framework outputs a single, interpretable quality value for each chromosome — which can be aggregated to score the whole genome as well. This enables direct ranking of assemblies and detection of major classes of structural variation, with future implications in detection of possible pathogenic variation and/or expansion of our approach to other species.

Building upon centeny maps to ascertain motif conservation across genomes, which effectively converts the occurrence between consecutive \textsc{cenp-b} boxes into distance values, we have been able to demonstrate their consistency across diverse \textsc{t\scht{2}t} assemblies while accounting for the substantial sequence-level variability underpinning the centromeric locus. To achieve this, each chromosomes’ centromeres was transformed into a numerical representation by extracting consecutive inter-motif distances and expressing them as normalized discrete histograms. When applied to the five \textsc{t\scht{2}t} genome assemblies selected in this work (\textsc{hg\scht{002}}, \textsc{i\scht{002}c}, \textsc{yao}, \textsc{h\scht{9}}, and \textsc{rpe\scht{1}}) along with the two human references \textsc{grc}h\textsc{\scht{38}} and \textsc{chm\scht{13}}, we reveal that those distance values are consistent across individuals and haplotypes. The former provides for a sound validation for the use of these distances as a biologically grounded fingerprint of centromeric organization.

By establishing a benchmark experiment, we tested and confirmed \textsc{kl} divergence to be the most representative measure among the others tested in this study for computing chromosomal distance profiles. To this end, we compared our metric with other similarity measures, in particular we looked at: Euclidean distance, a histogram-adapted Jaccard distance, and a deep sequence encoder baseline (Chronos-2). Overall, across haplotypes and genomes, symmetric \textsc{kl} yields the clearest diagonal structure and the strongest separation between homologous and non-homologous chromosomes (\Cref{fig:3_figure}), consistent with centromeric identity. \textsc{auroc} analysis supports near-perfect separation for all tested metrics, with \textsc{kl} providing the most discriminative and stable scoring. 
Additionally, in order to more directly answer the principal question about scoring genome assemblies employing our \textsc{kl} measure, we introduced synthetic noise in all five \textsc{t\scht{2}t} genomes part of the study. This has been fundamentally important to also comprehend how and to what extent our framework was able to classify structural rearrangements in the form of insertions and deletions greater than 49 bps, which are among the most common structural events at centromeres. While deletions are not well tolerated, as they factually change the probability distribution of a symmetric \textsc{kl}, and as such can be easily detected when \enquote{artificially} added; insertions, on the other hand, do not have the same effect and our approach has a limited capacity in characterizing this type of events. This likely reflects type of distances inserted, which were coming from the same chromosome and do not fundamentally alter the underlying \textsc{kl} distribution; indeed, when distances found in other chromosomes are inserted, the selected one scores poorly with respect to its unaltered counterpart. This testifies that our framework works well in cases of aberrant distances inserted, and it can classify translocation events as well, but remains blind of either short-range mutational events or other alterations that do not affect the overall structural organization and expected chromosome specificity. Furthermore, our ranking is currently based on just a handful of chromosome-level genomes, but also analyses so far were reliant on a comparison against a baseline truth distribution build from centeny map distances for \textsc{chm\scht{13}}. Therefore, to substantiate the applicability of \textsc{kl} in situations where the reference is different we produced chromosome rankings for all chromosomes using a population-level baseline distribution demonstrating, with a leave-one-out experiments, that top-ranking genomes remain close to it (after being processed through the framework) even when excluded from baseline construction. We thus provide a consensus score to benchmark future genomes against. However, continue refinement and improvement of the scoring metric, as more correctly assembled genomes with intact centromeres for all chromosomes become available, will be possible in a near future.

The completeness and accuracy of centromeres should be an essential feature of \textsc{t\scht{2}t} genomes, like telomeres. Here, we bypass sequences and turn genomes into numerical rendering, providing a rapid and mathematically tested framework to compare those numbers at chromosome and whole-genome level. Because we use a conserved and chromosome-specific architecture of numbers, we have been able to define a gold-standard numerical reference for human centromere assessment. As more genomes of high quality are produced, this reference can be further refined, becoming more representative and accurate. Our proof-of-concept thus addresses a real need in the genomic field and can be indefinitely expanded to other motifs, queries, difficult-to-assemble regions, species, and integrated with multiple other approaches. Altogether, this work represents a gateway to the future genomic and centromere assessment, enabling genome-to-genome comparison for years to come.

\newpage

\paragraph{Acknowledgments}
This work was possible thanks to the University of Rome “La Sapienza”, supported by grants RICERCA\_SAPIENZA, ERC, AIRC and FIS. We thank the High Performance Computing (HPC) Core Terastat2 and Terastat3 from Umberto Ferraro Petrillo of the Statistics Department, University of Rome Sapienza. We acknowledge all the members of the Giunta lab, especially Valentina Liguori for administrative support, Mauricio Orantes Bonilla and Ferdinando Randisi for critical reading and help with editing. S.G. and members of her lab are supported by Italian Foundation for Cancer Research (AIRC Start-Up Grant 2020 ID. \# 25189), the ERC CENTROFUN Starting Grant \#101078838, Fondo Italiano per la Scienza FIS2 \# 2023-02742, Sapienza University of Rome (Ricerca Sapienza Progetti di Eccellenza, Grant No. B83C24001270005), and the Rita Levi-Montalcini program from the Italian Ministry of University and Research (MIUR) to S.G.
 
\paragraph{Fundings}
S.G. is supported by the Fondazione AIRC per la Ricerca sul Cancro (AIRC Start-Up Grant 2020, Grant ID 2518 to S.G.); the European Research Council (ERC) under the European Union’s Horizon Europe research and innovation programme (CENTROFUN Starting Grant, Grant Agreement No. 101078838 to S.G.); the Ministero dell'Università e della Ricerca (MUR) through the Fondo Italiano per la Scienza (FIS2, Grant No. 2023-02742 to S.G.); and Sapienza University of Rome (Ricerca Sapienza Progetti di Eccellenza, Grant No. B83C24001270005 to S.G.).

\paragraph{Author contributions}
S.G., F.G., L.F., M.M., M.T.U., and E.C. designed the study and wrote the manuscript; L.F., M.M., M.T.U., and E.C. performed all computational analyses; L.C. provided help with centeny maps analyses. All authors have reviewed and approved the manuscript. 

\paragraph{Declaration of interests}
S.G. is an inventor on a patent (application no. 102023000028302) by University of Rome “La Sapienza,” filed on 28 December 2023, with international extension PCT/IT/2024/050269. 

\paragraph{Data and materials availability} 
All data are available in the manuscript or the supplementary materials. All code generated is available at \url{https://github. com/GiuntaLab/KL-GenomeEvaluation}

\paragraph{Declaration of generative AI and AI-assisted technologies in the manuscript preparation process}
During the editing, some of the authors used AI services (i.e. Claude, Gemini, etc.) to check for spelling and grammar, and for suggestions on condensing specific sentences. After this, the authors reviewed the text in its entirety and thus take full responsibility for the content of the published article.

\paragraph{Code availability}
Custom scripts used in this study and processed data will be available at the GitHub repository 
\url{https://github.com/PINlabSapienza} and \url{https://github.com/GiuntaLab/GCP-Centeny}.

\paragraph{Methods}
Centeny maps and centromere annotations using GCP pipeline models
The whole-genome and centromere-specific characterizations were performed using our previously described Genomic Centromere Profiling (GCP) pipeline, using CENP-B boxes-based annotation of the motif position and orientation (blue = forward; red = reverse complement sequence found within the same DNA strand) to define the centeny maps. Additional annotations of centromere structure were added using GCP pipeline Models 1 and Models 2, showing the inter-motif distances within centromeres (Model 1) and the organization of the motif within the alpha-satellite monomers (Model 2).\supercite{corda2025chromosome} More information and code can be found at \url{https://github.com/GiuntaLab/GCP-Centeny}

\newpage
\begin{refcontext}[sorting=nyt]
\printbibliography[title=References]
\end{refcontext}

\newpage

\begin{figure}[htbp]
\centering
\begin{subfigure}{.5\textwidth}
  \centering
  \includegraphics[width=\textwidth]{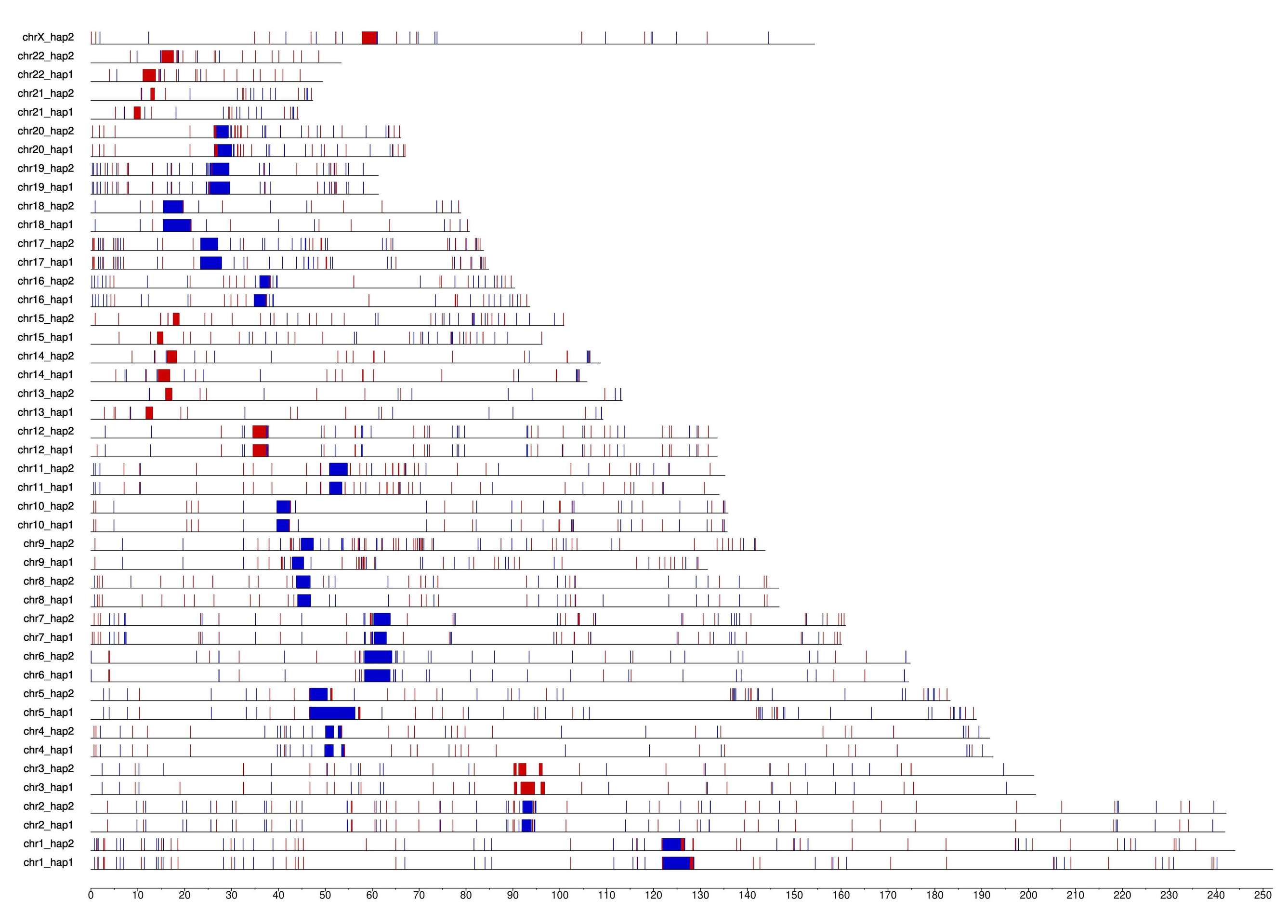}
  \caption{\textsc{hg}\oldstylenums{002} centeny map}
\end{subfigure}%
\begin{subfigure}{.5\textwidth}
  \centering
  \includegraphics[width=\textwidth]{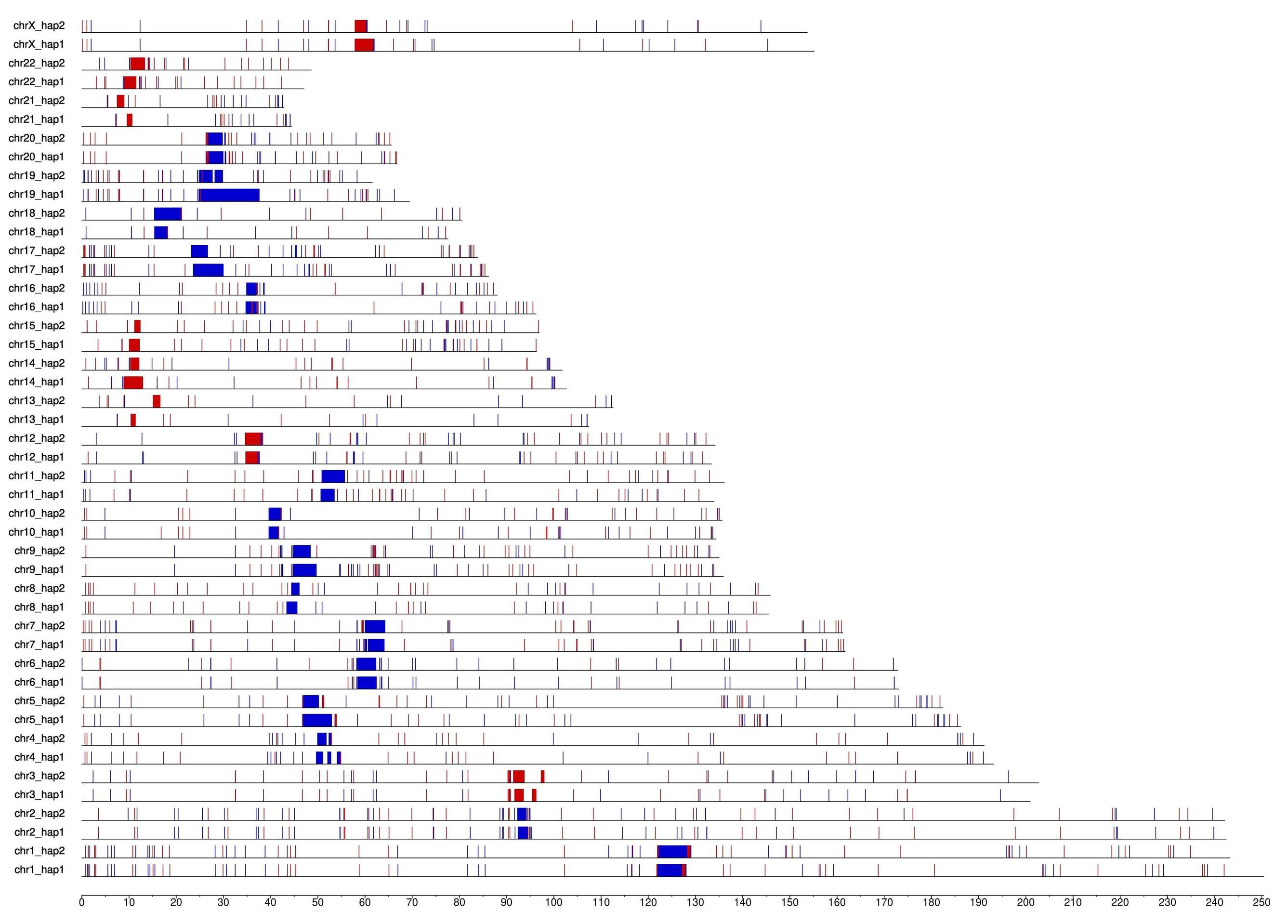}
  \caption{\textsc{h}\oldstylenums{9} centeny map}
\end{subfigure}
\vspace{1em}
\begin{subfigure}{.5\textwidth}
  \centering
  \includegraphics[width=\textwidth]{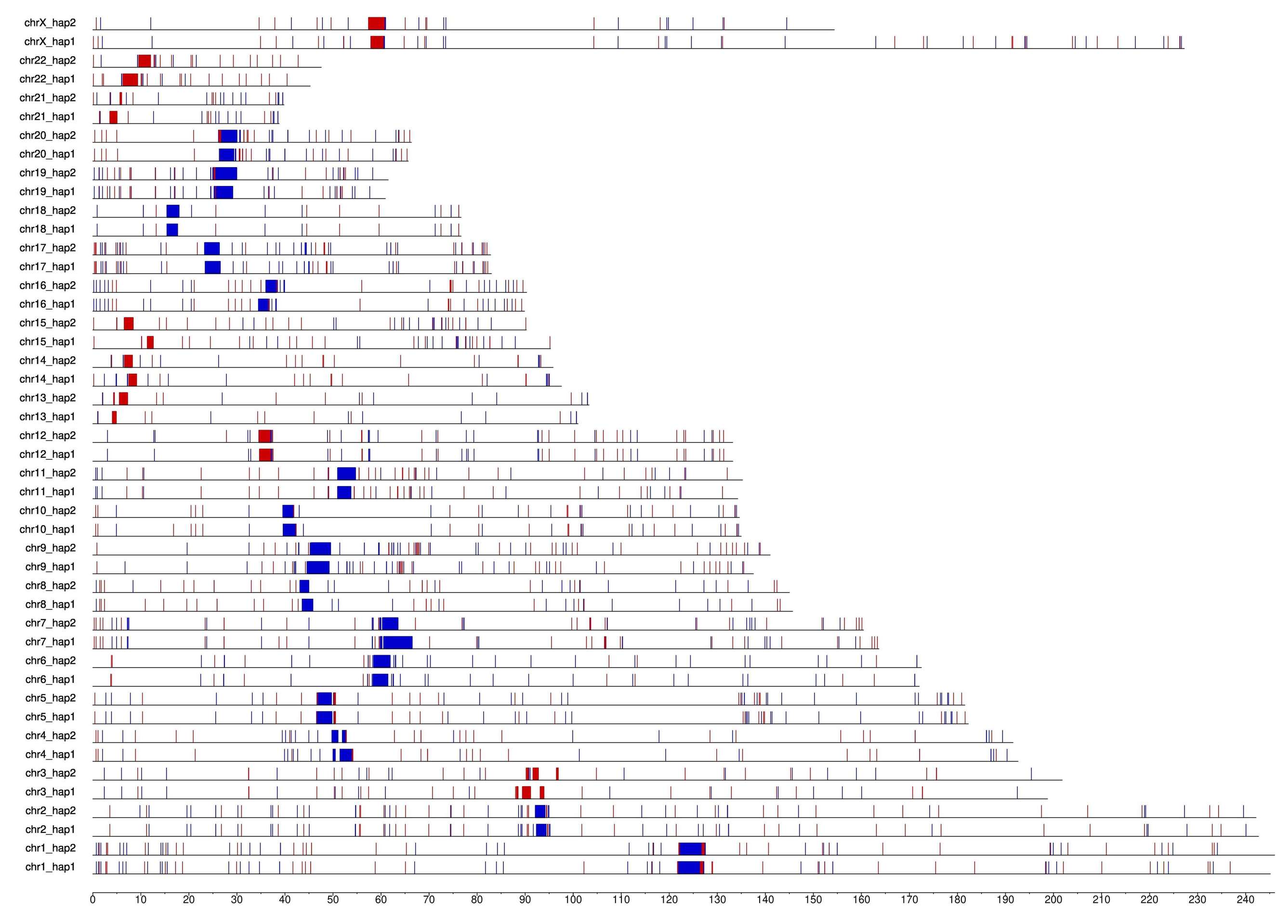}
  \caption{\textsc{rpe}\oldstylenums{1} centeny map}
\end{subfigure}%
\begin{subfigure}{.5\textwidth}
  \centering
  \includegraphics[width=\textwidth]{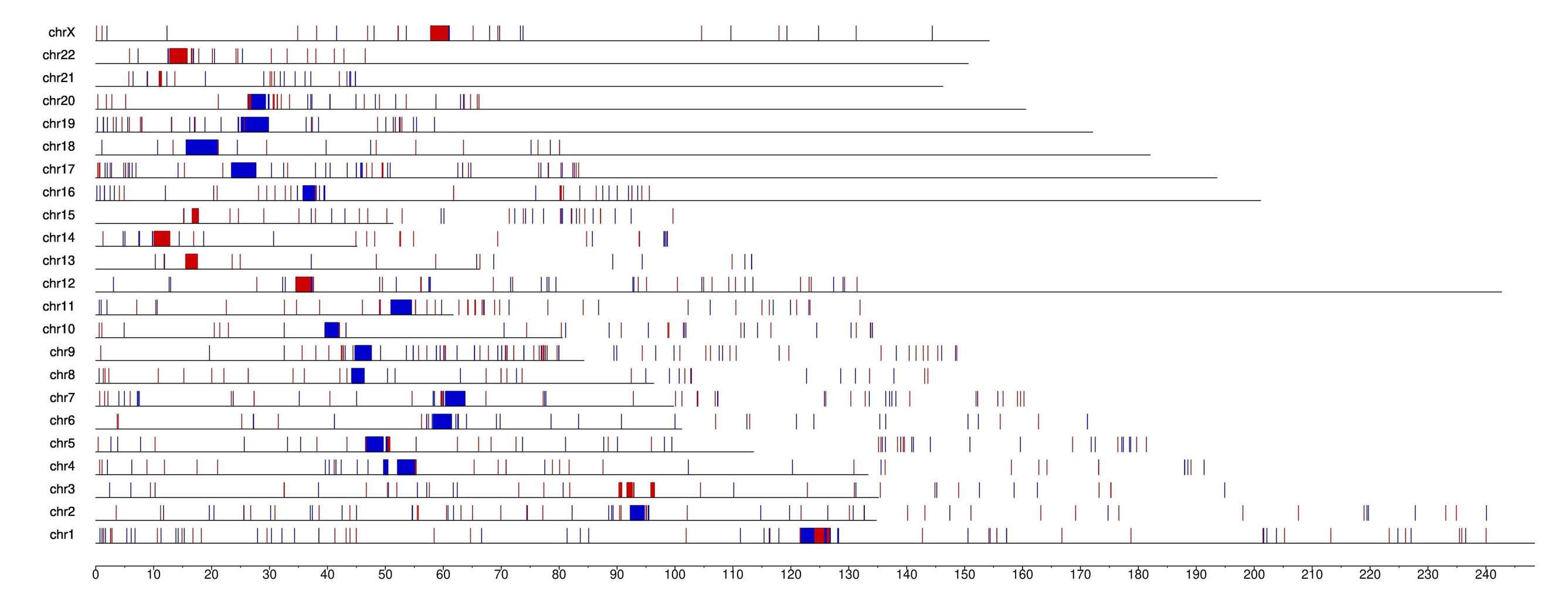}
  \caption{\textsc{chm}\oldstylenums{13} centeny map}
\end{subfigure}
\vspace{1em}
\begin{subfigure}{\textwidth}
  \centering
  \includegraphics[width=\textwidth]{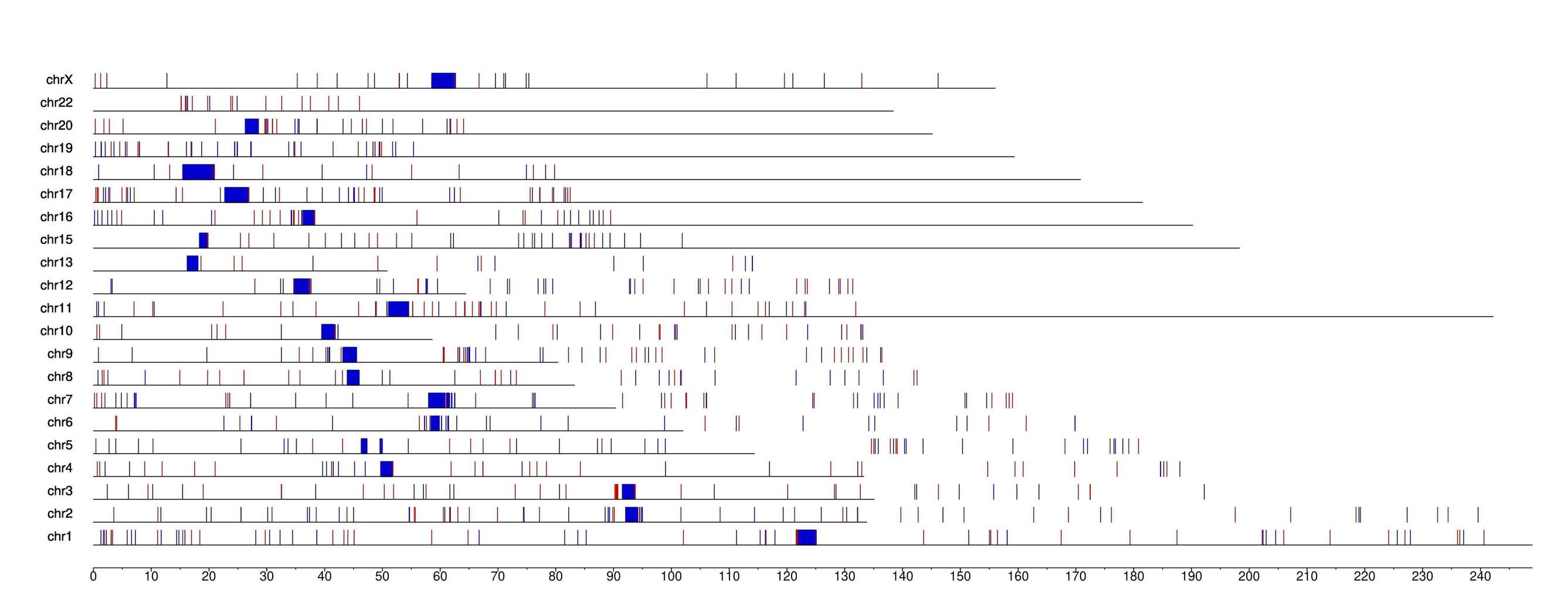}
  \caption{\textsc{grc}h\oldstylenums{38} centeny map}
\end{subfigure}
\caption{Expanded version of Figure 1 in the main text reporting the details of each sample's centeny map. The script is designed to plot chromosomes only when at least 50 \textsc{cenp-b} box hits are found; for this reason, chr14 and chr21 of \textsc{grc}h\oldstylenums{38} are not displayed. chrY is notoriously small and \textsc{cenp-b}-depleted, hence why it is not present in male genomes (\textsc{hg}\oldstylenums{002}) and haploid references (\textsc{chm}\oldstylenums{13} and \textsc{grc}h\oldstylenums{38}). The notation is otherwise the same as in main Figure 1.}
\label{fig:S1}
\end{figure}

\newpage

\begin{figure}[htbp]
\centering
\begin{subfigure}{\textwidth}
  \centering
  \includegraphics[width=\textwidth]{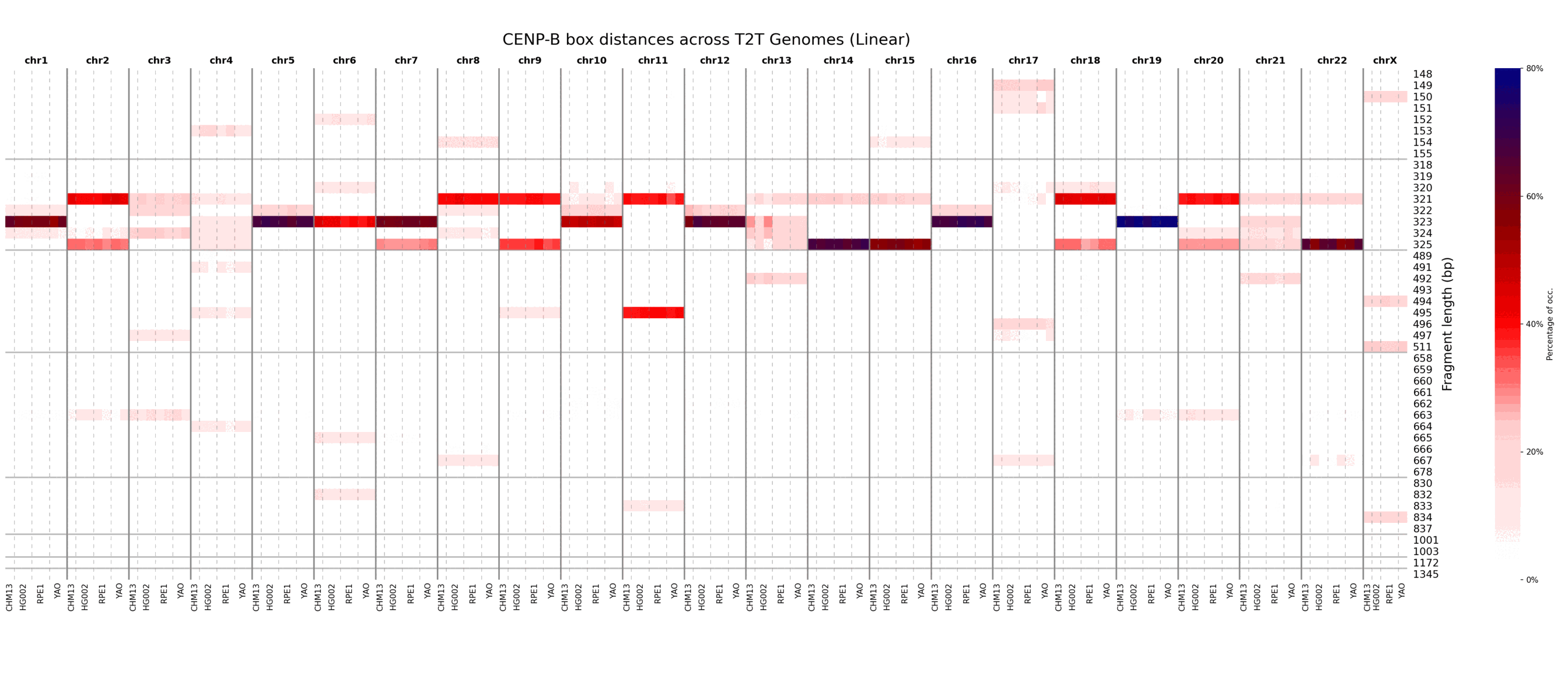}
  \caption{Linear values expressed as percentages of similarity}
\end{subfigure}%

\begin{subfigure}{.49\textwidth}
  \centering
  \includegraphics[width=\textwidth]{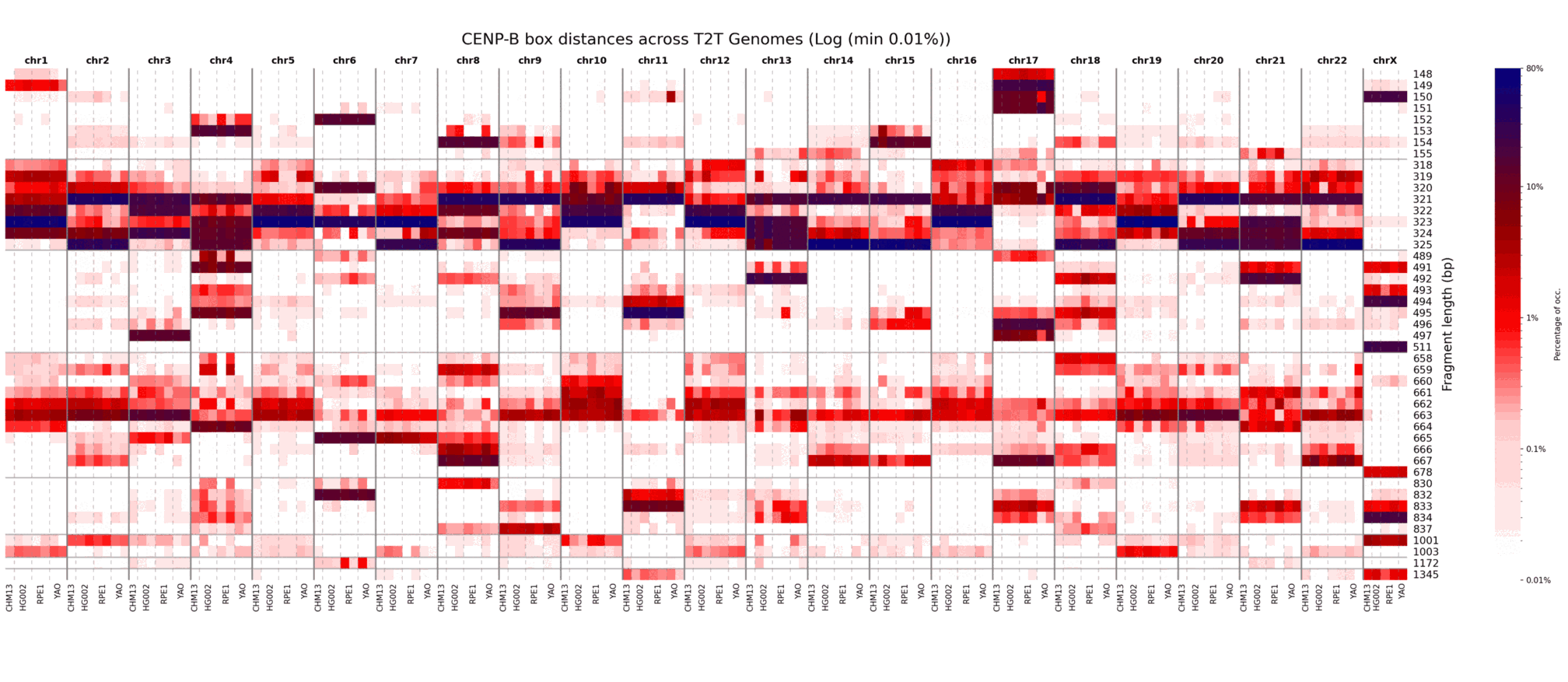}
  \caption{Log-scaled thresholds retained only if over 0.01\%}
\end{subfigure}
\begin{subfigure}{.49\textwidth}
  \centering
  \includegraphics[width=\textwidth]{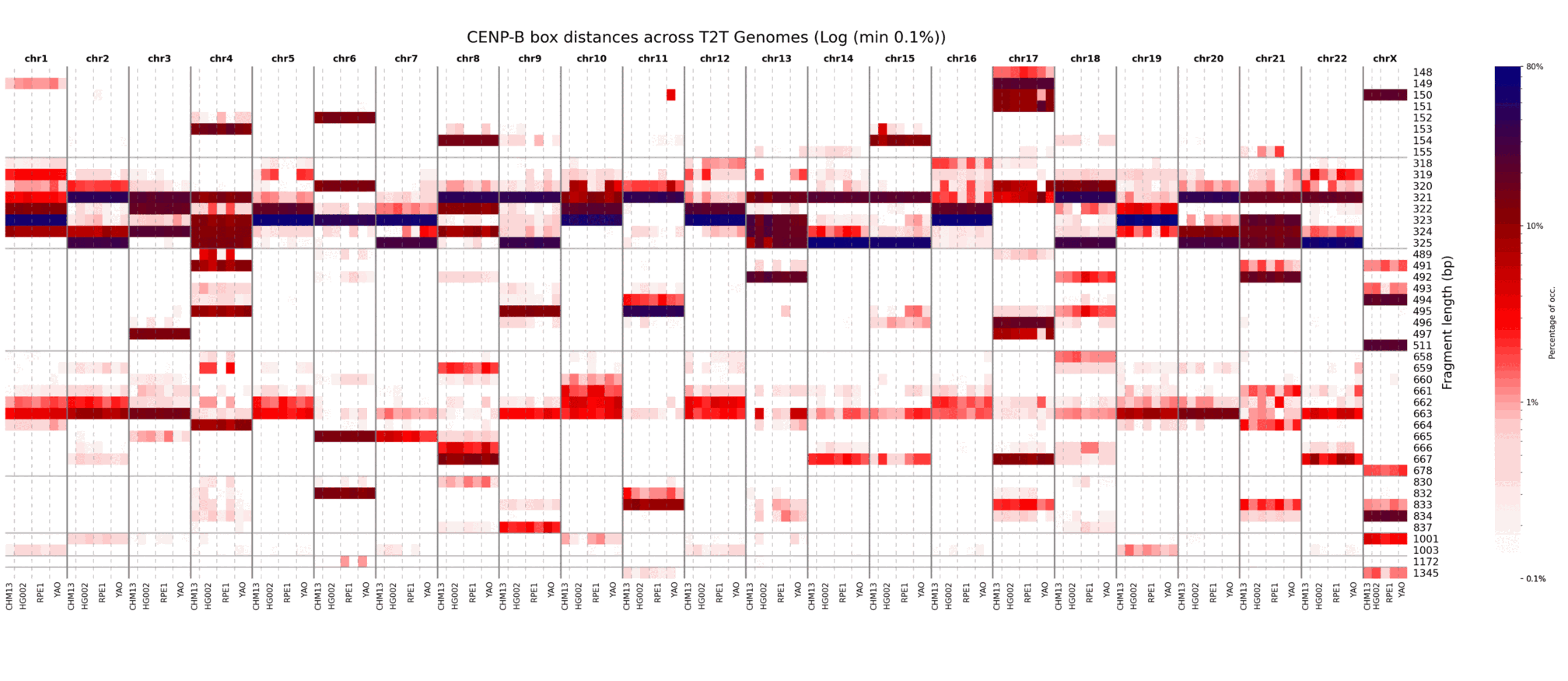}
  \caption{Log-scaled thresholds retained only if over 0.1\%}
\end{subfigure}
\caption{Heatmap view of different thresholds of frequency values for discrete inter-\textsc{cenpb} box distances.}
\label{fig:S2}
\end{figure}

\newpage

\begin{figure}[htbp]
\centering
\begin{subfigure}{.5\textwidth}
  \centering
  \includegraphics[width=\textwidth]{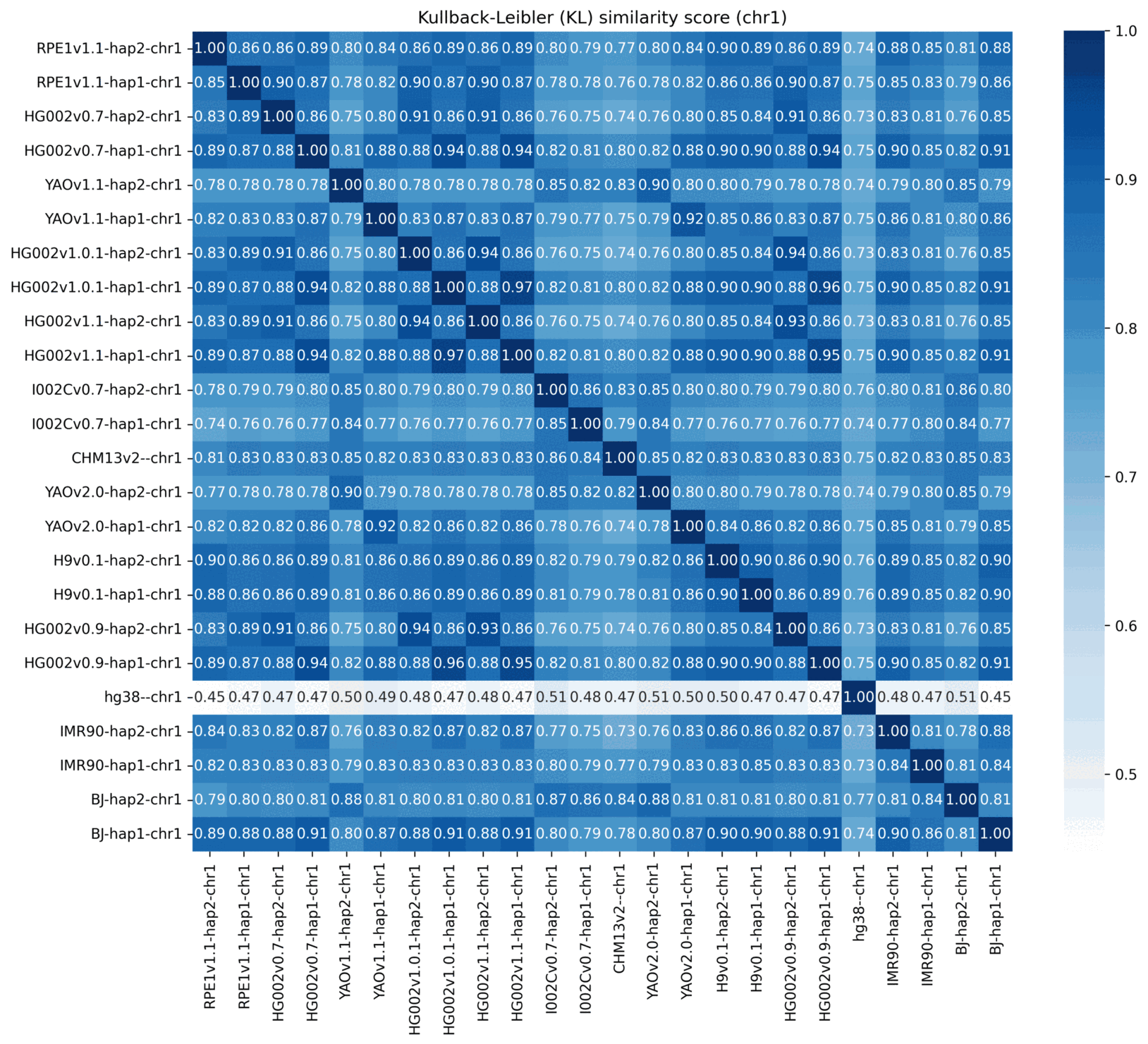}
  \caption{\textsc{kl} similarity matrix of chromosome 1}
\end{subfigure}\hfill
\begin{subfigure}{.5\textwidth}
  \centering
  \includegraphics[width=\textwidth]{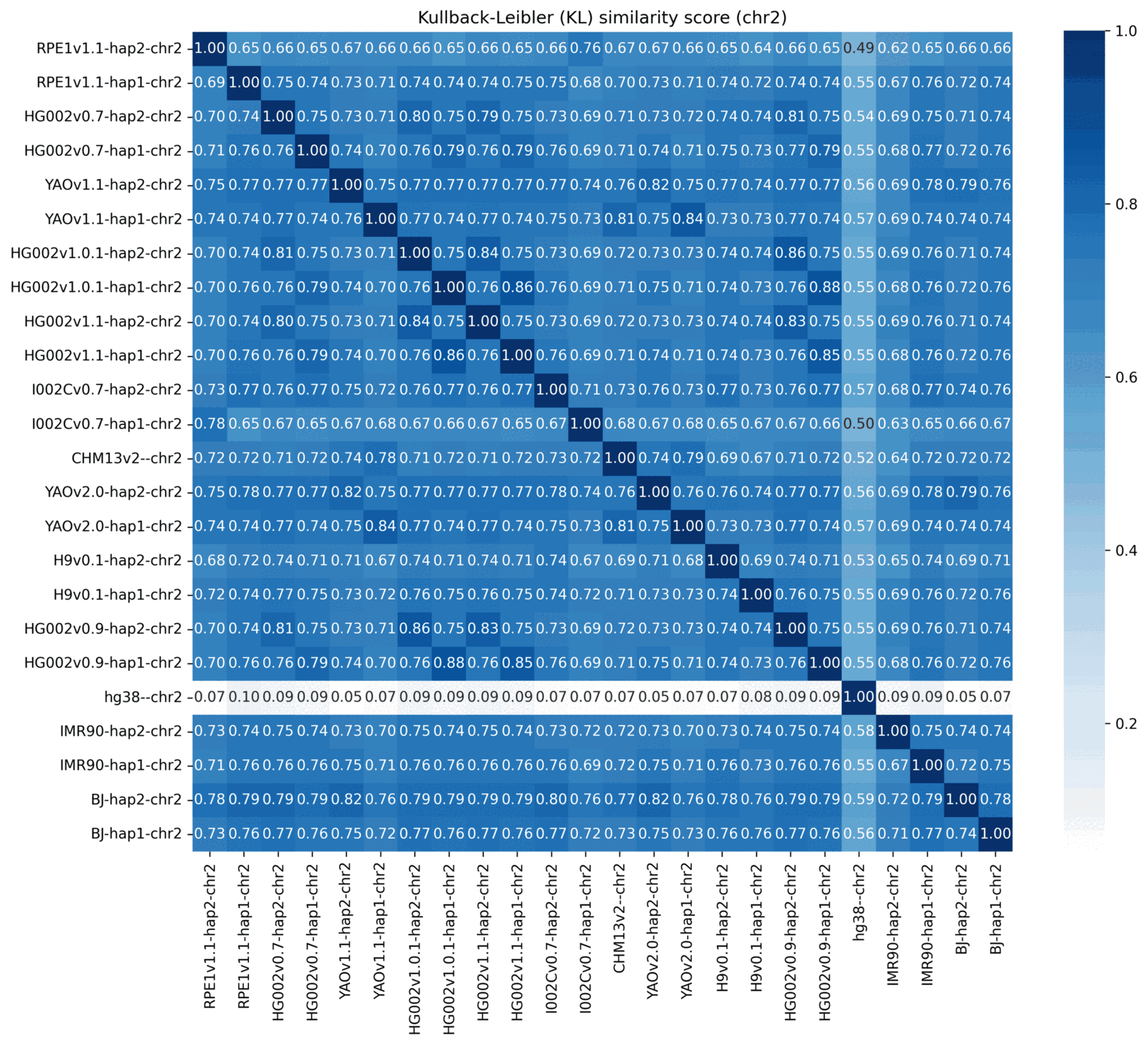}
  \caption{\textsc{kl} similarity matrix of chromosome 2}
\end{subfigure}

\vspace{1em}
\begin{subfigure}{.5\textwidth}
  \centering
  \includegraphics[width=\textwidth]{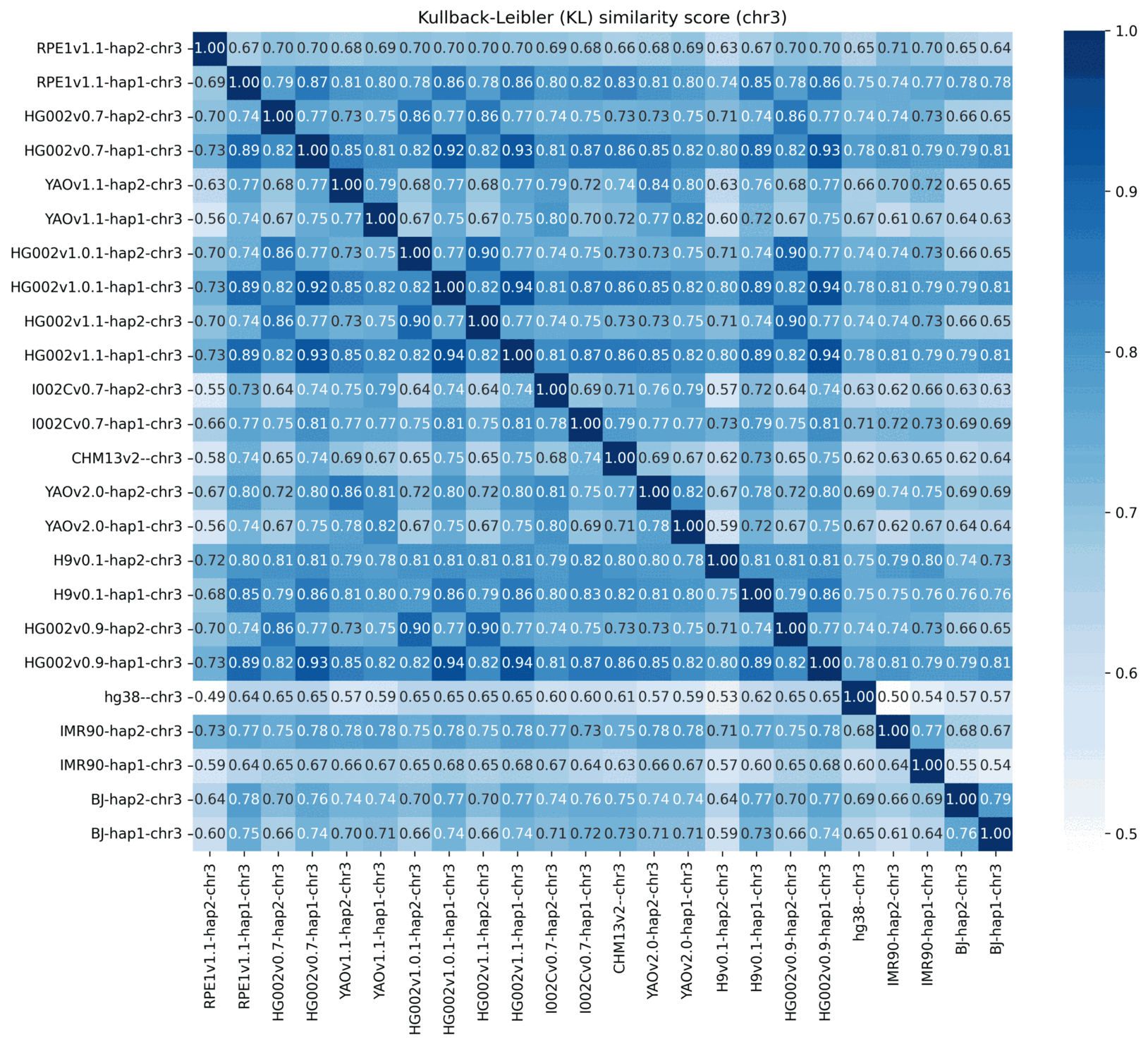}
  \caption{\textsc{kl} similarity matrix of chromosome 3}
\end{subfigure}\hfill
\begin{subfigure}{.5\textwidth}
  \centering
  \includegraphics[width=\textwidth]{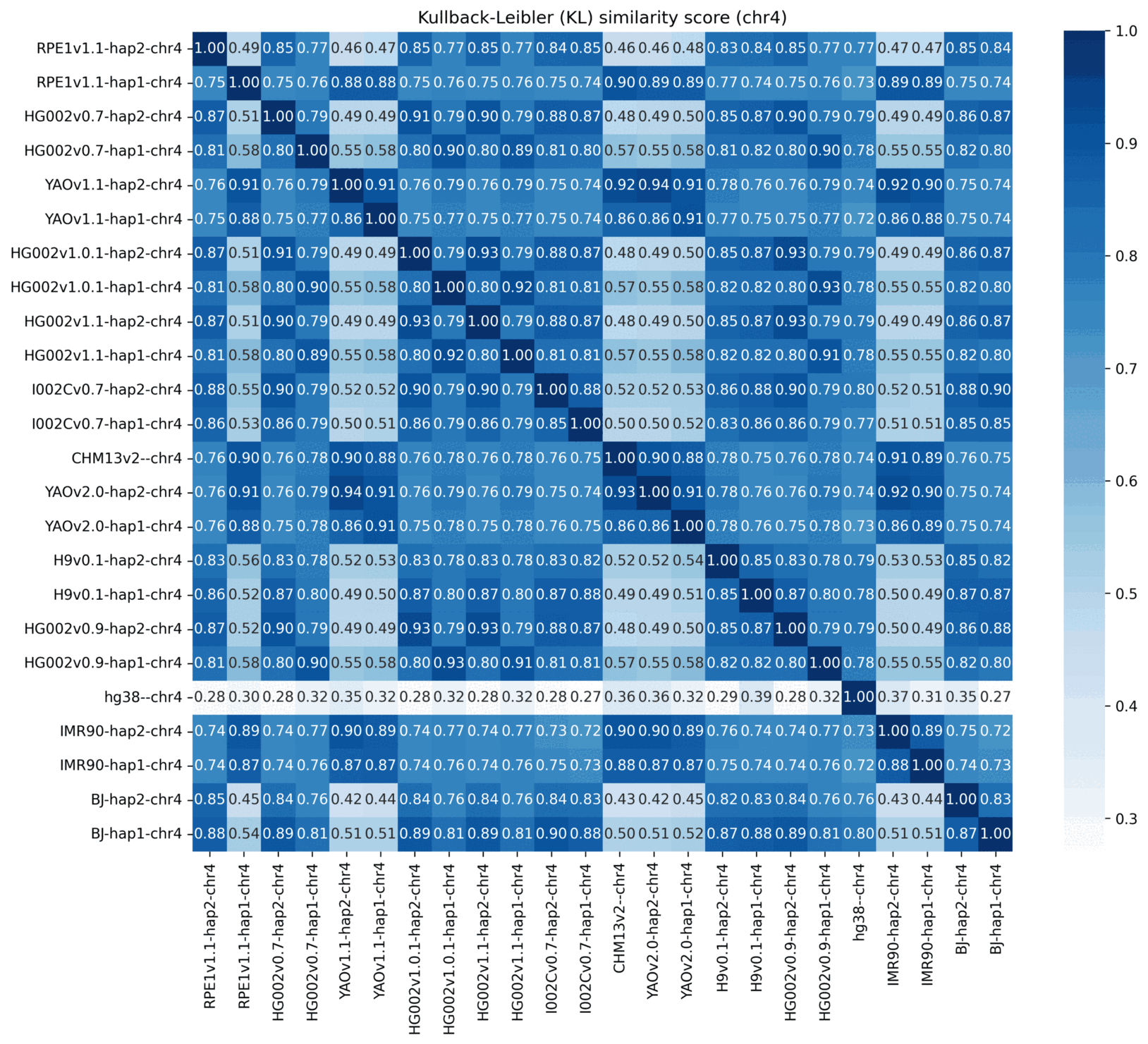}
  \caption{\textsc{kl} similarity matrix of chromosome 4}
\end{subfigure}
\caption{\textsc{kl} similarity heatmaps for chromosomes 1, 2, 3, and 4 evaluated among \textsc{rpe}\oldstylenums{1}v\oldstylenums{1.1}, \textsc{hg}\oldstylenums{002} (versions 0.7, 0.9, 1.0.1, and 1.1), \textsc{yao} (versions 1.1 and 2.0), \textsc{i}\oldstylenums{002}\textsc{c}, \textsc{chm}\oldstylenums{13}, \textsc{h}\oldstylenums{9}, \textsc{grc}h\oldstylenums{38}, \textsc{imr}\oldstylenums{90}, \textsc{bj} and their respective haplotypes.}
\end{figure}

\newpage

\begin{figure}[htbp]
\centering
\begin{subfigure}{.5\textwidth}
  \centering
  \includegraphics[width=\textwidth]{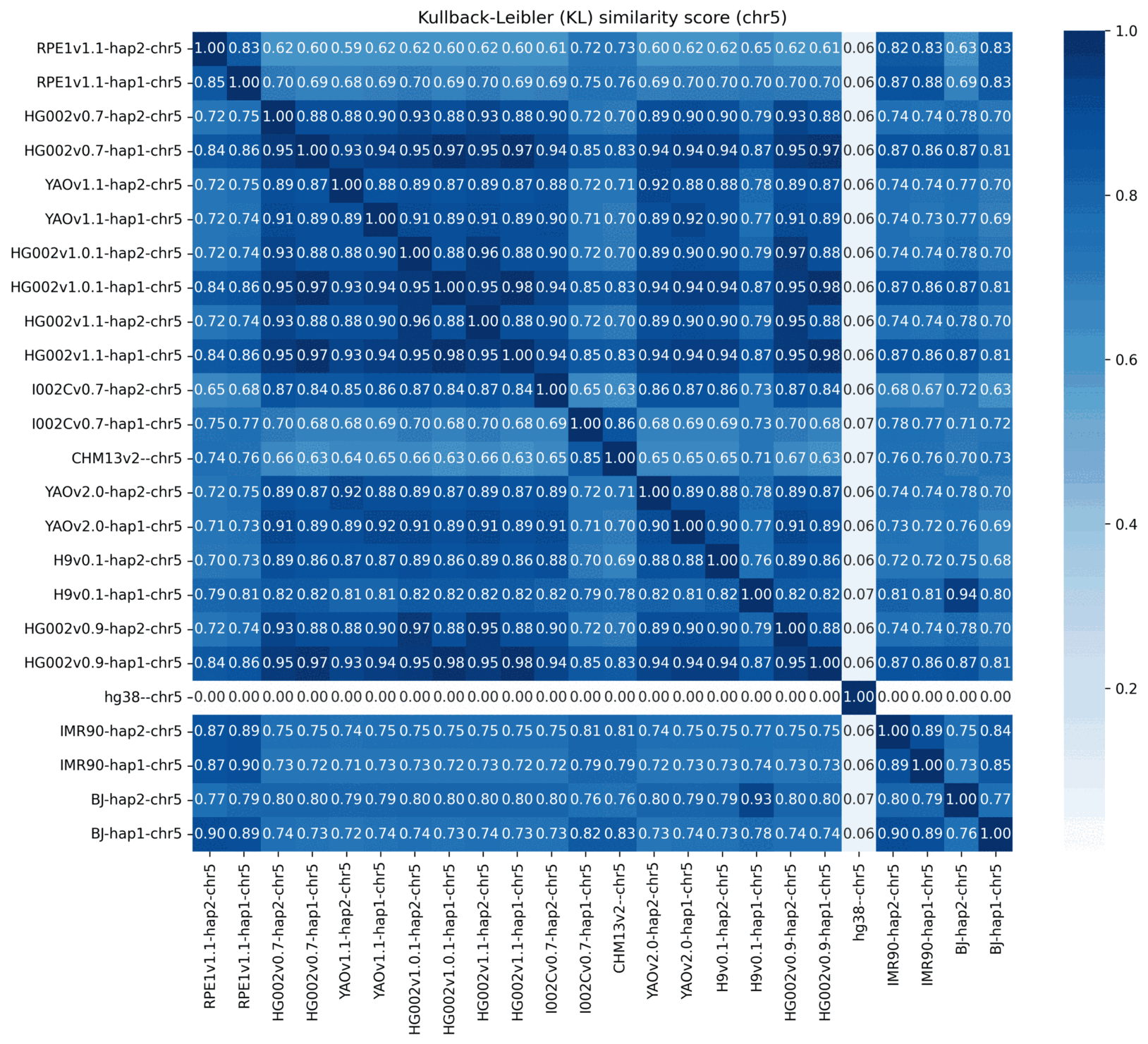}
  \caption{\textsc{kl} similarity matrix of chromosome 5}
\end{subfigure}\hfill
\begin{subfigure}{.5\textwidth}
  \centering
  \includegraphics[width=\textwidth]{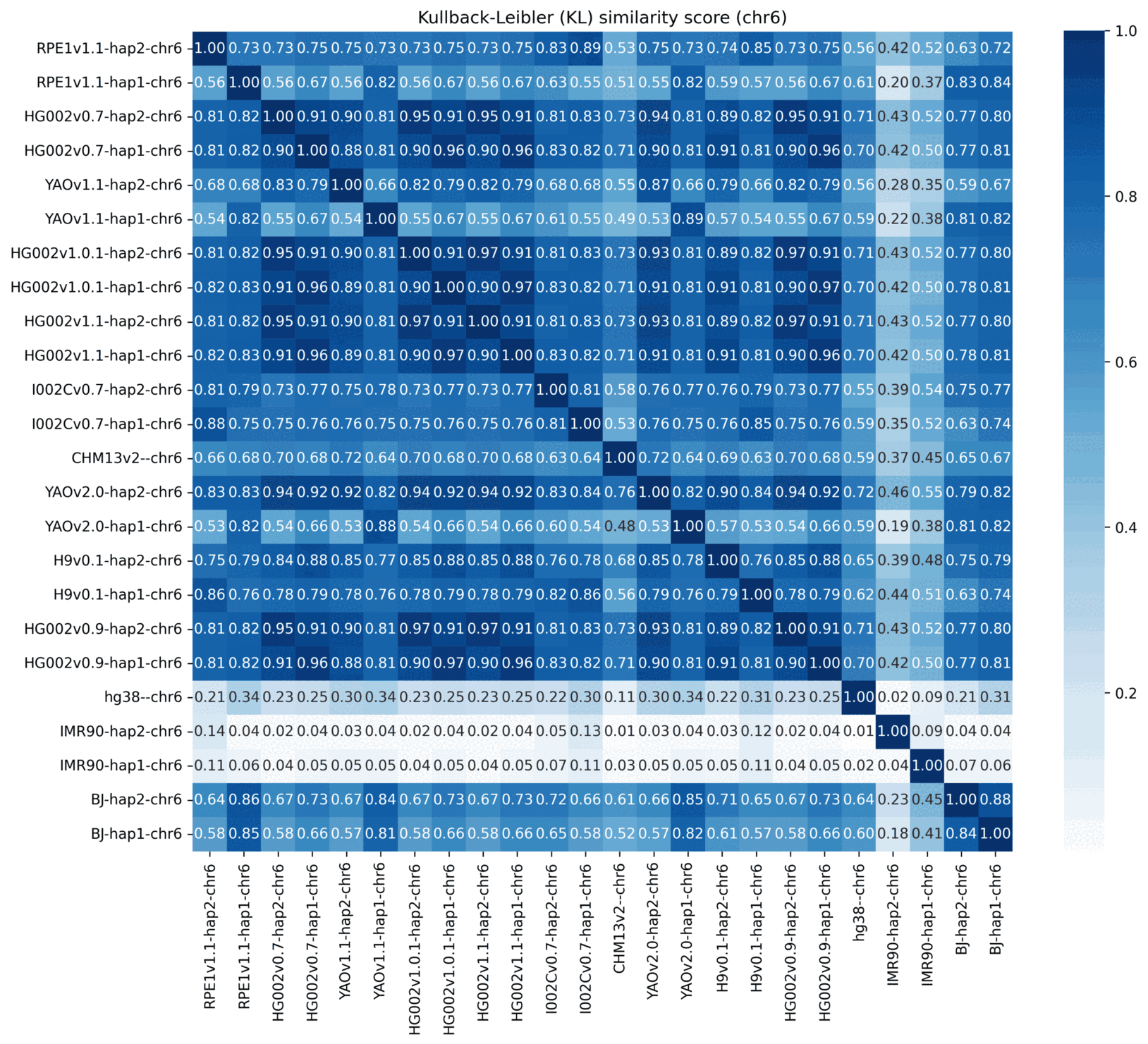}
  \caption{\textsc{kl} similarity matrix of chromosome 6}
\end{subfigure}

\vspace{1em}
\begin{subfigure}{.5\textwidth}
  \centering
  \includegraphics[width=\textwidth]{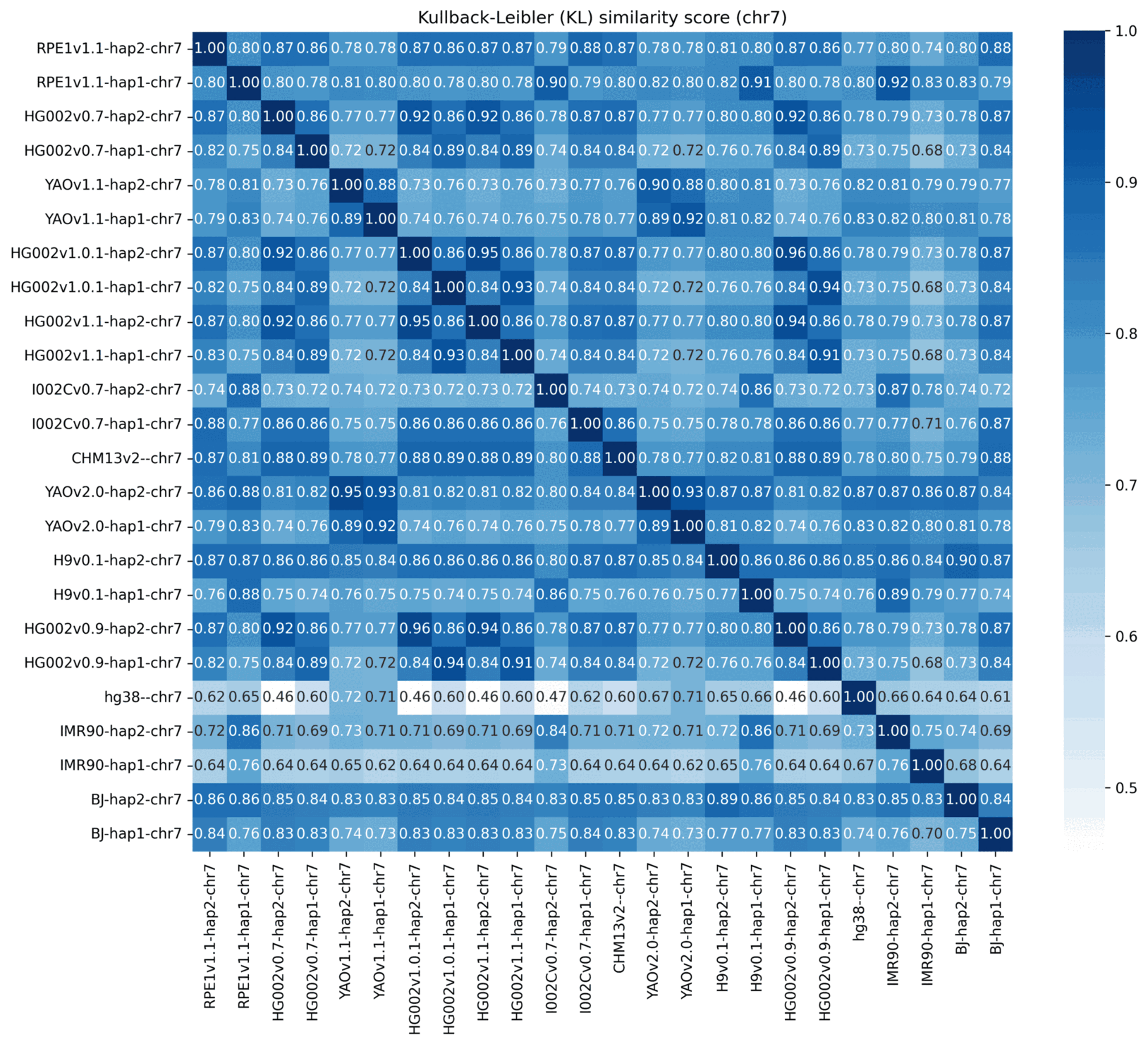}
  \caption{\textsc{kl} similarity matrix of chromosome 7}
\end{subfigure}\hfill
\begin{subfigure}{.5\textwidth}
  \centering
  \includegraphics[width=\textwidth]{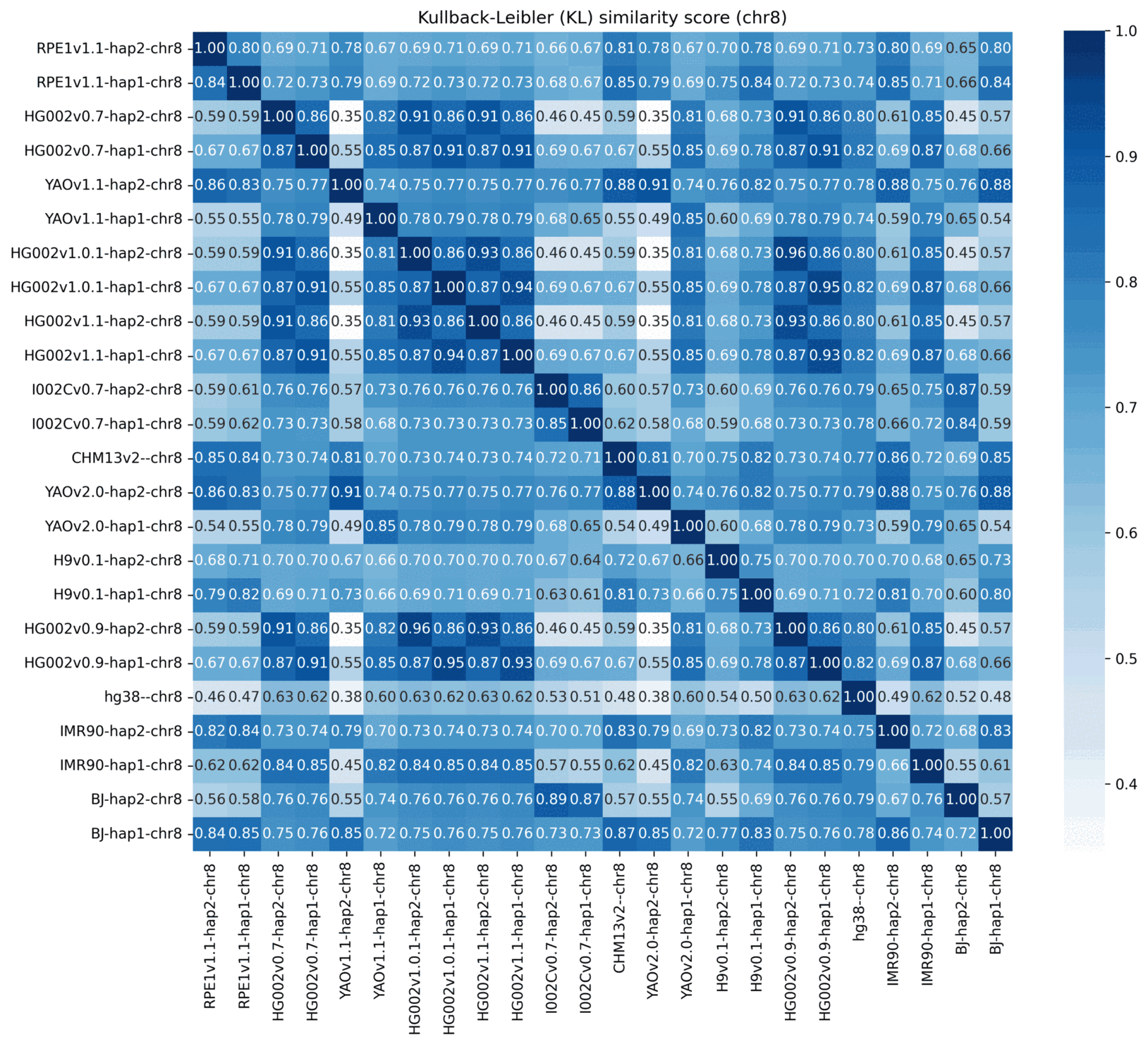}
  \caption{\textsc{kl} similarity matrix of chromosome 8}
\end{subfigure}
\caption{\textsc{kl} similarity heatmaps for chromosomes 5, 6, 7, and 8 evaluated among \textsc{rpe}\oldstylenums{1}v\oldstylenums{1.1}, \textsc{hg}\oldstylenums{002} (versions 0.7, 0.9, 1.0.1, and 1.1), \textsc{yao} (versions 1.1 and 2.0), \textsc{i}\oldstylenums{002}\textsc{c}, \textsc{chm}\oldstylenums{13}, \textsc{h}\oldstylenums{9}, \textsc{grc}h\oldstylenums{38}, \textsc{imr}\oldstylenums{90}, \textsc{bj} and their respective haplotypes.}
\end{figure}

\newpage

\begin{figure}[htbp]
\centering
\begin{subfigure}{.5\textwidth}
  \centering
  \includegraphics[width=\textwidth]{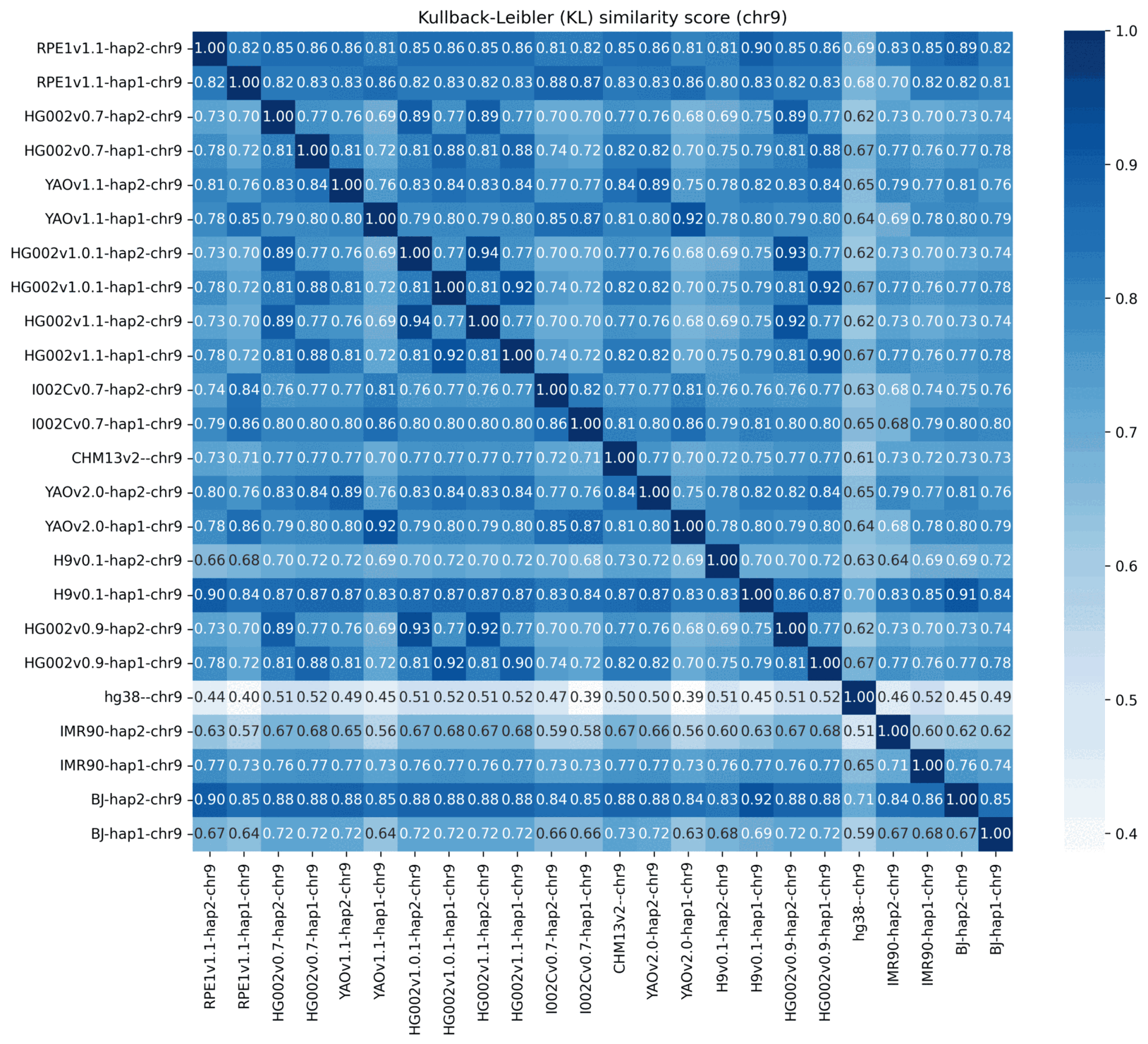}
  \caption{\textsc{kl} similarity matrix of chromosome 9}
\end{subfigure}\hfill
\begin{subfigure}{.5\textwidth}
  \centering
  \includegraphics[width=\textwidth]{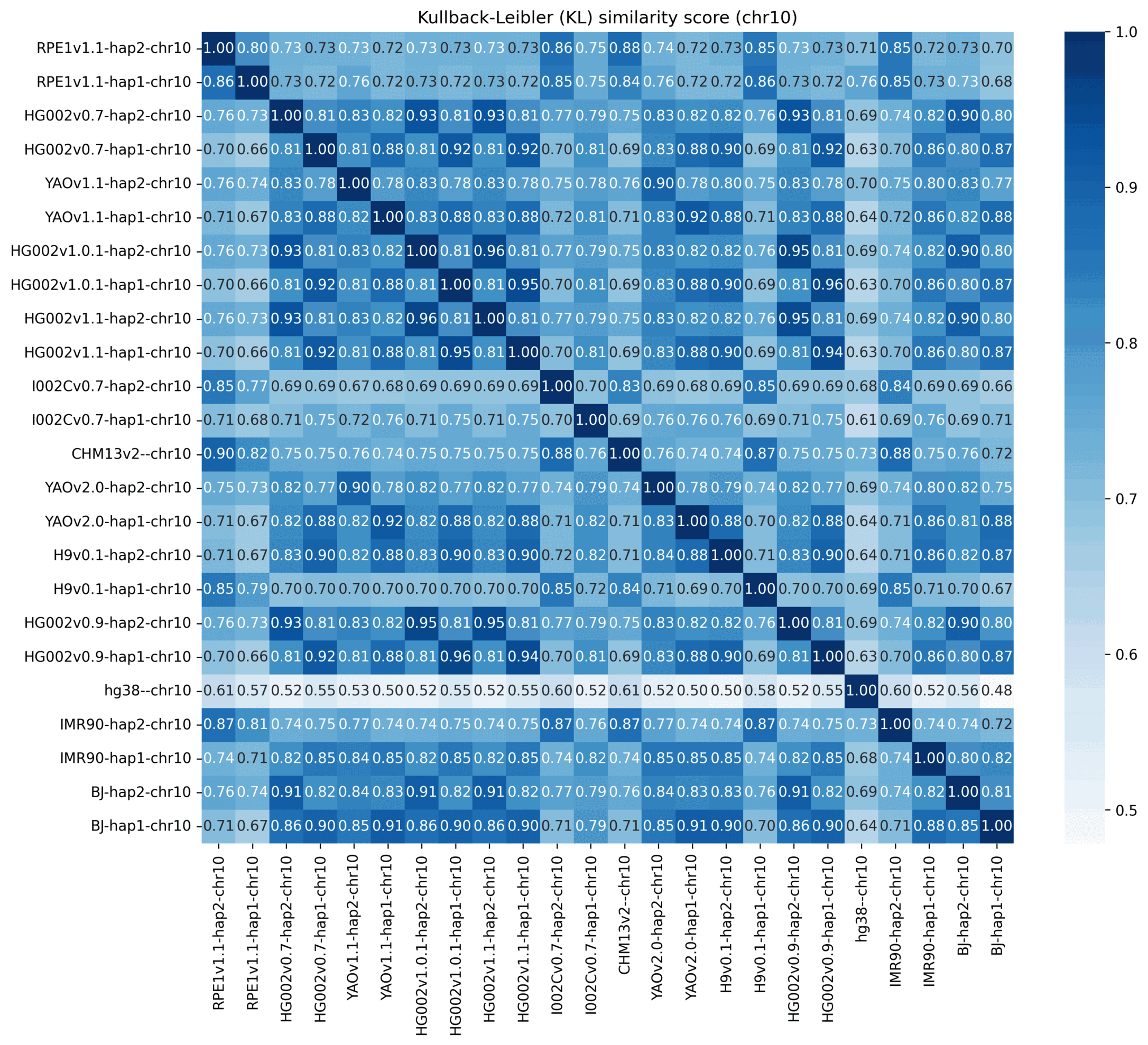}
  \caption{\textsc{kl} similarity matrix of chromosome 10}
\end{subfigure}

\vspace{1em}
\begin{subfigure}{.5\textwidth}
  \centering
  \includegraphics[width=\textwidth]{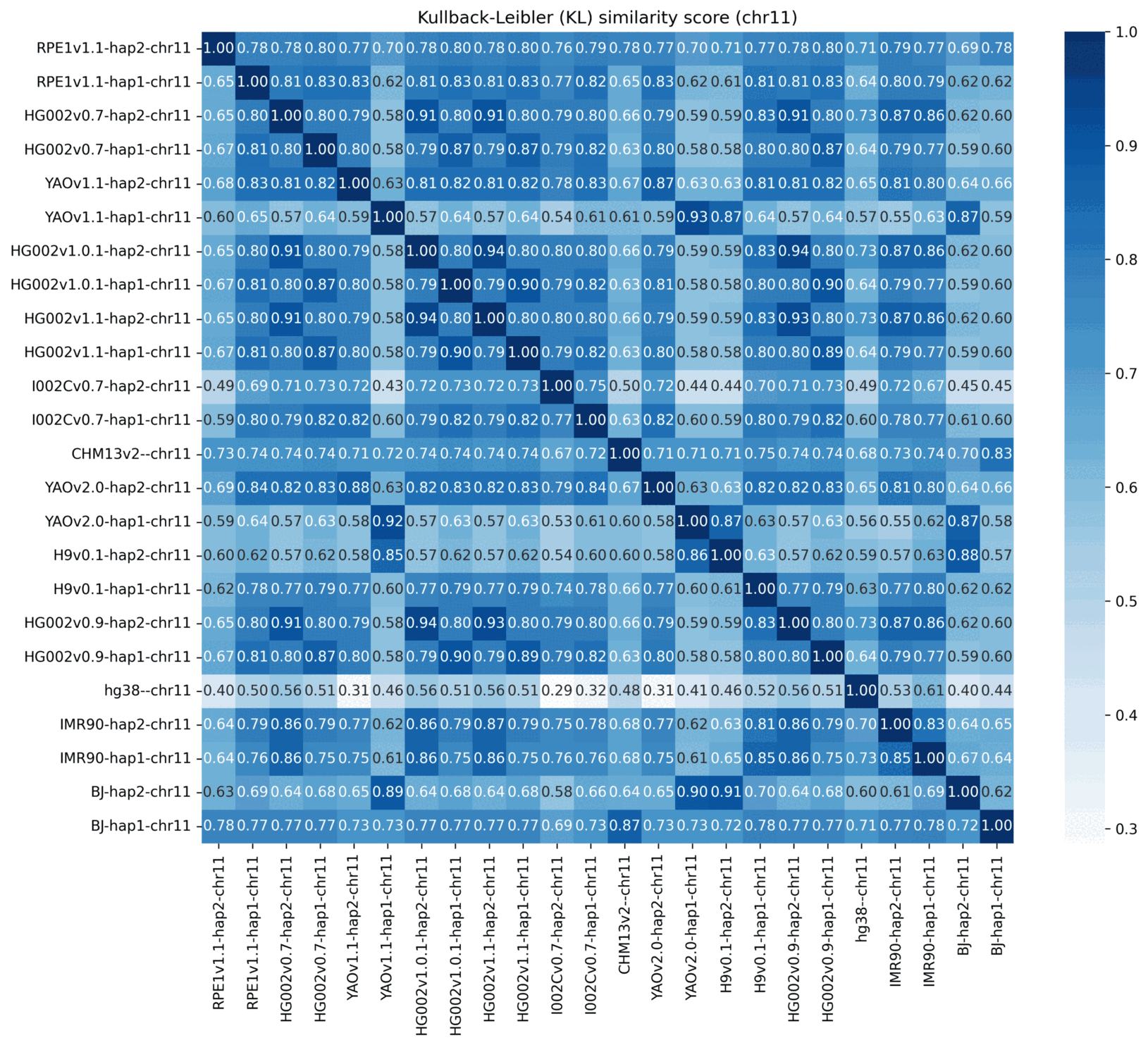}
  \caption{\textsc{kl} similarity matrix of chromosome 11}
\end{subfigure}\hfill
\begin{subfigure}{.5\textwidth}
  \centering
  \includegraphics[width=\textwidth]{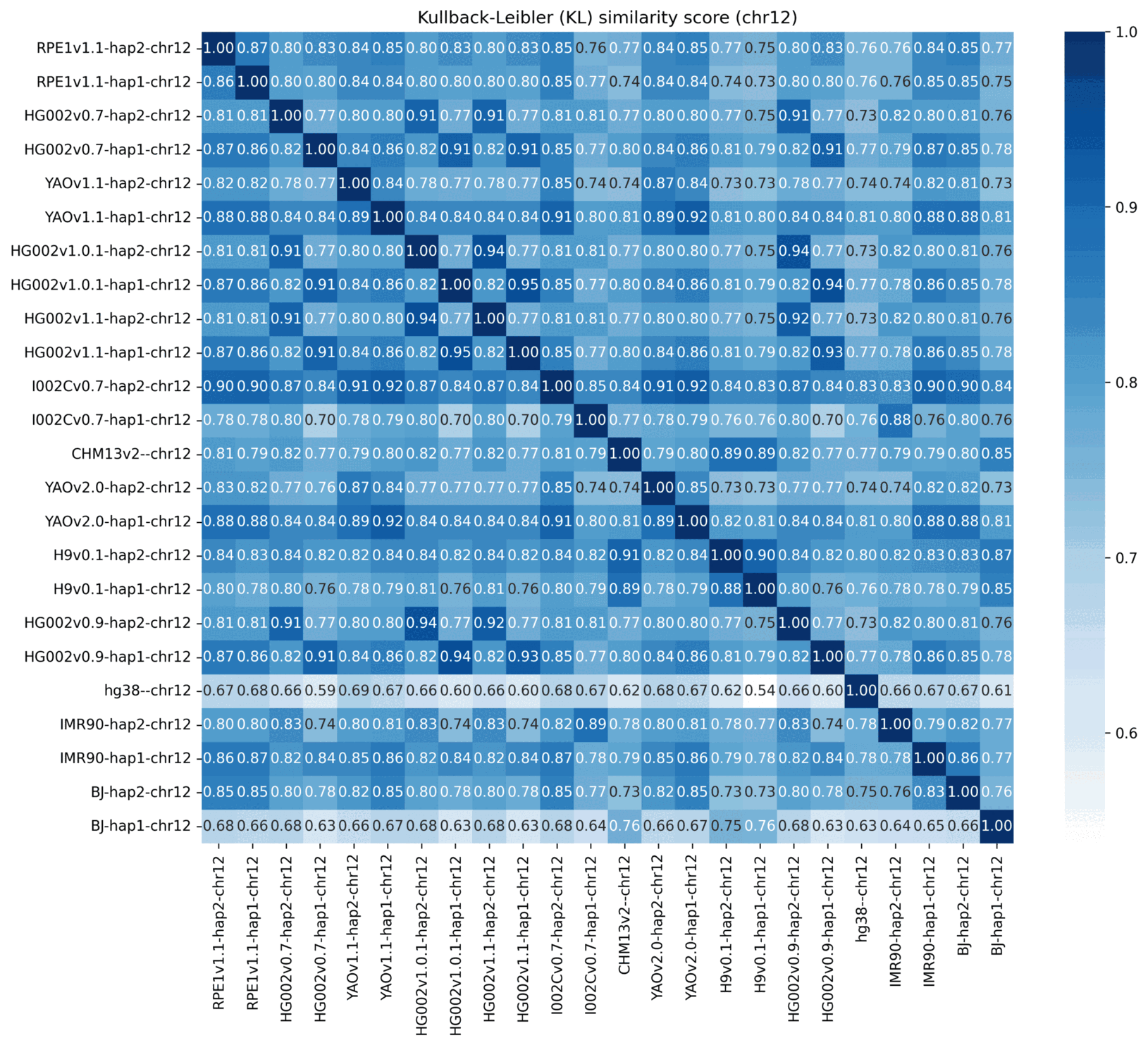}
  \caption{\textsc{kl} similarity matrix of chromosome 12}
\end{subfigure}
\caption{\textsc{kl} similarity heatmaps for chromosomes 9, 10, 11, and 12 evaluated among \textsc{rpe}\oldstylenums{1}v\oldstylenums{1.1}, \textsc{hg}\oldstylenums{002} (versions 0.7, 0.9, 1.0.1, and 1.1), \textsc{yao} (versions 1.1 and 2.0), \textsc{i}\oldstylenums{002}\textsc{c}, \textsc{chm}\oldstylenums{13}, \textsc{h}\oldstylenums{9}, \textsc{grc}h\oldstylenums{38}, \textsc{imr}\oldstylenums{90}, \textsc{bj} and their respective haplotypes.}
\end{figure}

\newpage

\begin{figure}[htbp]
\centering
\begin{subfigure}{.5\textwidth}
  \centering
  \includegraphics[width=\textwidth]{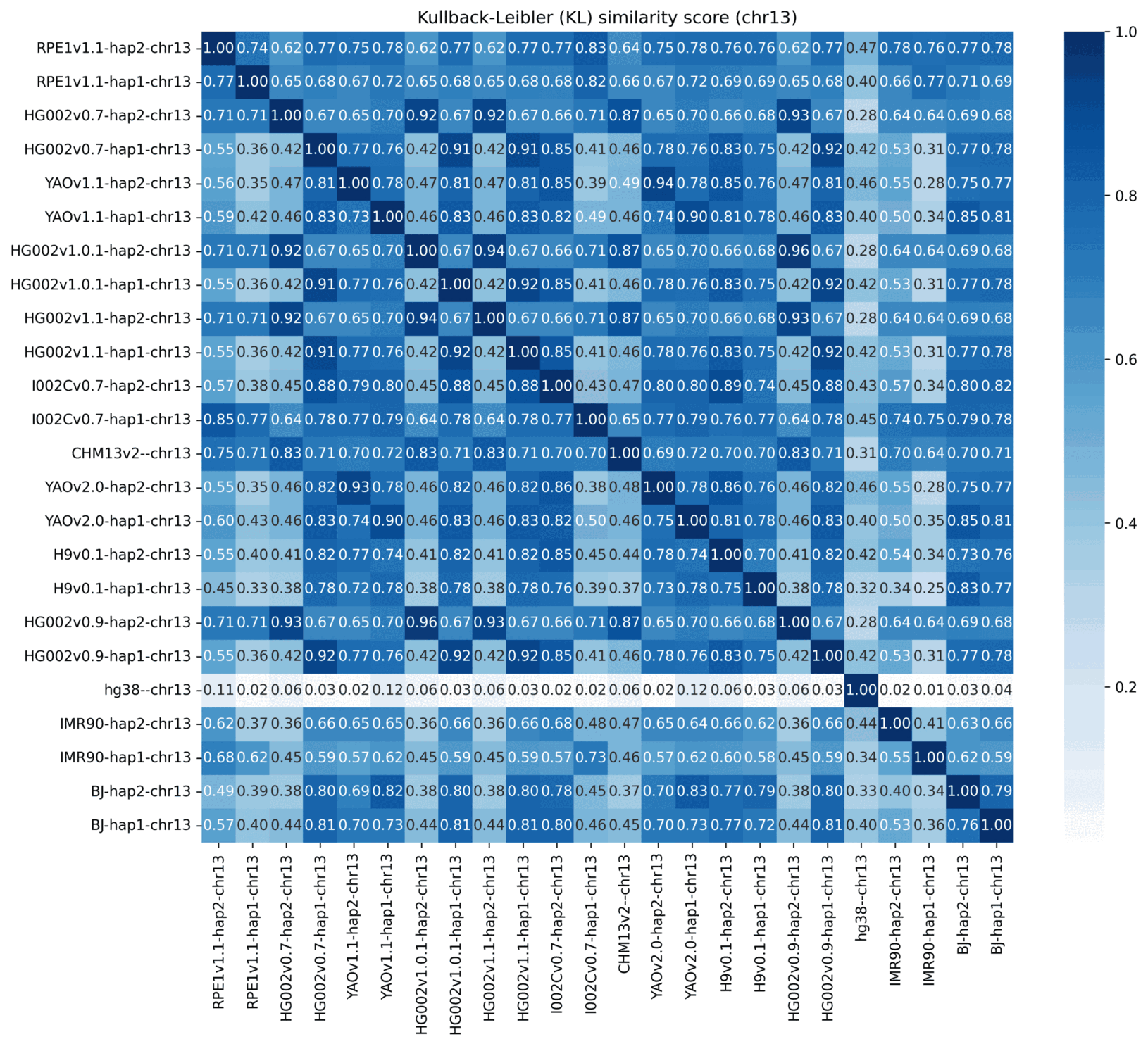}
  \caption{\textsc{kl} similarity matrix of chromosome 13}
\end{subfigure}\hfill
\begin{subfigure}{.5\textwidth}
  \centering
  \includegraphics[width=\textwidth]{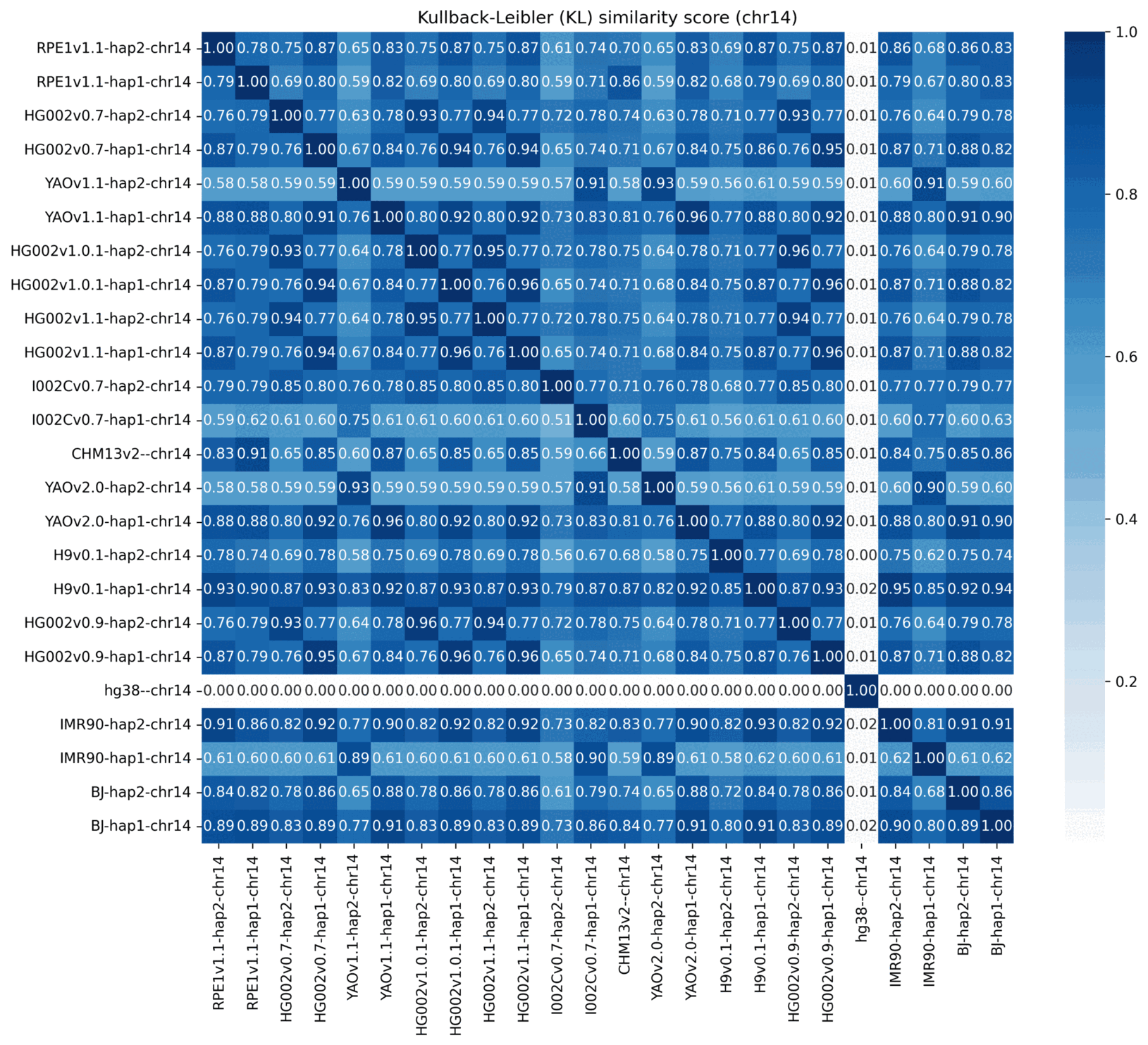}
  \caption{\textsc{kl} similarity matrix of chromosome 14}
\end{subfigure}

\vspace{1em}
\begin{subfigure}{.5\textwidth}
  \centering
  \includegraphics[width=\textwidth]{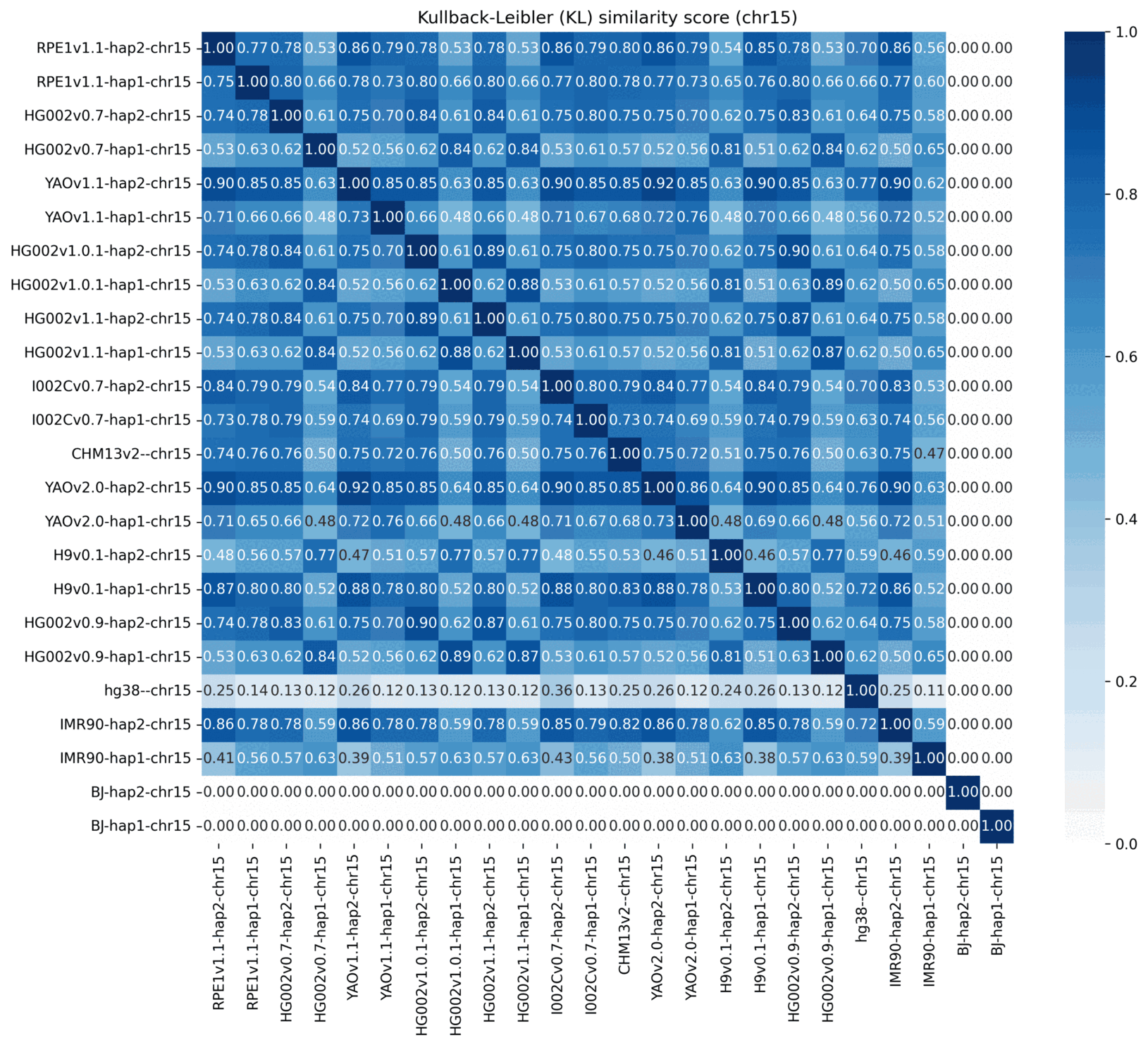}
  \caption{\textsc{kl} similarity matrix of chromosome 15}
\end{subfigure}\hfill
\begin{subfigure}{.5\textwidth}
  \centering
  \includegraphics[width=\textwidth]{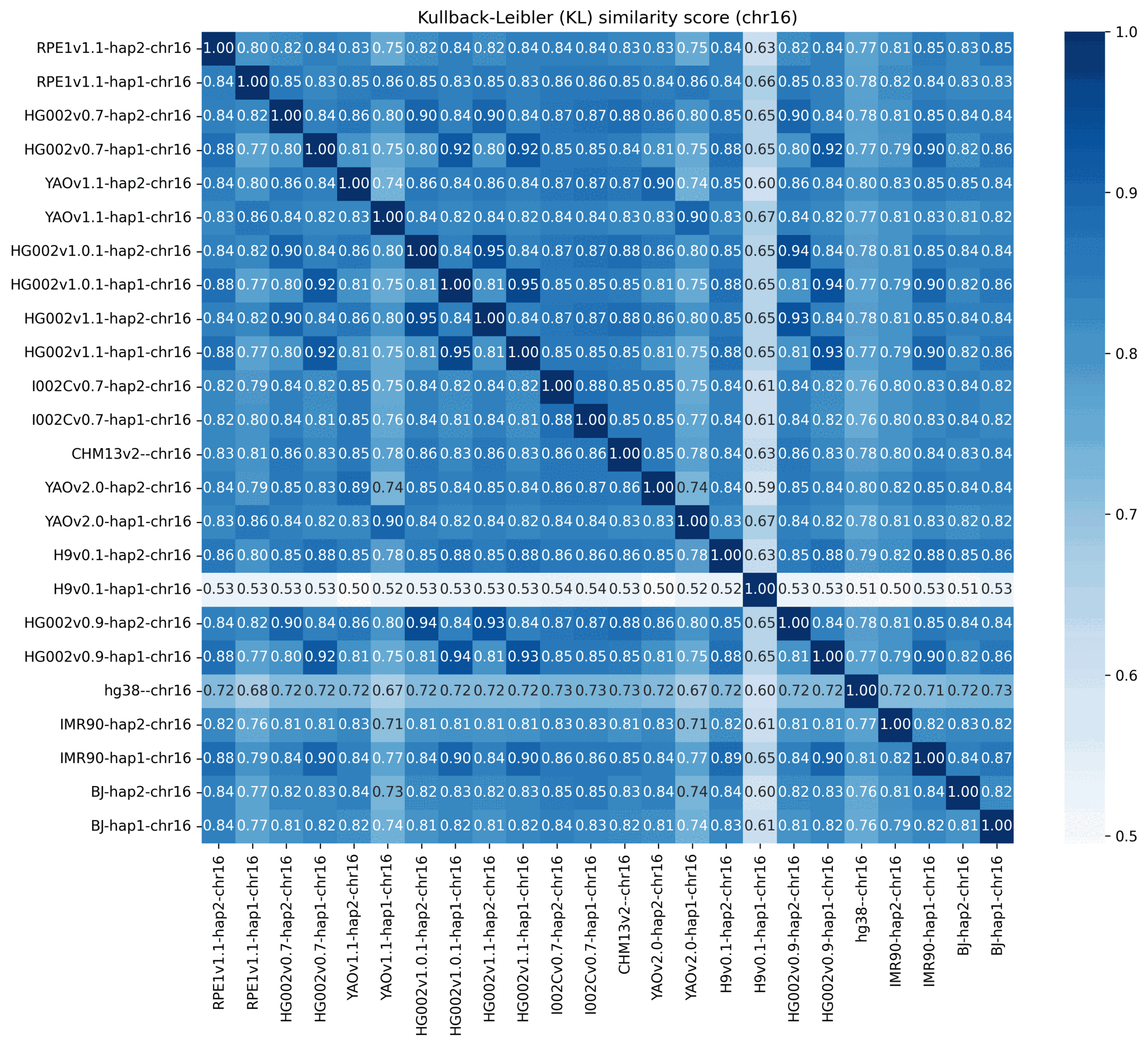}
  \caption{\textsc{kl} similarity matrix of chromosome 16}
\end{subfigure}
\caption{\textsc{kl} similarity heatmaps for chromosomes 13, 14, 15, and 16 evaluated among \textsc{rpe}\oldstylenums{1}v\oldstylenums{1.1}, \textsc{hg}\oldstylenums{002} (versions 0.7, 0.9, 1.0.1, and 1.1), \textsc{yao} (versions 1.1 and 2.0), \textsc{i}\oldstylenums{002}\textsc{c}, \textsc{chm}\oldstylenums{13}, \textsc{h}\oldstylenums{9}, \textsc{grc}h\oldstylenums{38}, \textsc{imr}\oldstylenums{90}, \textsc{bj} and their respective haplotypes.}
\end{figure}

\newpage

\begin{figure}[htbp]
\centering
\begin{subfigure}{.5\textwidth}
  \centering
  \includegraphics[width=\textwidth]{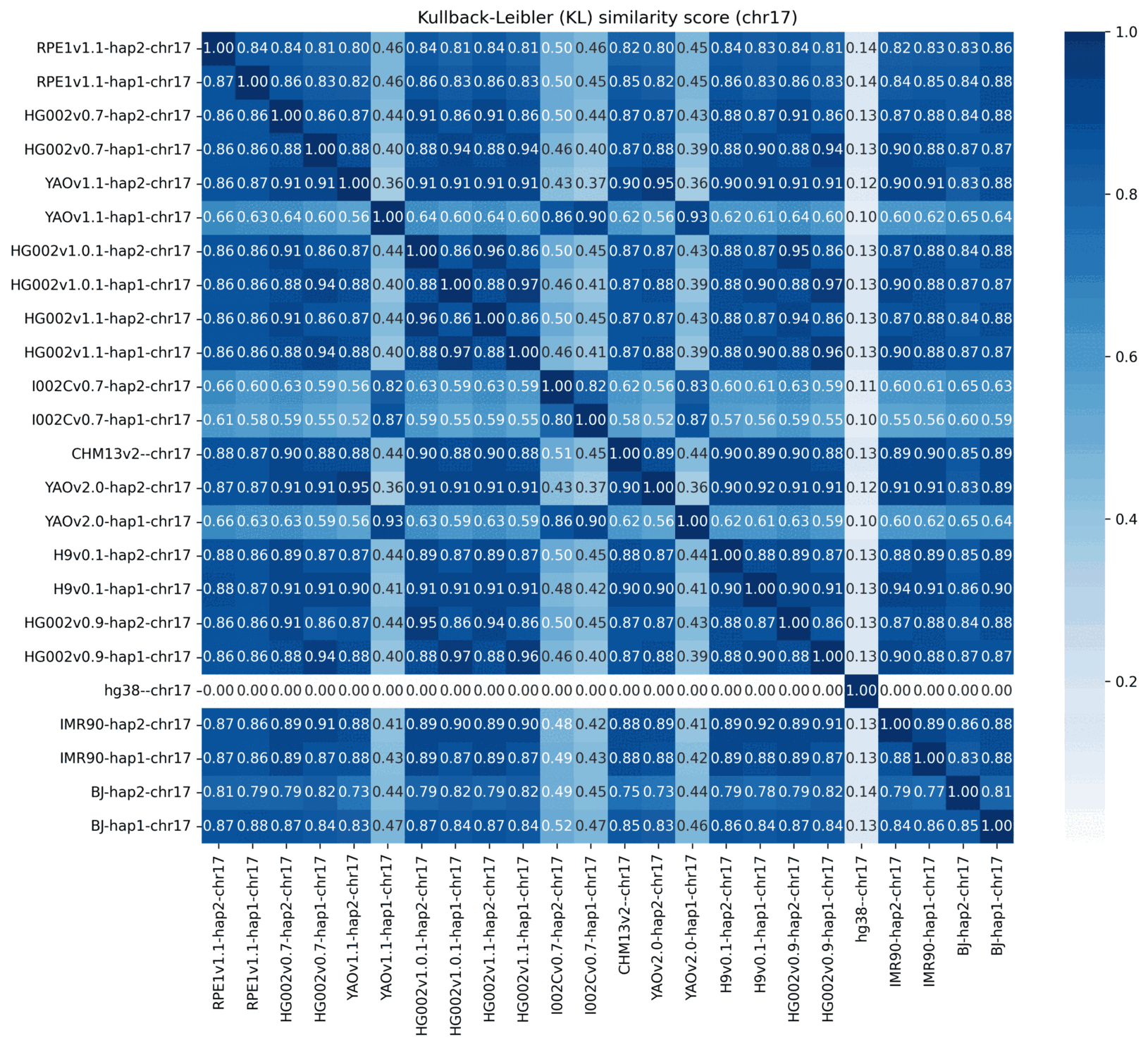}
  \caption{\textsc{kl} similarity matrix of chromosome 17}
\end{subfigure}\hfill
\begin{subfigure}{.5\textwidth}
  \centering
  \includegraphics[width=\textwidth]{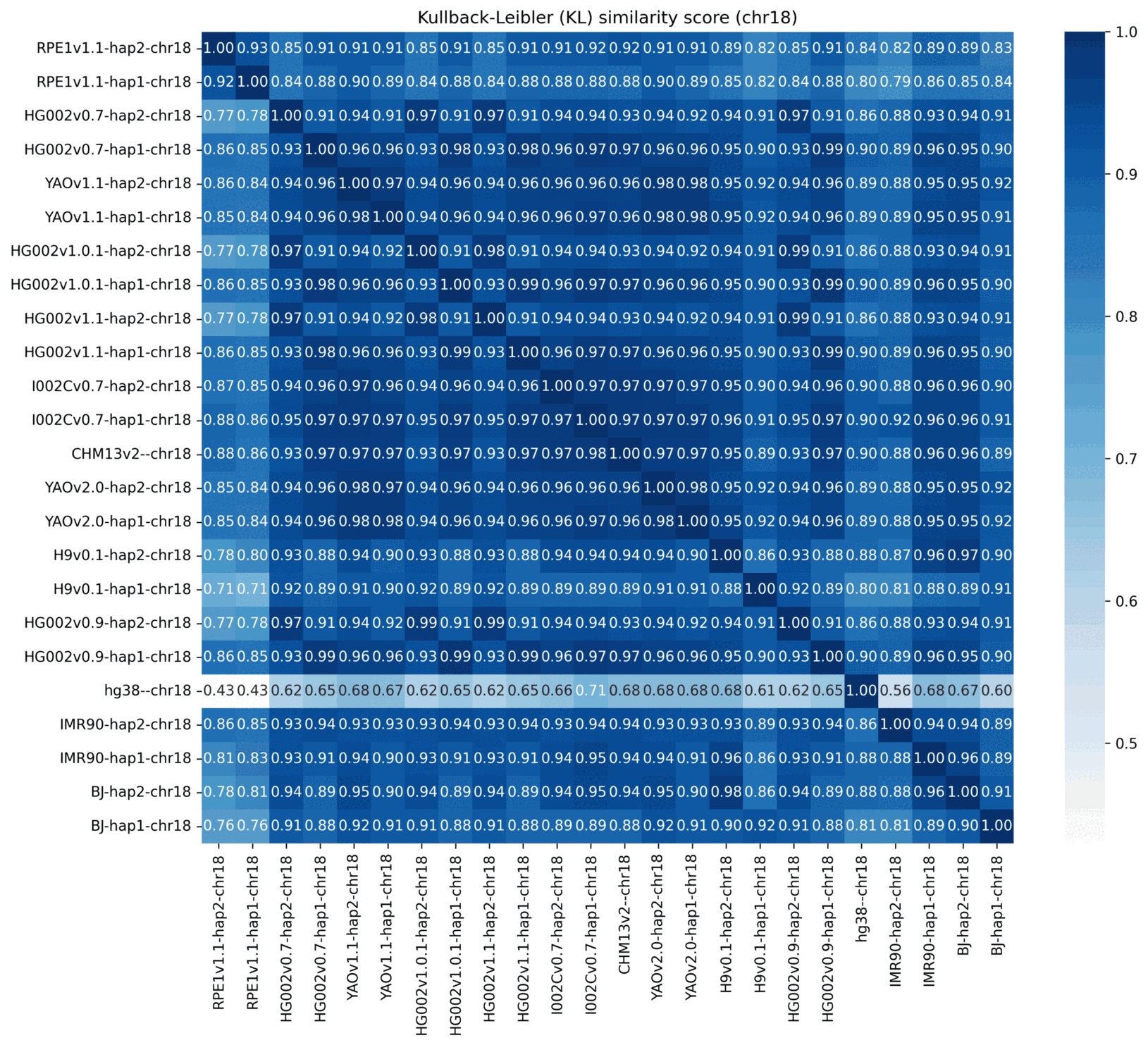}
  \caption{\textsc{kl} similarity matrix of chromosome 18}
\end{subfigure}

\vspace{1em}
\begin{subfigure}{.5\textwidth}
  \centering
  \includegraphics[width=\textwidth]{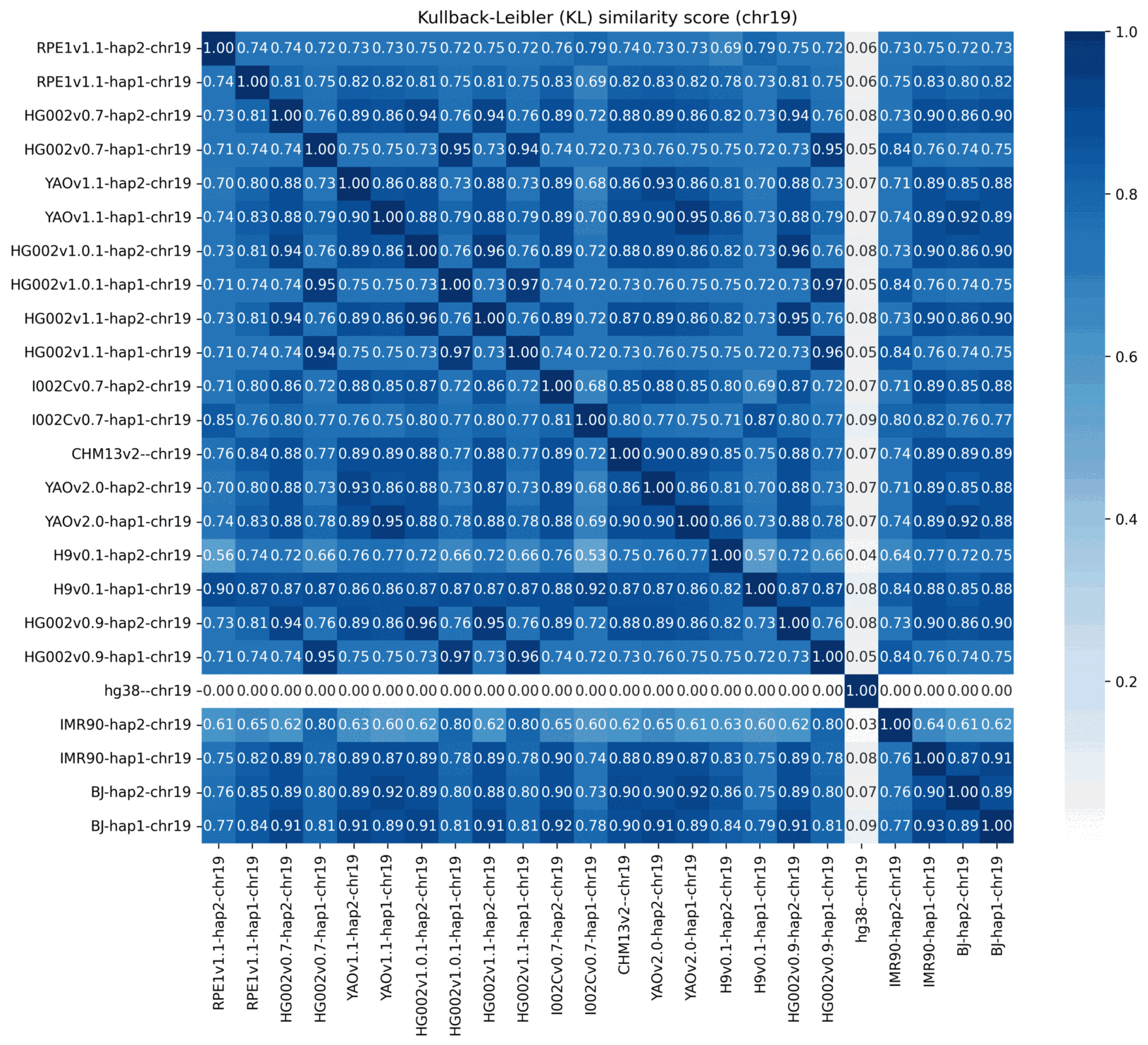}
  \caption{\textsc{kl} similarity matrix of chromosome 19}
\end{subfigure}\hfill
\begin{subfigure}{.5\textwidth}
  \centering
  \includegraphics[width=\textwidth]{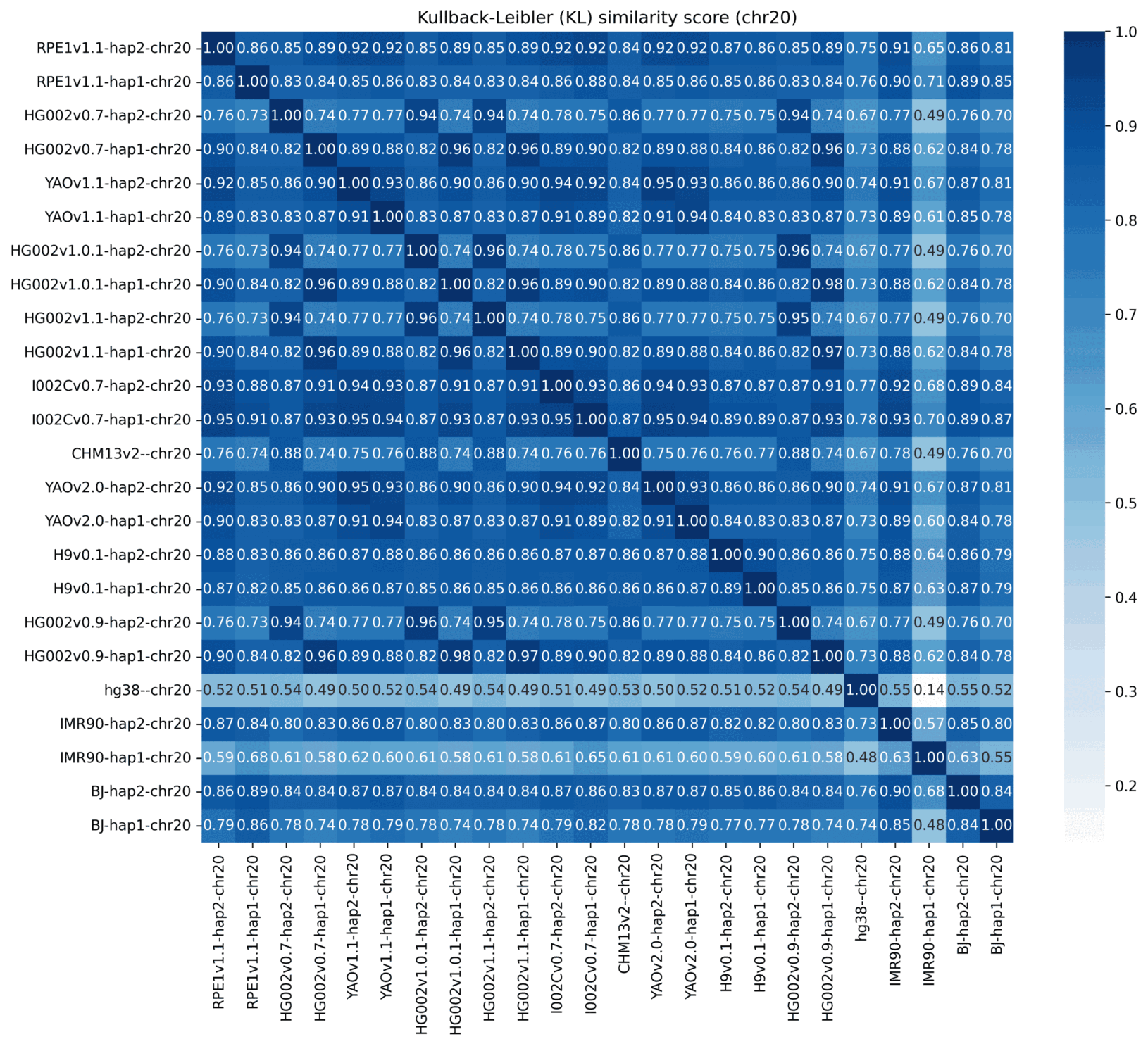}
  \caption{\textsc{kl} similarity matrix of chromosome 20}
\end{subfigure}
\caption{\textsc{kl} similarity heatmaps for chromosomes 17, 18, 19, and 20 evaluated among \textsc{rpe}\oldstylenums{1}v\oldstylenums{1.1}, \textsc{hg}\oldstylenums{002} (versions 0.7, 0.9, 1.0.1, and 1.1), \textsc{yao} (versions 1.1 and 2.0), \textsc{i}\oldstylenums{002}\textsc{c}, \textsc{chm}\oldstylenums{13}, \textsc{h}\oldstylenums{9}, \textsc{grc}h\oldstylenums{38}, \textsc{imr}\oldstylenums{90}, \textsc{bj} and their respective haplotypes.}
\end{figure}

\newpage

\begin{figure}[htbp]
\centering
\begin{subfigure}{.5\textwidth}
  \centering
  \includegraphics[width=\textwidth]{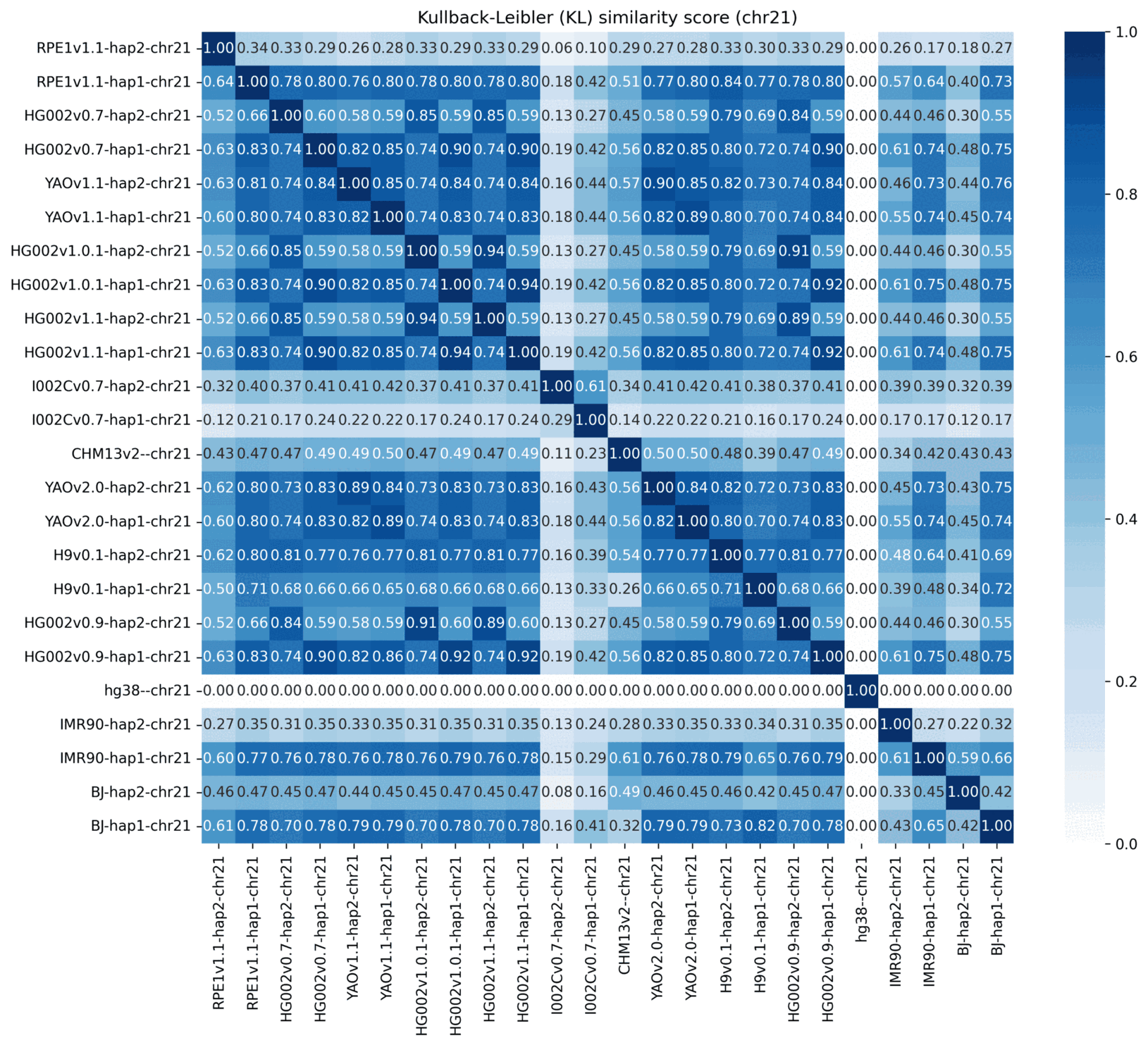}
  \caption{\textsc{kl} similarity matrix of chromosome 21}
\end{subfigure}\hfill
\begin{subfigure}{.5\textwidth}
  \centering
  \includegraphics[width=\textwidth]{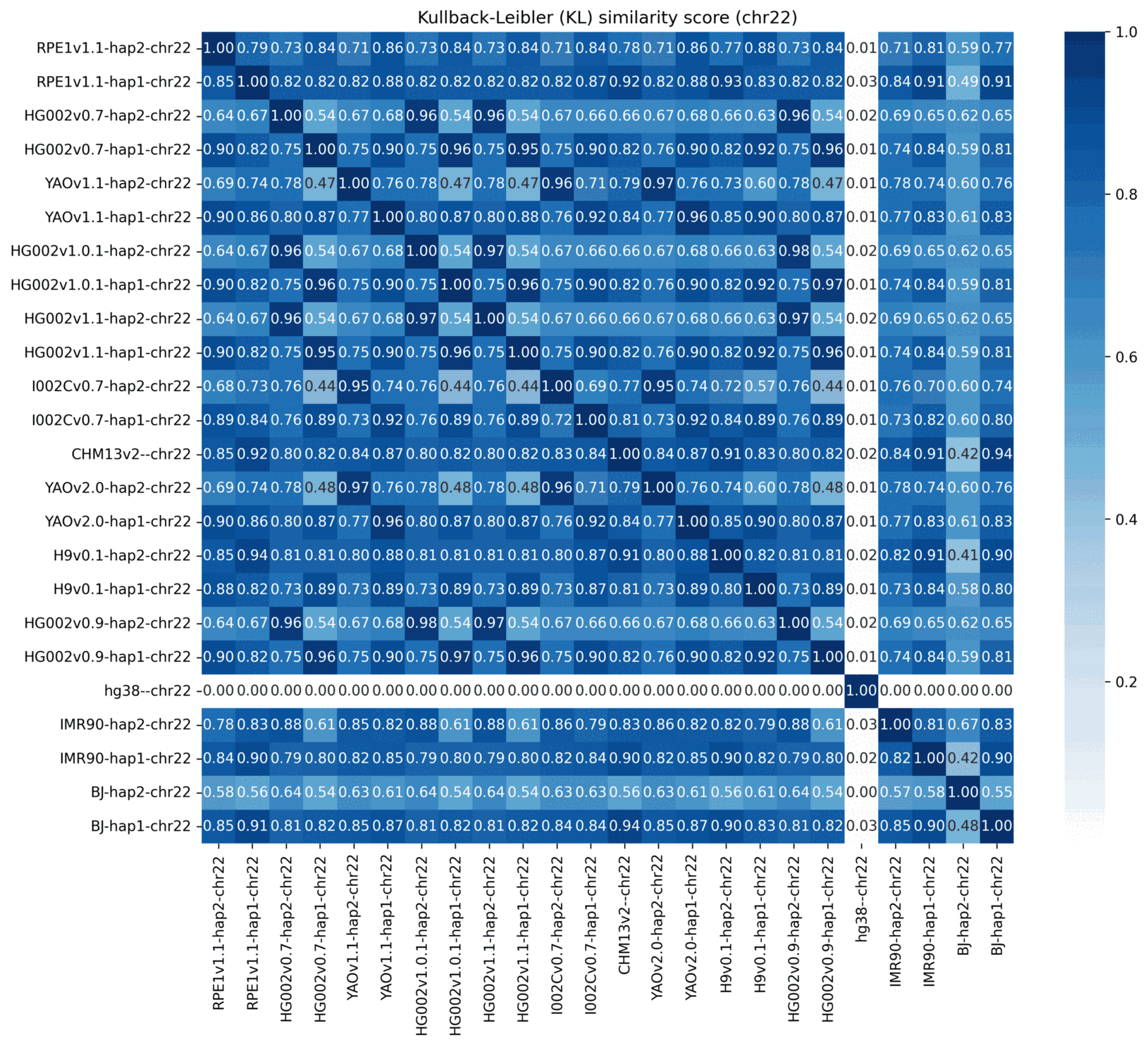}
  \caption{\textsc{kl} similarity matrix of chromosome 22}
\end{subfigure}
\caption{\textsc{kl} similarity heatmaps for chromosomes 21 and 22 evaluated among \textsc{rpe}\oldstylenums{1}v\oldstylenums{1.1}, \textsc{hg}\oldstylenums{002} (versions 0.7, 0.9, 1.0.1, and 1.1), \textsc{yao} (versions 1.1 and 2.0), \textsc{i}\oldstylenums{002}\textsc{c}, \textsc{chm}\oldstylenums{13}, \textsc{h}\oldstylenums{9}, \textsc{grc}h\oldstylenums{38}, \textsc{imr}\oldstylenums{90}, \textsc{bj} and their respective haplotypes.}
\end{figure}

\newpage

\begin{figure}[htbp]
\centering
\begin{subfigure}{.5\textwidth}
  \centering
  \includegraphics[width=\textwidth]{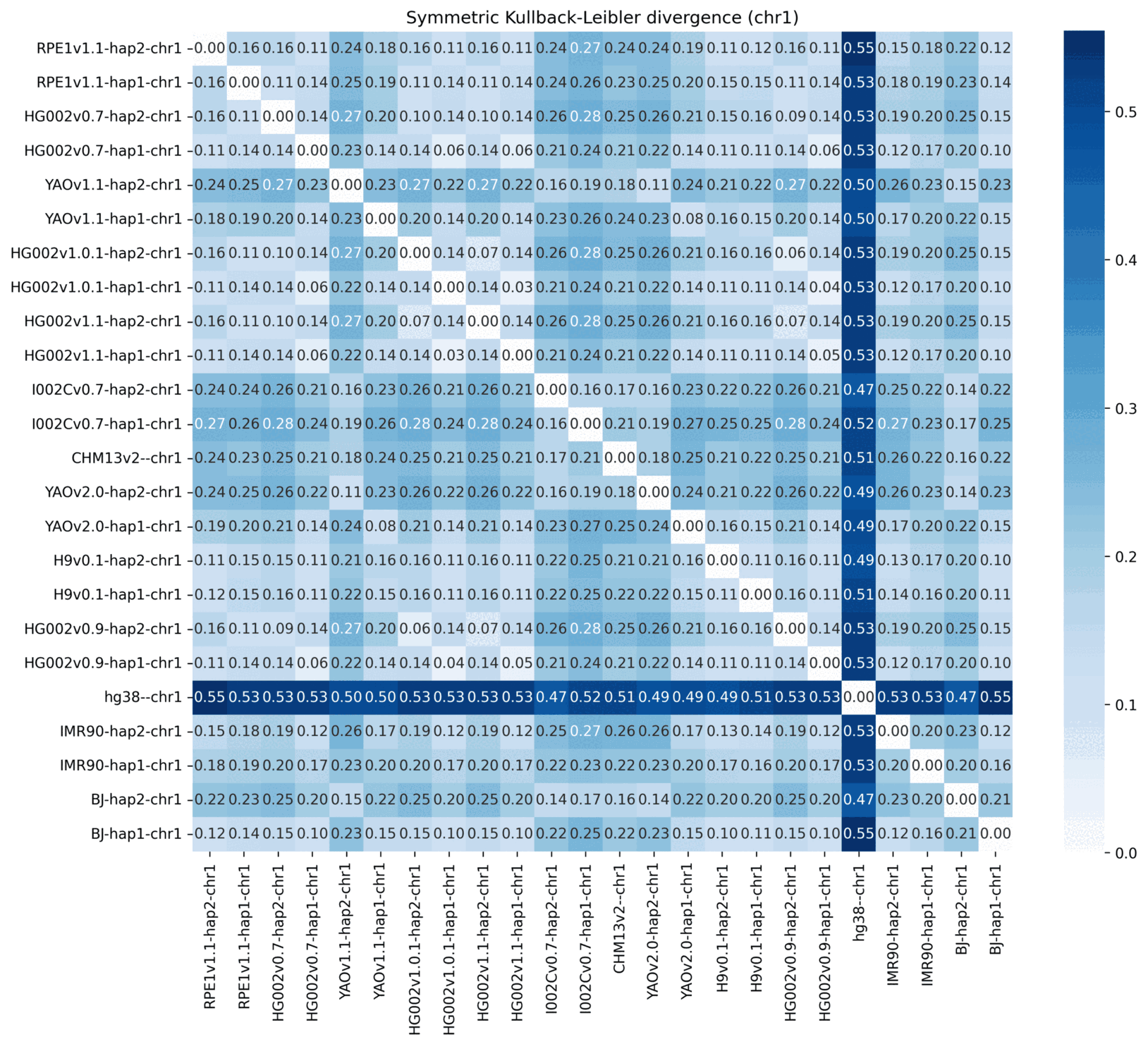}
  \caption{Symmetric \textsc{kl} heatmap chromosome 1}
\end{subfigure}\hfill
\begin{subfigure}{.5\textwidth}
  \centering
  \includegraphics[width=\textwidth]{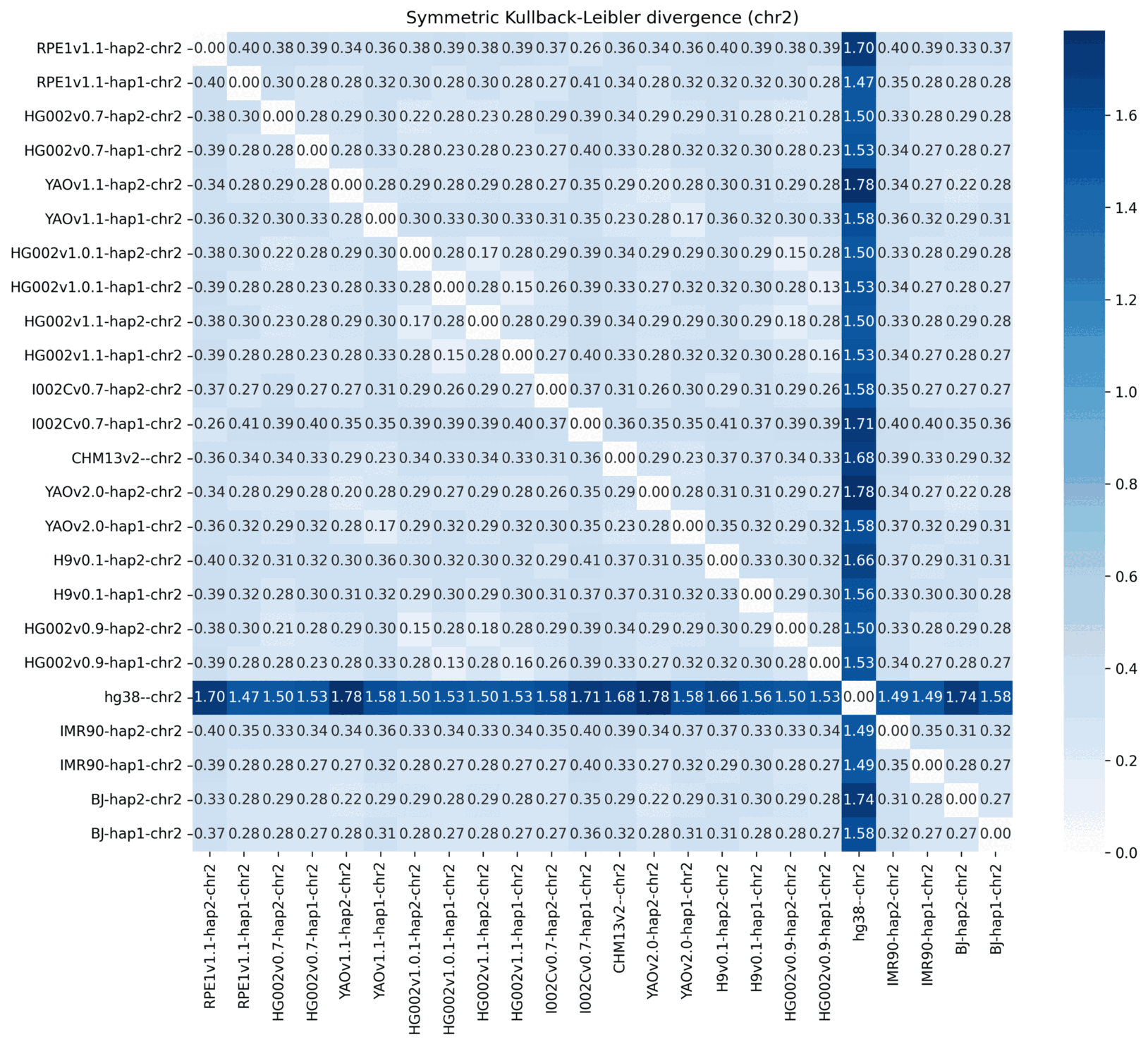}
  \caption{Symmetric \textsc{kl} heatmap chromosome 2}
\end{subfigure}

\vspace{1em}
\begin{subfigure}{.5\textwidth}
  \centering
  \includegraphics[width=\textwidth]{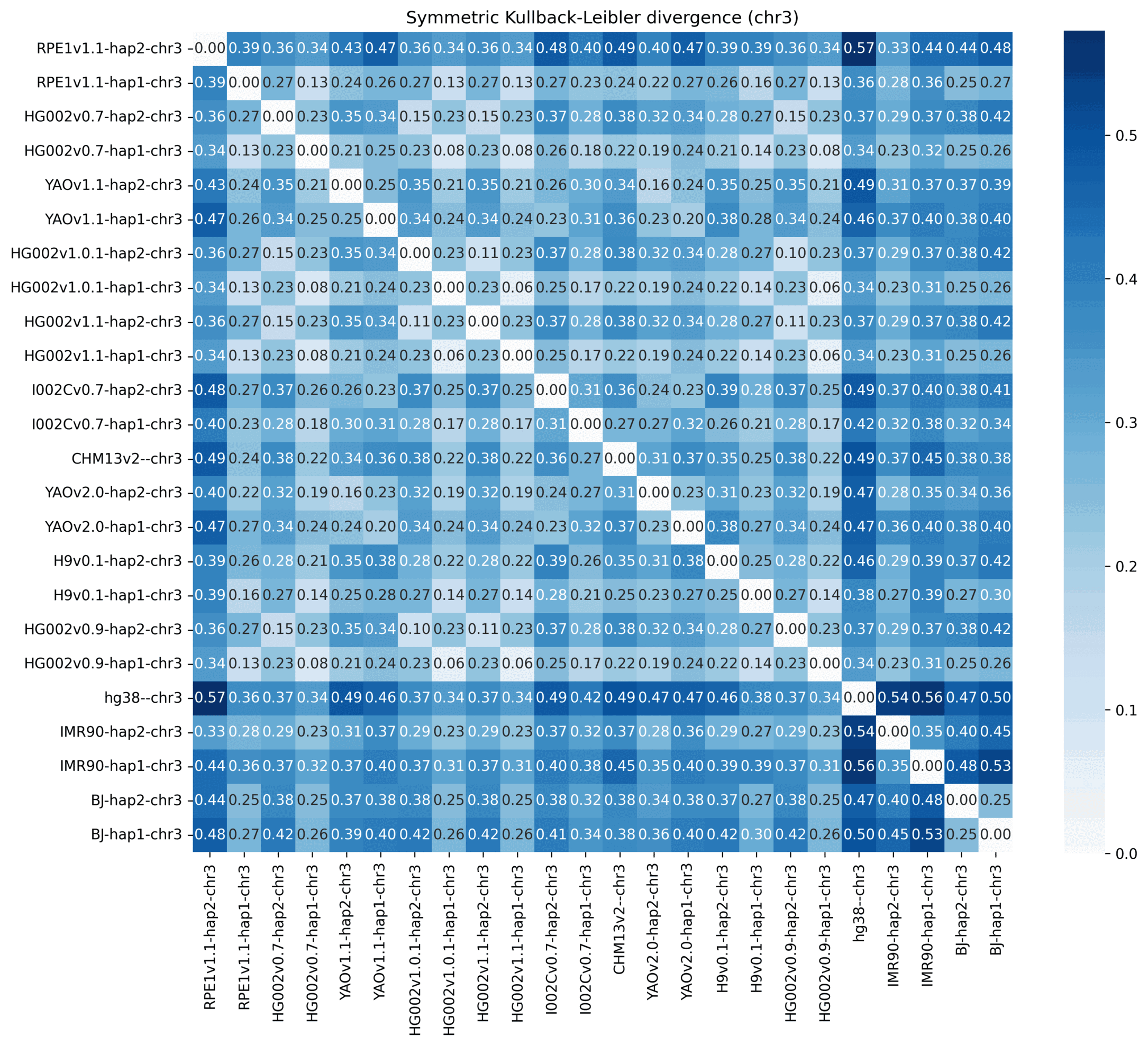}
  \caption{Symmetric \textsc{kl} heatmap chromosome 3}
\end{subfigure}\hfill
\begin{subfigure}{.5\textwidth}
  \centering
  \includegraphics[width=\textwidth]{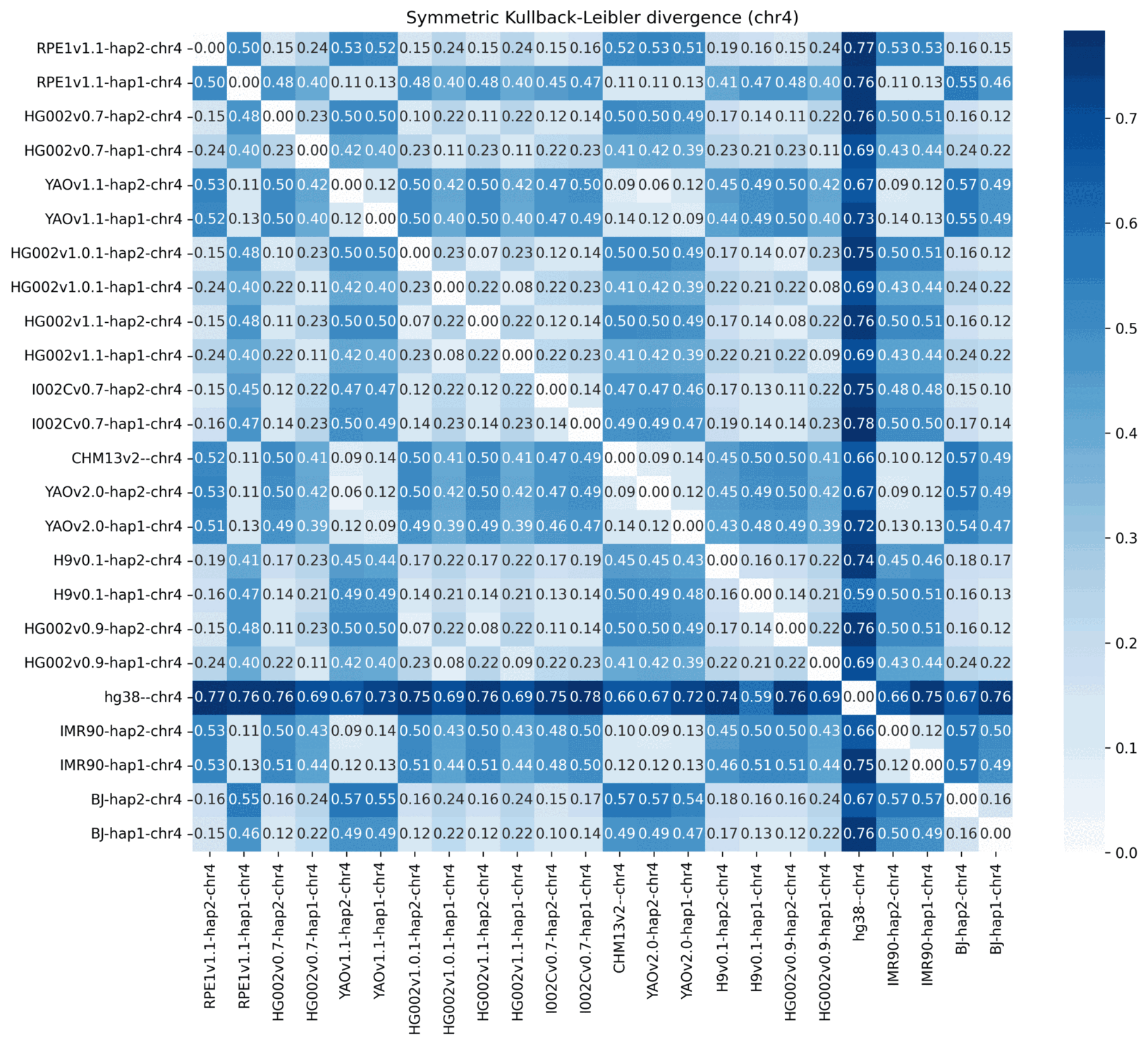}
  \caption{Symmetric \textsc{kl} heatmap chromosome 4}
\end{subfigure}
\caption{Symmetric \textsc{kl} heatmaps for chromosomes 1, 2, 3, and 4.}
\end{figure}

\newpage

\begin{figure}[htbp]
\centering
\begin{subfigure}{.5\textwidth}
  \centering
  \includegraphics[width=\textwidth]{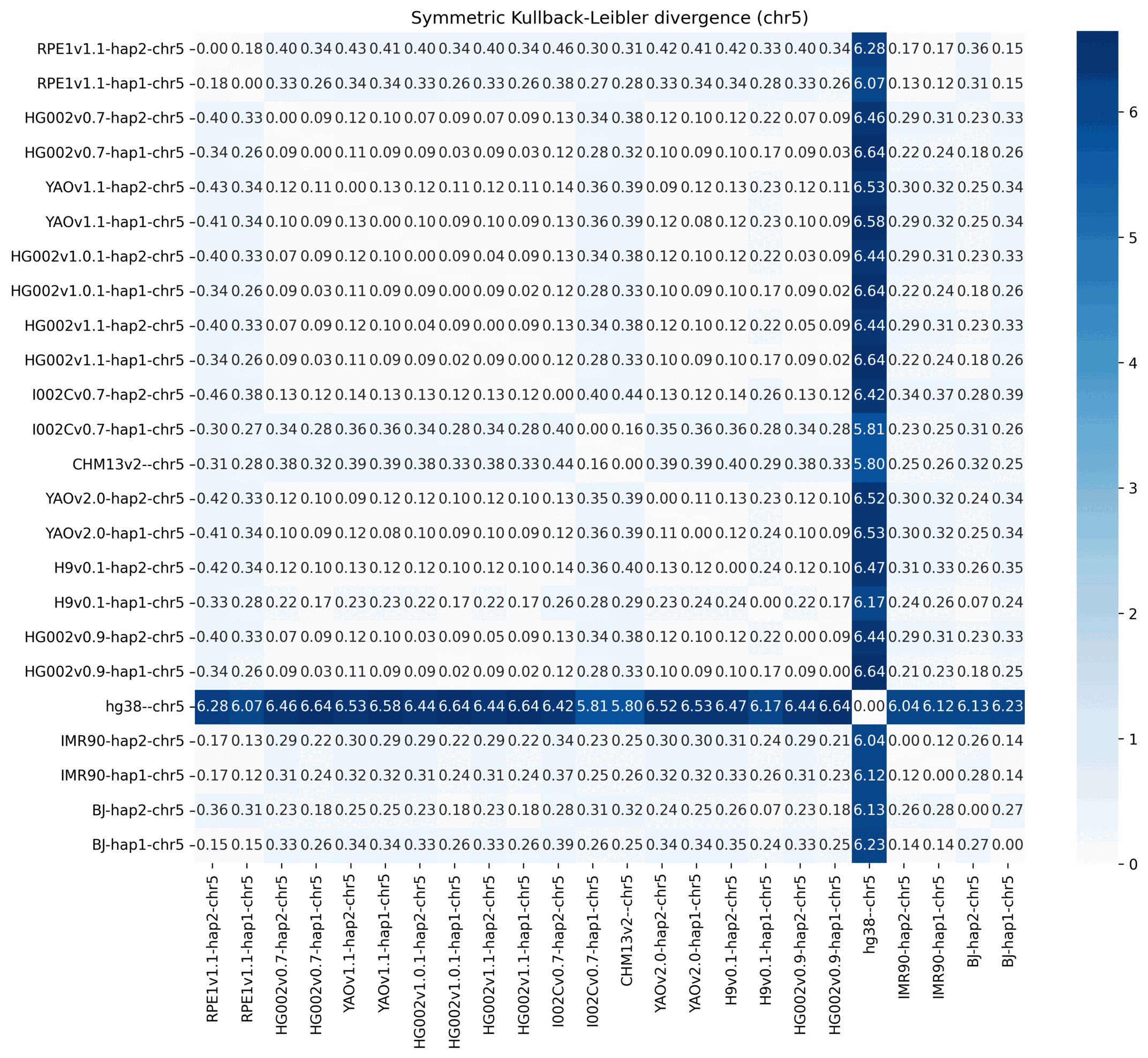}
  \caption{Symmetric \textsc{kl} heatmap chromosome 5}
\end{subfigure}\hfill
\begin{subfigure}{.5\textwidth}
  \centering
  \includegraphics[width=\textwidth]{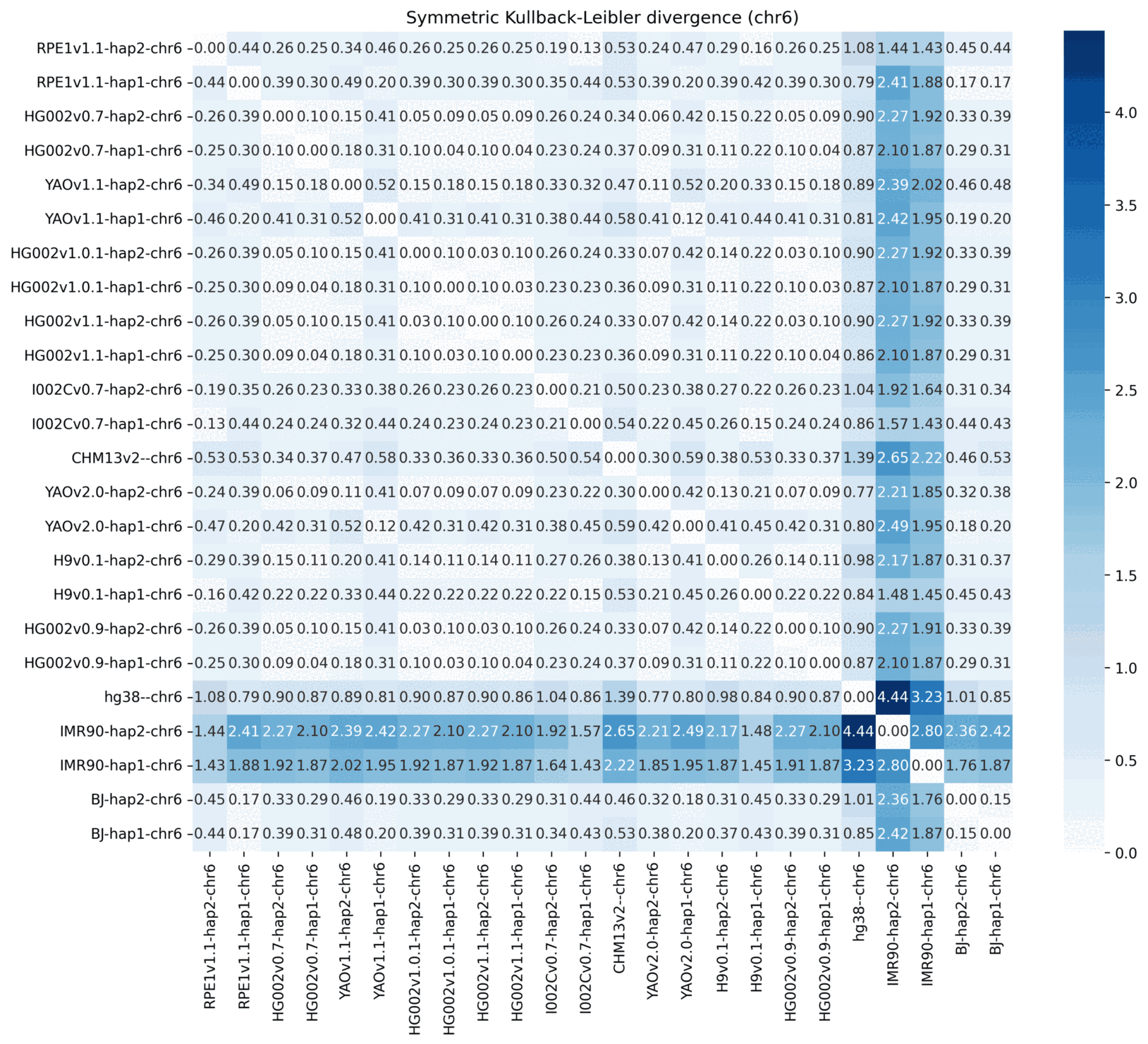}
  \caption{Symmetric \textsc{kl} heatmap chromosome 6}
\end{subfigure}

\vspace{1em}
\begin{subfigure}{.5\textwidth}
  \centering
  \includegraphics[width=\textwidth]{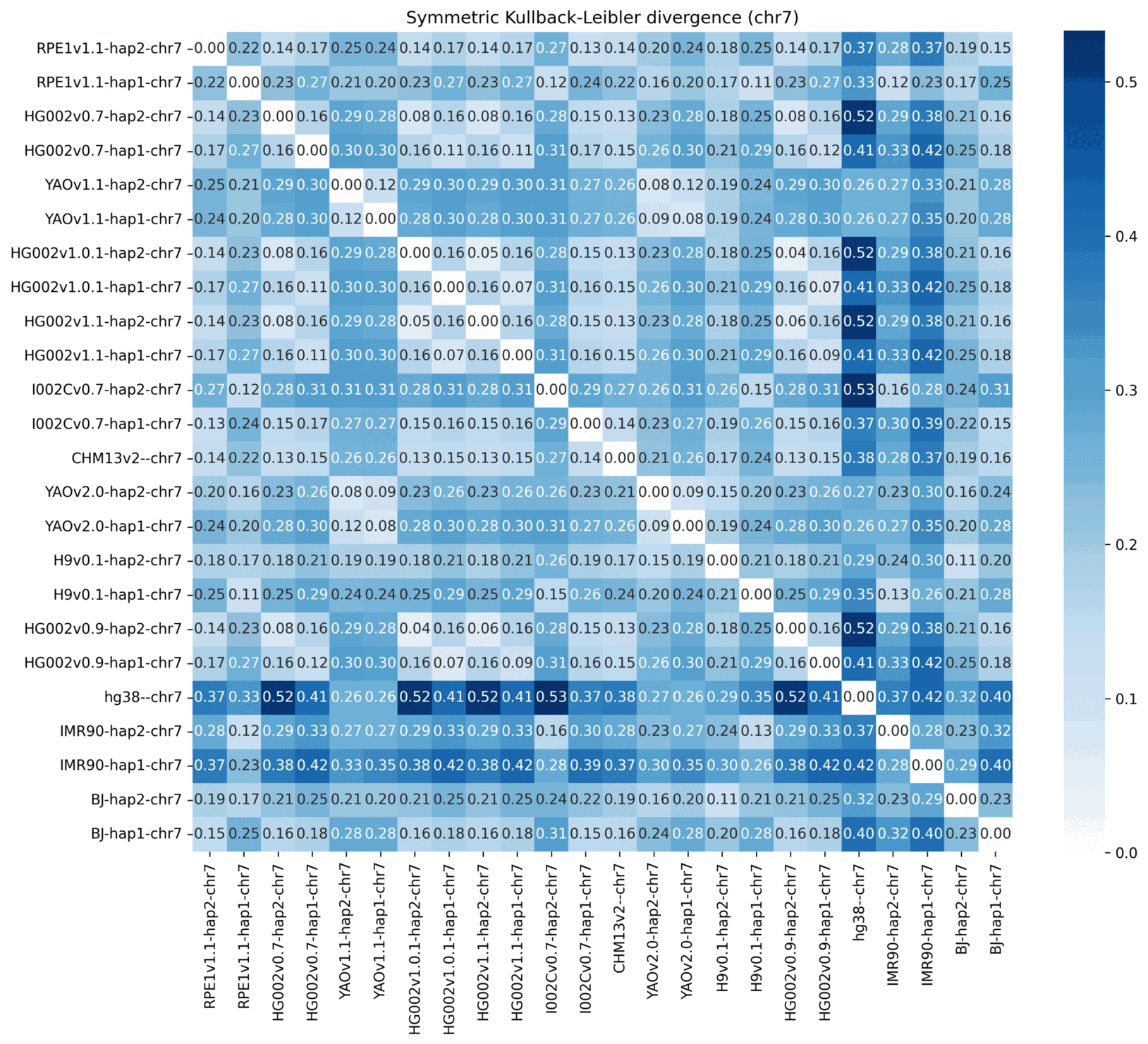}
  \caption{Symmetric \textsc{kl} heatmap chromosome 7}
\end{subfigure}\hfill
\begin{subfigure}{.5\textwidth}
  \centering
  \includegraphics[width=\textwidth]{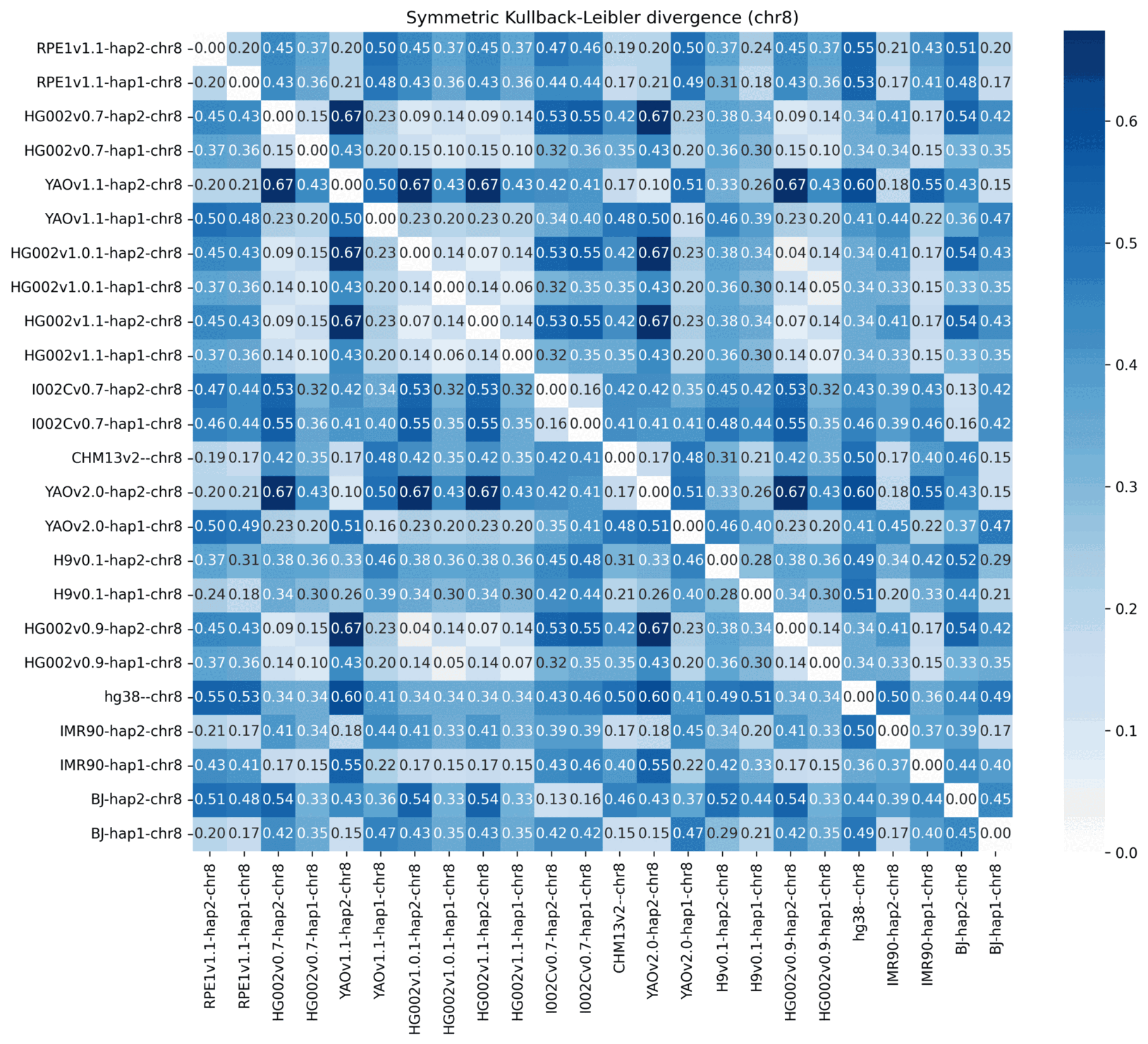}
  \caption{Symmetric \textsc{kl} heatmap chromosome 8}
\end{subfigure}
\caption{Symmetric \textsc{kl} heatmaps for chromosomes 5, 6, 7, and 8.}
\end{figure}

\newpage

\begin{figure}[htbp]
\centering
\begin{subfigure}{.5\textwidth}
  \centering
  \includegraphics[width=\textwidth]{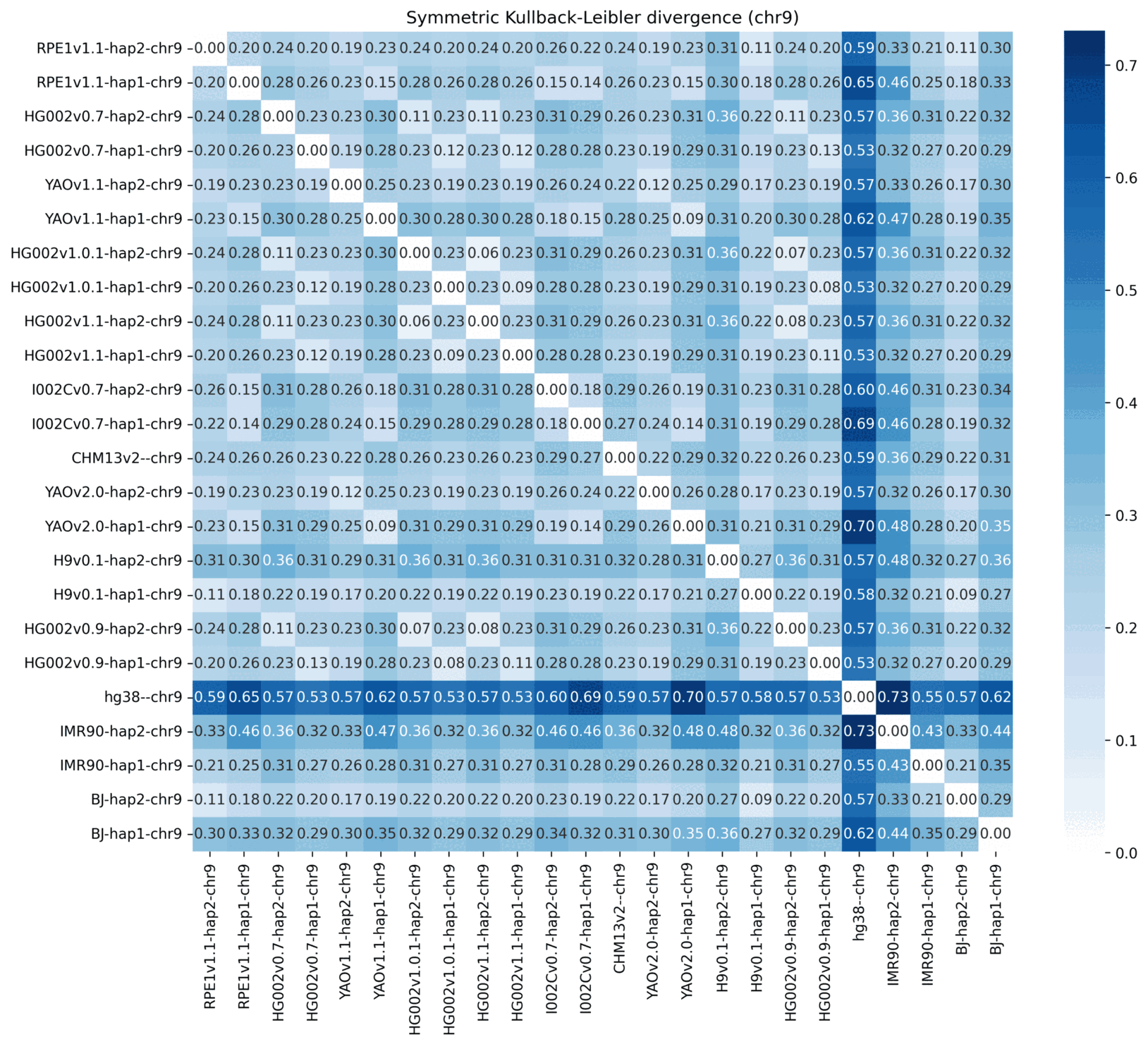}
  \caption{Symmetric \textsc{kl} heatmap chromosome 9}
\end{subfigure}\hfill
\begin{subfigure}{.5\textwidth}
  \centering
  \includegraphics[width=\textwidth]{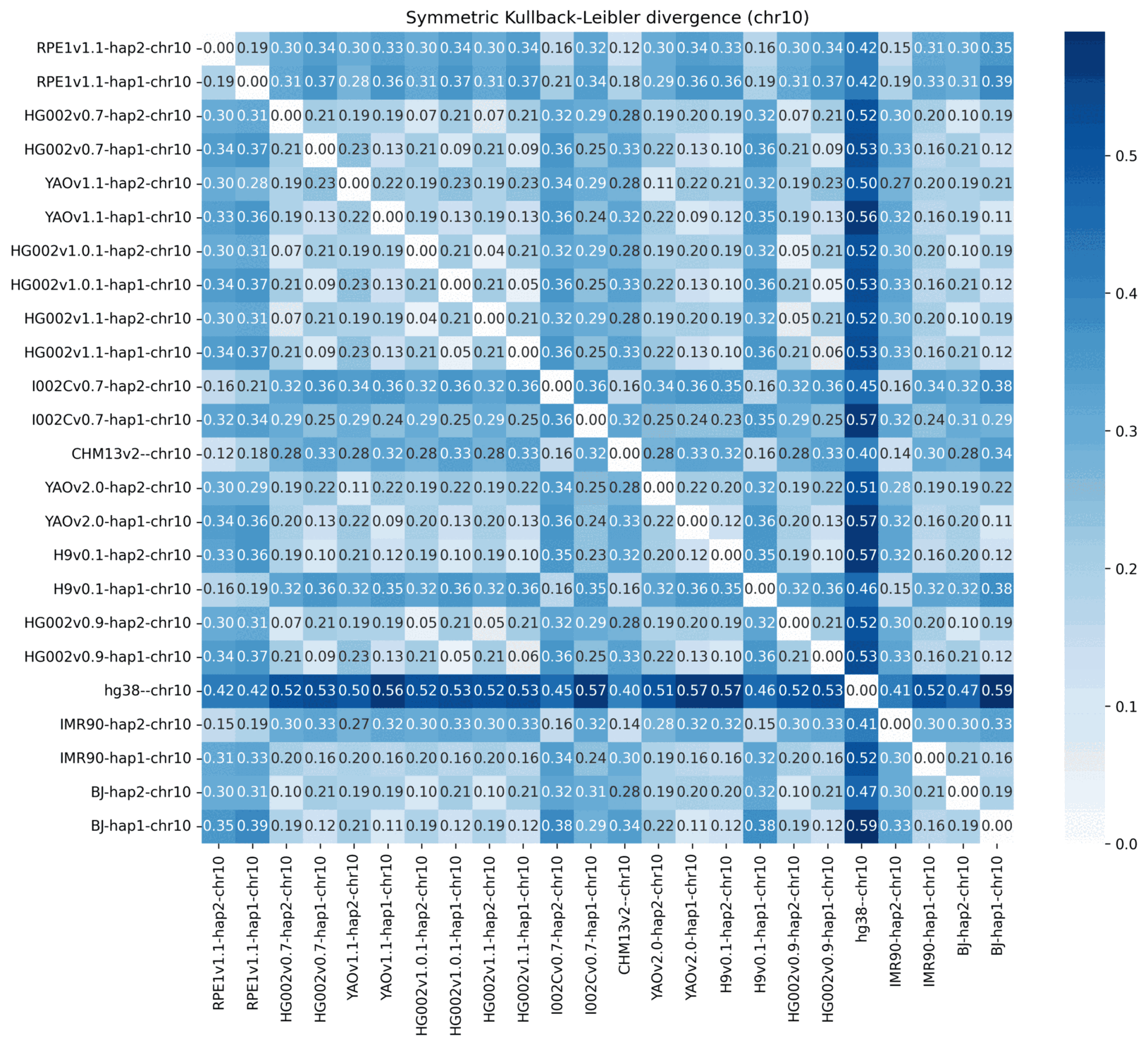}
  \caption{Symmetric \textsc{kl} heatmap chromosome 10}
\end{subfigure}

\vspace{1em}
\begin{subfigure}{.5\textwidth}
  \centering
  \includegraphics[width=\textwidth]{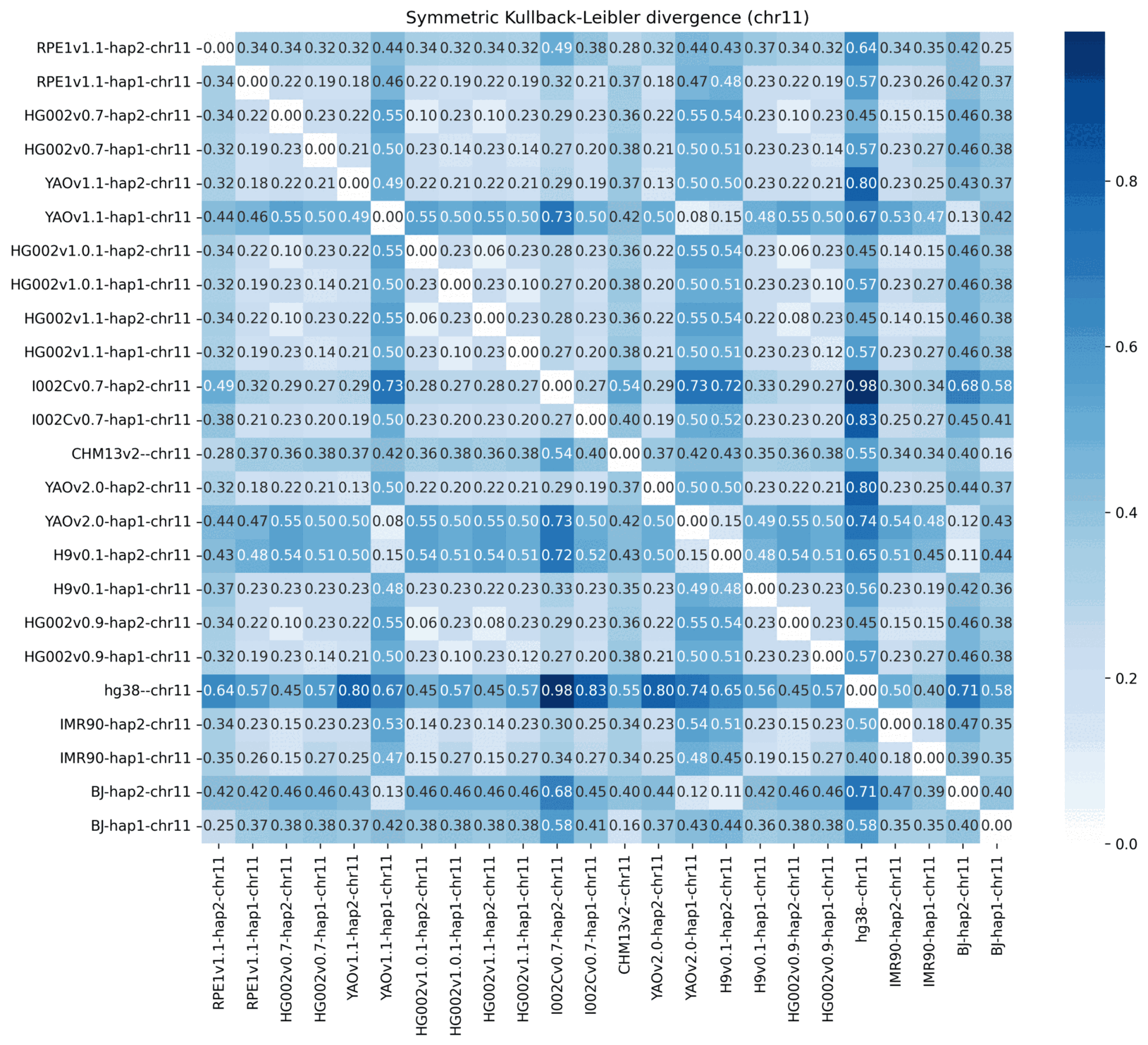}
  \caption{Symmetric \textsc{kl} heatmap chromosome 11}
\end{subfigure}\hfill
\begin{subfigure}{.5\textwidth}
  \centering
  \includegraphics[width=\textwidth]{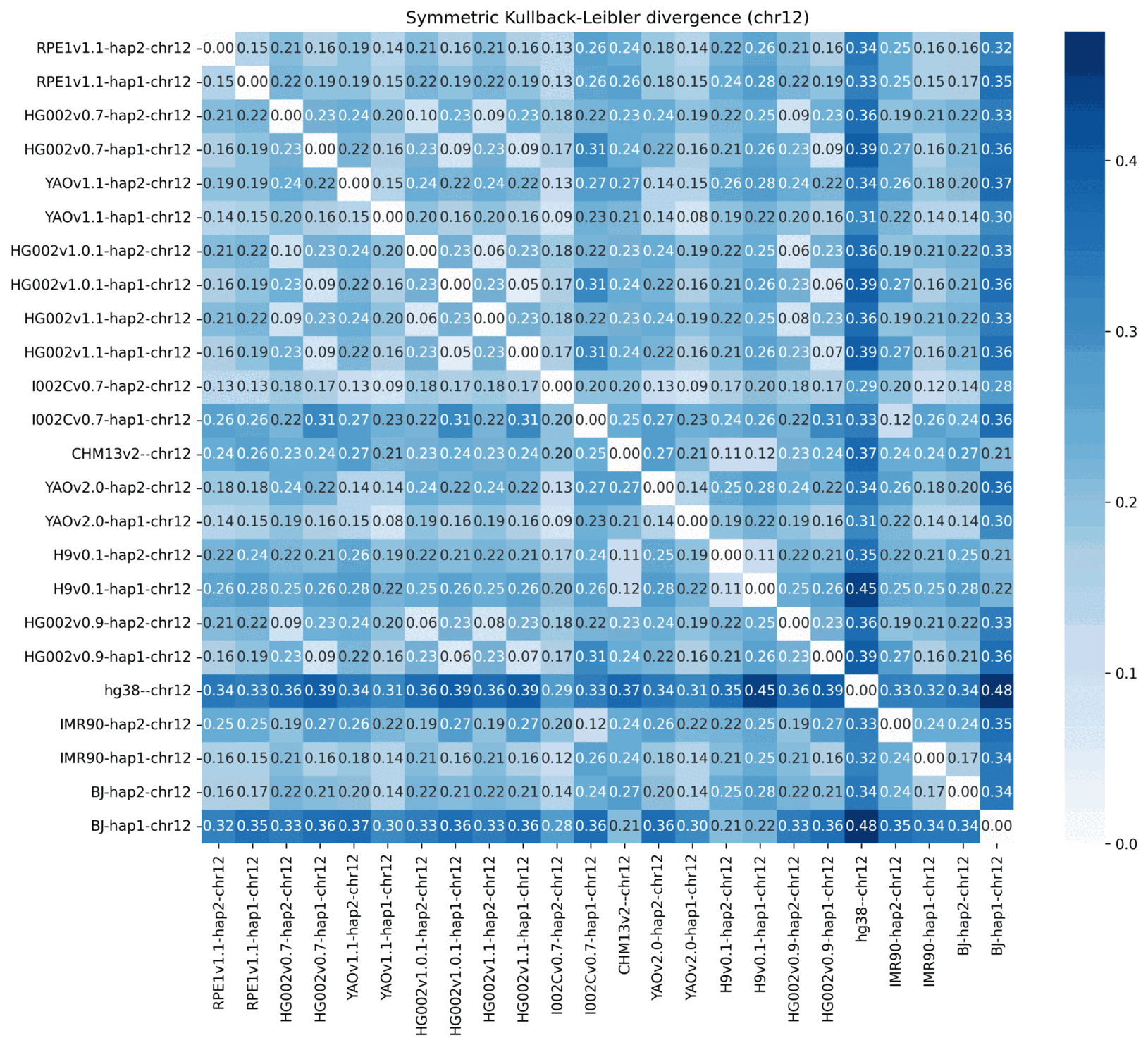}
  \caption{Symmetric \textsc{kl} heatmap chromosome 12}
\end{subfigure}
\caption{Symmetric \textsc{kl} heatmaps for chromosomes 9, 10, 11, and 12.}
\end{figure}

\newpage

\begin{figure}[htbp]
\centering
\begin{subfigure}{.5\textwidth}
  \centering
  \includegraphics[width=\textwidth]{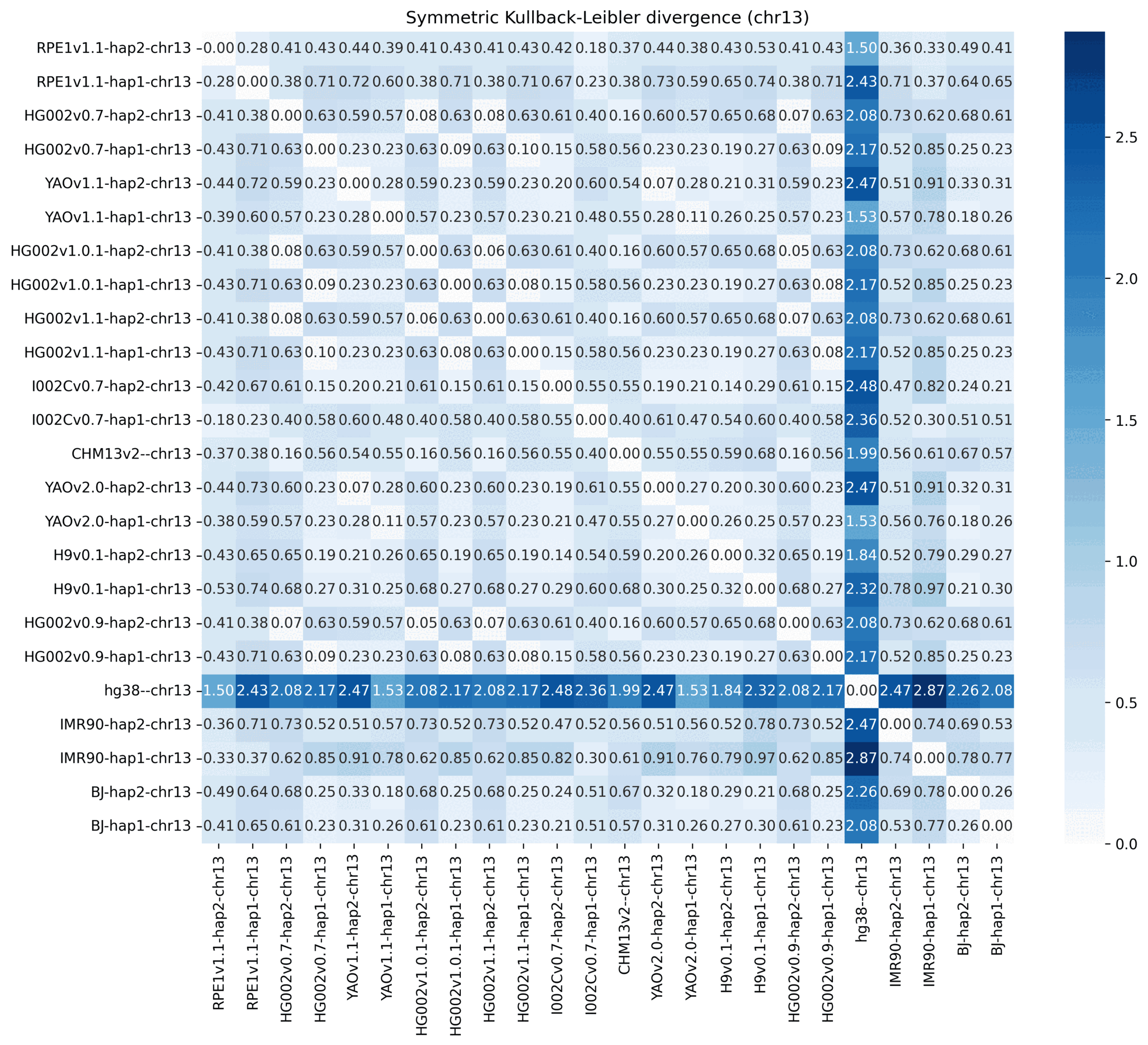}
  \caption{Symmetric \textsc{kl} heatmap chromosome 13}
\end{subfigure}\hfill
\begin{subfigure}{.5\textwidth}
  \centering
  \includegraphics[width=\textwidth]{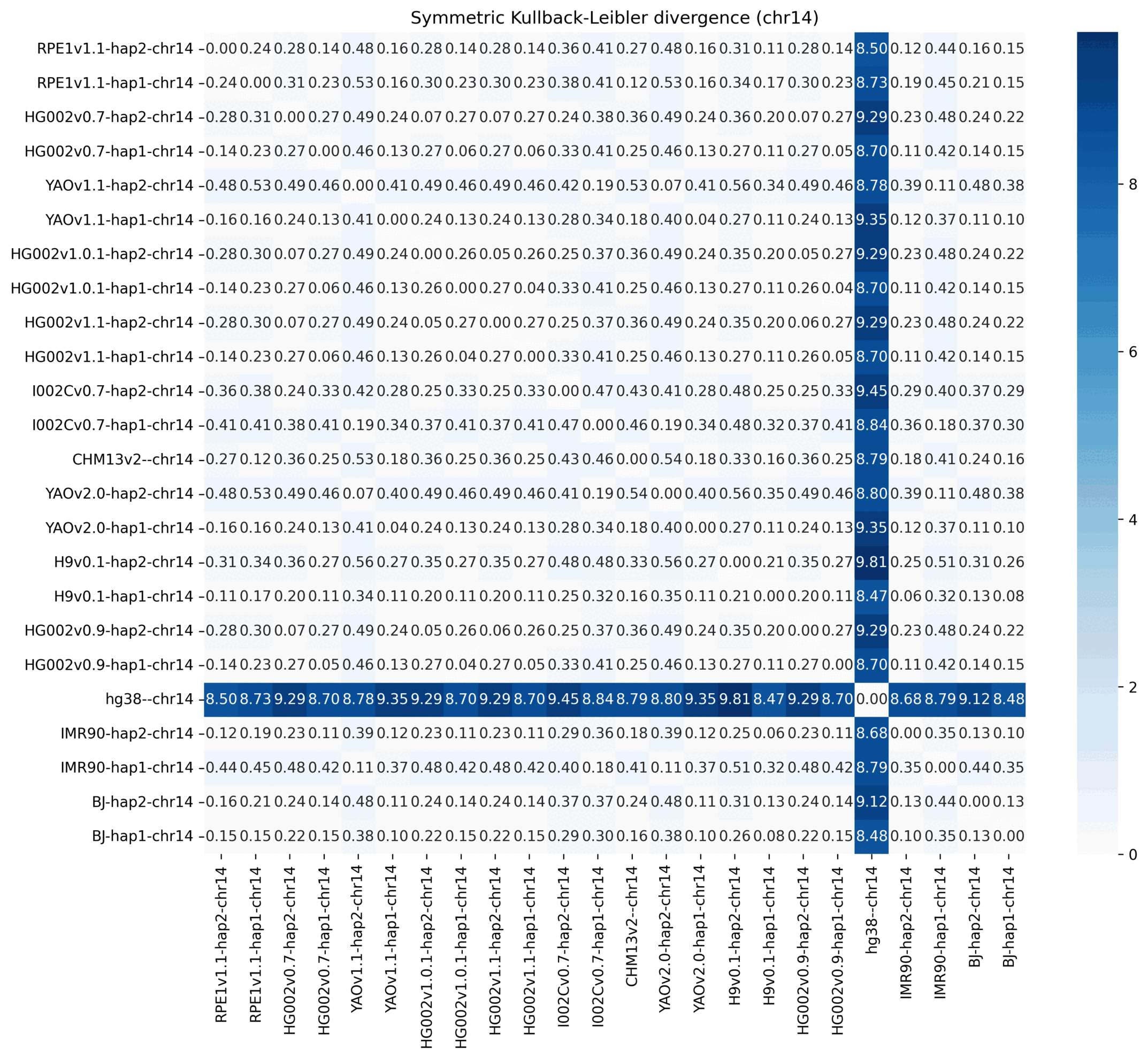}
  \caption{Symmetric \textsc{kl} heatmap chromosome 14}
\end{subfigure}

\vspace{1em}
\begin{subfigure}{.5\textwidth}
  \centering
  \includegraphics[width=\textwidth]{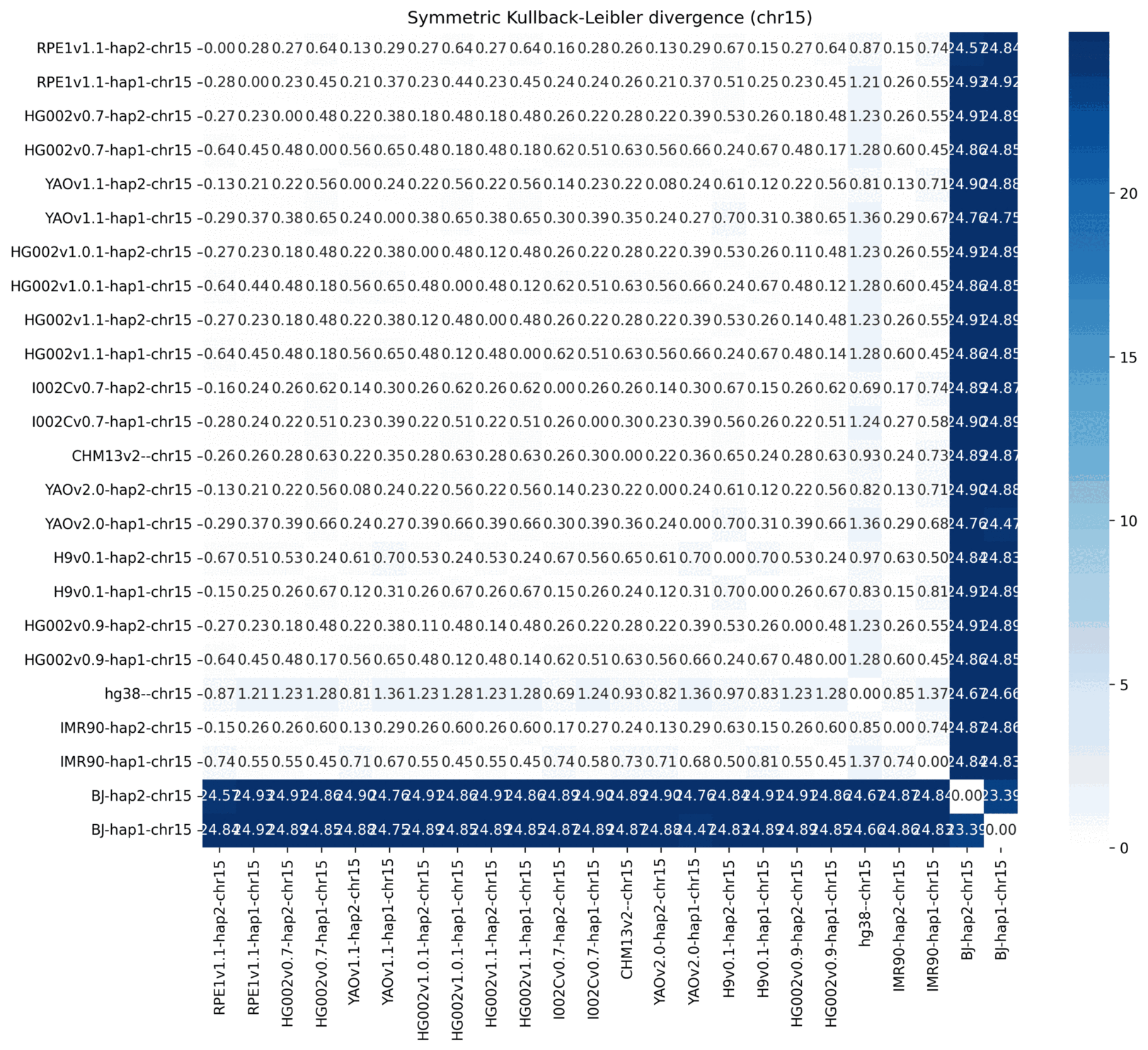}
  \caption{Symmetric \textsc{kl} heatmap chromosome 15}
\end{subfigure}\hfill
\begin{subfigure}{.5\textwidth}
  \centering
  \includegraphics[width=\textwidth]{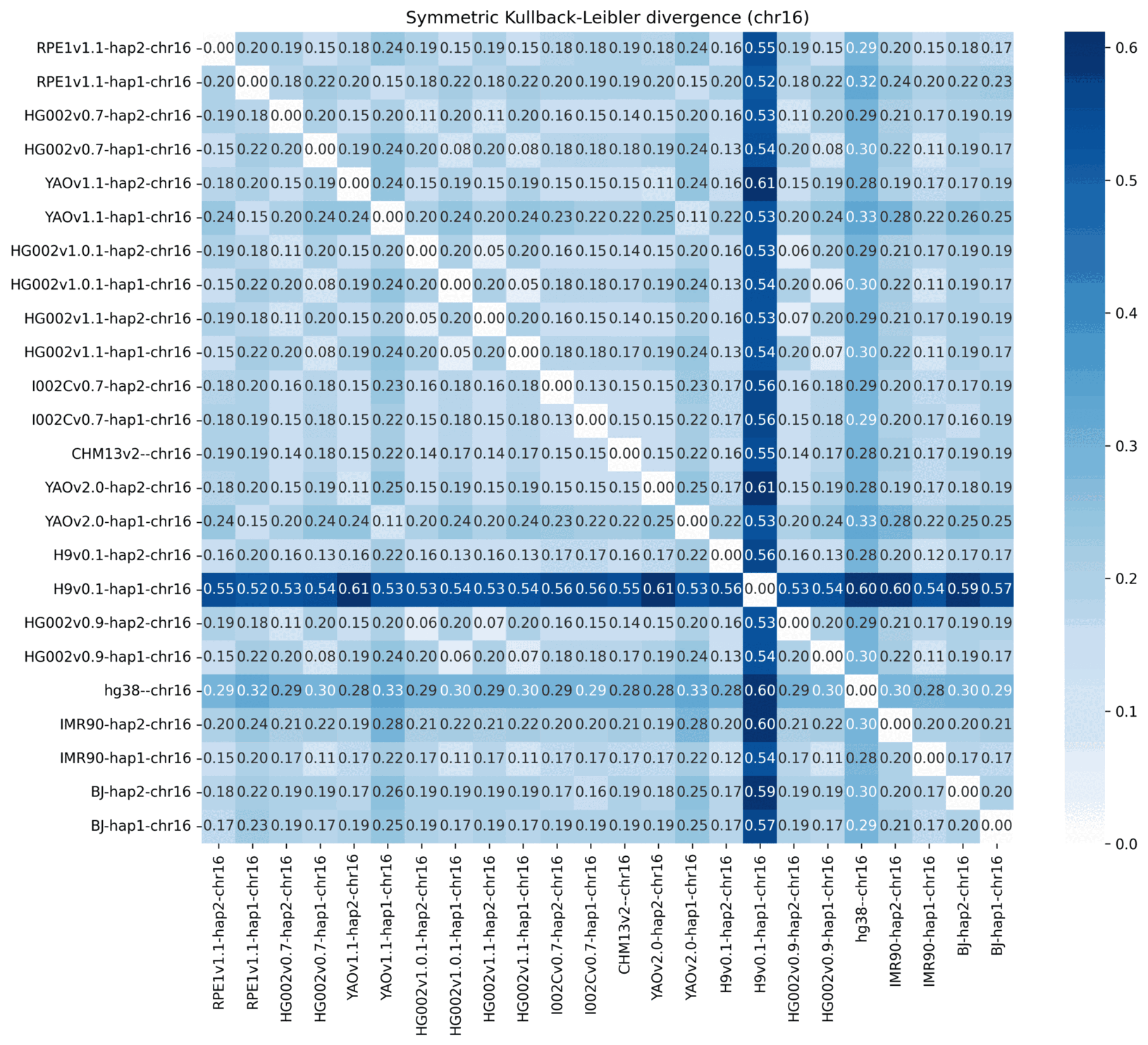}
  \caption{Symmetric \textsc{kl} heatmap chromosome 16}
\end{subfigure}
\caption{Symmetric \textsc{kl} heatmaps for chromosomes 13, 14, 15, and 16.}
\end{figure}

\newpage

\begin{figure}[htbp]
\centering
\begin{subfigure}{.5\textwidth}
  \centering
  \includegraphics[width=\textwidth]{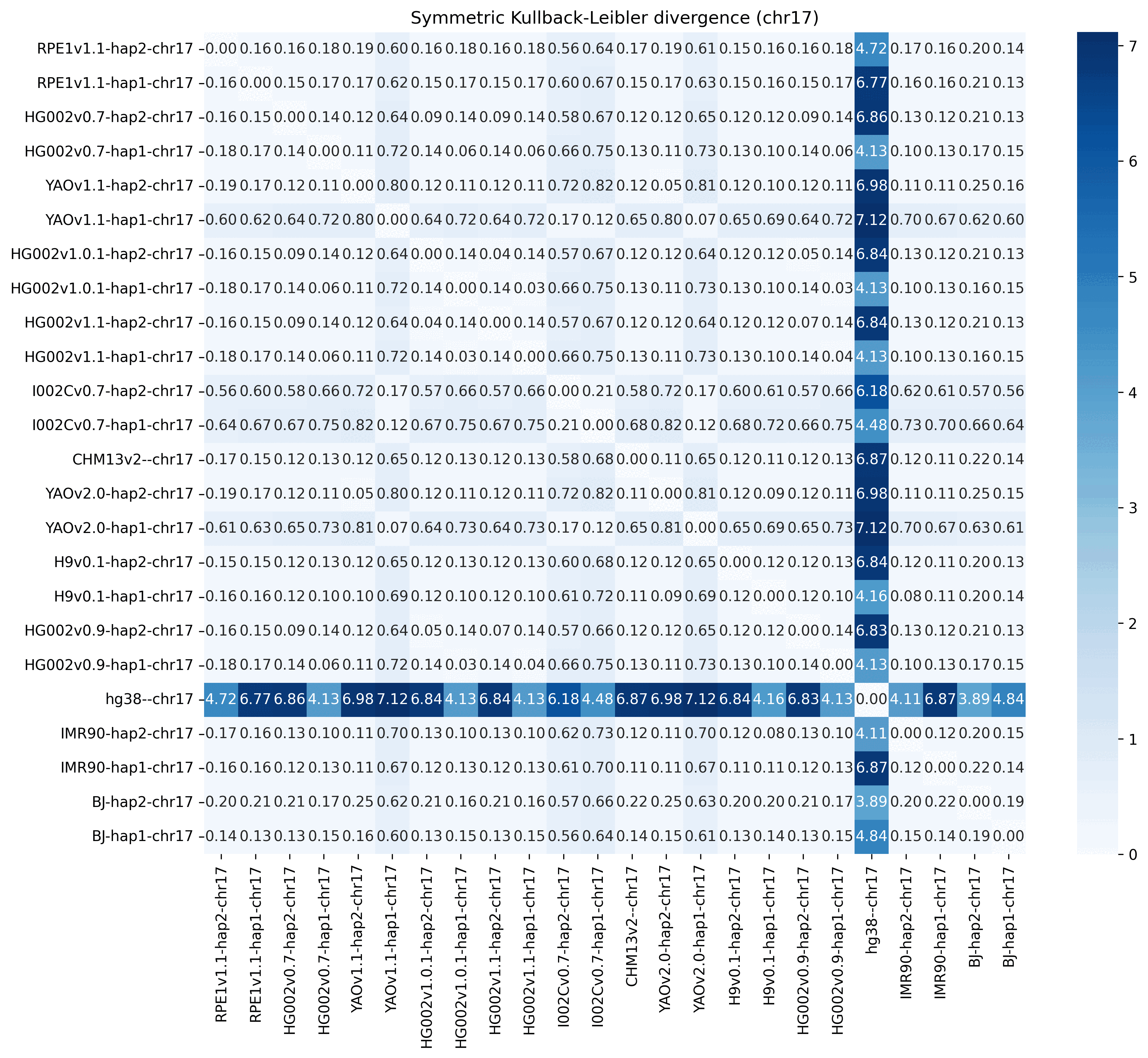}
  \caption{Symmetric \textsc{kl} heatmap chromosome 17}
\end{subfigure}\hfill
\begin{subfigure}{.5\textwidth}
  \centering
  \includegraphics[width=\textwidth]{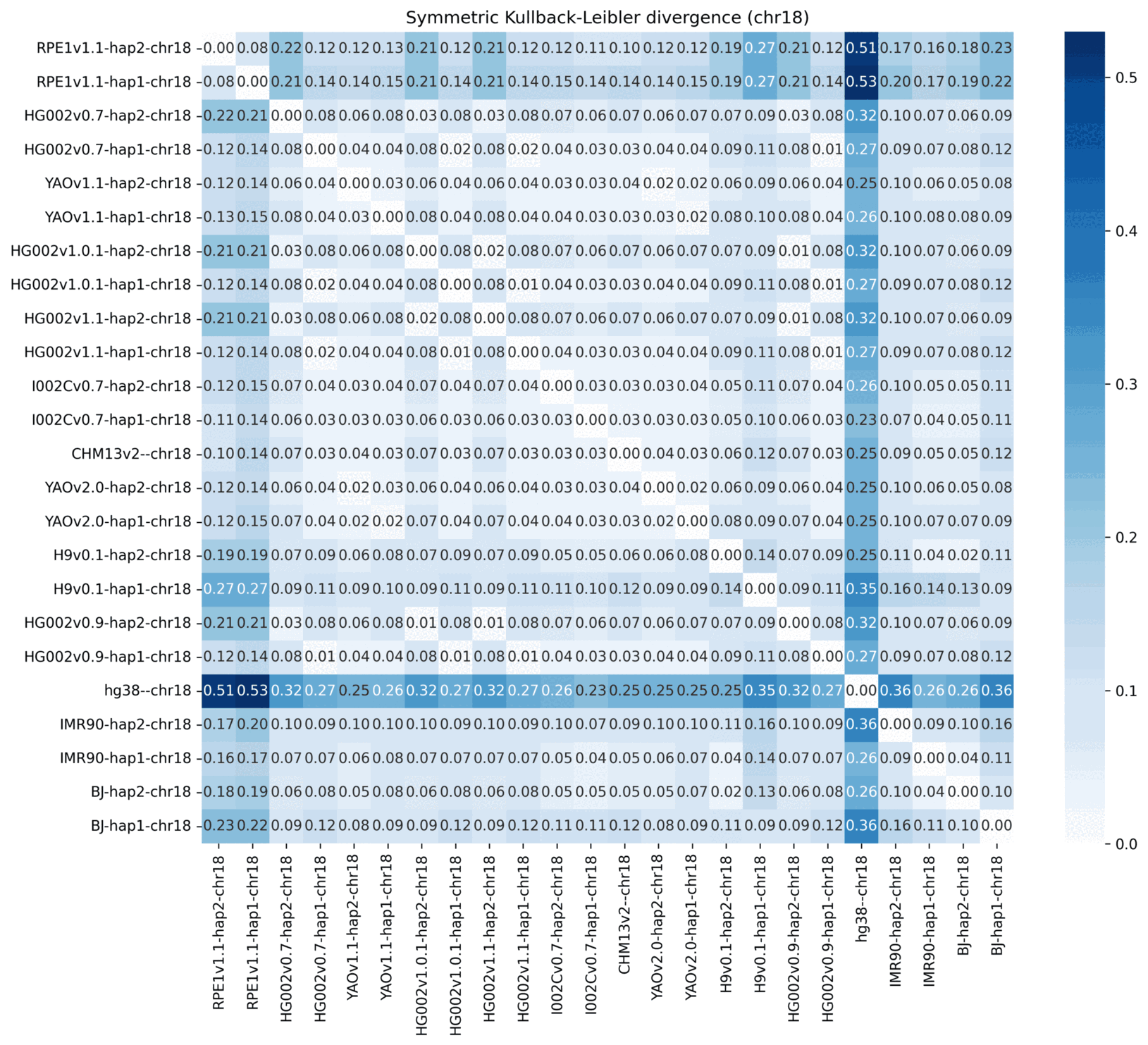}
  \caption{Symmetric \textsc{kl} heatmap chromosome 18}
\end{subfigure}

\vspace{1em}
\begin{subfigure}{.5\textwidth}
  \centering
  \includegraphics[width=\textwidth]{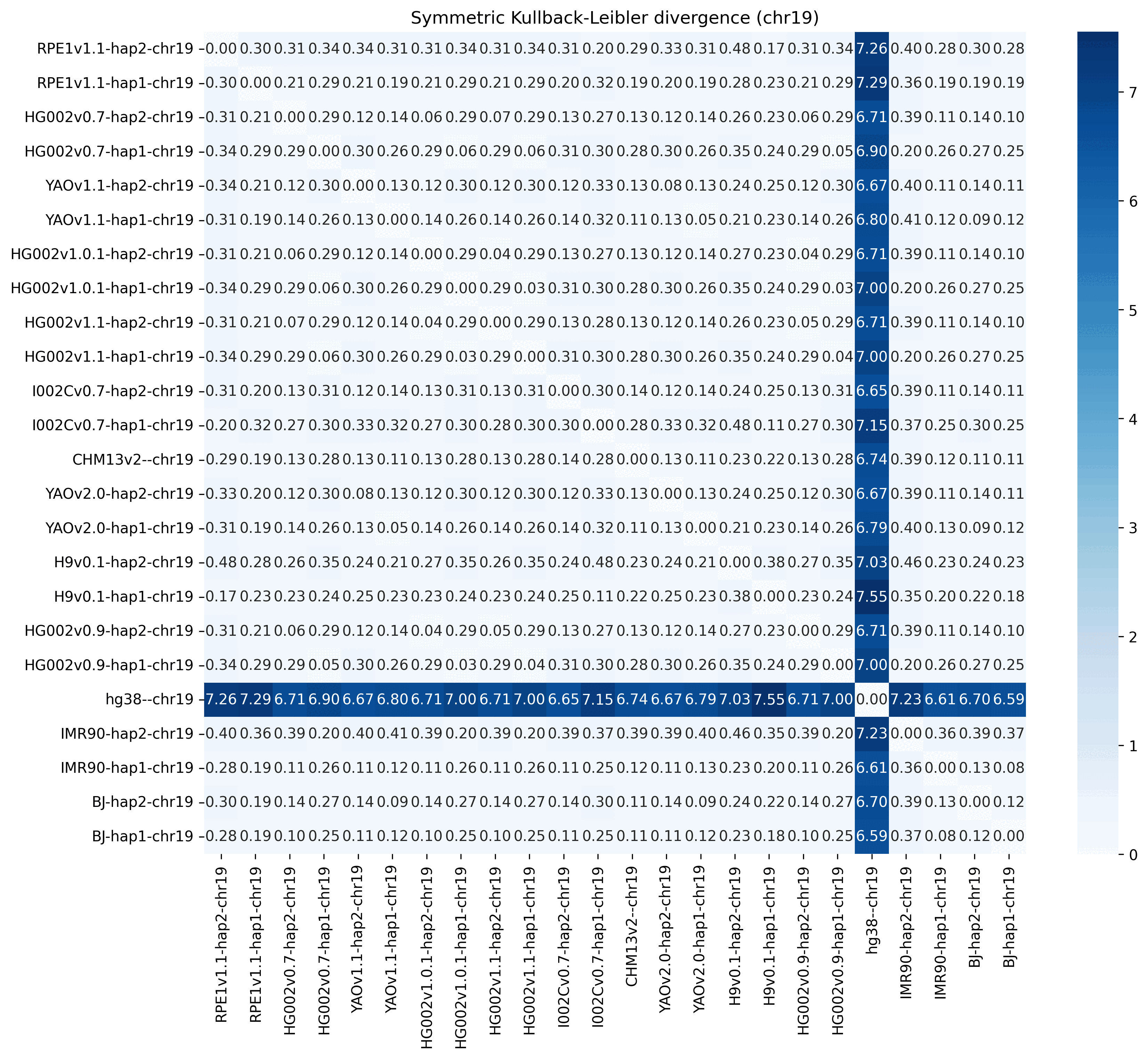}
  \caption{Symmetric \textsc{kl} heatmap chromosome 19}
\end{subfigure}\hfill
\begin{subfigure}{.5\textwidth}
  \centering
  \includegraphics[width=\textwidth]{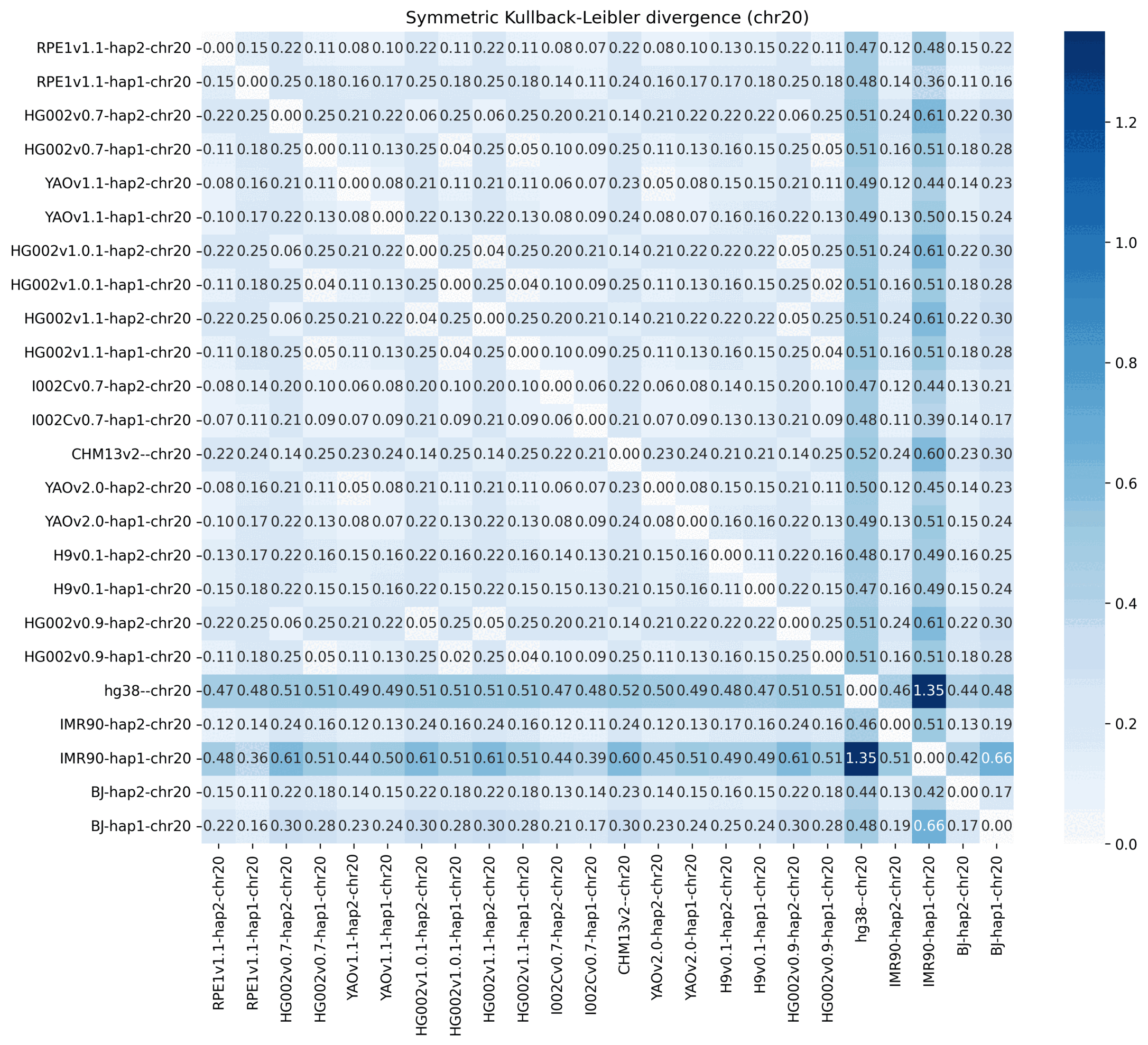}
  \caption{Symmetric \textsc{kl} heatmap chromosome 20}
\end{subfigure}
\caption{Symmetric \textsc{kl} heatmaps for chromosomes 17, 18, 19, and 20.}
\end{figure}

\newpage

\begin{figure}[htbp]
\centering
\begin{subfigure}{.5\textwidth}
  \centering
  \includegraphics[width=\textwidth]{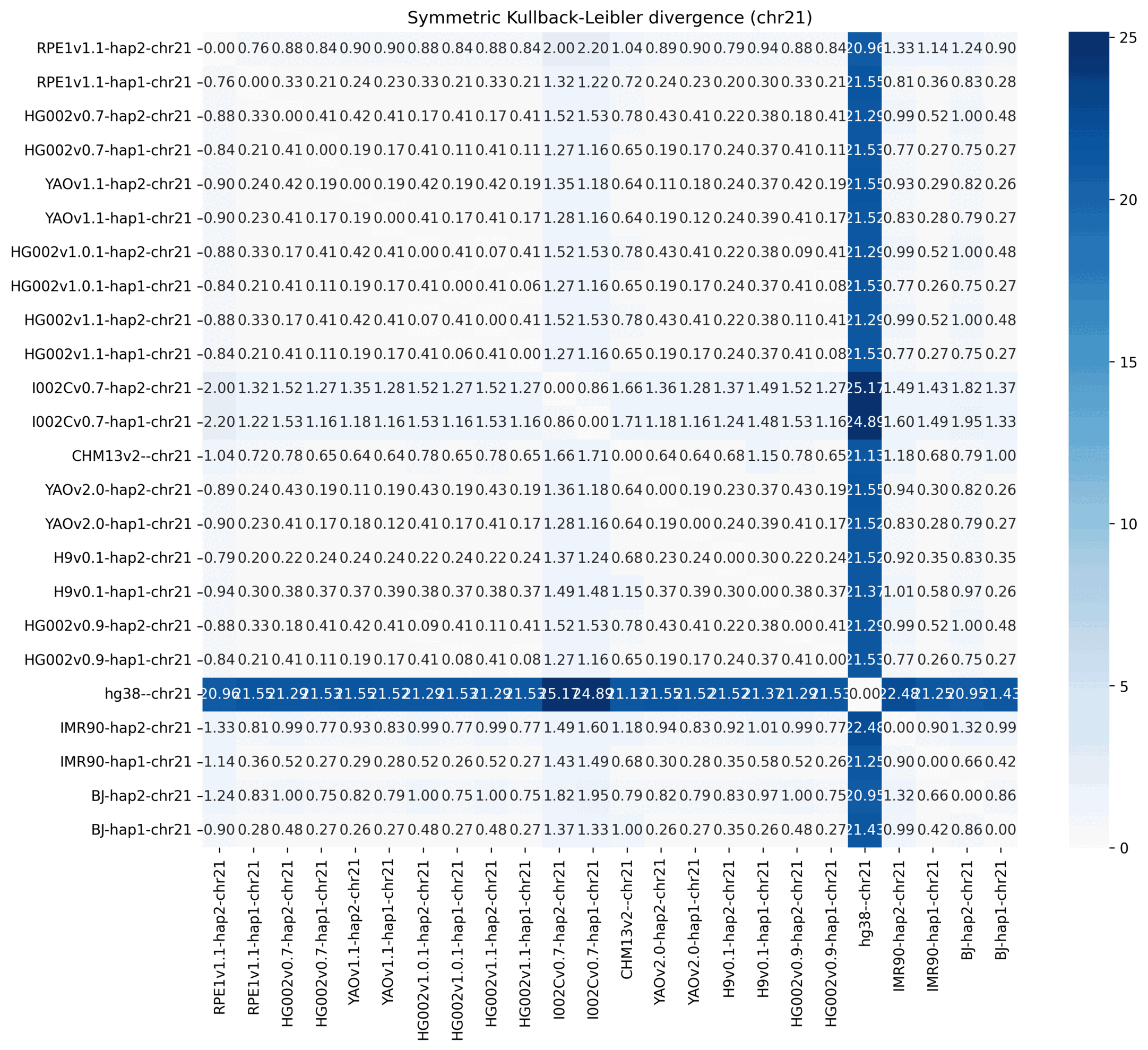}
  \caption{Symmetric \textsc{kl} heatmap chromosome 21}
\end{subfigure}\hfill
\begin{subfigure}{.5\textwidth}
  \centering
  \includegraphics[width=\textwidth]{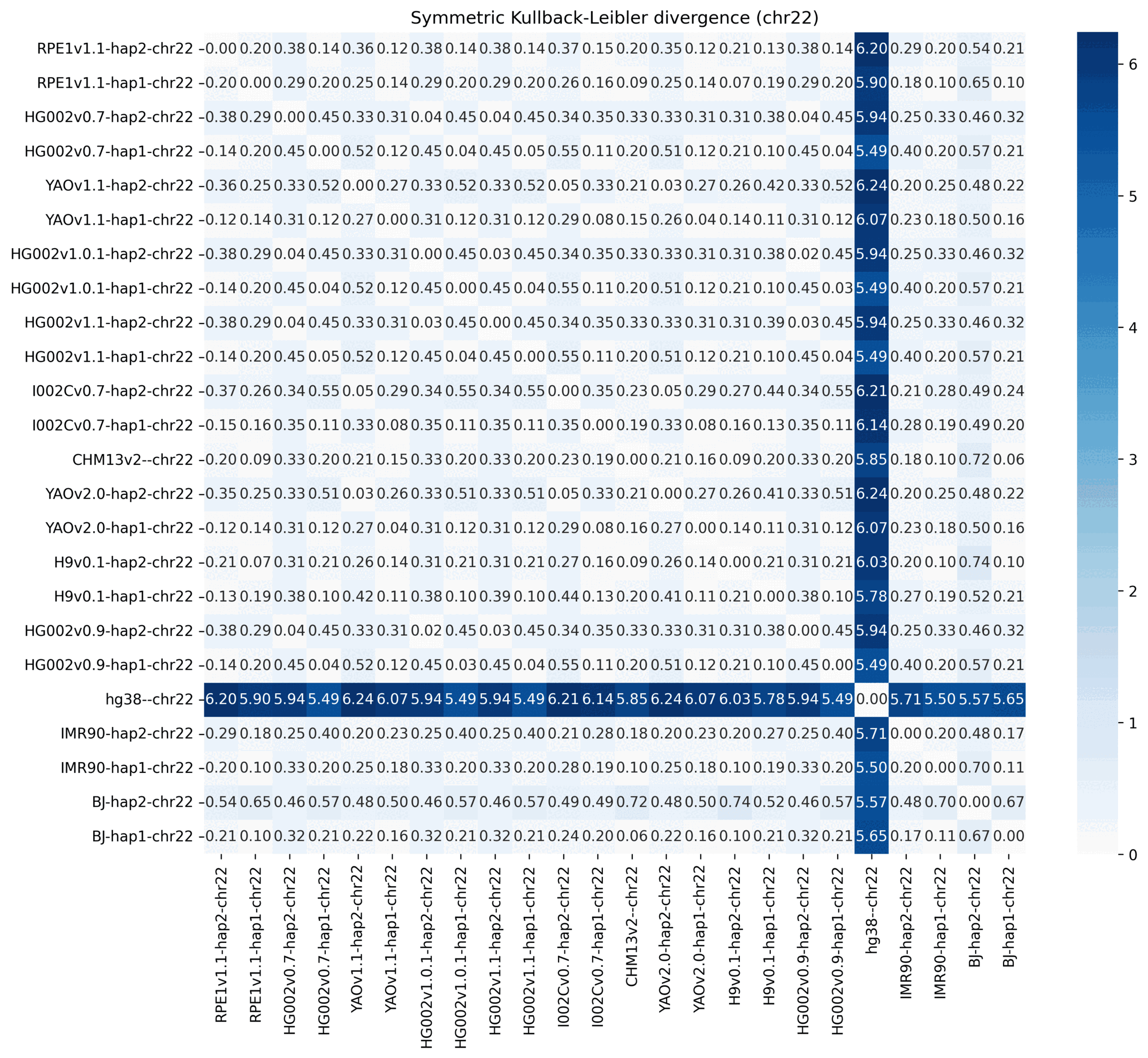}
  \caption{Symmetric \textsc{kl} heatmap chromosome 22}
\end{subfigure}
\caption{Symmetric \textsc{kl} heatmaps for chromosomes 21 and 22.}
\end{figure}

\newpage

\begin{figure}[htbp]
\centering
\begin{subfigure}{.5\textwidth}
   \centering
   \includegraphics[width=\textwidth]{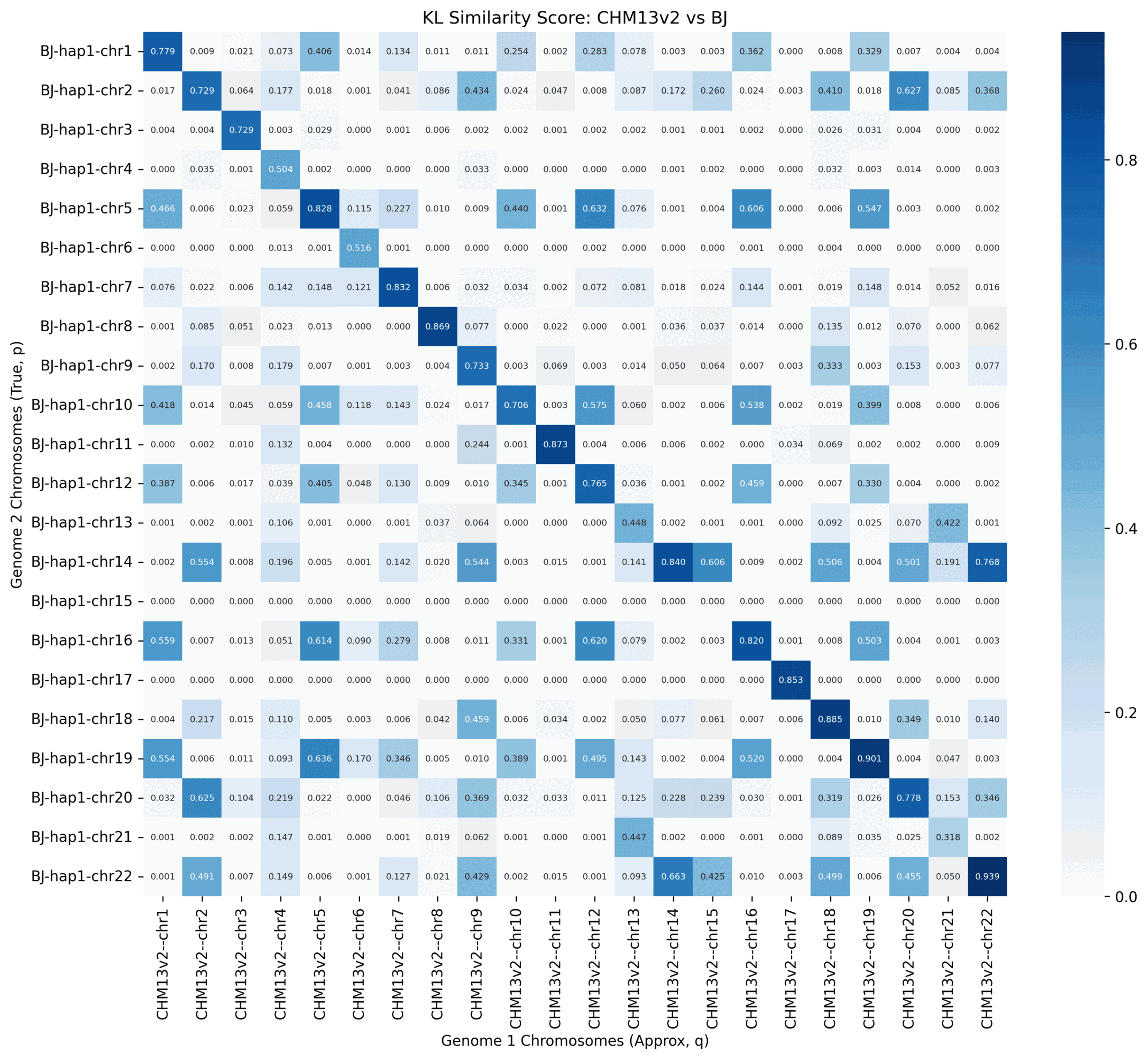}
   \caption{\textsc{chm}\oldstylenums{13}v\oldstylenums{2.0} vs \textsc{bj}}
\end{subfigure}\hfill
\begin{subfigure}{.5\textwidth}
   \centering
   \includegraphics[width=\textwidth]{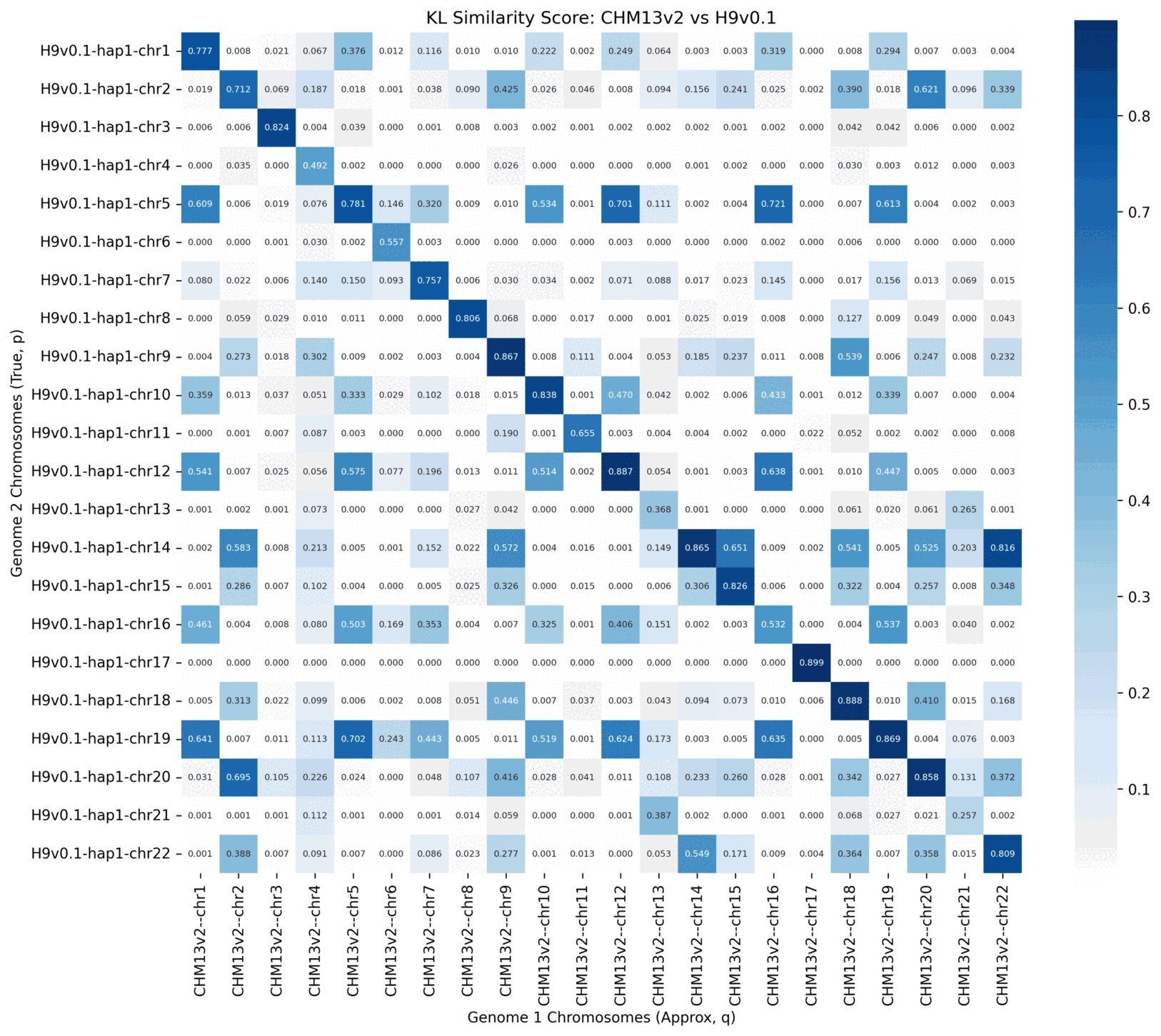}
   \caption{\textsc{chm}\oldstylenums{13}v\oldstylenums{2.0} vs \textsc{h}\oldstylenums{9}v\oldstylenums{0.1}}
\end{subfigure}

\vspace{1em}
\begin{subfigure}{.5\textwidth}
   \centering
   \includegraphics[width=\textwidth]{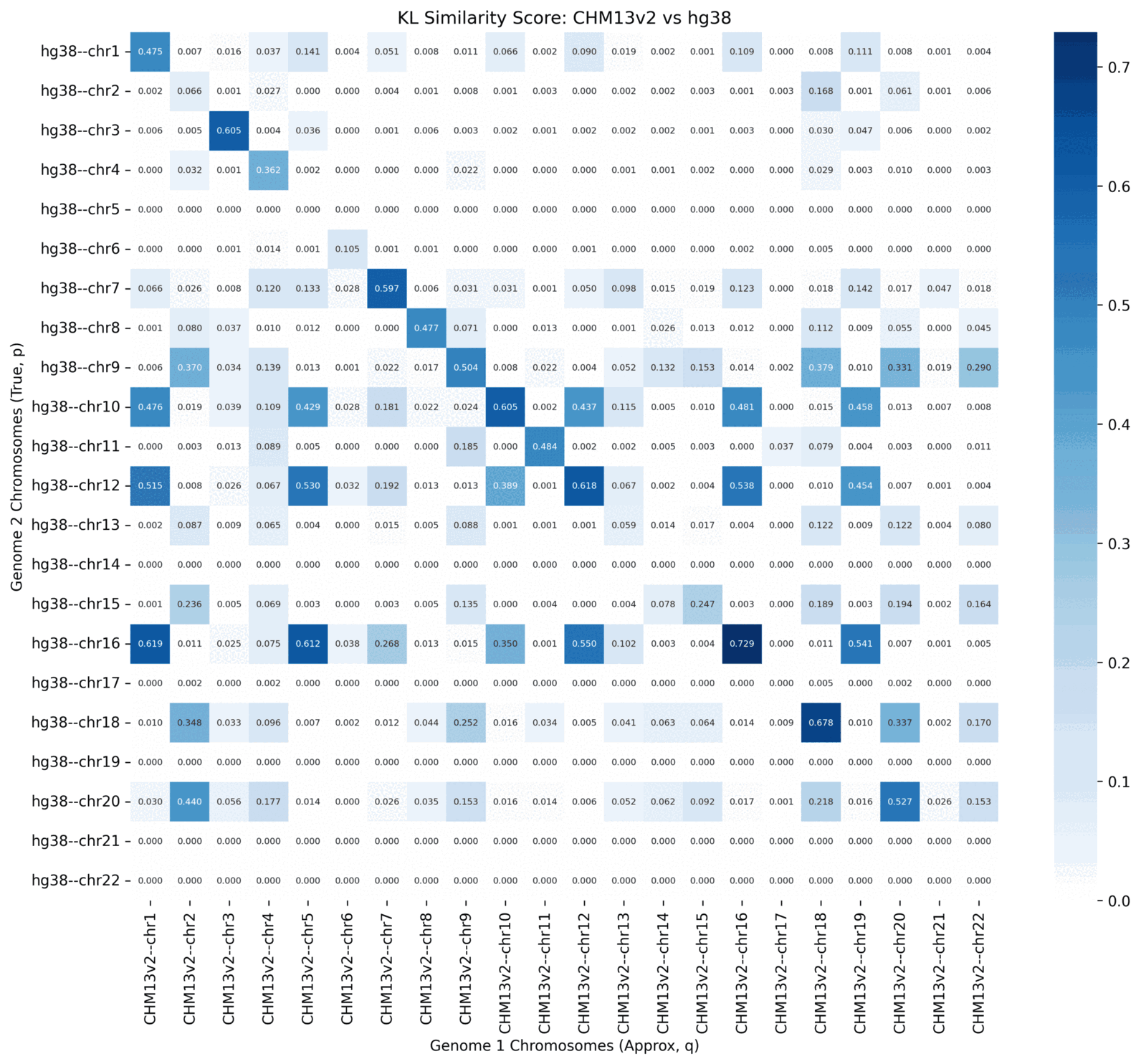}
   \caption{\textsc{chm}\oldstylenums{13}v\oldstylenums{2.0} vs \textsc{grc}h\oldstylenums{38}}
\end{subfigure}\hfill
\begin{subfigure}{.5\textwidth}
   \centering
   \includegraphics[width=\textwidth]{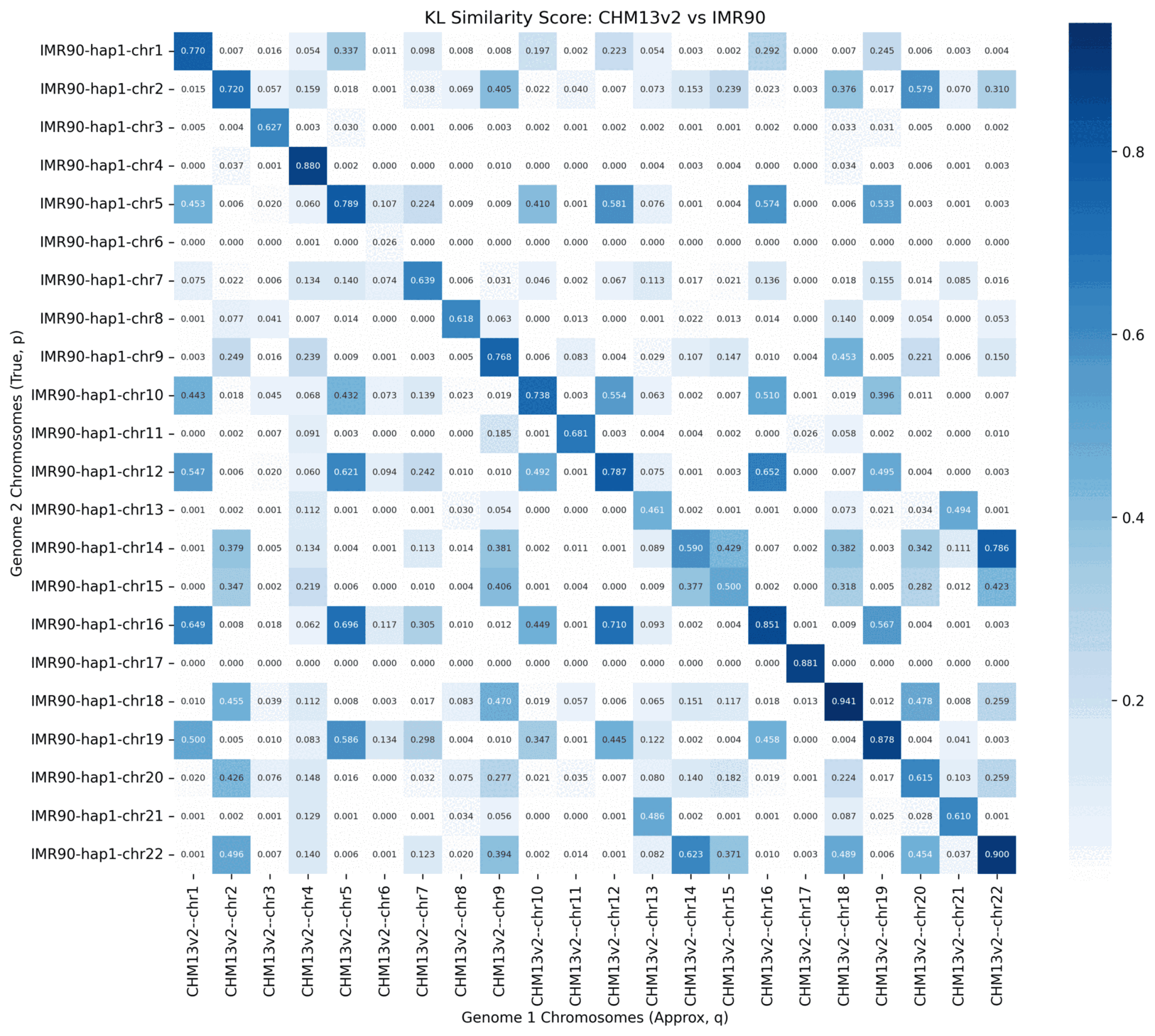}
   \caption{\textsc{chm}\oldstylenums{13}v\oldstylenums{2.0} vs \textsc{imr}\oldstylenums{90}}
\end{subfigure}
\caption{Pairwise \textsc{kl} similarity matrices used to score accuracy, entropy, InfoNCE \textemdash\space hap1 only.}
\end{figure}

\textsc{rpe}\oldstylenums{1}v\oldstylenums{1.1}, \textsc{hg}\oldstylenums{002} (versions 0.7, 0.9, 1.0.1, and 1.1), \textsc{yao} (versions 1.1 and 2.0), \textsc{i}\oldstylenums{002}\textsc{c}, \textsc{chm}\oldstylenums{13}, \textsc{h}\oldstylenums{9}, \textsc{grc}h\oldstylenums{38}, \textsc{imr}\oldstylenums{90}, \textsc{bj}

\newpage

\begin{figure}[htbp]
\centering
\begin{subfigure}{.5\textwidth}
   \centering
   \includegraphics[width=\textwidth]{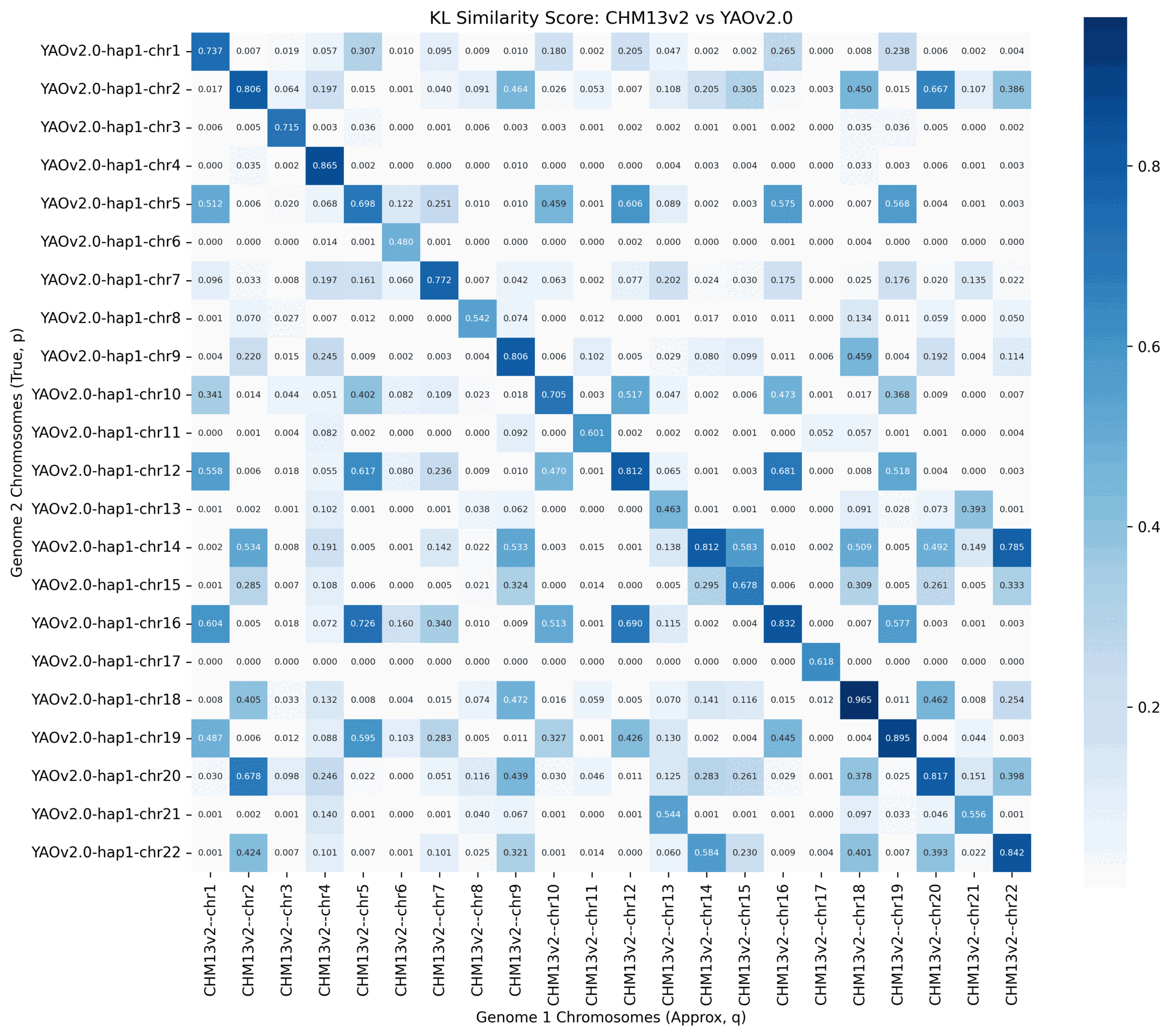}
   \caption{\textsc{chm}\oldstylenums{13}v\oldstylenums{2.0} vs \textsc{yao}v\oldstylenums{2.0}}
\end{subfigure}\hfill
\begin{subfigure}{.5\textwidth}
   \centering
   \includegraphics[width=\textwidth]{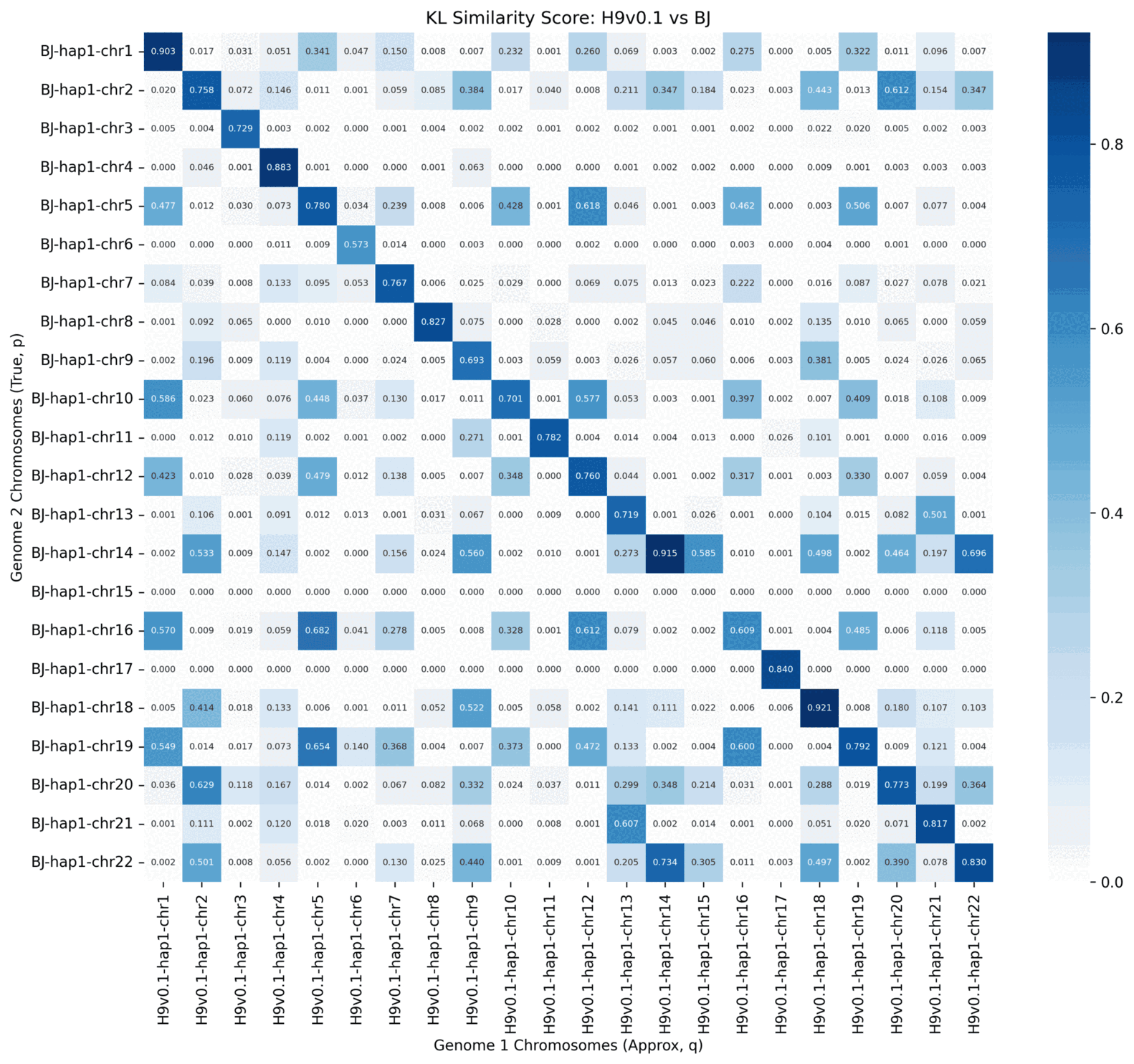}
   \caption{\textsc{h}\oldstylenums{9}v\oldstylenums{0.1} vs \textsc{bj}}
\end{subfigure}

\vspace{1em}
\begin{subfigure}{.5\textwidth}
   \centering
   \includegraphics[width=\textwidth]{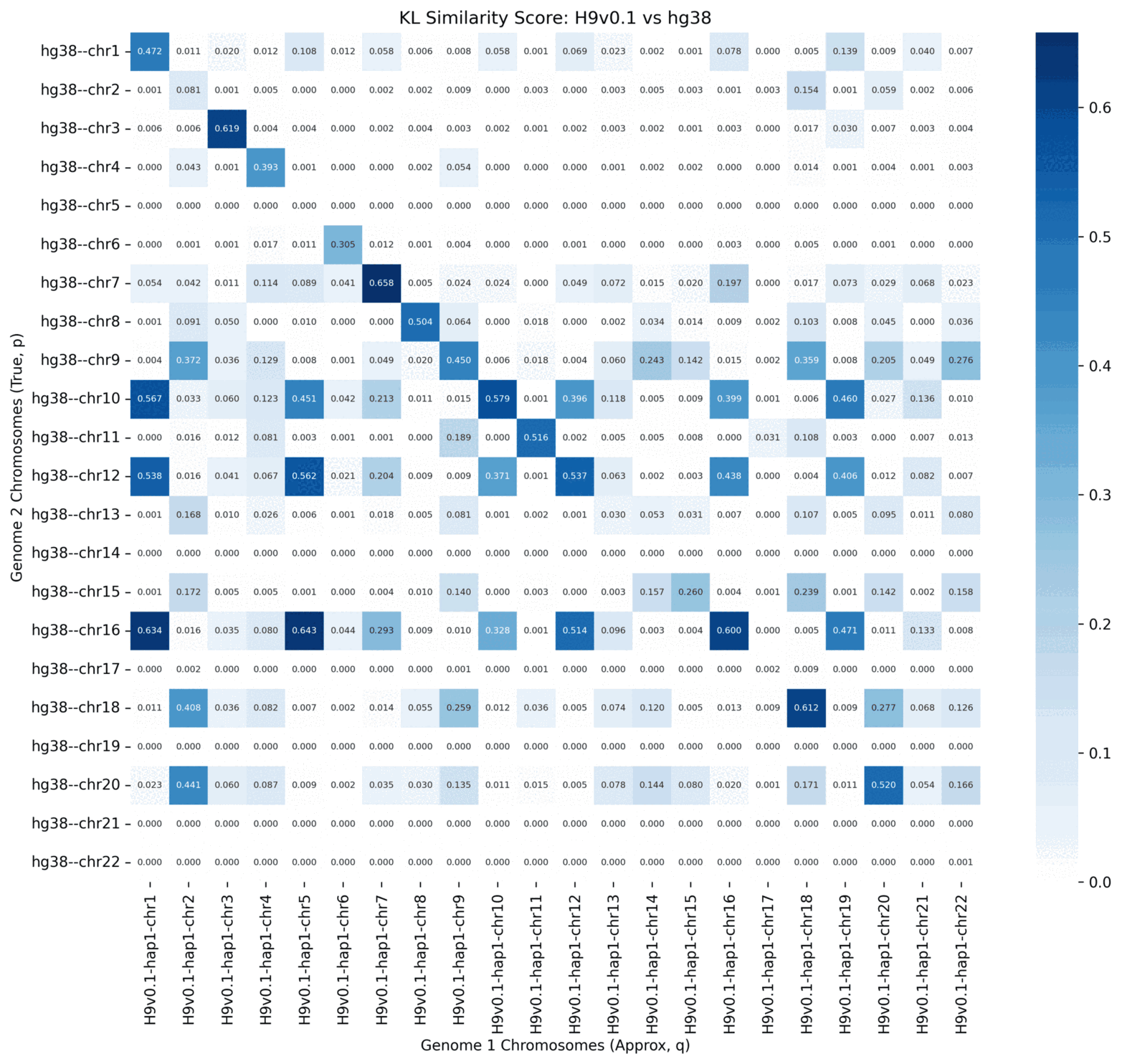}
   \caption{\textsc{h}\oldstylenums{9}v\oldstylenums{0.1} vs \textsc{grc}h\oldstylenums{38}}
\end{subfigure}\hfill
\begin{subfigure}{.5\textwidth}
   \centering
   \includegraphics[width=\textwidth]{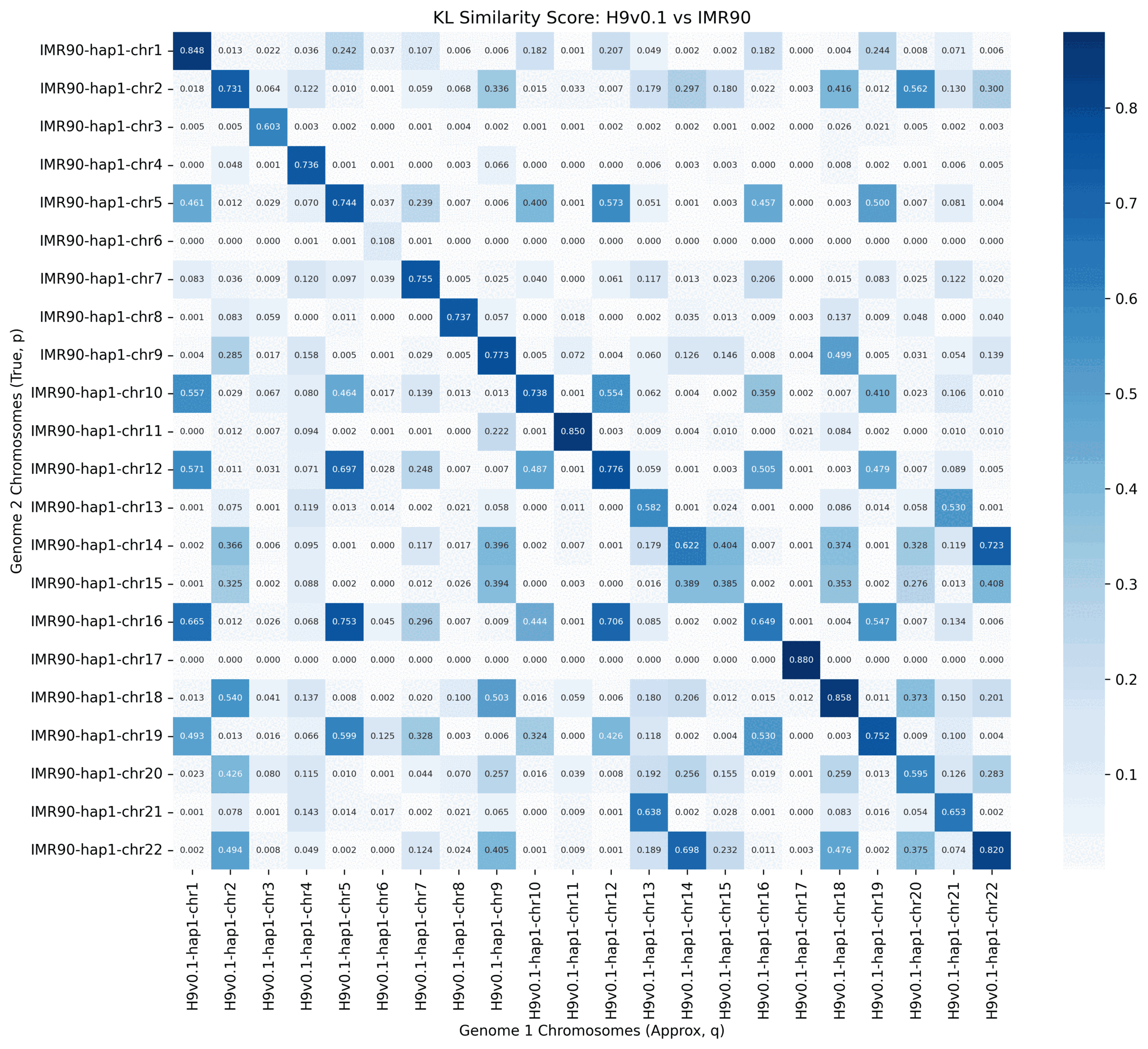}
   \caption{\textsc{h}\oldstylenums{9}v\oldstylenums{0.1} vs \textsc{imr}\oldstylenums{90}}
\end{subfigure}
\caption{Pairwise \textsc{kl} similarity matrices used to score accuracy, entropy, InfoNCE \textemdash\space hap1 only.}
\end{figure}

\newpage

\begin{figure}[htbp]
\centering
\begin{subfigure}{.5\textwidth}
   \centering
   \includegraphics[width=\textwidth]{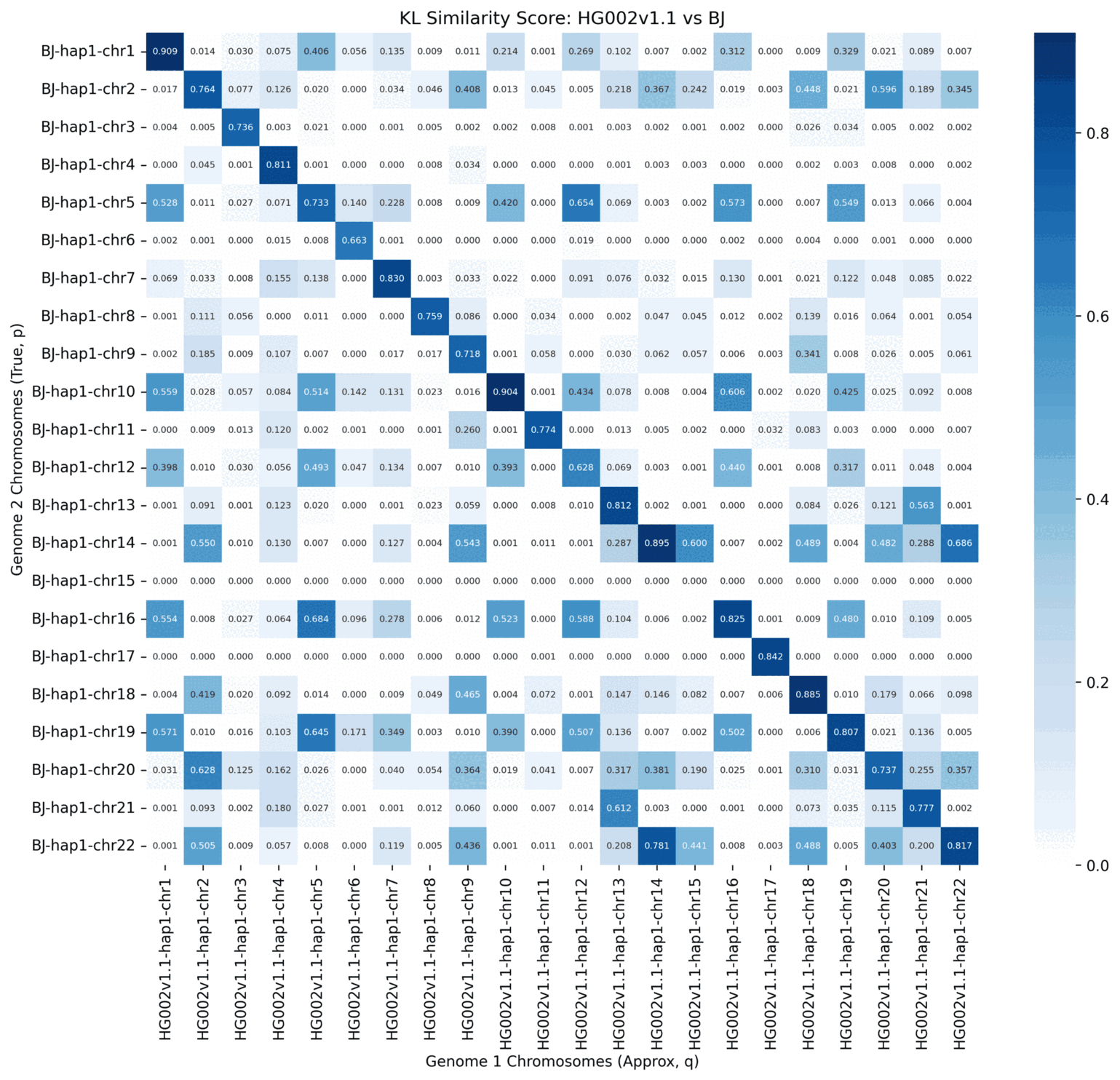}
   \caption{\textsc{hg}\oldstylenums{002}v\oldstylenums{1.1} vs \textsc{bj}}
\end{subfigure}\hfill
\begin{subfigure}{.5\textwidth}
   \centering
   \includegraphics[width=\textwidth]{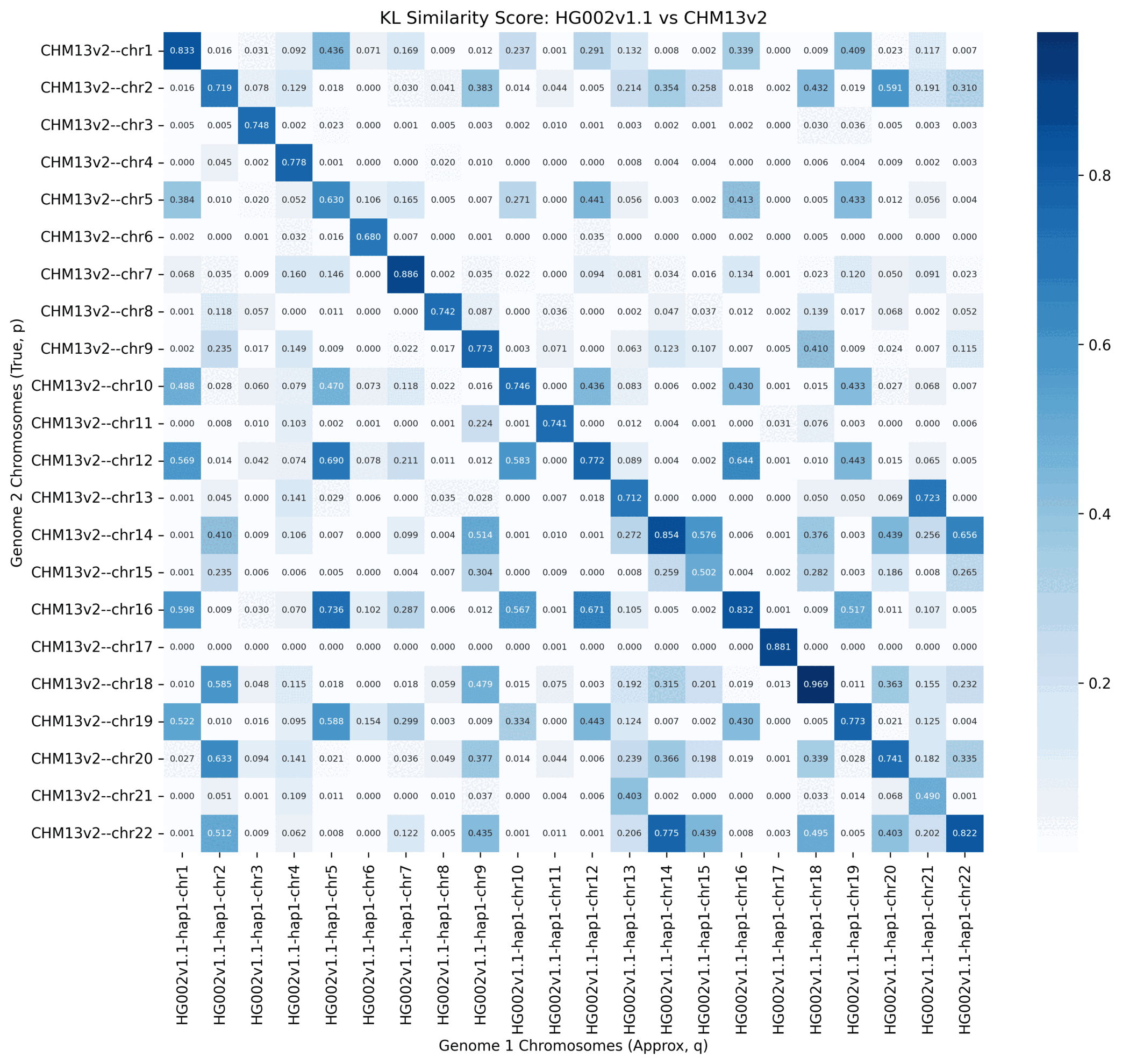}
   \caption{\textsc{hg}\oldstylenums{002}v\oldstylenums{1.1} vs \textsc{chm}\oldstylenums{13}v\oldstylenums{2.0}}
\end{subfigure}

\vspace{1em}
\begin{subfigure}{.5\textwidth}
   \centering
   \includegraphics[width=\textwidth]{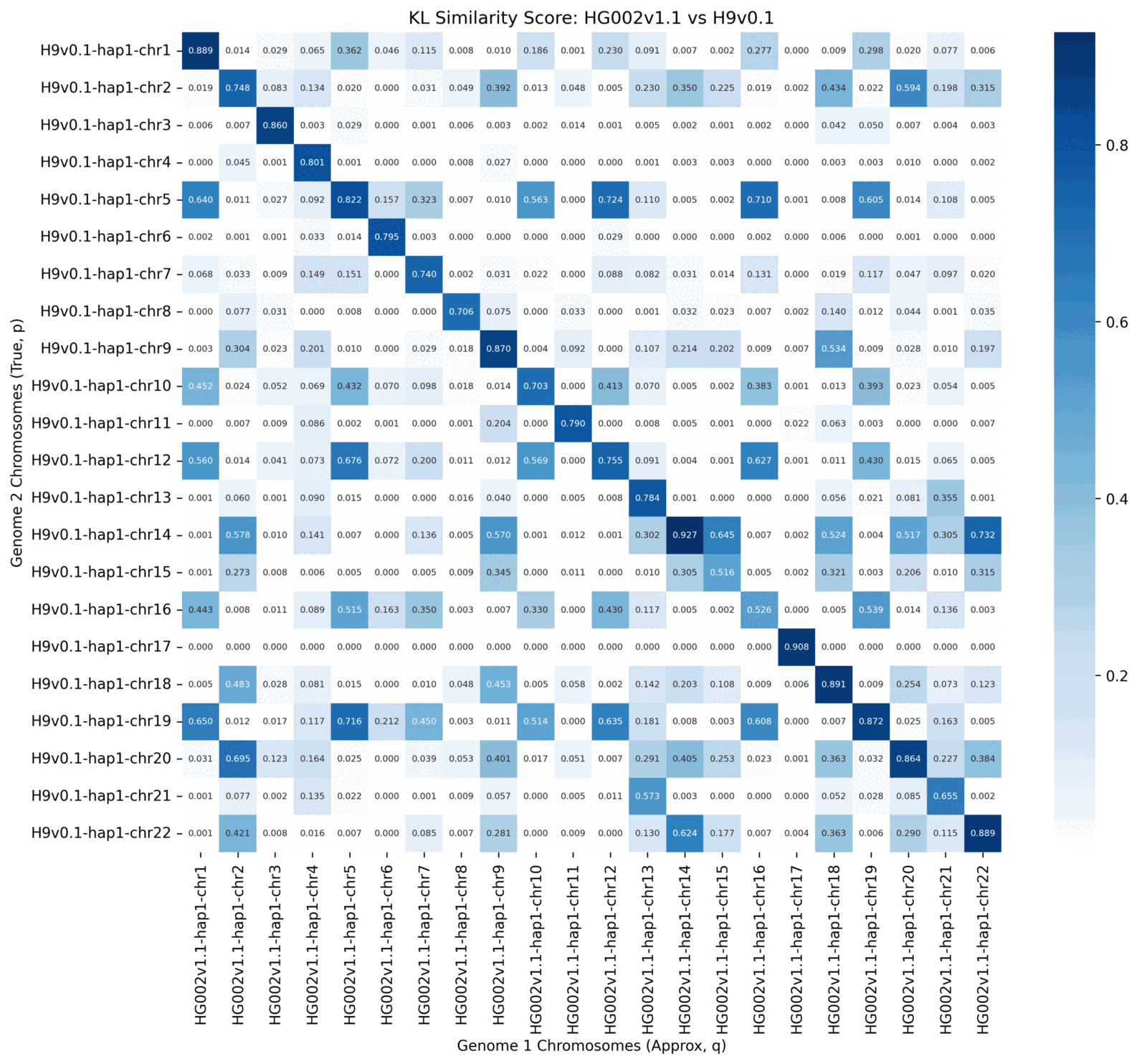}
   \caption{\textsc{hg}\oldstylenums{002}v\oldstylenums{1.1} vs \textsc{h}\oldstylenums{9}v\oldstylenums{0.1}}
\end{subfigure}\hfill
\begin{subfigure}{.5\textwidth}
   \centering
   \includegraphics[width=\textwidth]{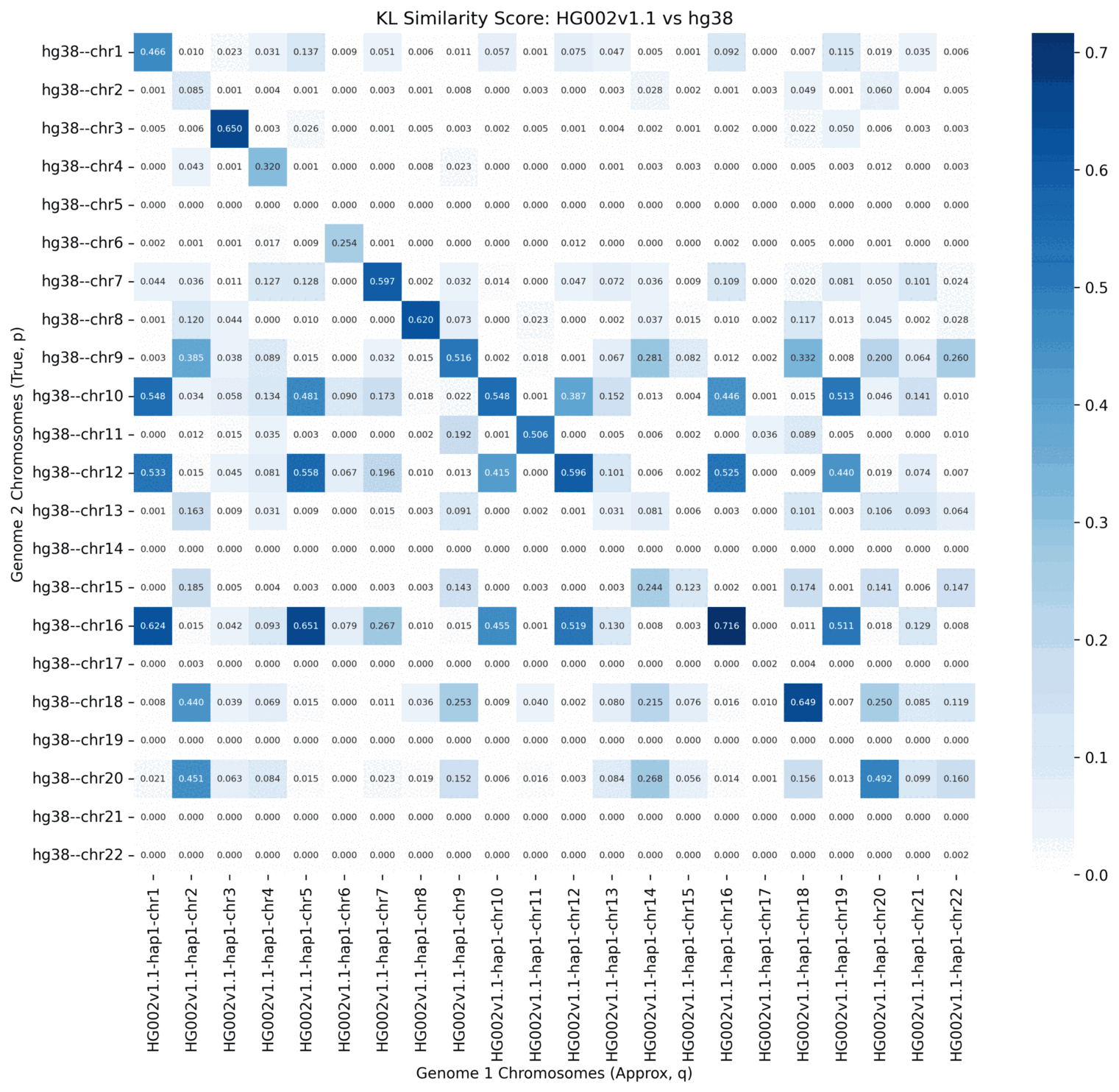}
   \caption{\textsc{hg}\oldstylenums{002}v\oldstylenums{1.1} vs \textsc{grc}h\oldstylenums{38}}
\end{subfigure}
\caption{Pairwise \textsc{kl} similarity matrices used to score accuracy, entropy, InfoNCE \textemdash\space hap1 only.}
\end{figure}

\newpage

\begin{figure}[htbp]
\centering
\begin{subfigure}{.5\textwidth}
   \centering
   \includegraphics[width=\textwidth]{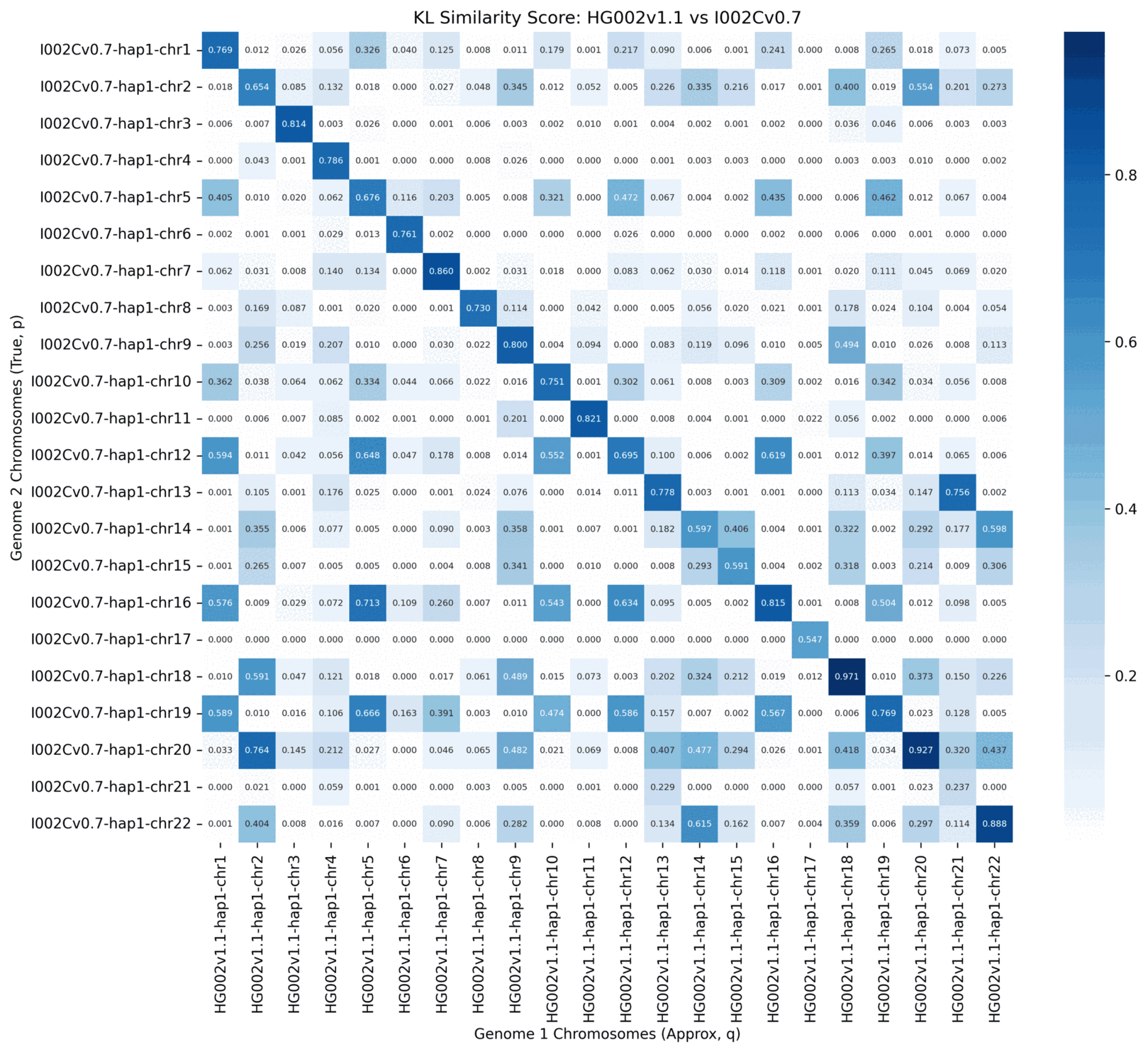}
   \caption{\textsc{hg}\oldstylenums{002}v\oldstylenums{1.1} vs \textsc{i}\oldstylenums{002}\textsc{c}v\oldstylenums{0.7}}
\end{subfigure}\hfill
\begin{subfigure}{.5\textwidth}
   \centering
   \includegraphics[width=\textwidth]{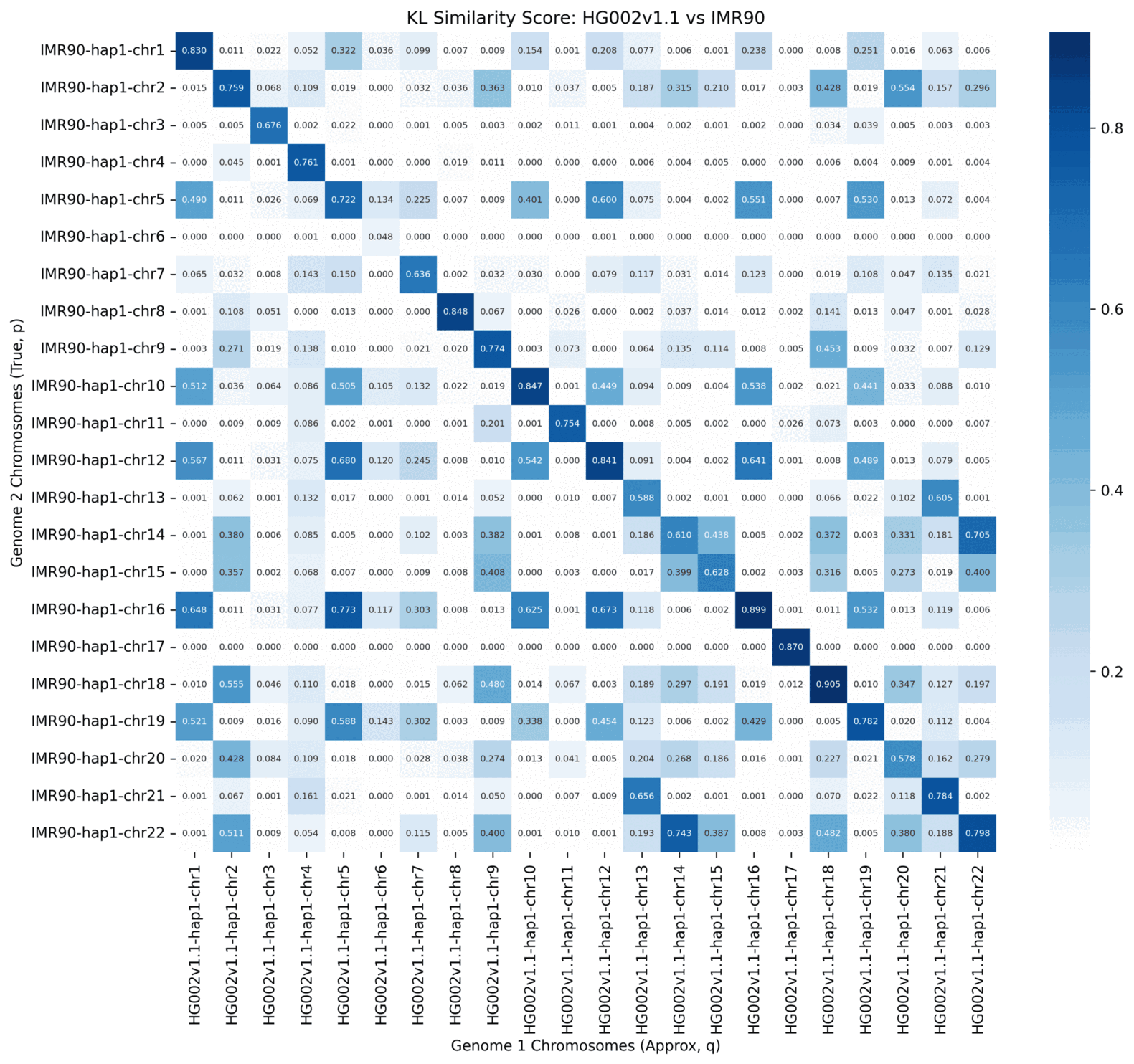}
   \caption{\textsc{hg}\oldstylenums{002}v\oldstylenums{1.1} vs \textsc{imr}\oldstylenums{90}}
\end{subfigure}

\vspace{1em}
\begin{subfigure}{.5\textwidth}
   \centering
   \includegraphics[width=\textwidth]{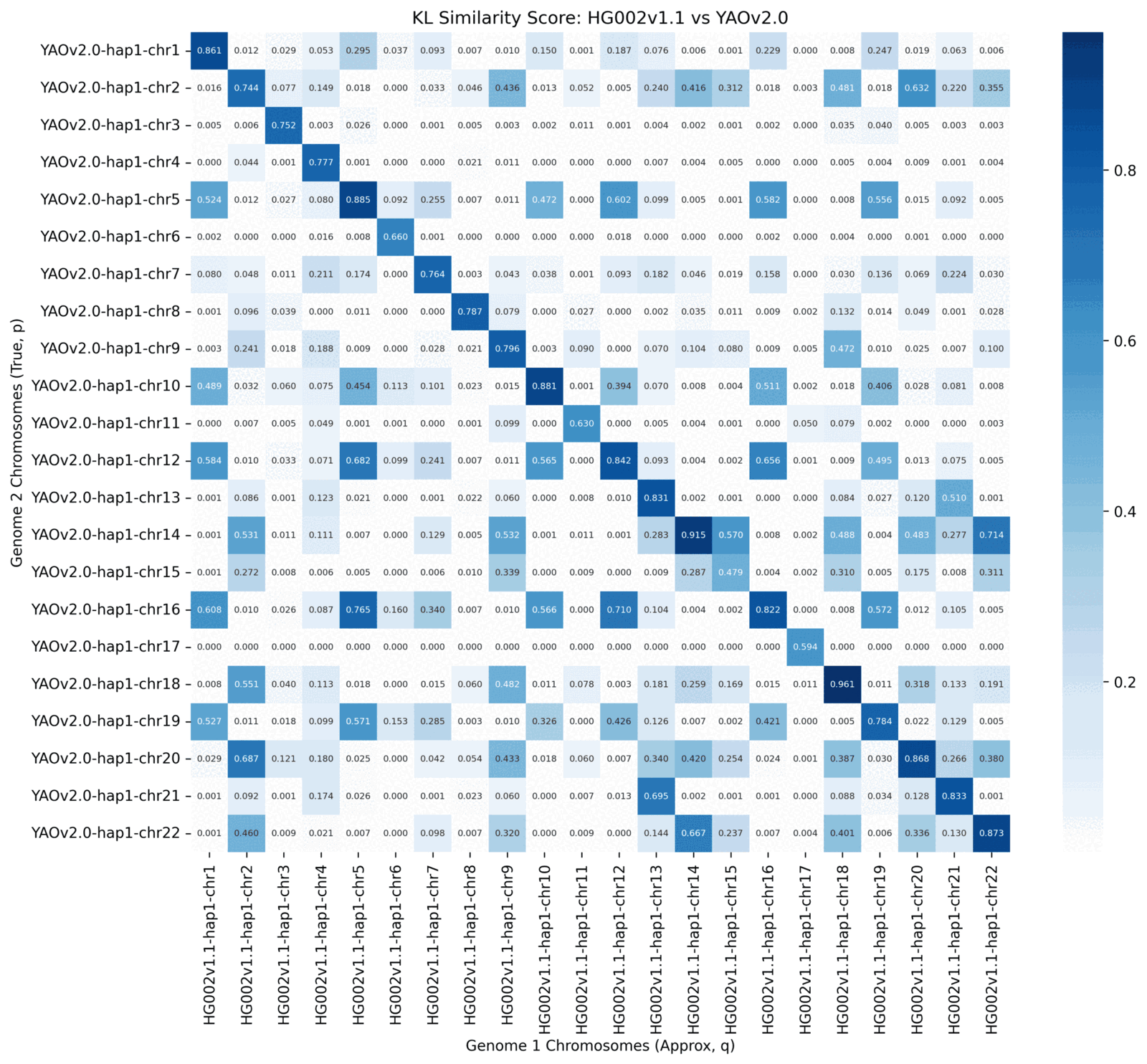}
   \caption{\textsc{hg}\oldstylenums{002}v\oldstylenums{1.1} vs \textsc{yao}v\oldstylenums{2.0}}
\end{subfigure}\hfill
\begin{subfigure}{.5\textwidth}
   \centering
   \includegraphics[width=\textwidth]{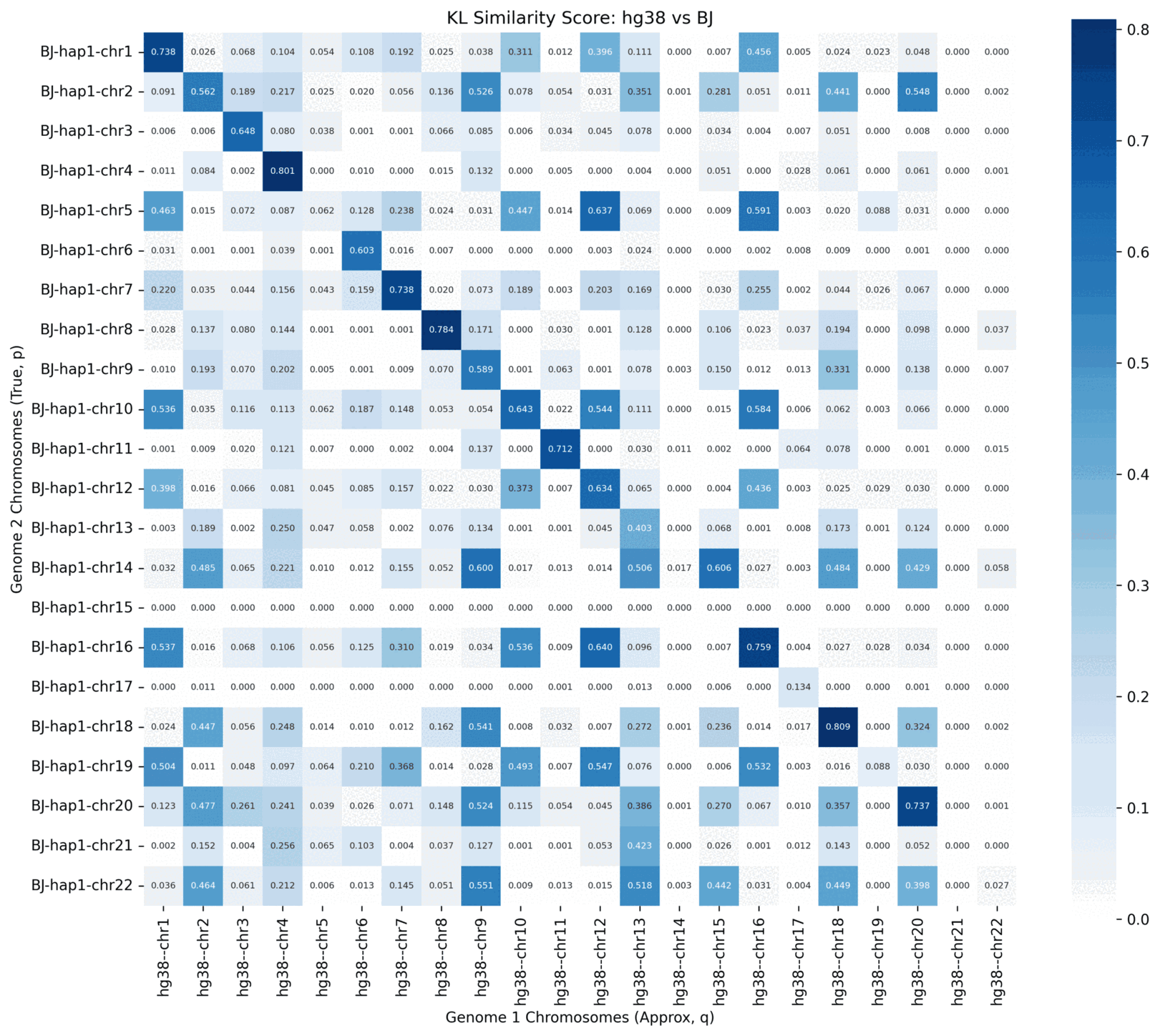}
   \caption{\textsc{grc}h\oldstylenums{38} vs \textsc{bj}}
\end{subfigure}
\caption{Pairwise \textsc{kl} similarity matrices used to score accuracy, entropy, InfoNCE \textemdash\space hap1 only.}
\end{figure}

\newpage

\begin{figure}[htbp]
\centering
\begin{subfigure}{.5\textwidth}
   \centering
   \includegraphics[width=\textwidth]{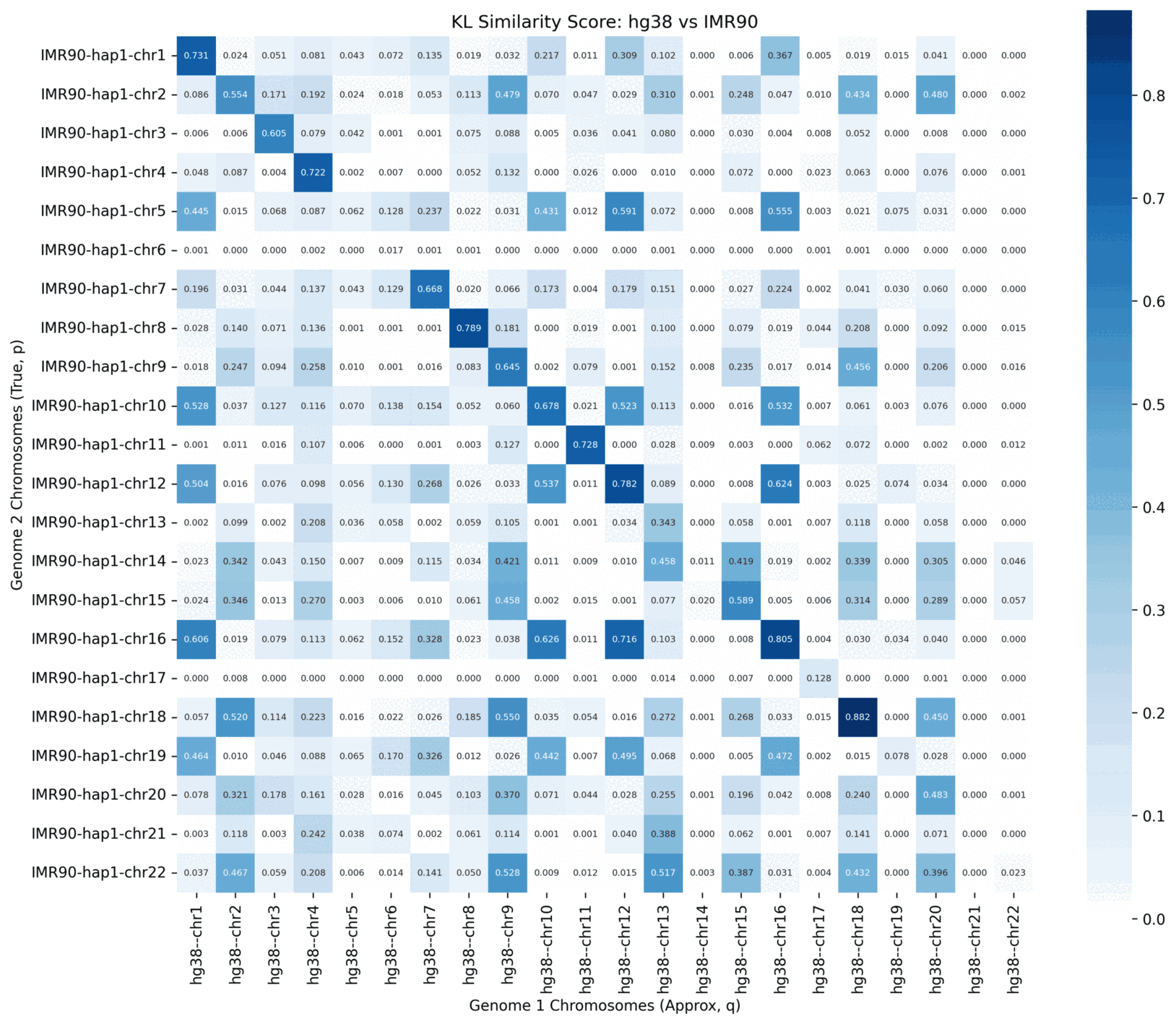}
   \caption{\textsc{grc}h\oldstylenums{38} vs \textsc{imr}\oldstylenums{90}}
\end{subfigure}\hfill
\begin{subfigure}{.5\textwidth}
   \centering
   \includegraphics[width=\textwidth]{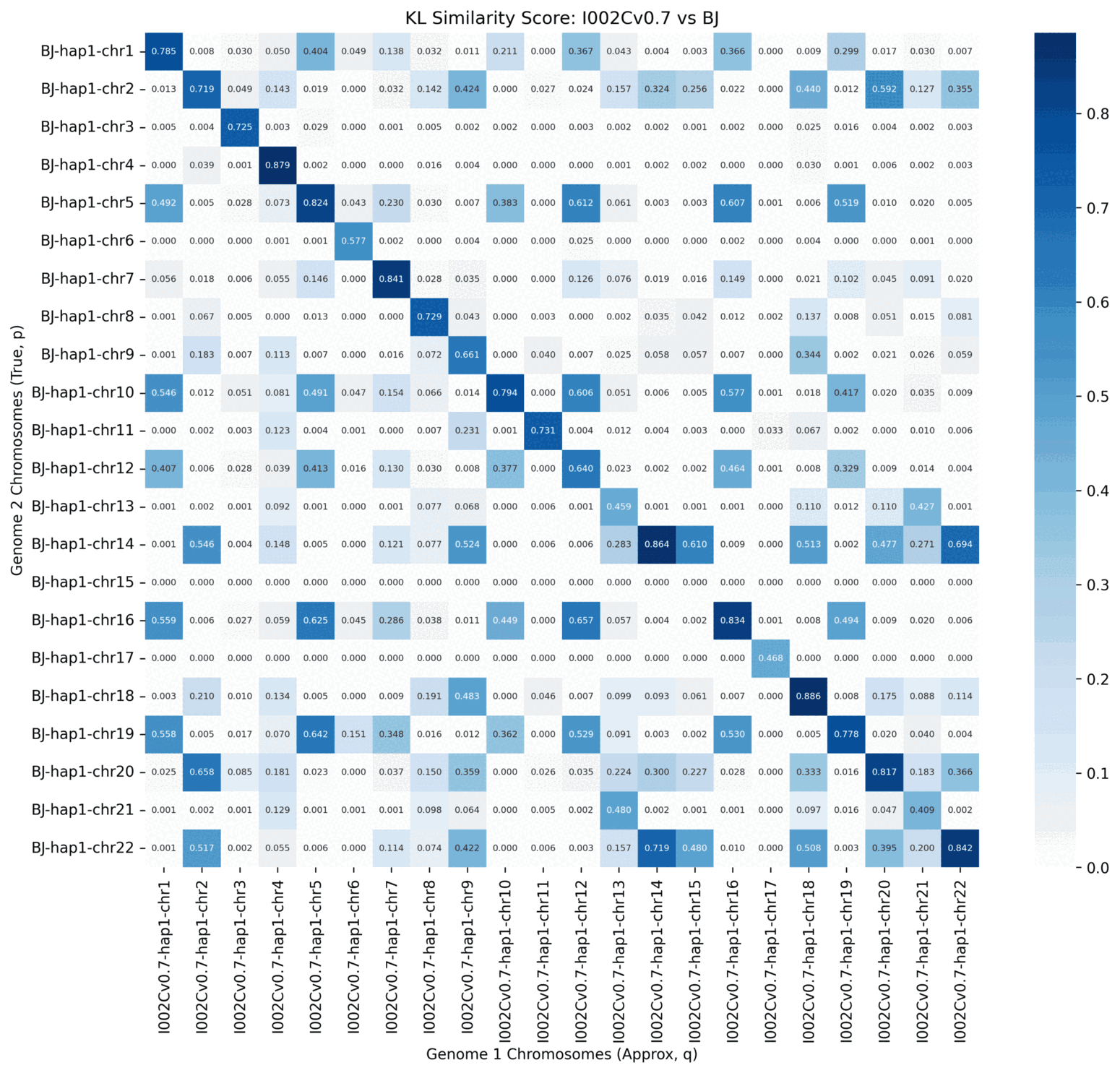}
   \caption{\textsc{i}\oldstylenums{002}\textsc{c}v\oldstylenums{0.7} vs \textsc{bj}}
\end{subfigure}

\vspace{1em}
\begin{subfigure}{.5\textwidth}
   \centering
   \includegraphics[width=\textwidth]{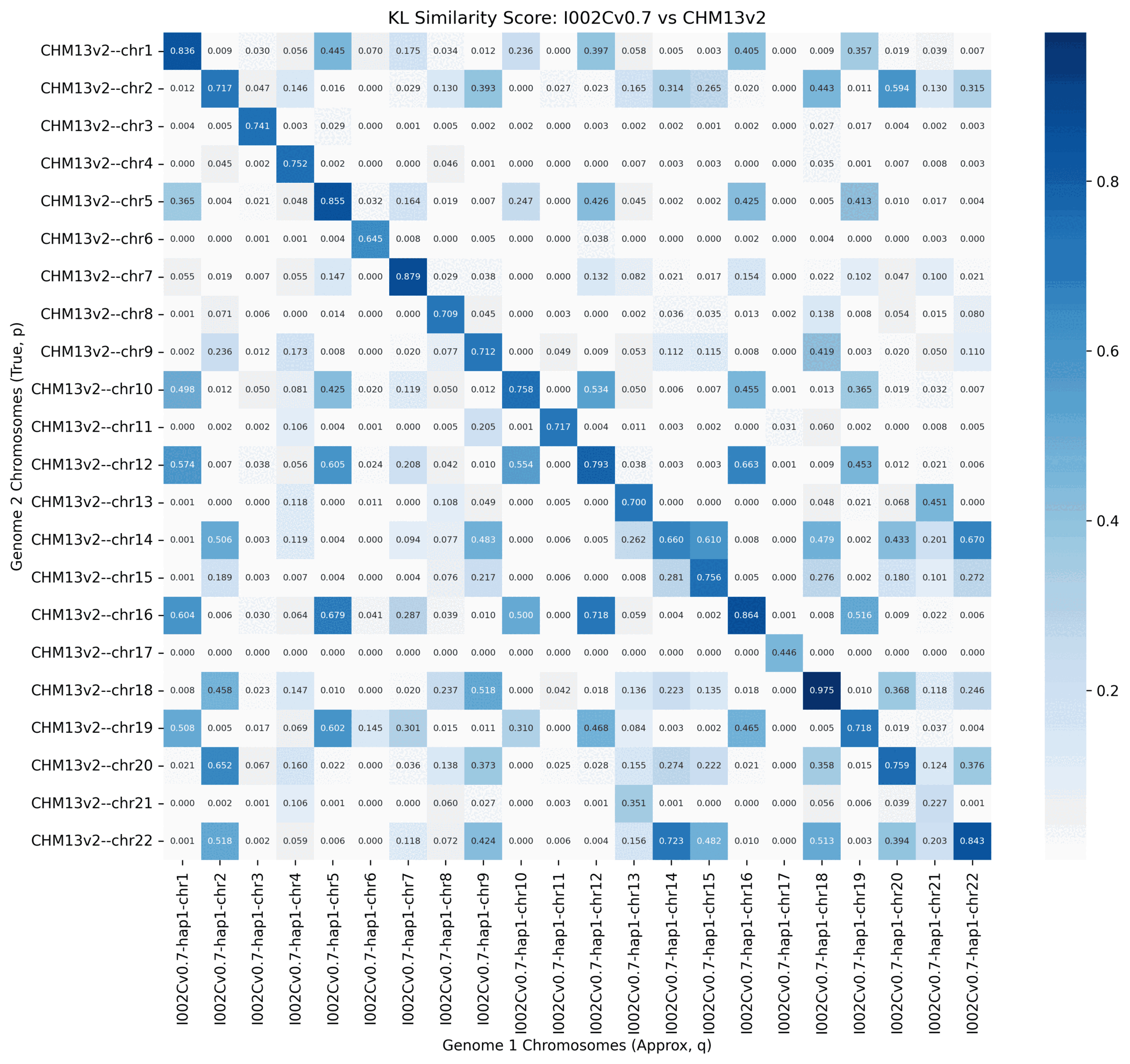}
   \caption{\textsc{i}\oldstylenums{002}\textsc{c}v\oldstylenums{0.7} vs \textsc{chm}\oldstylenums{13}v\oldstylenums{2.0}}
\end{subfigure}\hfill
\begin{subfigure}{.5\textwidth}
   \centering
   \includegraphics[width=\textwidth]{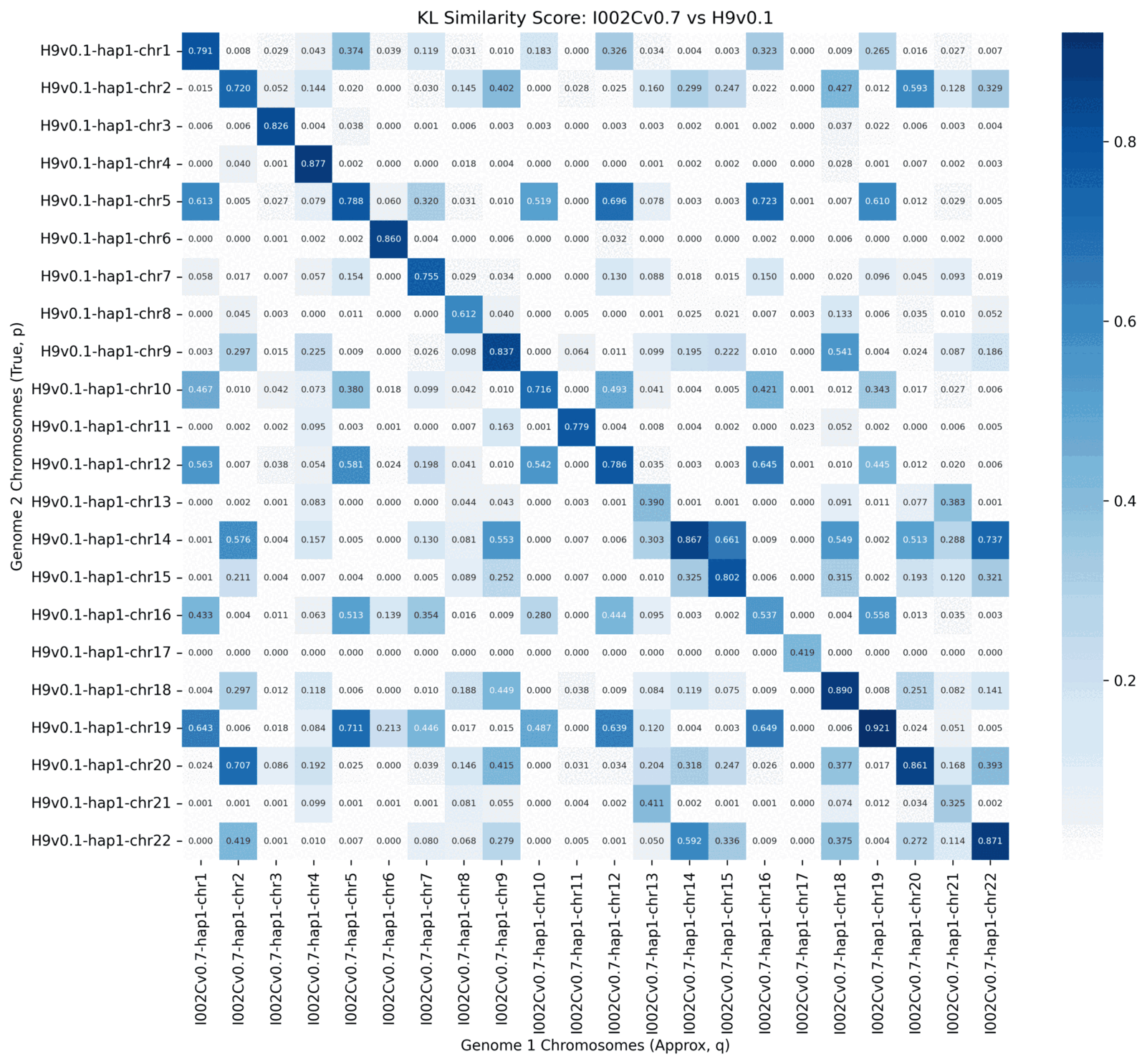}
   \caption{\textsc{i}\oldstylenums{002}\textsc{c}v\oldstylenums{0.7} vs \textsc{h}\oldstylenums{9}v\oldstylenums{0.1}}
\end{subfigure}
\caption{Pairwise \textsc{kl} similarity matrices used to score accuracy, entropy, InfoNCE \textemdash\space hap1 only.}
\end{figure}

\newpage

\begin{figure}[htbp]
\centering
\begin{subfigure}{.5\textwidth}
   \centering
   \includegraphics[width=\textwidth]{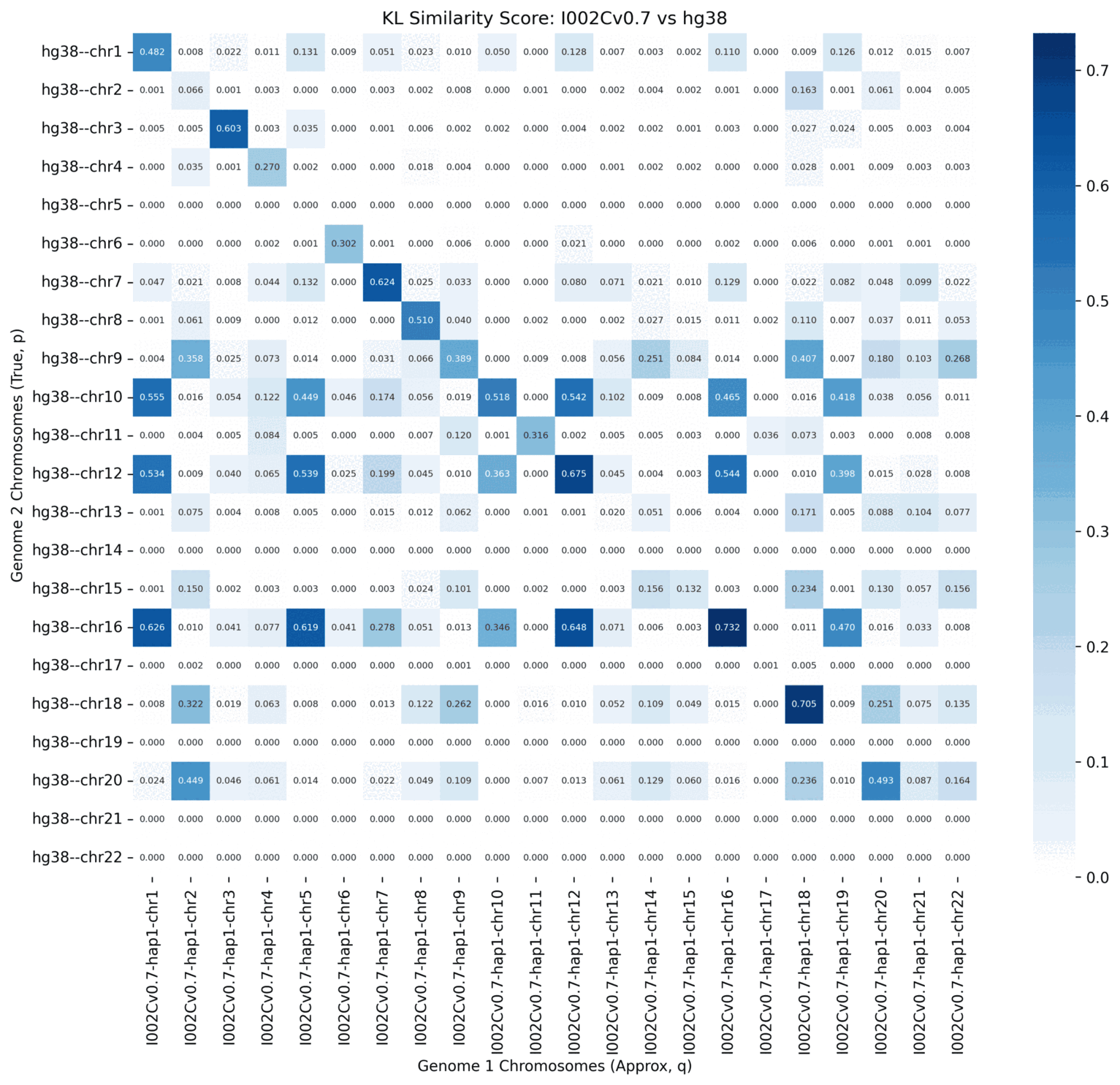}
   \caption{\textsc{i}\oldstylenums{002}\textsc{c}v\oldstylenums{0.7} vs \textsc{grc}h\oldstylenums{38}}
\end{subfigure}\hfill
\begin{subfigure}{.5\textwidth}
   \centering
   \includegraphics[width=\textwidth]{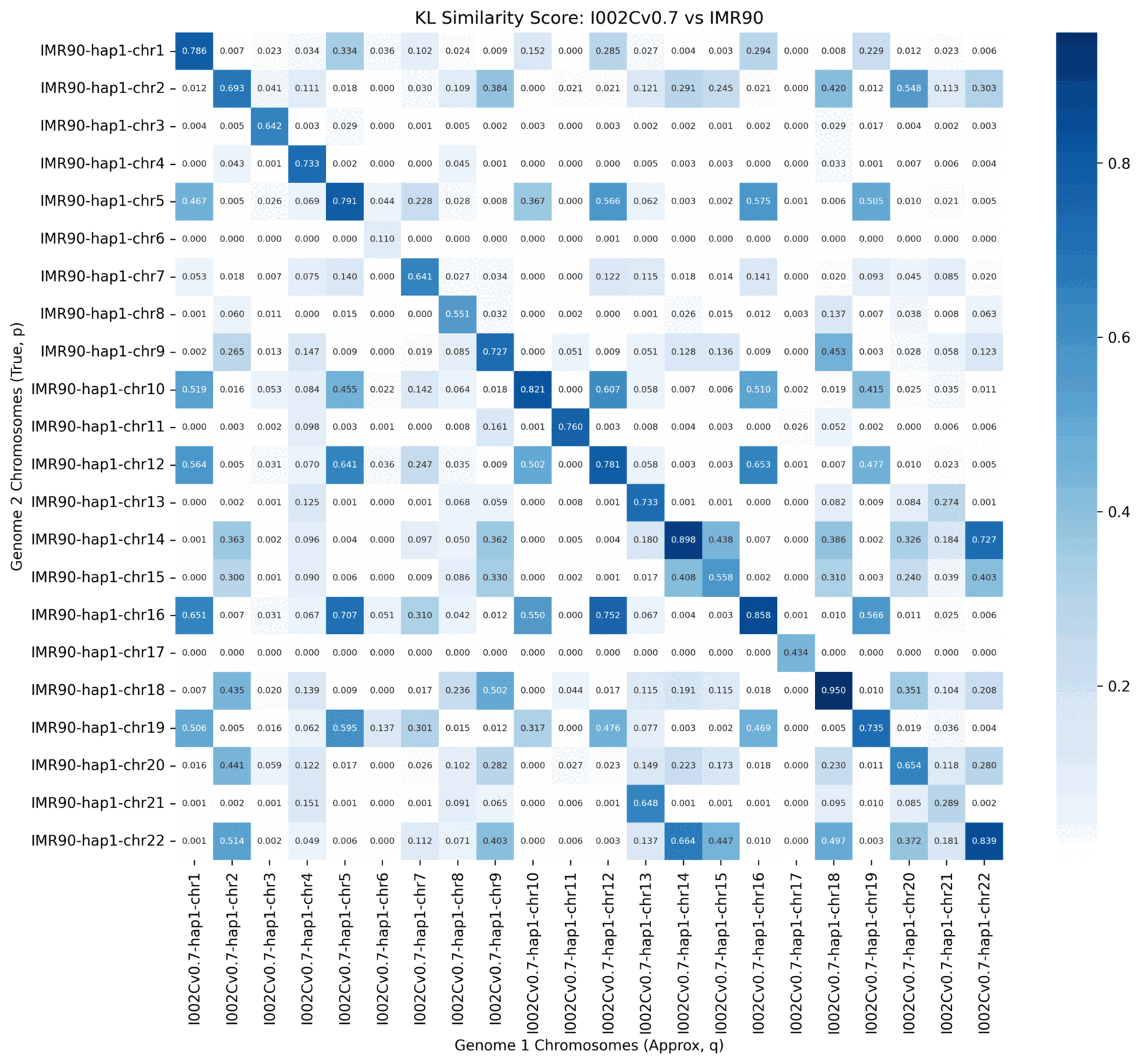}
   \caption{\textsc{i}\oldstylenums{002}\textsc{c}v\oldstylenums{0.7} vs \textsc{imr}\oldstylenums{90}}
\end{subfigure}

\vspace{1em}
\begin{subfigure}{.5\textwidth}
   \centering
   \includegraphics[width=\textwidth]{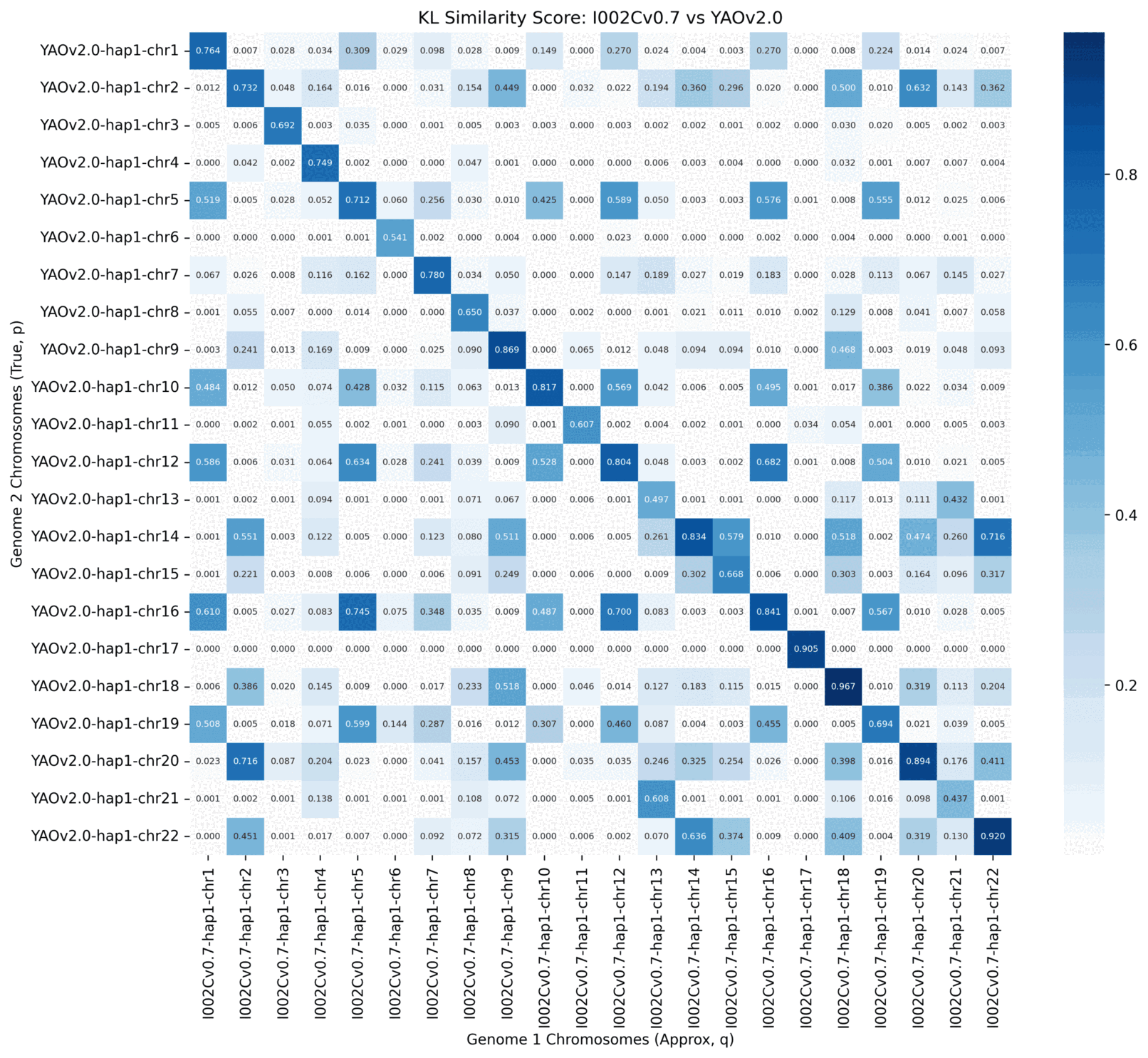}
   \caption{\textsc{i}\oldstylenums{002}\textsc{c}v\oldstylenums{0.7} vs \textsc{yao}v\oldstylenums{2.0}}
\end{subfigure}\hfill
\begin{subfigure}{.5\textwidth}
   \centering
   \includegraphics[width=\textwidth]{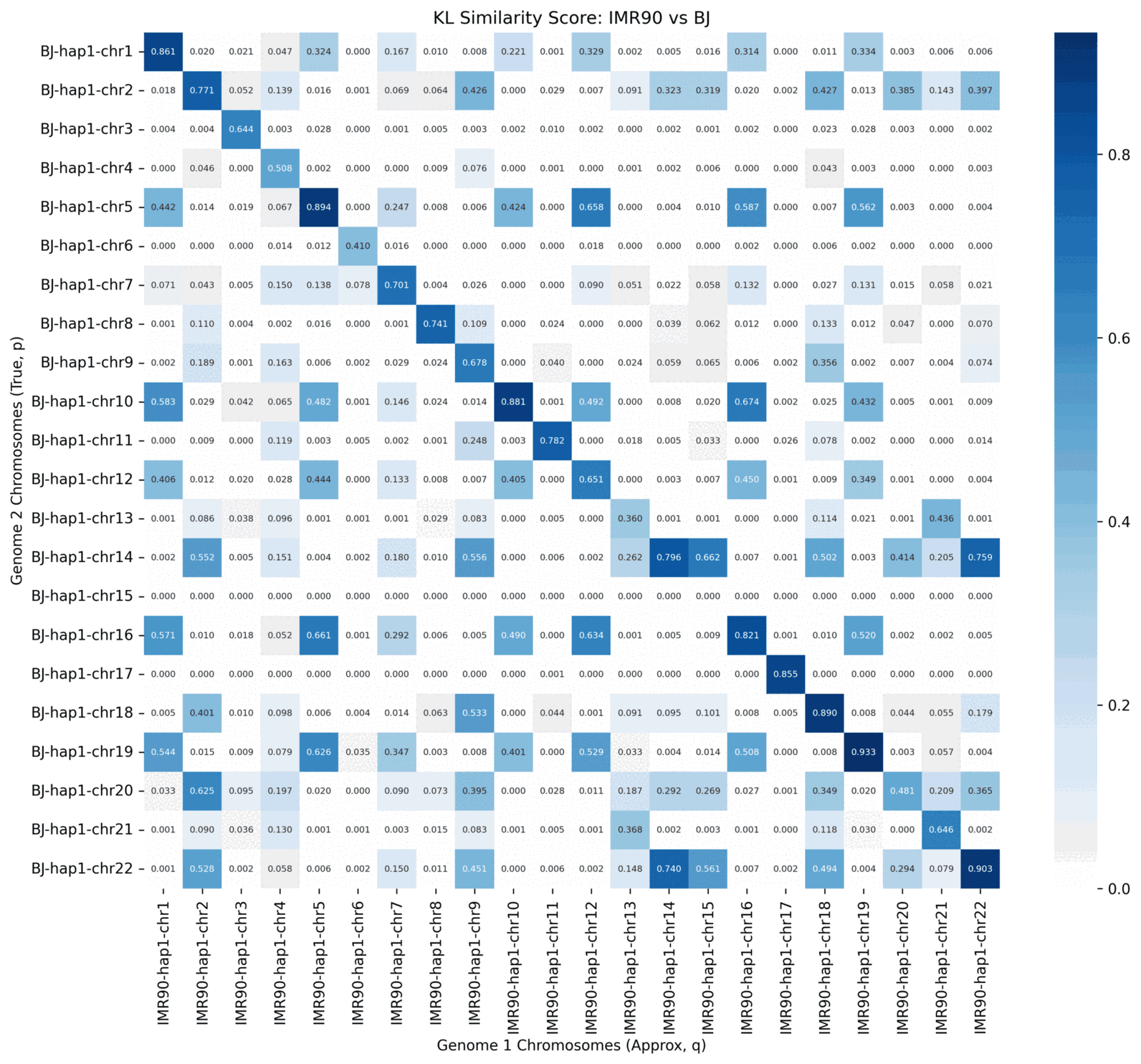}
   \caption{\textsc{imr}\oldstylenums{90} vs \textsc{bj}}
\end{subfigure}
\caption{Pairwise \textsc{kl} similarity matrices used to score accuracy, entropy, InfoNCE \textemdash\space hap1 only.}
\end{figure}

\newpage

\begin{figure}[htbp]
\centering
\begin{subfigure}{.5\textwidth}
   \centering
   \includegraphics[width=\textwidth]{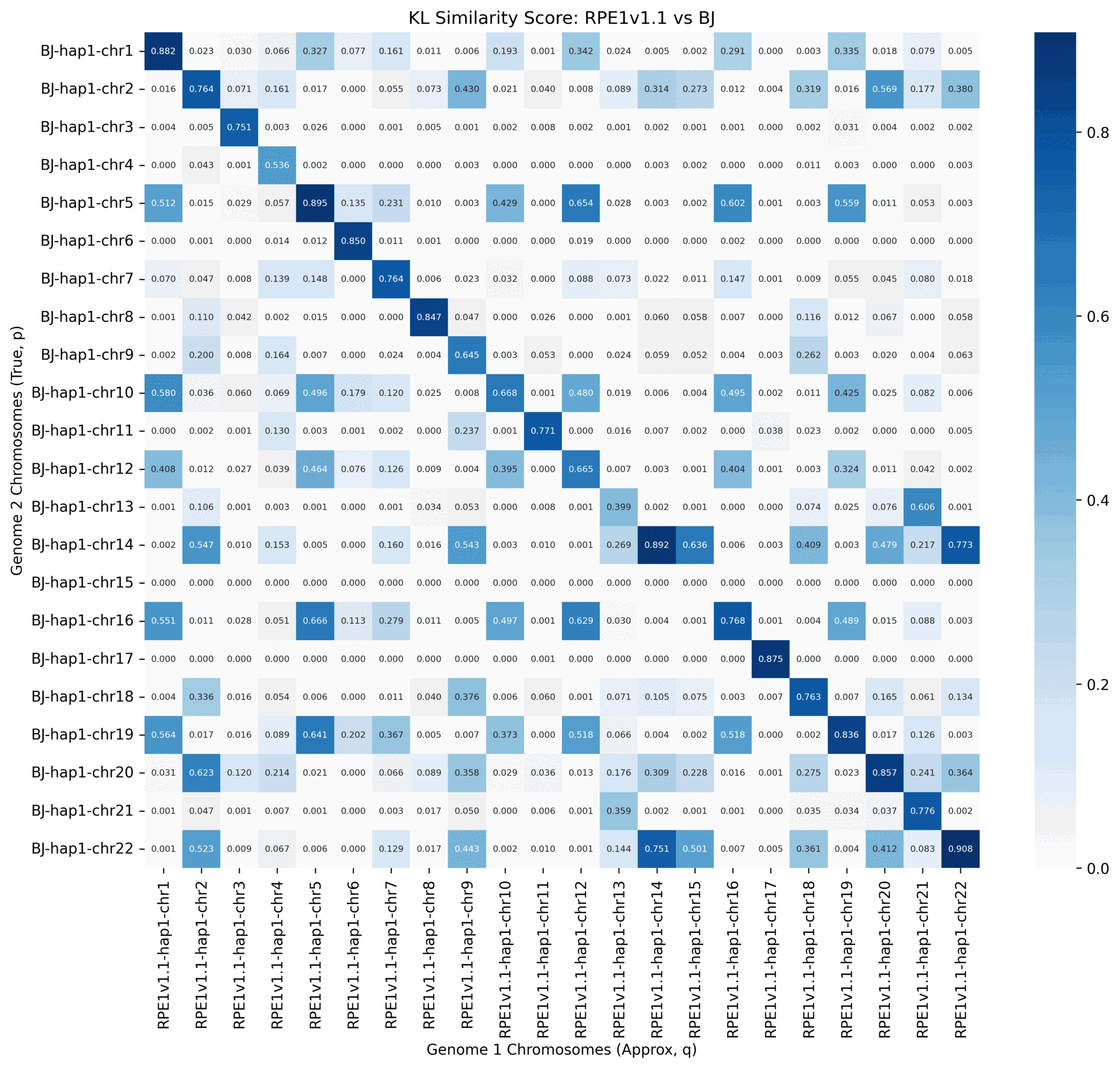}
   \caption{\textsc{rpe}\oldstylenums{1}v\oldstylenums{1.1} vs \textsc{bj}}
\end{subfigure}\hfill
\begin{subfigure}{.5\textwidth}
   \centering
   \includegraphics[width=\textwidth]{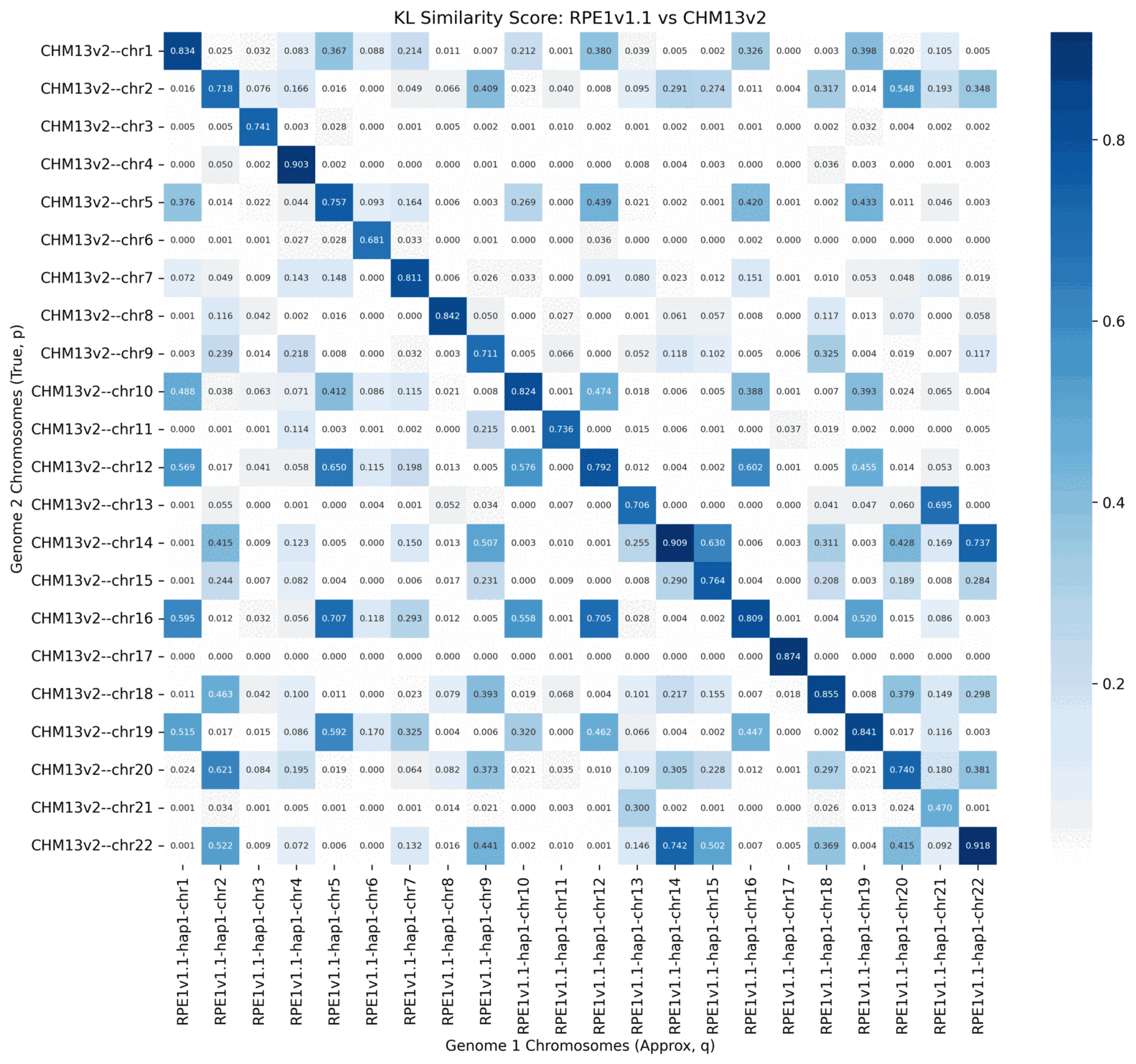}
   \caption{\textsc{rpe}\oldstylenums{1}v\oldstylenums{1.1} vs \textsc{chm}\oldstylenums{13}v\oldstylenums{2.0}}
\end{subfigure}

\vspace{1em}
\begin{subfigure}{.5\textwidth}
   \centering
   \includegraphics[width=\textwidth]{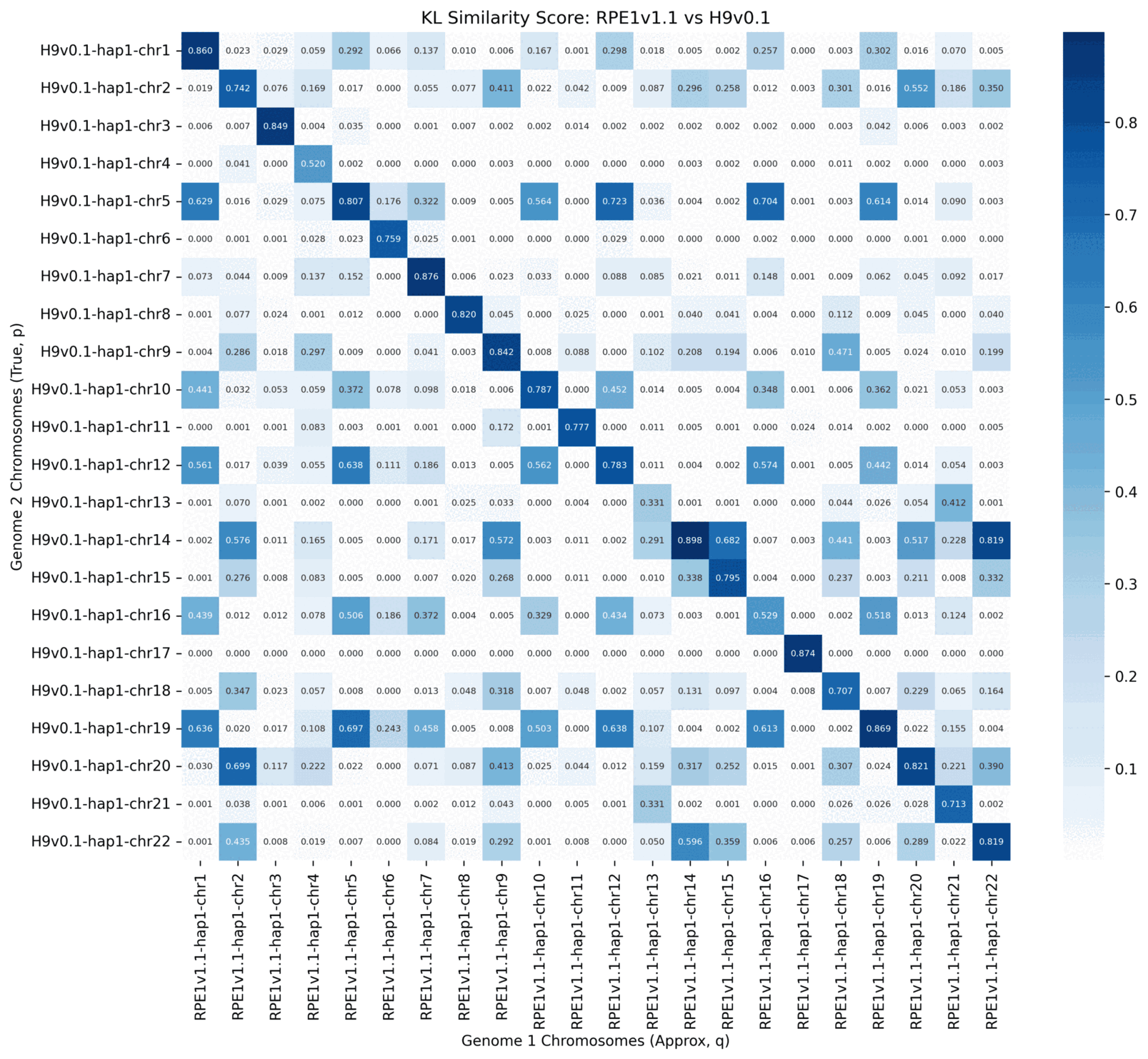}
   \caption{\textsc{rpe}\oldstylenums{1}v\oldstylenums{1.1} vs \textsc{h}\oldstylenums{9}v\oldstylenums{0.1}}
\end{subfigure}\hfill
\begin{subfigure}{.5\textwidth}
   \centering
   \includegraphics[width=\textwidth]{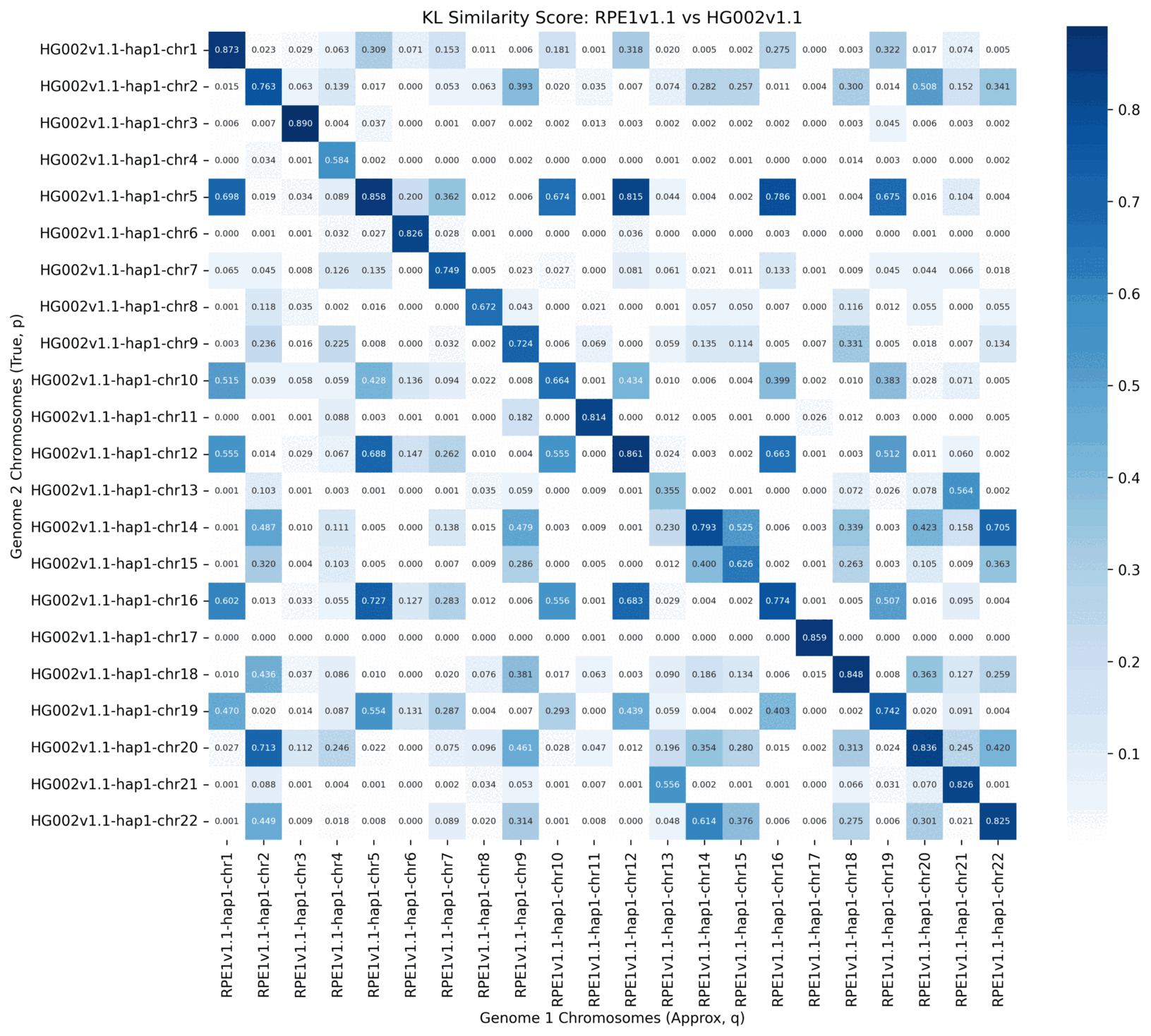}
   \caption{\textsc{rpe}\oldstylenums{1}v\oldstylenums{1.1} vs \textsc{hg}\oldstylenums{002}v\oldstylenums{1.1}}
\end{subfigure}
\caption{Pairwise \textsc{kl} similarity matrices used to score accuracy, entropy, InfoNCE \textemdash\space hap1 only.}
\end{figure}

\newpage

\begin{figure}[htbp]
\centering
\begin{subfigure}{.5\textwidth}
   \centering
   \includegraphics[width=\textwidth]{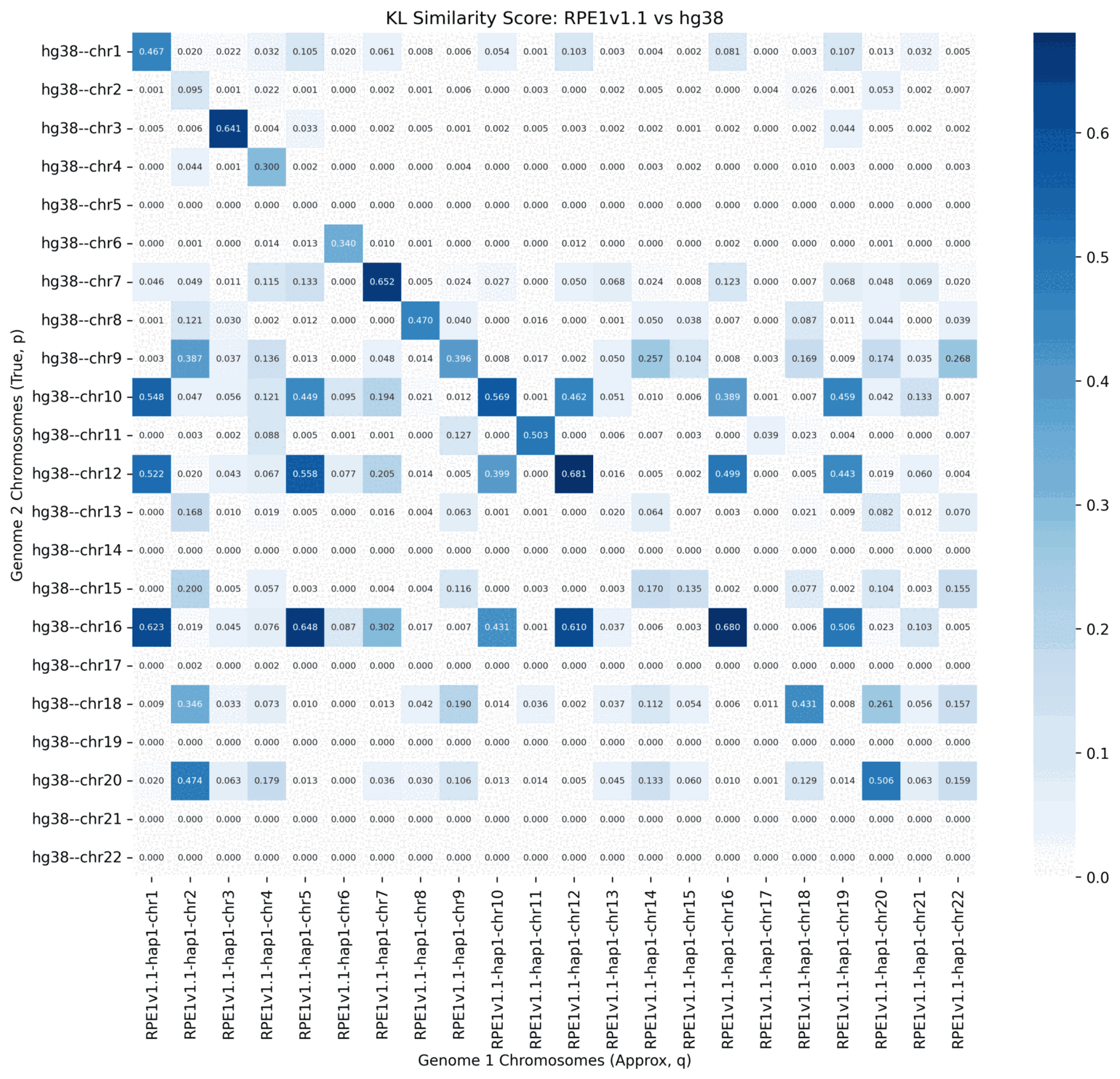}
   \caption{\textsc{rpe}\oldstylenums{1}v\oldstylenums{1.1} vs \textsc{grc}h\oldstylenums{38}}
\end{subfigure}\hfill
\begin{subfigure}{.5\textwidth}
   \centering
   \includegraphics[width=\textwidth]{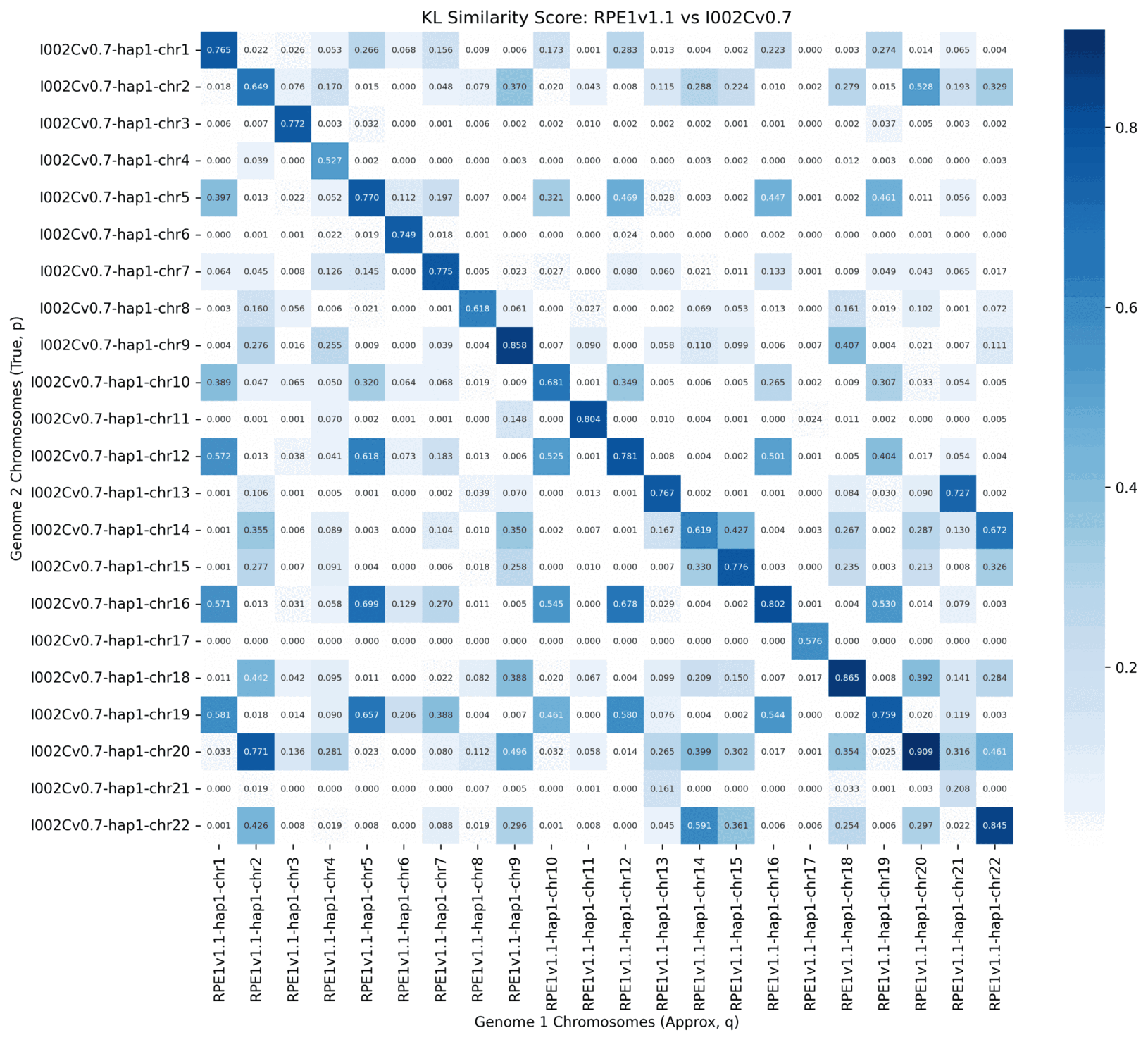}
   \caption{\textsc{rpe}\oldstylenums{1}v\oldstylenums{1.1} vs \textsc{i}\oldstylenums{002}\textsc{c}v\oldstylenums{0.7}}
\end{subfigure}

\vspace{1em}
\begin{subfigure}{.5\textwidth}
   \centering
   \includegraphics[width=\textwidth]{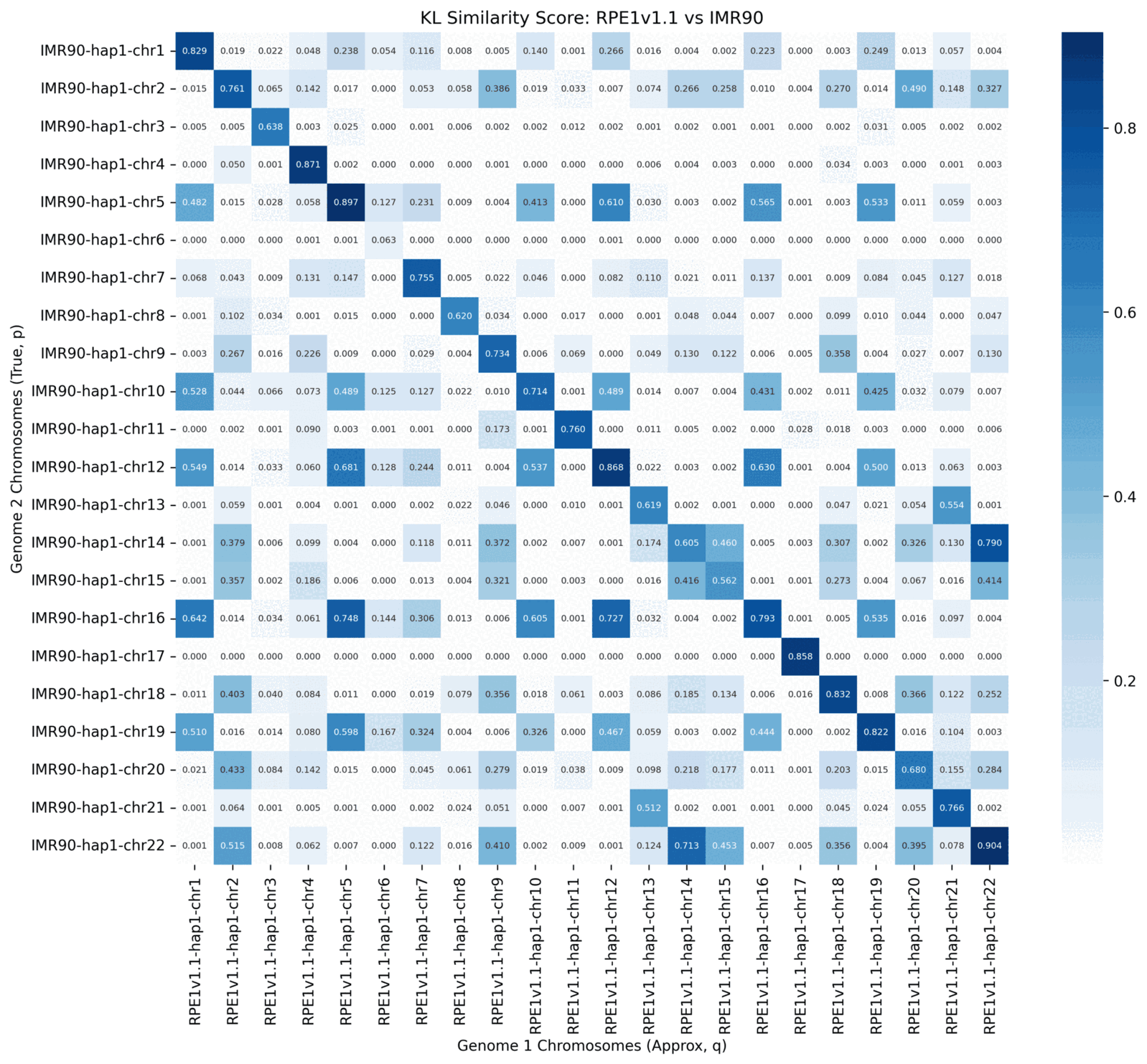}
   \caption{\textsc{rpe}\oldstylenums{1}v\oldstylenums{1.1} vs \textsc{imr}\oldstylenums{90}}
\end{subfigure}\hfill
\begin{subfigure}{.5\textwidth}
   \centering
   \includegraphics[width=\textwidth]{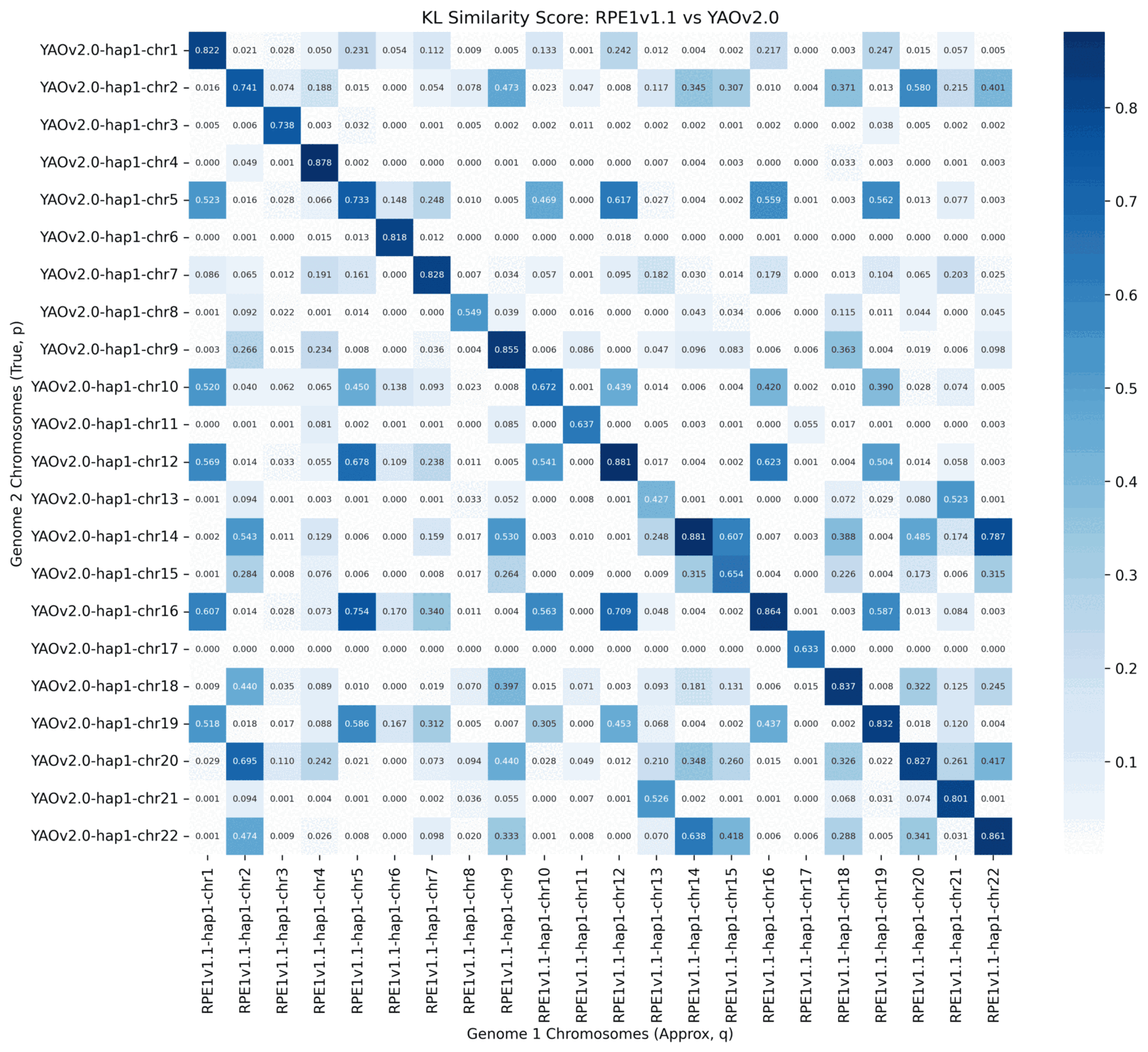}
   \caption{\textsc{rpe}\oldstylenums{1}v\oldstylenums{1.1} vs \textsc{yao}v\oldstylenums{2.0}}
\end{subfigure}
\caption{Pairwise \textsc{kl} similarity matrices used to score accuracy, entropy, InfoNCE \textemdash\space hap1 only.}
\end{figure}

\newpage

\begin{figure}[htbp]
\centering
\begin{subfigure}{.5\textwidth}
   \centering
   \includegraphics[width=\textwidth]{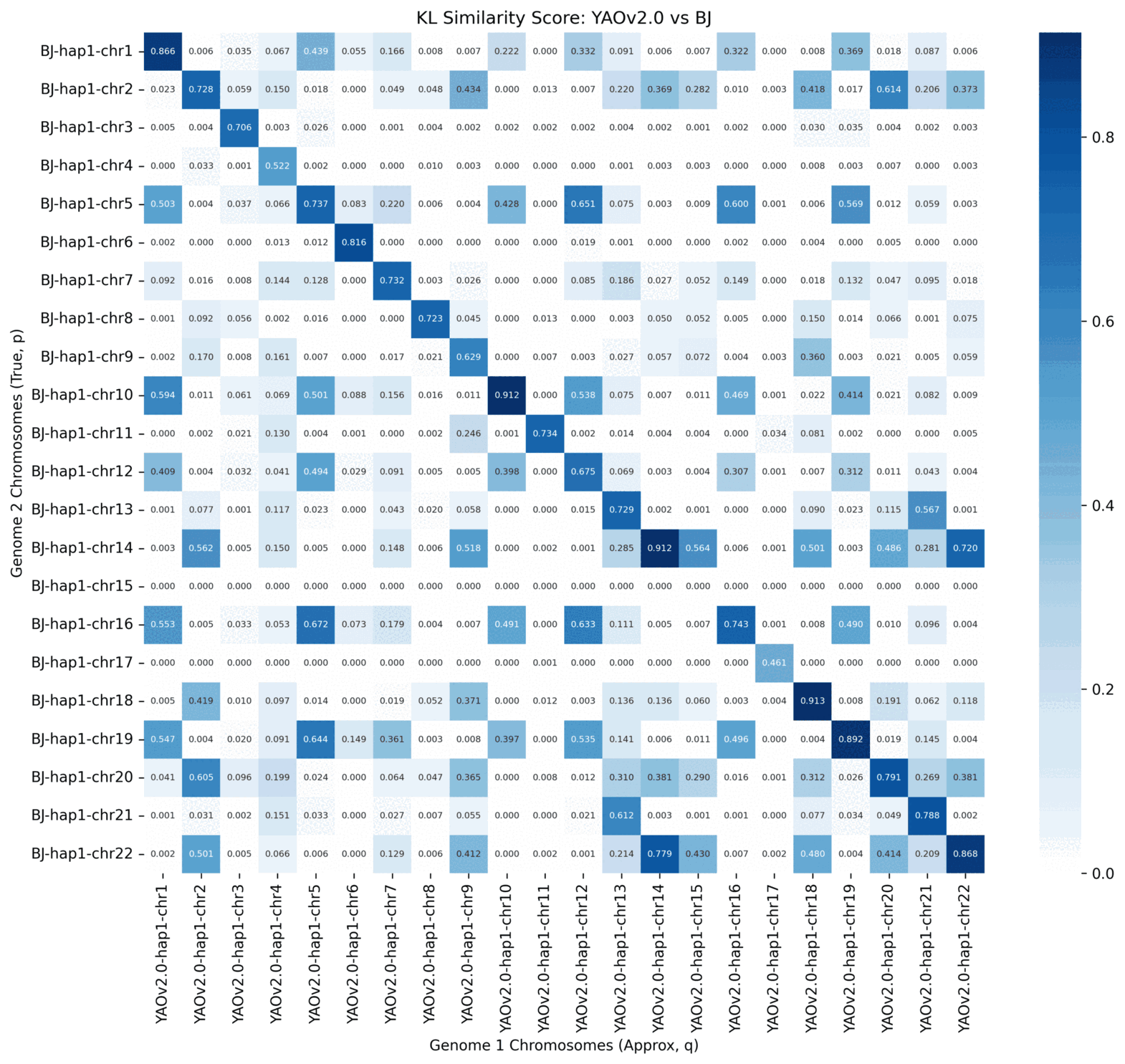}
   \caption{\textsc{yao}v\oldstylenums{2.0} vs \textsc{bj}}
\end{subfigure}\hfill
\begin{subfigure}{.5\textwidth}
   \centering
   \includegraphics[width=\textwidth]{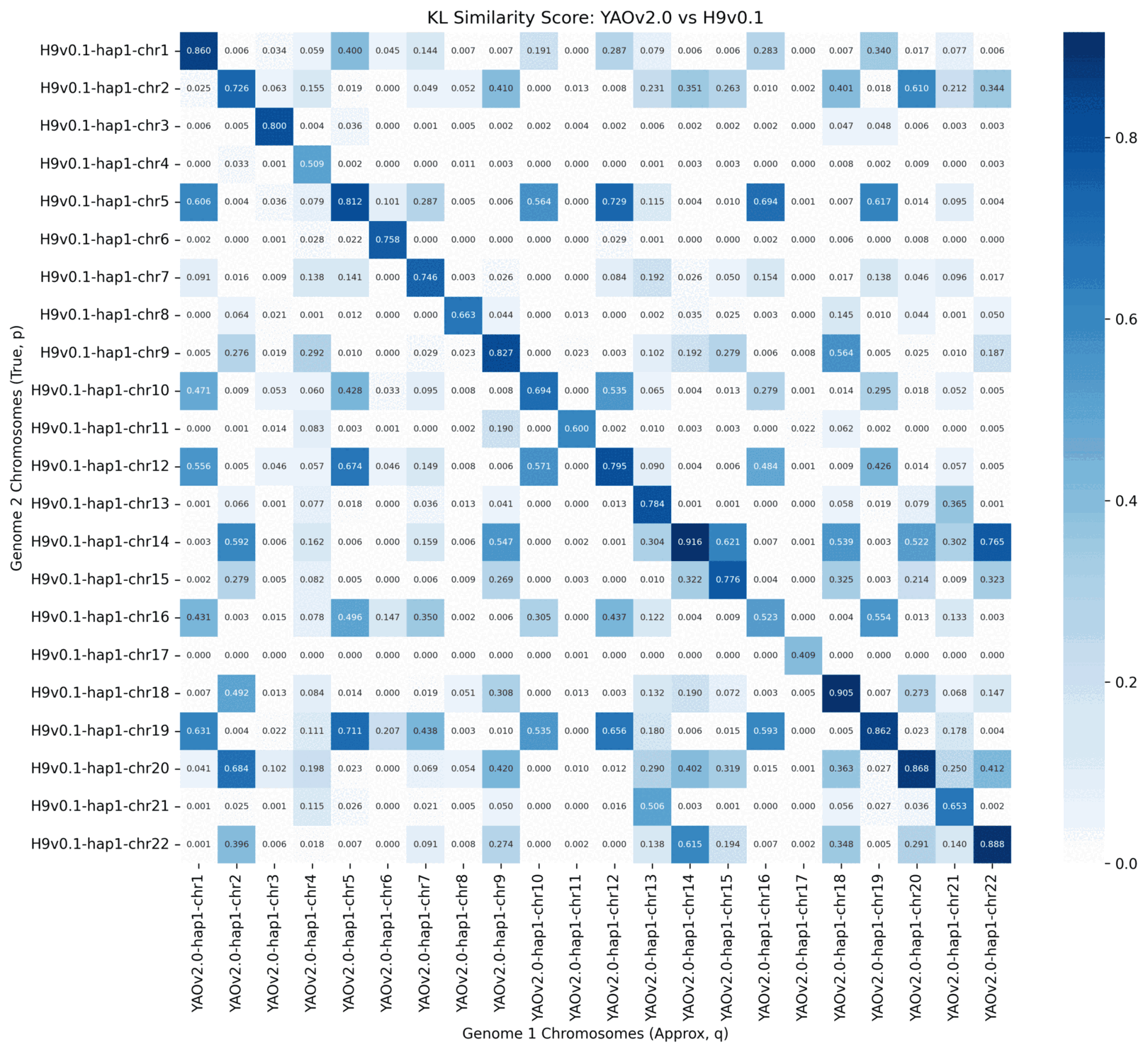}
   \caption{\textsc{yao}v\oldstylenums{2.0} vs \textsc{h}\oldstylenums{9}v\oldstylenums{0.1}}
\end{subfigure}

\vspace{1em}
\begin{subfigure}{.5\textwidth}
   \centering
   \includegraphics[width=\textwidth]{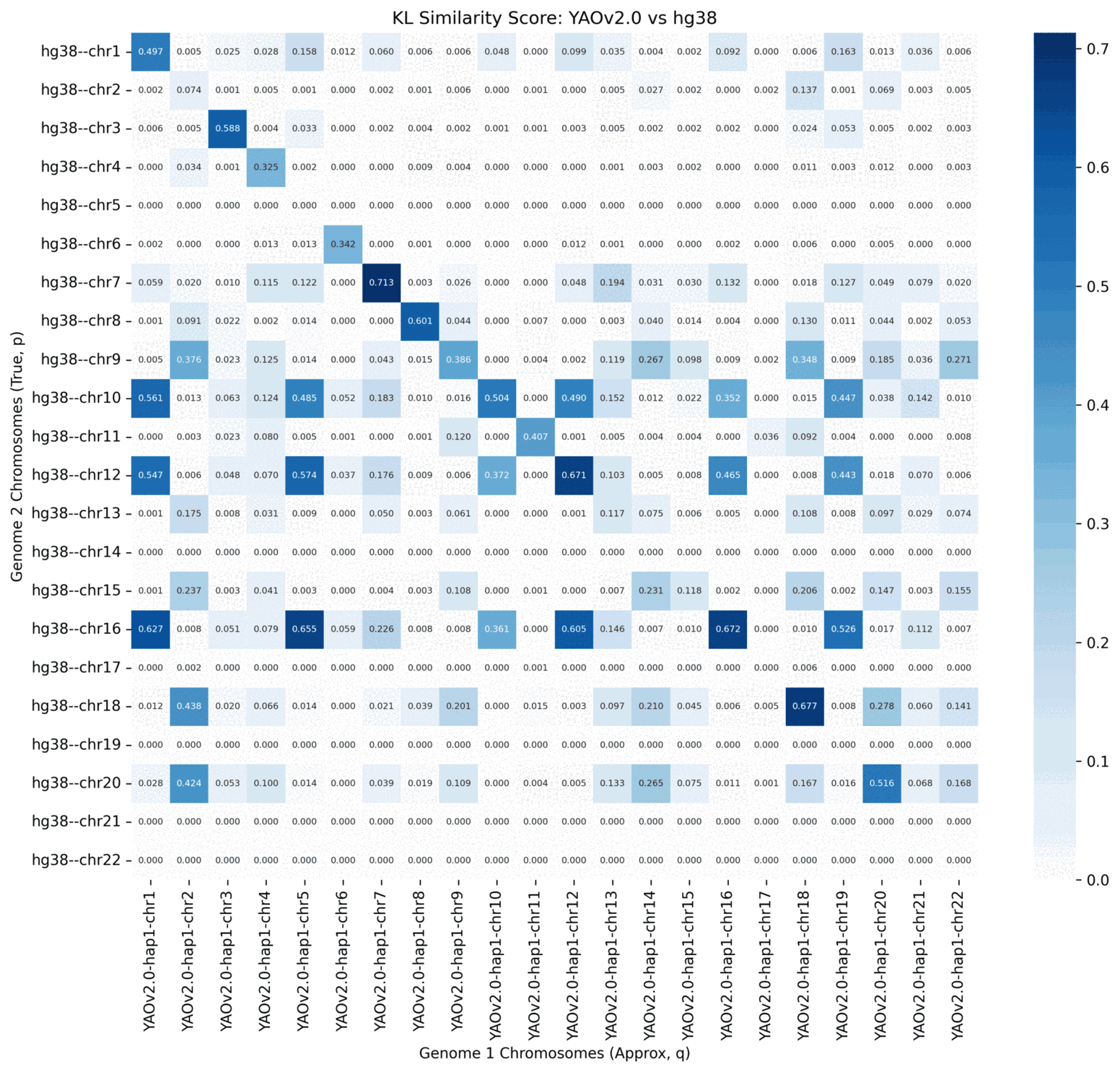}
   \caption{\textsc{yao}v\oldstylenums{2.0} vs \textsc{grc}h\oldstylenums{38}}
\end{subfigure}\hfill
\begin{subfigure}{.5\textwidth}
   \centering
   \includegraphics[width=\textwidth]{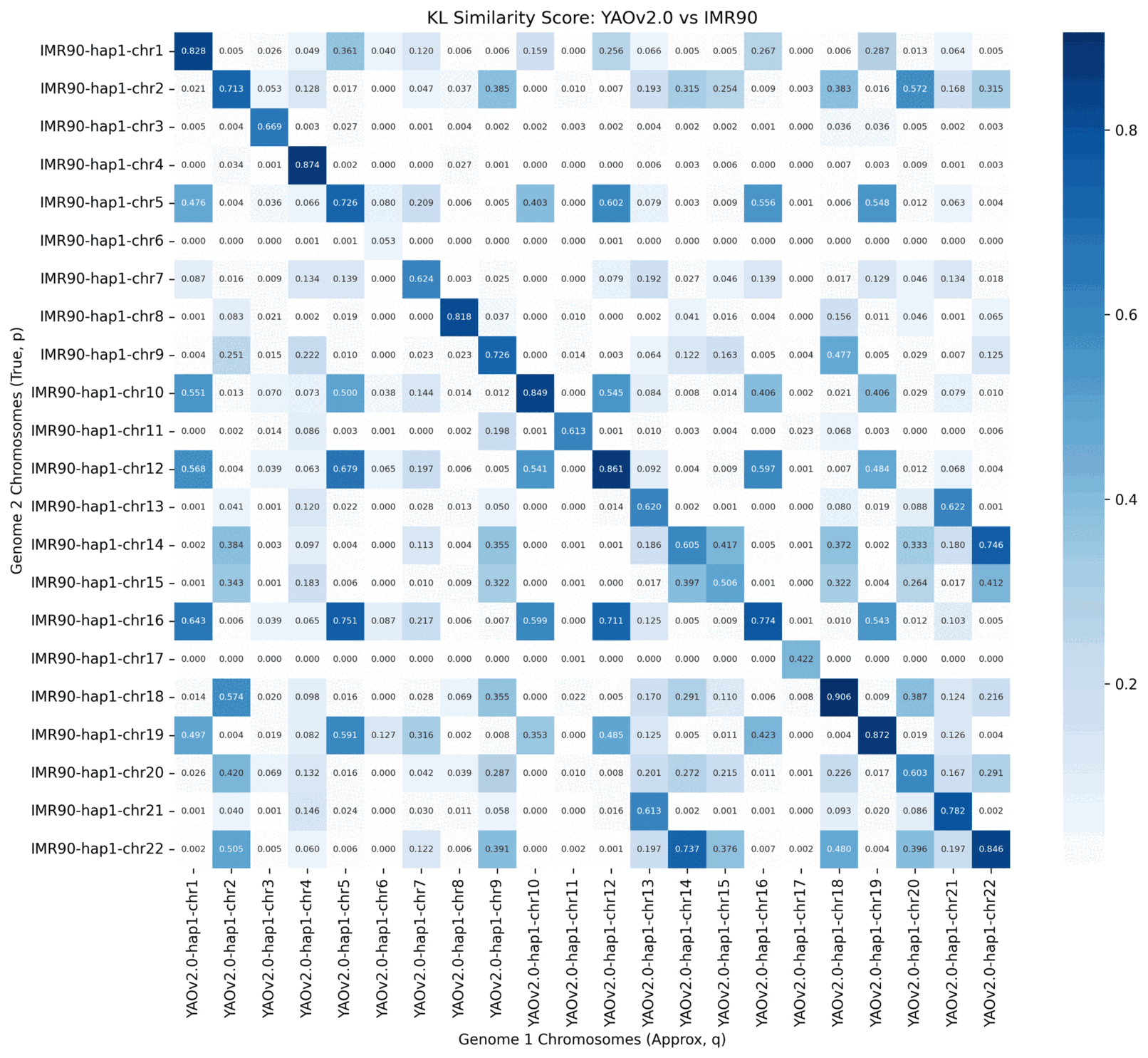}
   \caption{\textsc{yao}v\oldstylenums{2.0} vs \textsc{imr}\oldstylenums{90}}
\end{subfigure}
\caption{Pairwise \textsc{kl} similarity matrices used to score accuracy, entropy, InfoNCE \textemdash\space hap1 only.}
\end{figure}

\begin{figure}[htbp]
\centering
\begin{subfigure}{.4\textwidth}
  \centering
  \includegraphics[width=\textwidth]{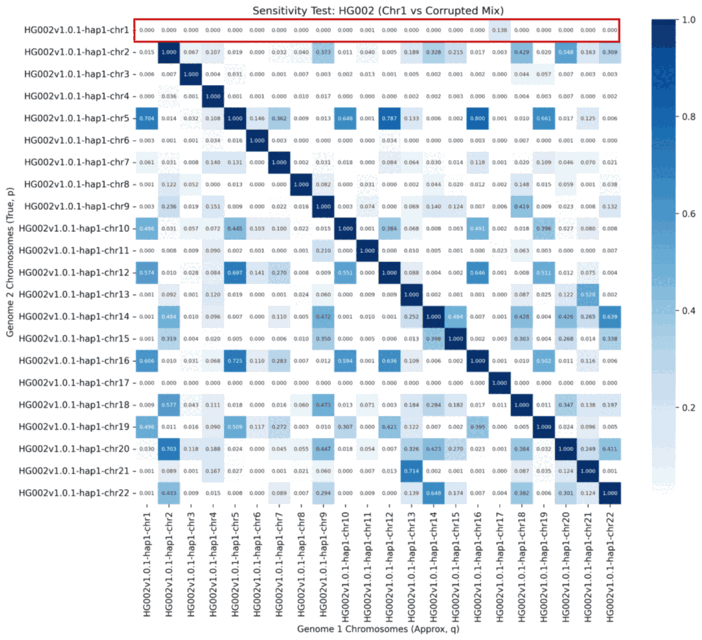}
  \caption{Contamination of chromosome 1 by chromosome 17 of \textsc{hg}\oldstylenums{002}}
\end{subfigure}

\begin{subfigure}{.4\textwidth}
  \centering
  \includegraphics[width=\textwidth]{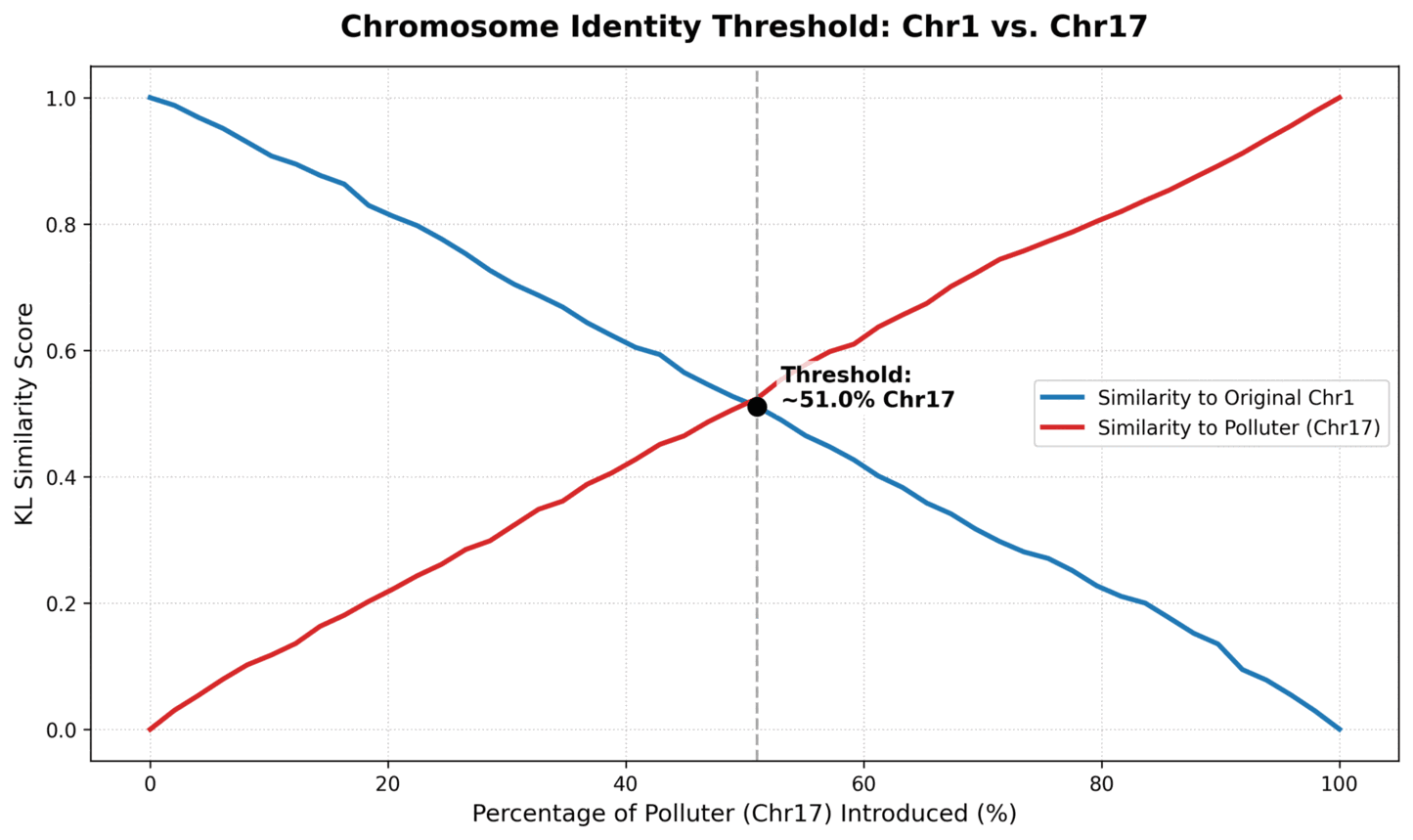}
  \caption{Threshold analysis of the amount of chromosome 17 needed to be injected in chromosome 1 to reflect a change in the \textsc{kl} score}
\end{subfigure}

\begin{subfigure}{.4\textwidth}
  \centering
  \includegraphics[width=\textwidth]{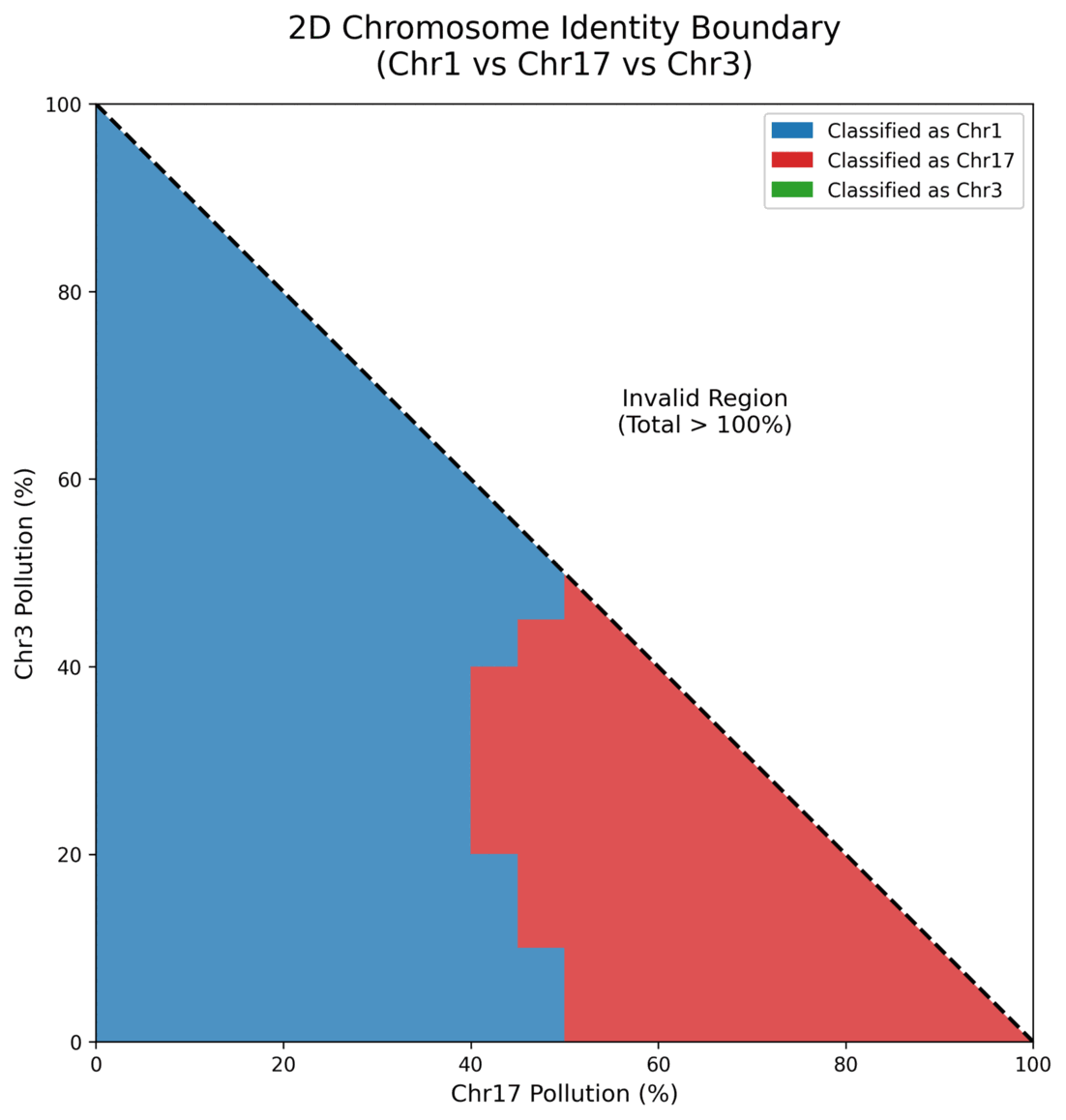}
  \caption{2D threshold analysis of the amount of chromosome 17 and chromosome 3 needed to be injected in chromosome 1 to reflect a change in the \textsc{kl} score}
\end{subfigure}
\caption{Overall corruption analysis and discrimination thresholds.}
\end{figure}
\newpage

\begin{figure}[htbp]
\centering
\begin{subfigure}{.65\textwidth}
  \centering
  \includegraphics[width=\textwidth]{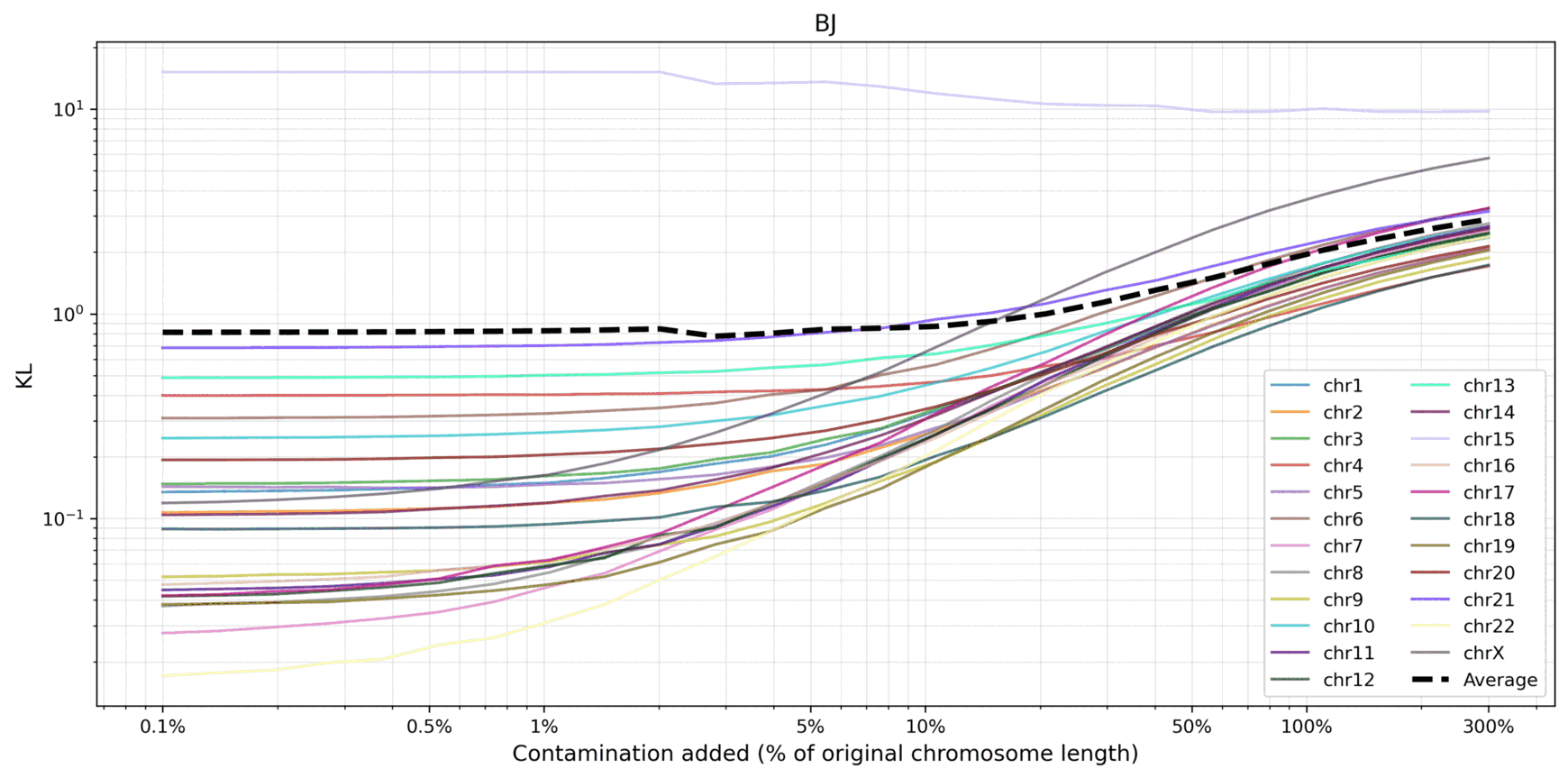}
  \caption{Contamination analysis \textsc{bj}}
\end{subfigure}

\begin{subfigure}{.65\textwidth}
  \centering
  \includegraphics[width=\textwidth]{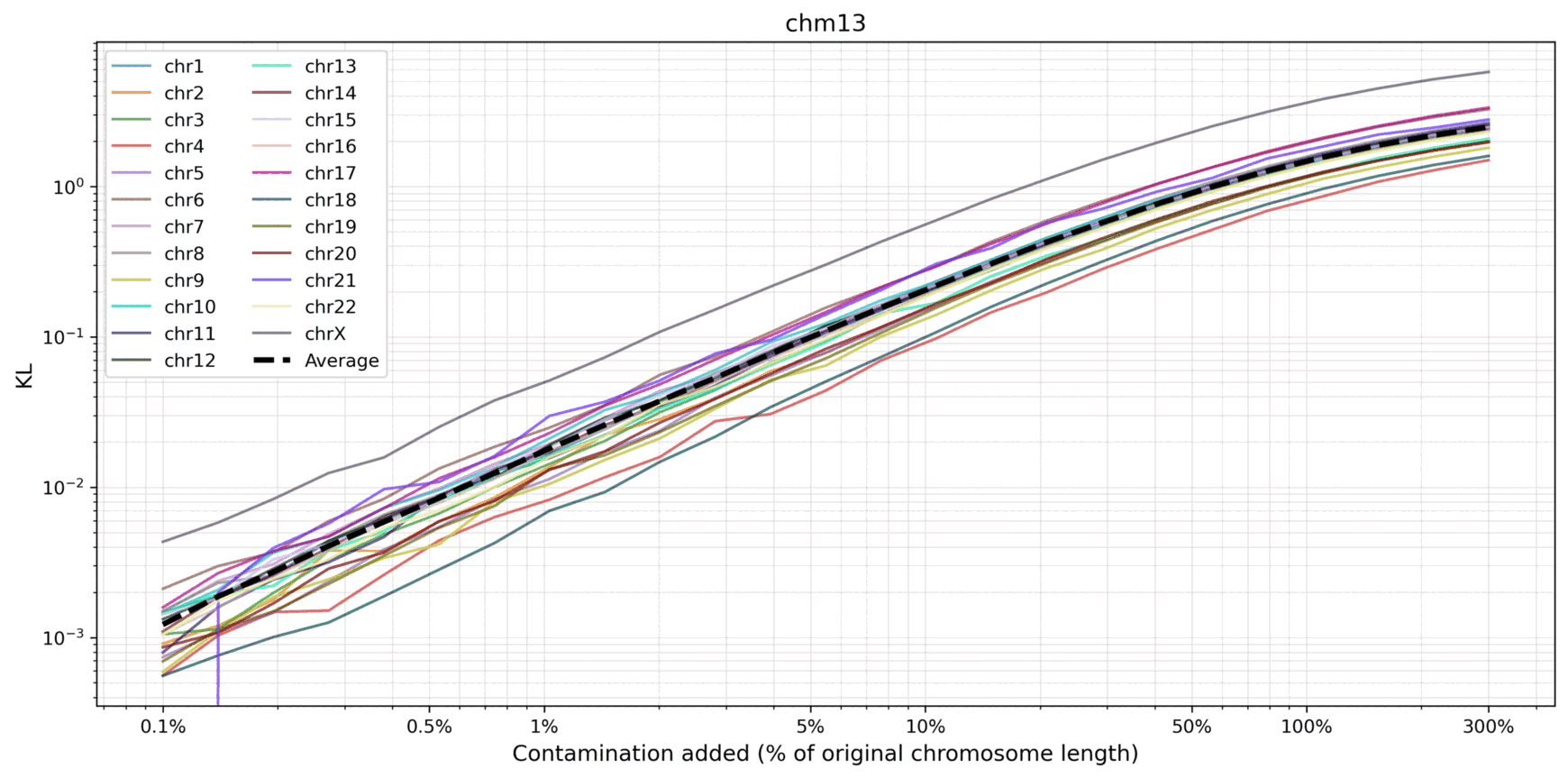}
  \caption{Contamination analysis \textsc{chm}\oldstylenums{13}}
\end{subfigure}

\begin{subfigure}{.65\textwidth}
  \centering
  \includegraphics[width=\textwidth]{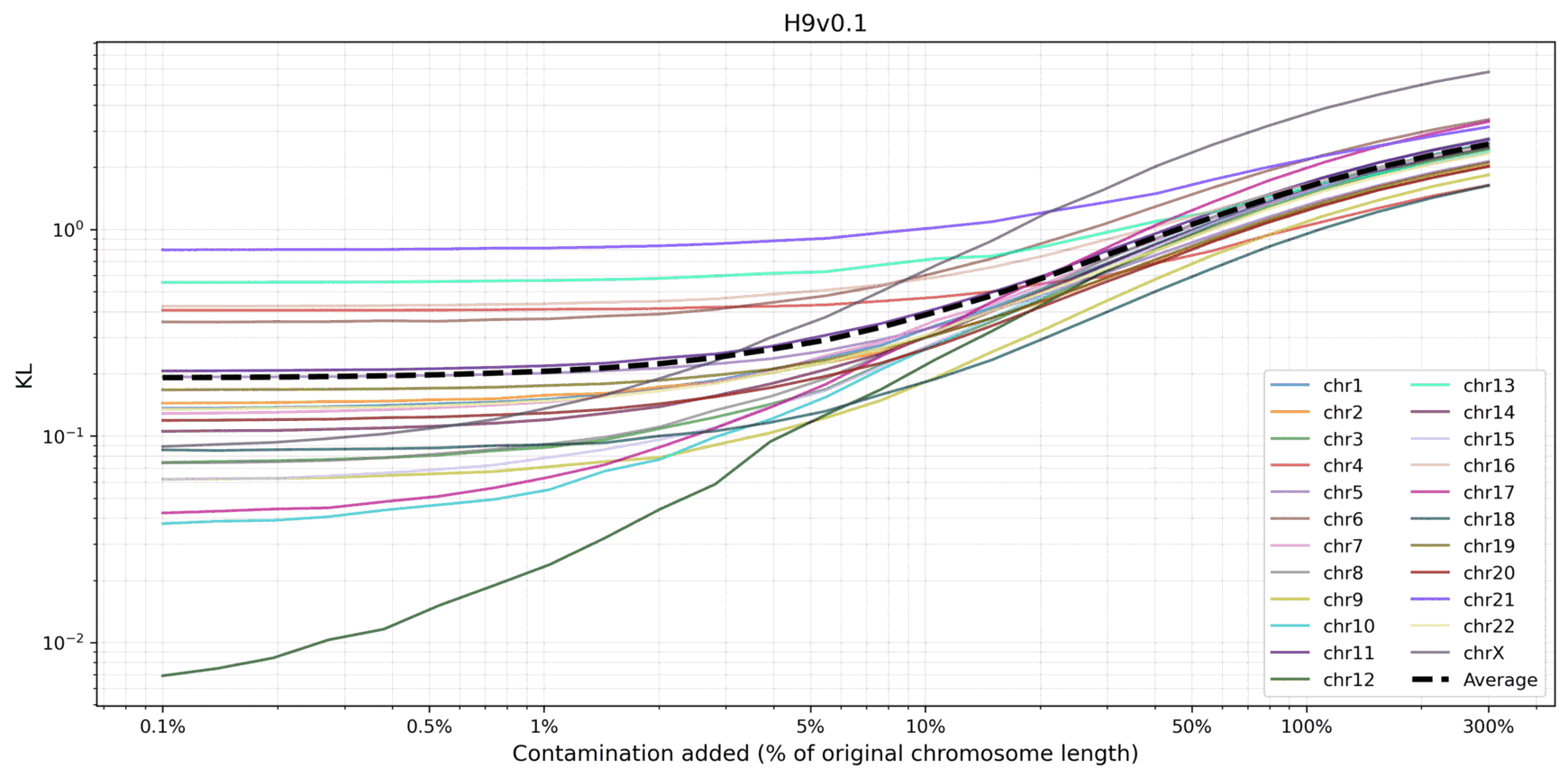}
  \caption{Contamination analysis \textsc{h}\oldstylenums{9}v\oldstylenums{0.1}}
\end{subfigure}
\caption{Contamination analysis for \textsc{bj}, \textsc{chm}\oldstylenums{13}, and \textsc{h}\oldstylenums{9}v\oldstylenums{0.1}.}
\end{figure}
\newpage

\begin{figure}[htbp]
\centering
\begin{subfigure}{.65\textwidth}
  \centering
  \includegraphics[width=\textwidth]{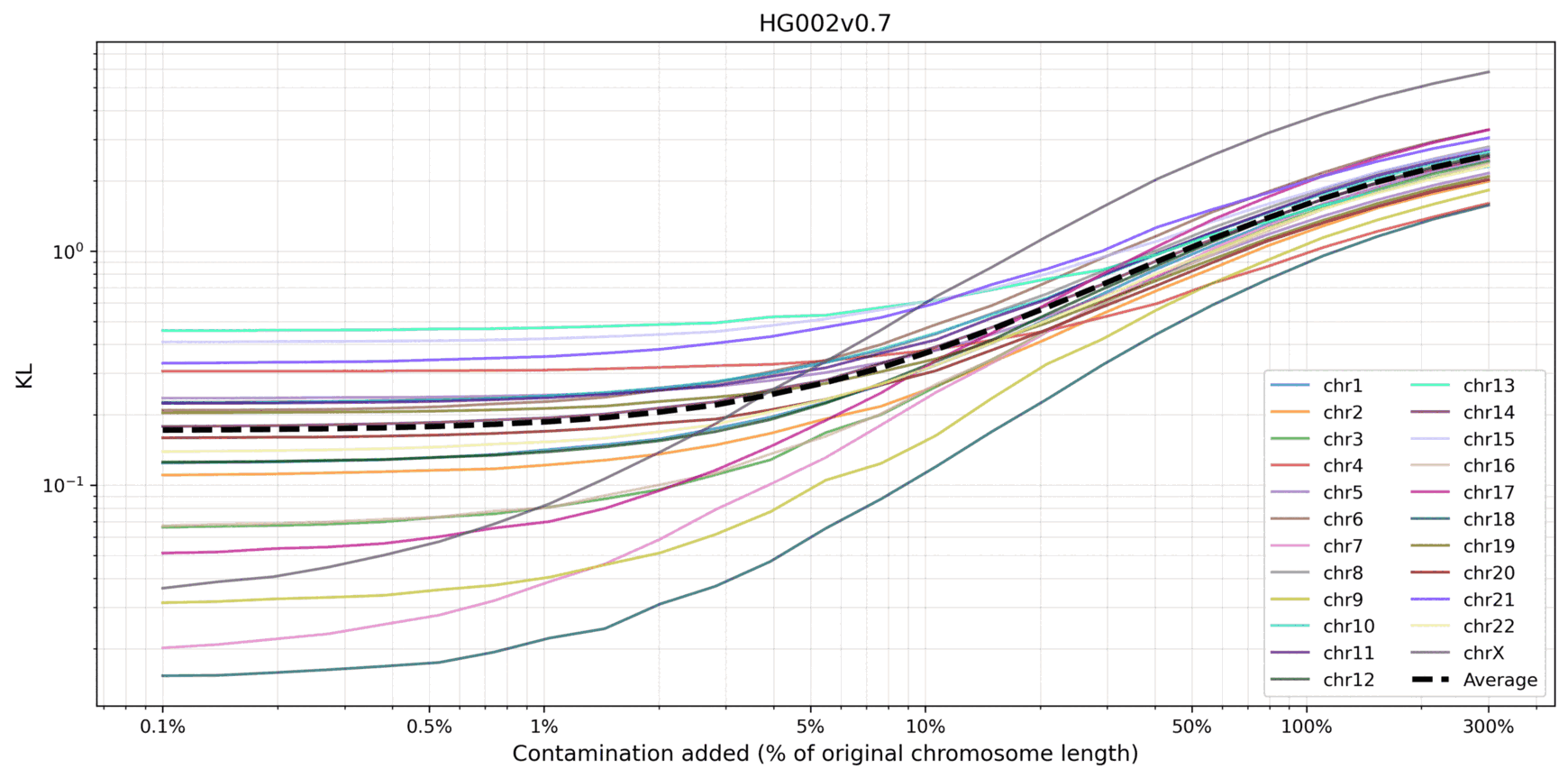}
  \caption{Contamination analysis \textsc{hg}\oldstylenums{002}v\oldstylenums{0.7}}
\end{subfigure}

\begin{subfigure}{.65\textwidth}
  \centering
  \includegraphics[width=\textwidth]{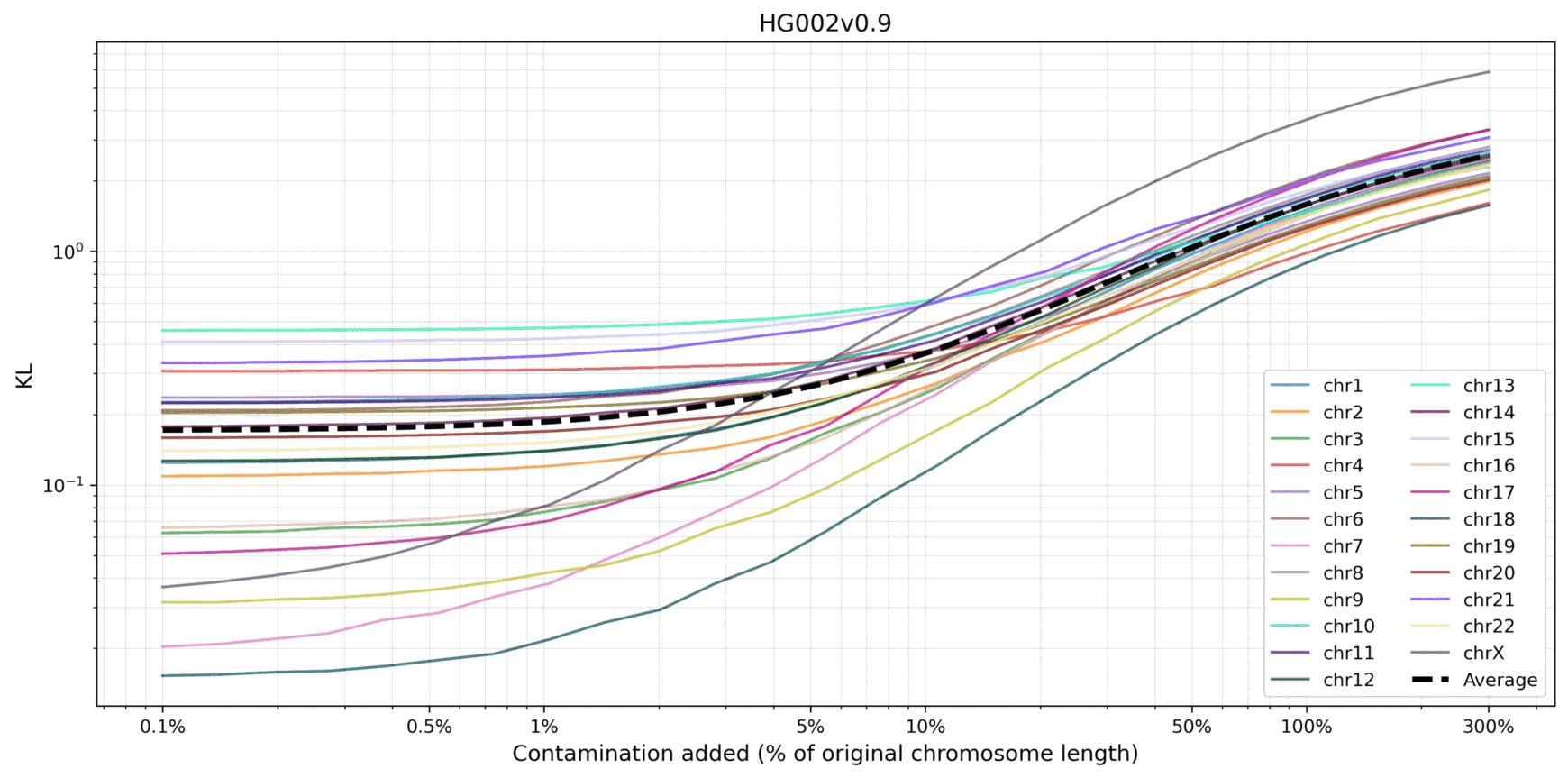}
  \caption{Contamination analysis \textsc{hg}\oldstylenums{002}v\oldstylenums{0.9}}
\end{subfigure}

\begin{subfigure}{.65\textwidth}
  \centering
  \includegraphics[width=\textwidth]{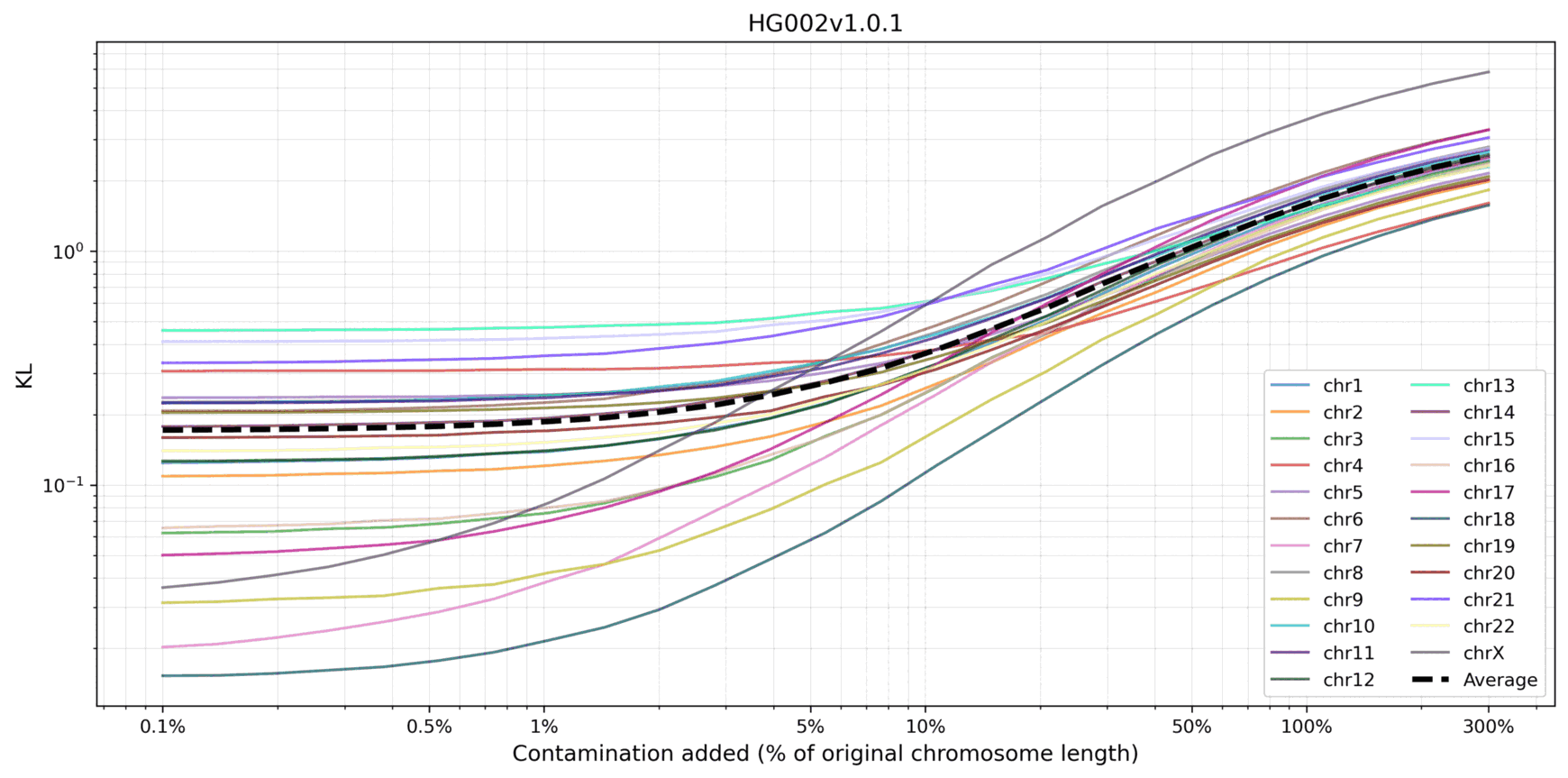}
  \caption{Contamination analysis \textsc{hg}\oldstylenums{002}v\oldstylenums{1.0.1}}
\end{subfigure}
\caption{Contamination analysis for \textsc{hg}\oldstylenums{002} versions 0.7, 0.9, and 1.0.1.}
\end{figure}
\newpage

\begin{figure}[htbp]
\centering
\begin{subfigure}{.65\textwidth}
  \centering
  \includegraphics[width=\textwidth]{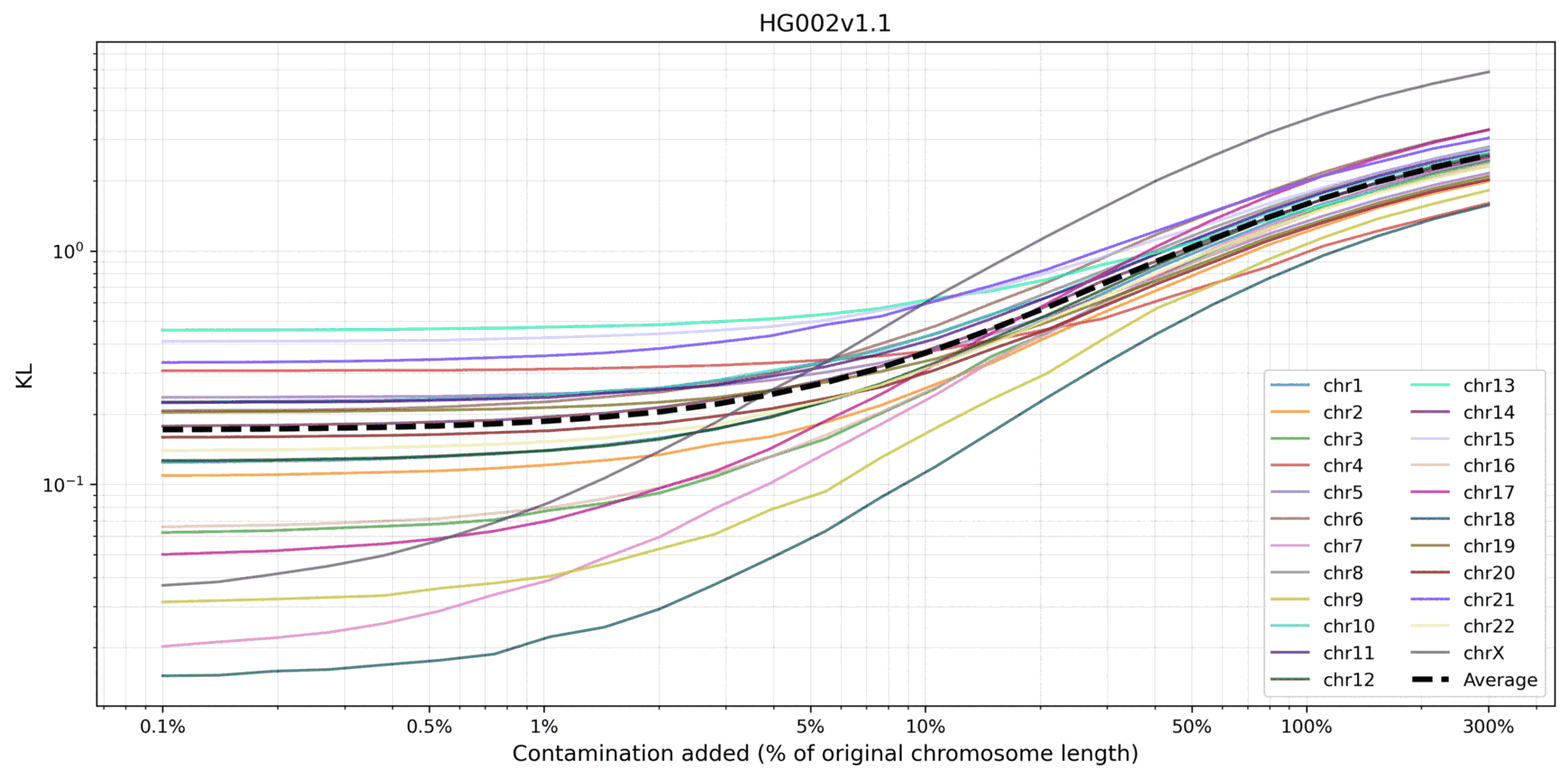}
  \caption{Contamination analysis \textsc{hg}\oldstylenums{002}v\oldstylenums{1.1}}
\end{subfigure}

\begin{subfigure}{.65\textwidth}
  \centering
  \includegraphics[width=\textwidth]{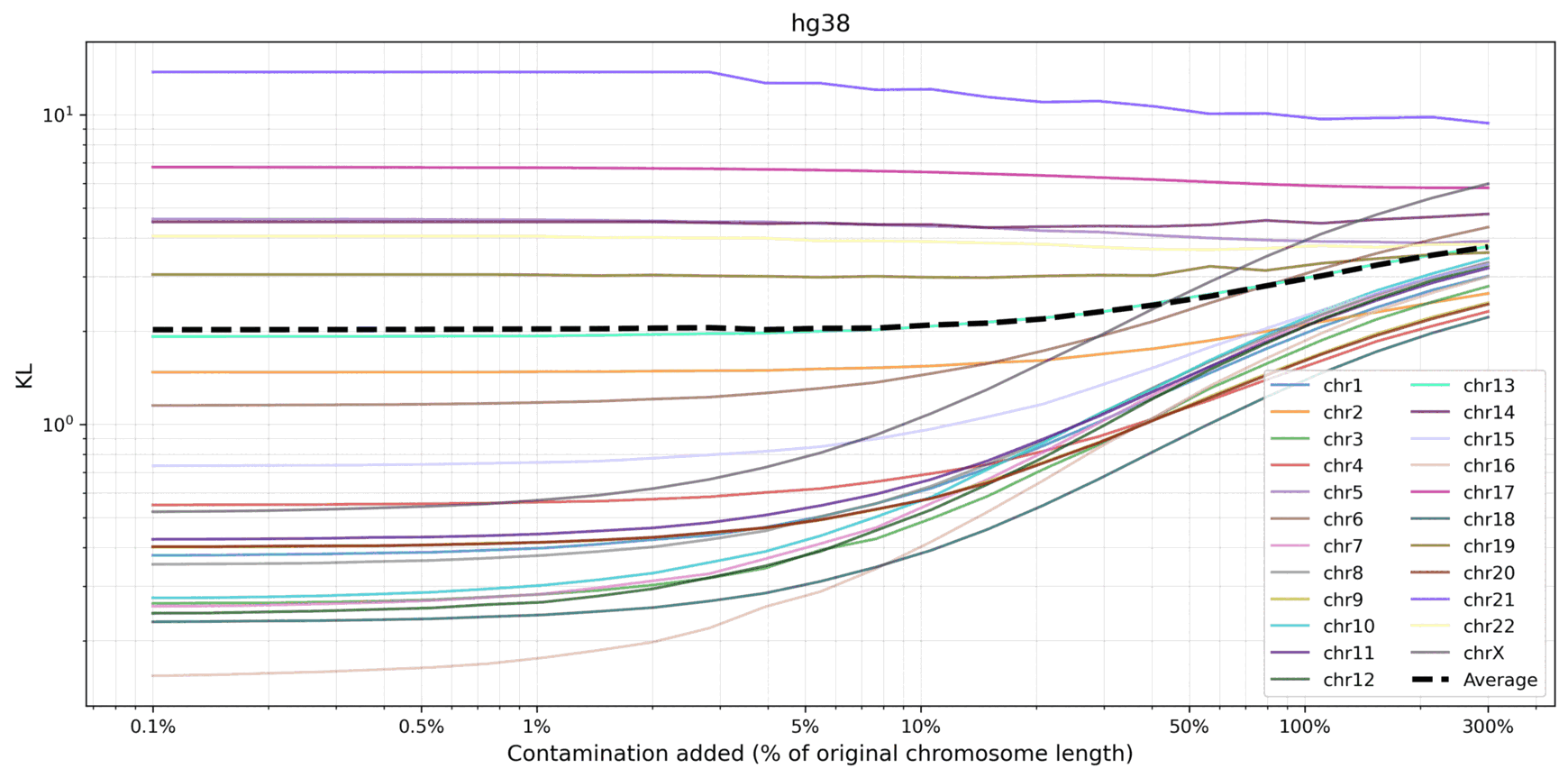}
  \caption{Contamination analysis \textsc{grc}h\oldstylenums{38}}
\end{subfigure}

\begin{subfigure}{.65\textwidth}
  \centering
  \includegraphics[width=\textwidth]{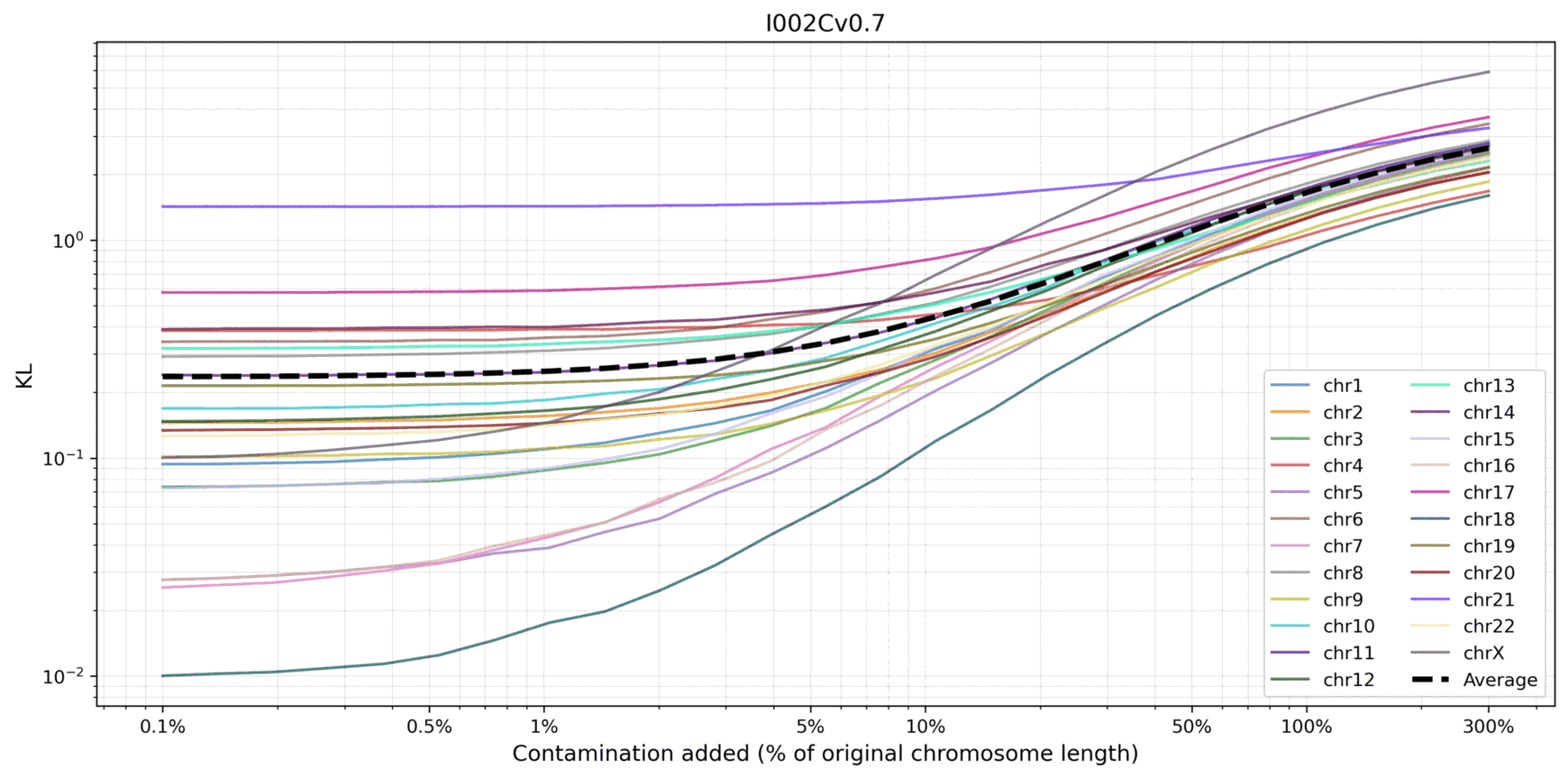}
  \caption{Contamination analysis \textsc{i}\oldstylenums{002}v\oldstylenums{0.7}}
\end{subfigure}
\caption{Contamination analysis for \textsc{hg}\oldstylenums{002}v\oldstylenums{1.1}, \textsc{grc}h\oldstylenums{38}, and \textsc{i}\oldstylenums{002}v\oldstylenums{0.7}.}
\end{figure}
\newpage

\begin{figure}[htbp]
\centering
\begin{subfigure}{.5\textwidth}
  \centering
  \includegraphics[width=\textwidth]{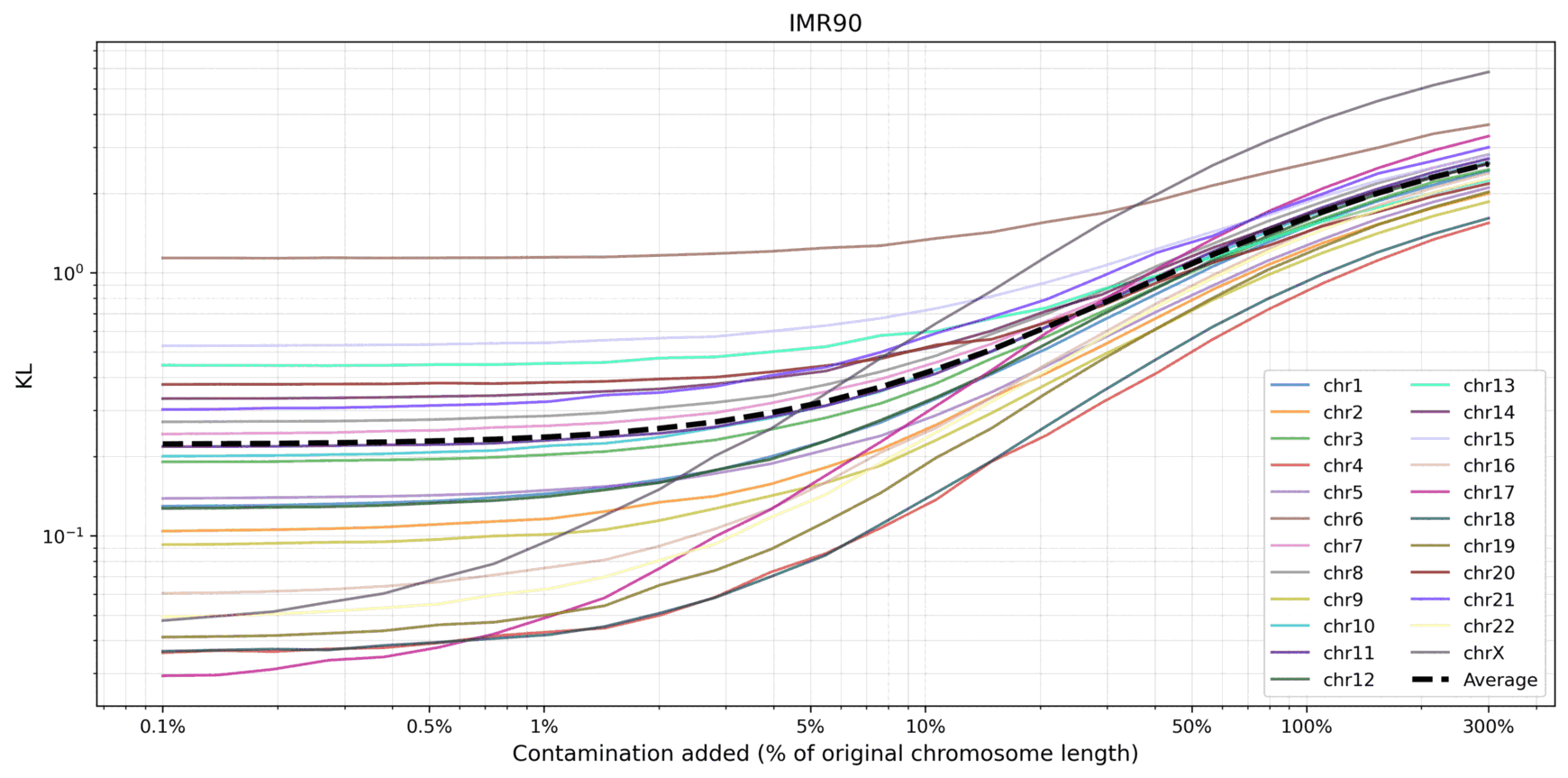}
  \caption{Contamination analysis \textsc{imr}\oldstylenums{90}}
\end{subfigure}

\begin{subfigure}{.5\textwidth}
  \centering
  \includegraphics[width=\textwidth]{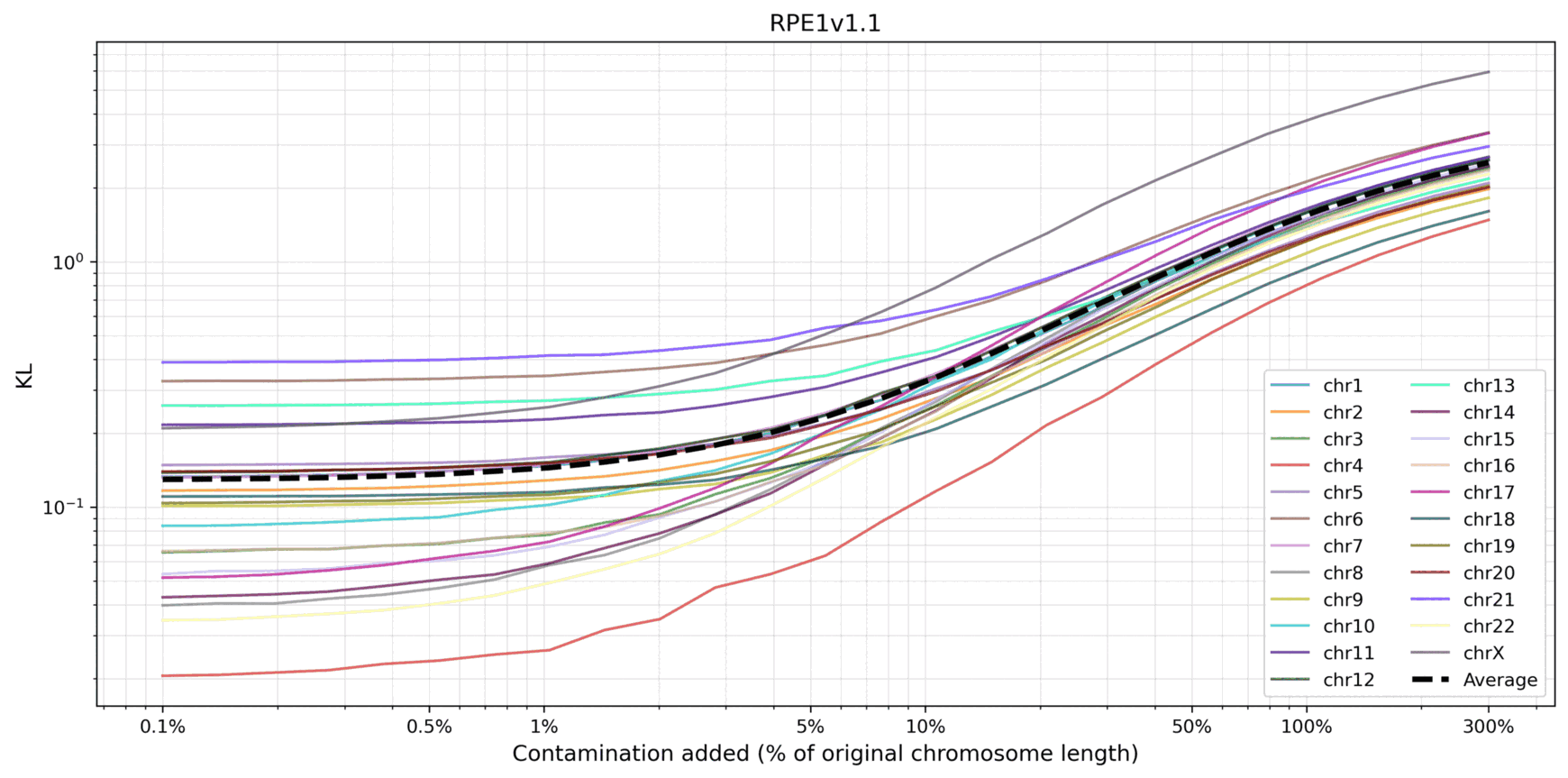}
  \caption{Contamination analysis \textsc{rpe}\oldstylenums{1}v\oldstylenums{1.1}}
\end{subfigure}

\begin{subfigure}{.5\textwidth}
  \centering
  \includegraphics[width=\textwidth]{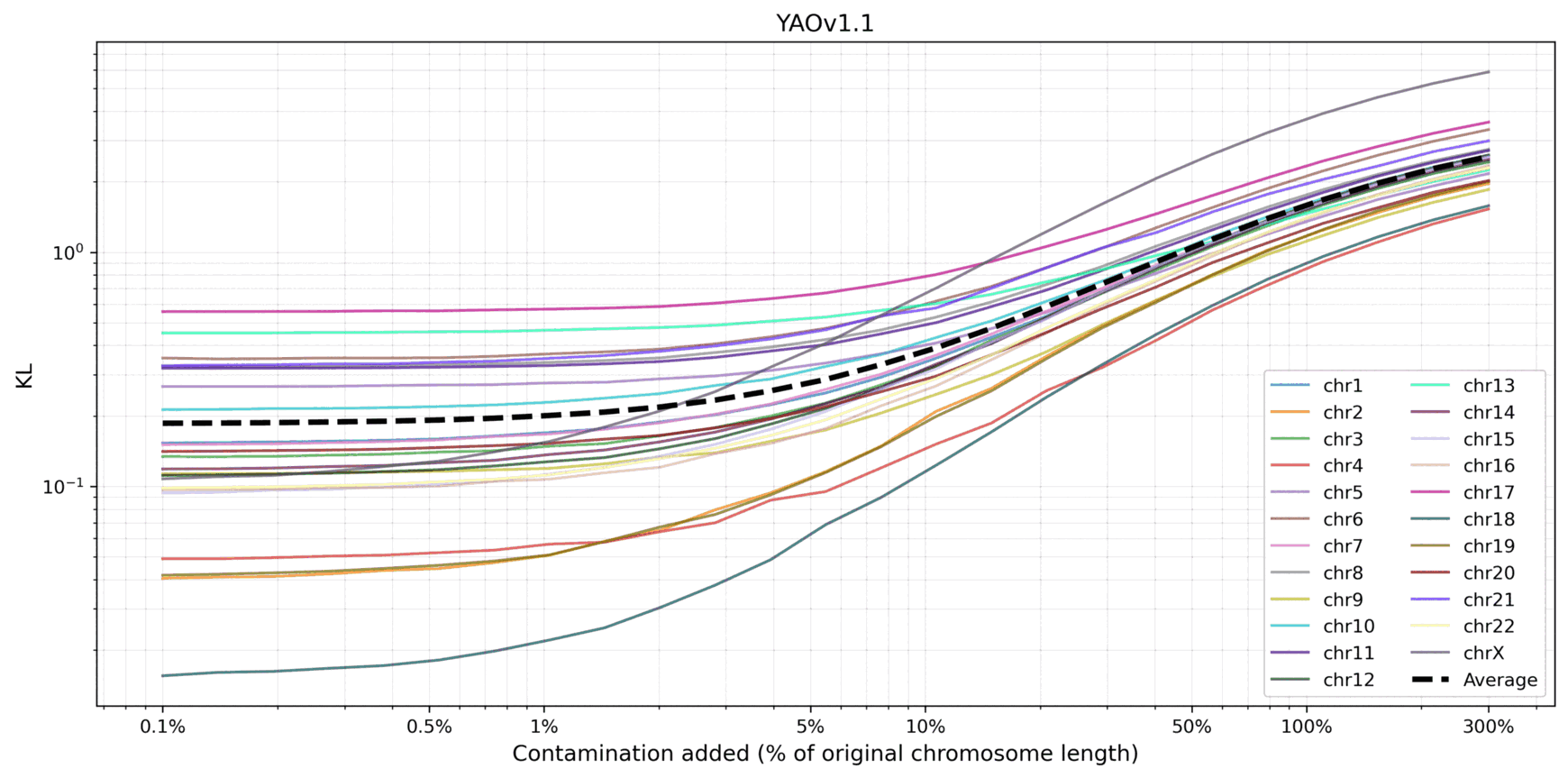}
  \caption{Contamination analysis \textsc{yao}v\oldstylenums{1.1}}
\end{subfigure}

\begin{subfigure}{.5\textwidth}
  \centering
  \includegraphics[width=\textwidth]{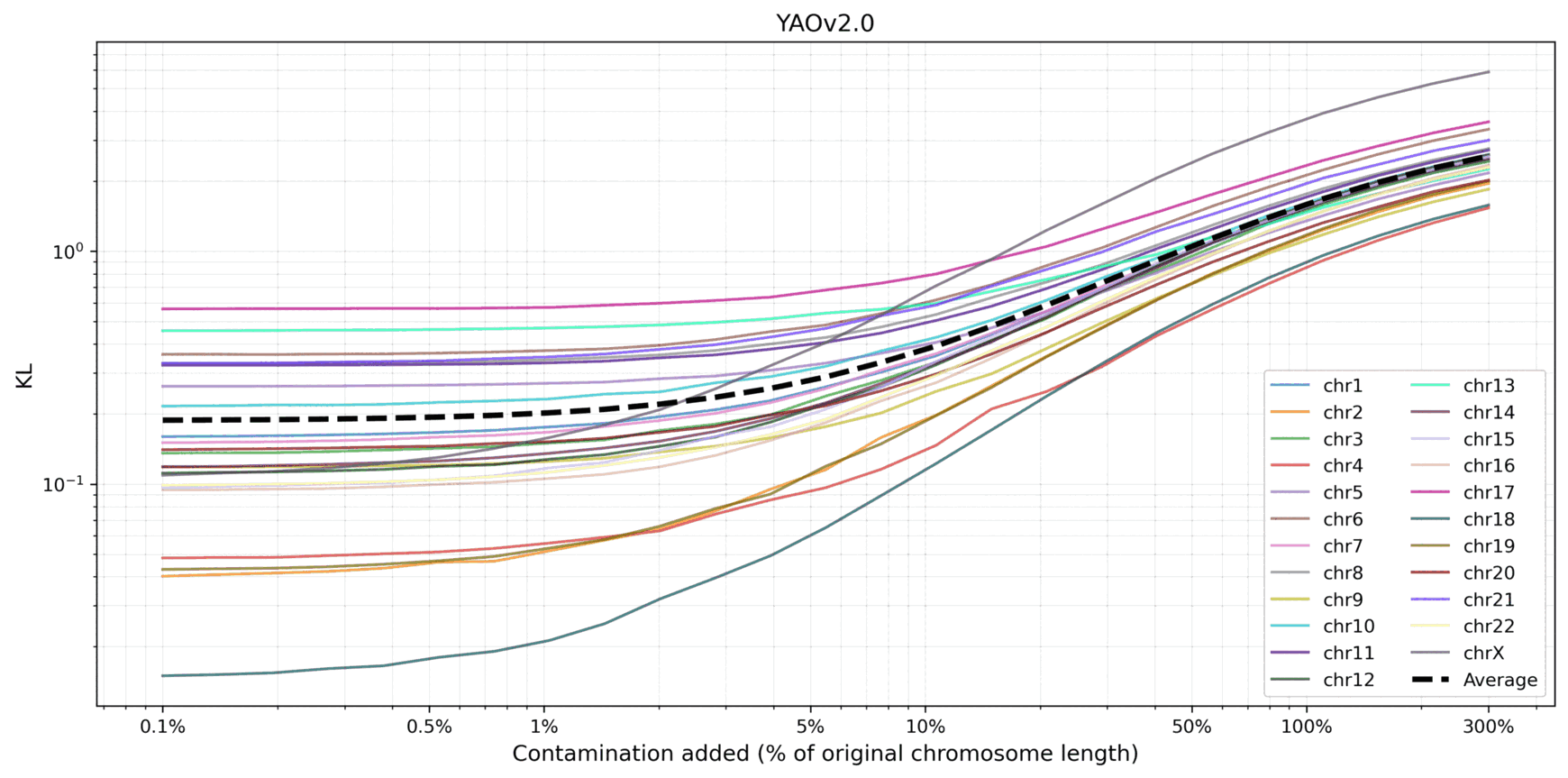}
  \caption{Contamination analysis \textsc{yao}v\oldstylenums{2.0}}
\end{subfigure}
\caption{Contamination analysis for \textsc{imr}\oldstylenums{90}, \textsc{rpe}\oldstylenums{1}, and \textsc{yao} versions 1.1 and 2.0.}
\end{figure}

\newpage

\begin{figure}[htbp]
\centering
\begin{subfigure}{.5\textwidth}
  \centering
  \includegraphics[width=\textwidth]{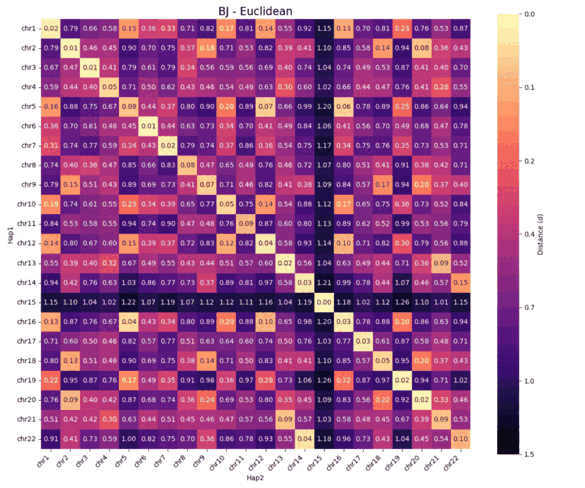}
  \caption{\textsc{bj} Euclidean}
\end{subfigure}

\begin{subfigure}{.5\textwidth}
  \centering
  \includegraphics[width=\textwidth]{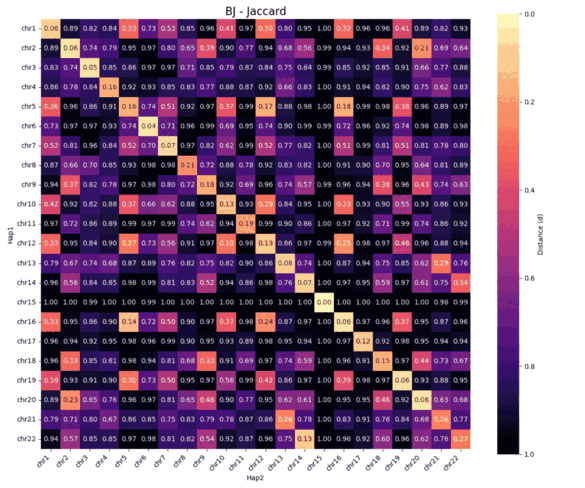}
  \caption{\textsc{bj} Jaccard}
\end{subfigure}

\begin{subfigure}{.5\textwidth}
  \centering
  \includegraphics[width=\textwidth]{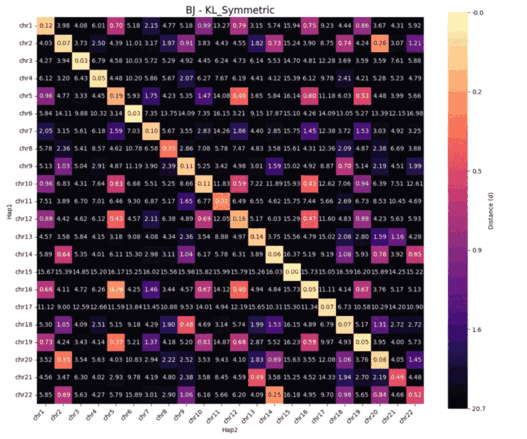}
  \caption{\textsc{bj} Symmetric \textsc{kl} }
\end{subfigure}
\caption{Distance heatmaps for the \textsc{bj} cell line.}
\end{figure}
\newpage

\begin{figure}[htbp]
\centering
\begin{subfigure}{.5\textwidth}
  \centering
  \includegraphics[width=\textwidth]{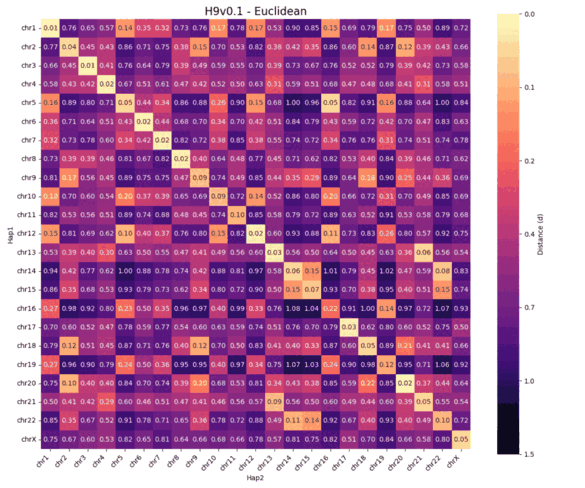}
  \caption{\textsc{h}\oldstylenums{9}v\oldstylenums{0.1} Euclidean}
\end{subfigure}

\begin{subfigure}{.5\textwidth}
  \centering
  \includegraphics[width=\textwidth]{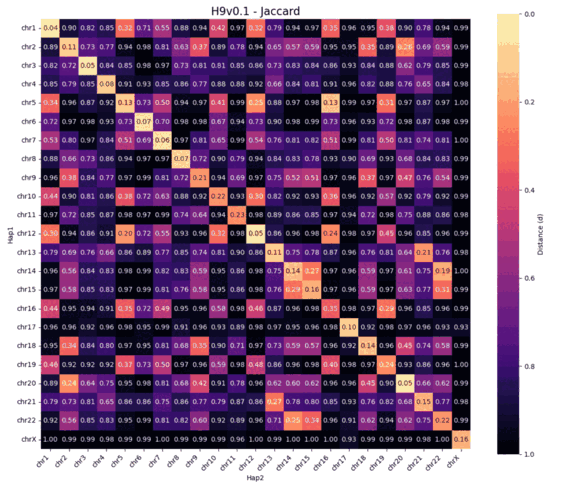}
  \caption{\textsc{h}\oldstylenums{9}v\oldstylenums{0.1} Jaccard}
\end{subfigure}

\begin{subfigure}{.5\textwidth}
  \centering
  \includegraphics[width=\textwidth]{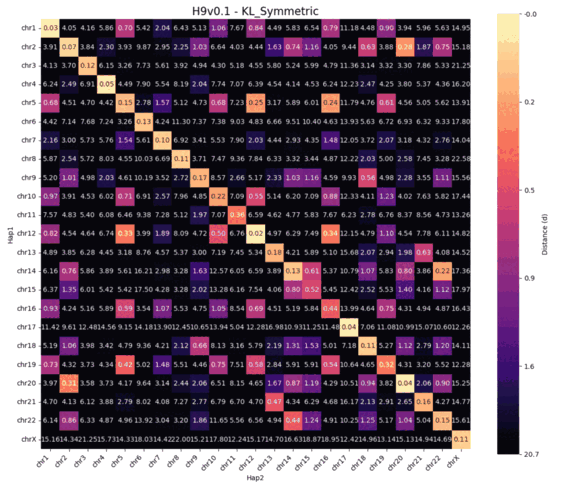}
  \caption{\textsc{h}\oldstylenums{9}v\oldstylenums{0.1} Symmetric \textsc{kl} }
\end{subfigure}
\caption{Distance heatmaps for the \textsc{h}\oldstylenums{9}v\oldstylenums{0.1} cell line.}
\end{figure}
\newpage

\begin{figure}[htbp]
\centering
\begin{subfigure}{.5\textwidth}
  \centering
  \includegraphics[width=\textwidth]{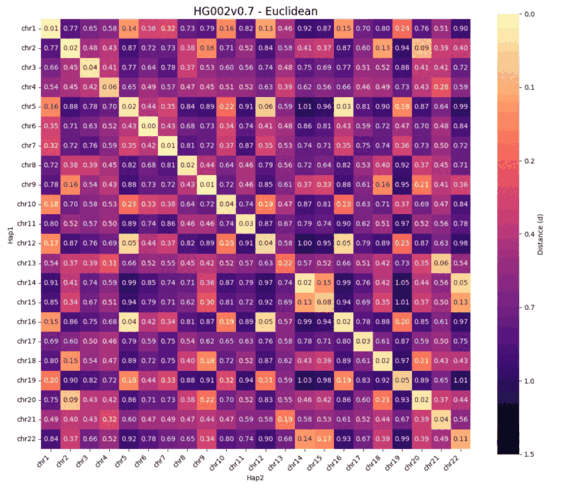}
  \caption{\textsc{hg}\oldstylenums{002}v\oldstylenums{0.7} Euclidean}
\end{subfigure}

\begin{subfigure}{.5\textwidth}
  \centering
  \includegraphics[width=\textwidth]{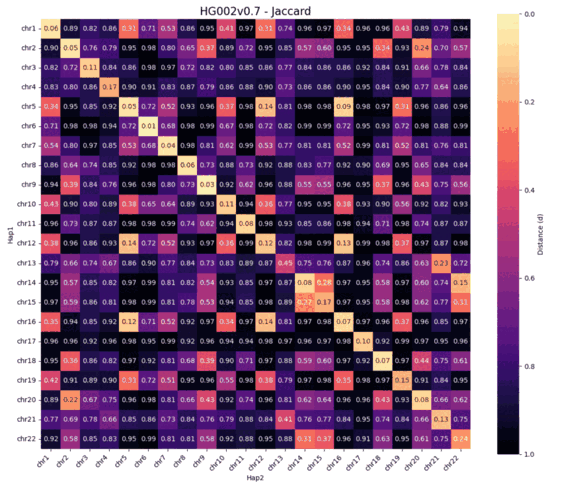}
  \caption{\textsc{hg}\oldstylenums{002}v\oldstylenums{0.7} Jaccard}
\end{subfigure}

\begin{subfigure}{.5\textwidth}
  \centering
  \includegraphics[width=\textwidth]{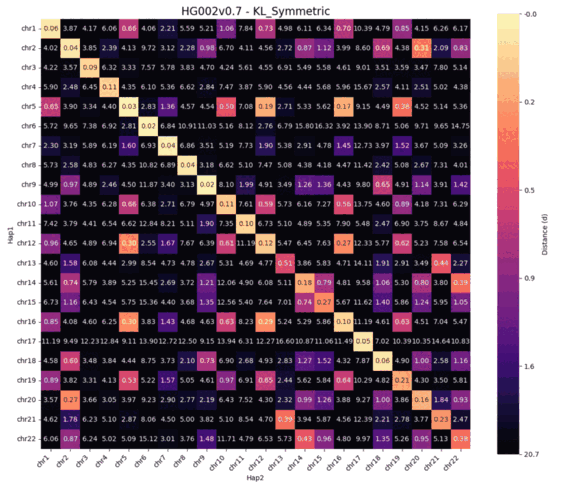}
  \caption{\textsc{hg}\oldstylenums{002}v\oldstylenums{0.7} Symmetric \textsc{kl} }
\end{subfigure}
\caption{Distance heatmaps for \textsc{hg}\oldstylenums{002}v\oldstylenums{0.7}.}
\end{figure}
\newpage

\begin{figure}[htbp]
\centering
\begin{subfigure}{.5\textwidth}
  \centering
  \includegraphics[width=\textwidth]{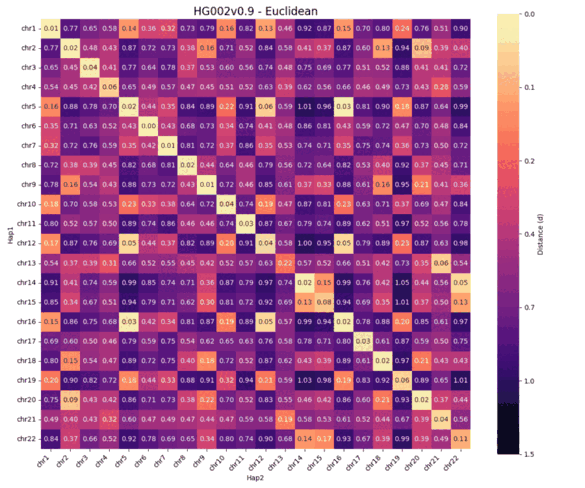}
  \caption{\textsc{hg}\oldstylenums{002}v\oldstylenums{0.9} Euclidean}
\end{subfigure}

\begin{subfigure}{.5\textwidth}
  \centering
  \includegraphics[width=\textwidth]{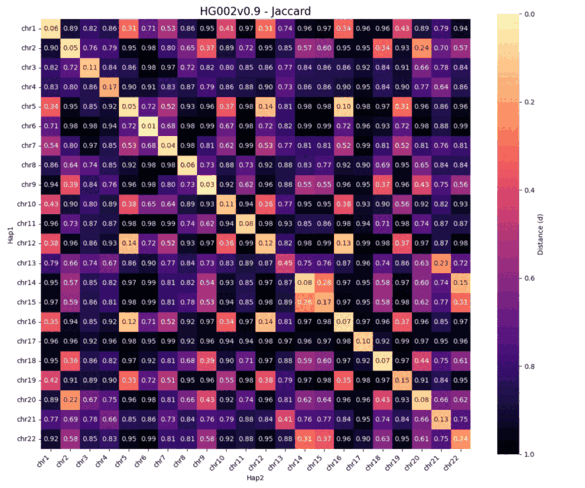}
  \caption{\textsc{hg}\oldstylenums{002}v\oldstylenums{0.9} Jaccard}
\end{subfigure}

\begin{subfigure}{.5\textwidth}
  \centering
  \includegraphics[width=\textwidth]{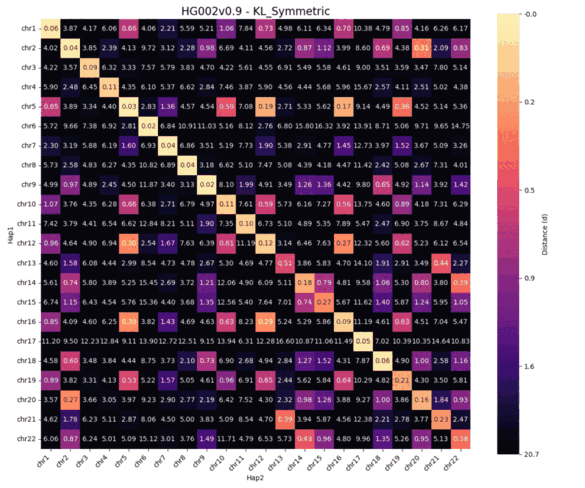}
  \caption{\textsc{hg}\oldstylenums{002}v\oldstylenums{0.9} Symmetric \textsc{kl} }
\end{subfigure}
\caption{Distance heatmaps for \textsc{hg}\oldstylenums{002}v\oldstylenums{0.9}.}
\end{figure}
\newpage

\begin{figure}[htbp]
\centering
\begin{subfigure}{.5\textwidth}
  \centering
  \includegraphics[width=\textwidth]{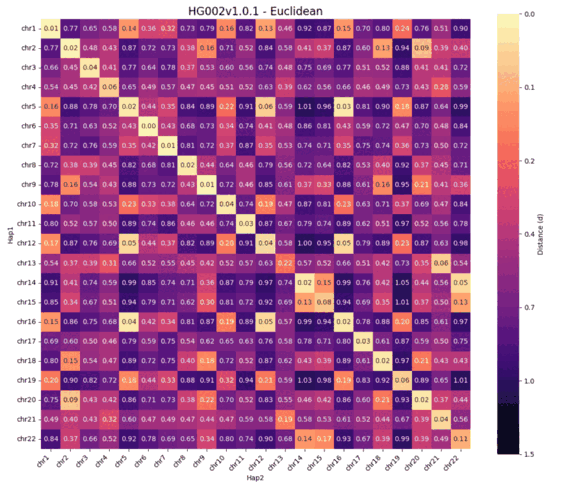}
  \caption{\textsc{hg}\oldstylenums{002}v\oldstylenums{1.0.1} Euclidean}
\end{subfigure}

\begin{subfigure}{.5\textwidth}
  \centering
  \includegraphics[width=\textwidth]{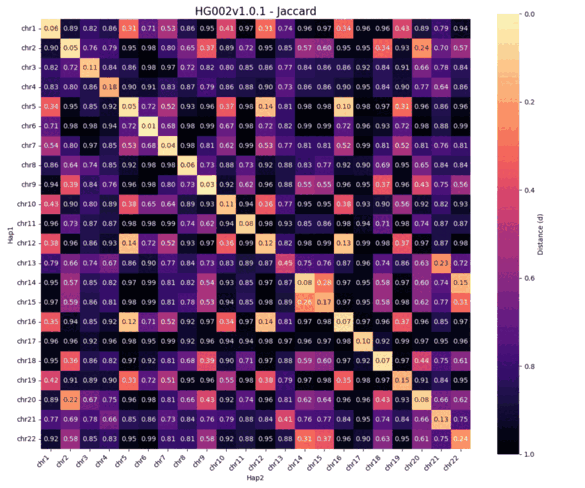}
  \caption{\textsc{hg}\oldstylenums{002}v\oldstylenums{1.0.1} Jaccard}
\end{subfigure}

\begin{subfigure}{.5\textwidth}
  \centering
  \includegraphics[width=\textwidth]{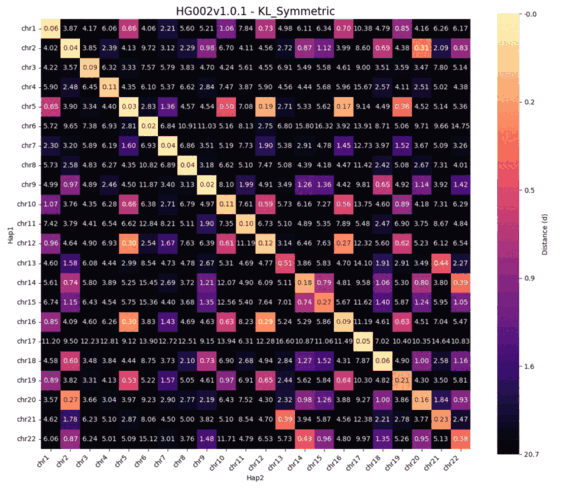}
  \caption{\textsc{hg}\oldstylenums{002}v\oldstylenums{1.0.1} Symmetric \textsc{kl} }
\end{subfigure}
\caption{Distance heatmaps for \textsc{hg}\oldstylenums{002}v\oldstylenums{1.0.1}.}
\end{figure}
\newpage

\begin{figure}[htbp]
\centering
\begin{subfigure}{.5\textwidth}
  \centering
  \includegraphics[width=\textwidth]{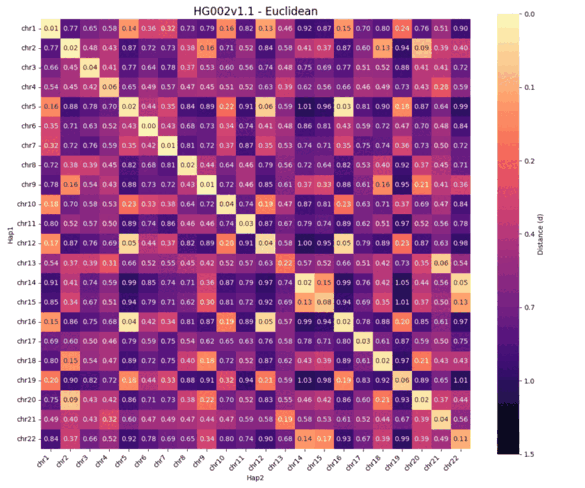}
  \caption{\textsc{hg}\oldstylenums{002}v\oldstylenums{1.1} Euclidean}
\end{subfigure}

\begin{subfigure}{.5\textwidth}
  \centering
  \includegraphics[width=\textwidth]{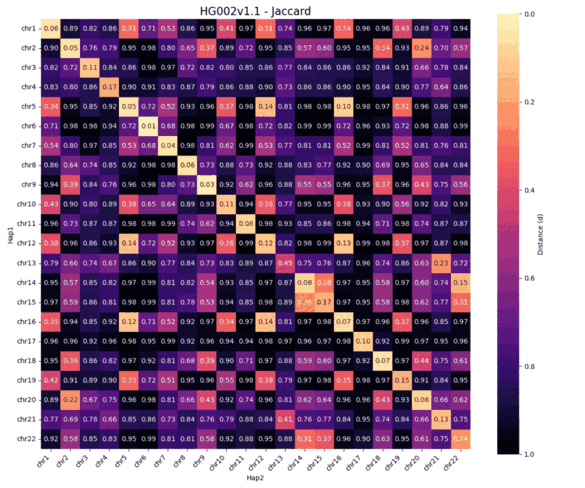}
  \caption{\textsc{hg}\oldstylenums{002}v\oldstylenums{1.1} Jaccard}
\end{subfigure}

\begin{subfigure}{.5\textwidth}
  \centering
  \includegraphics[width=\textwidth]{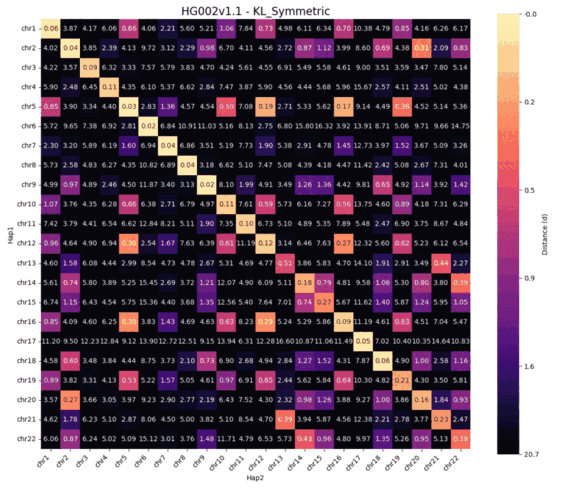}
  \caption{\textsc{hg}\oldstylenums{002}v\oldstylenums{1.1} Symmetric \textsc{kl} }
\end{subfigure}
\caption{Distance heatmaps for \textsc{hg}\oldstylenums{002}v\oldstylenums{1.1}.}
\end{figure}
\newpage

\begin{figure}[htbp]
\centering
\begin{subfigure}{.5\textwidth}
  \centering
  \includegraphics[width=\textwidth]{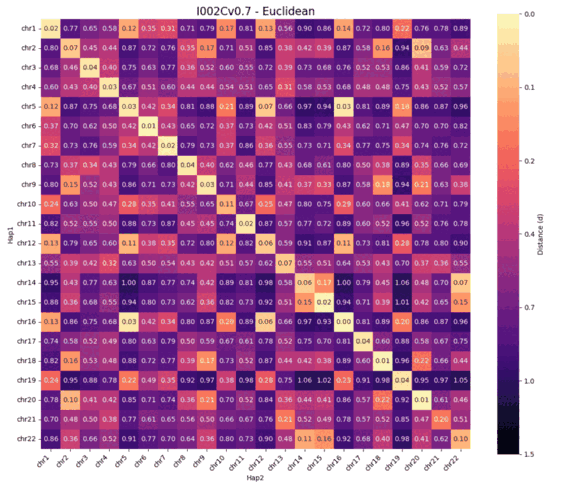}
  \caption{\textsc{i}\oldstylenums{002}v\oldstylenums{0.7} Euclidean}
\end{subfigure}

\begin{subfigure}{.5\textwidth}
  \centering
  \includegraphics[width=\textwidth]{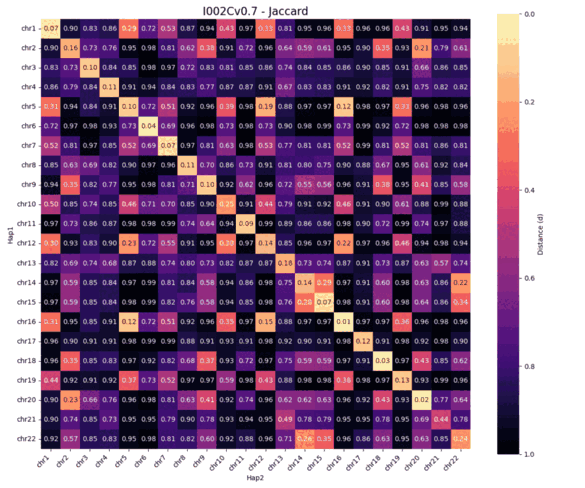}
  \caption{\textsc{i}\oldstylenums{002}v\oldstylenums{0.7} Jaccard}
\end{subfigure}

\begin{subfigure}{.5\textwidth}
  \centering
  \includegraphics[width=\textwidth]{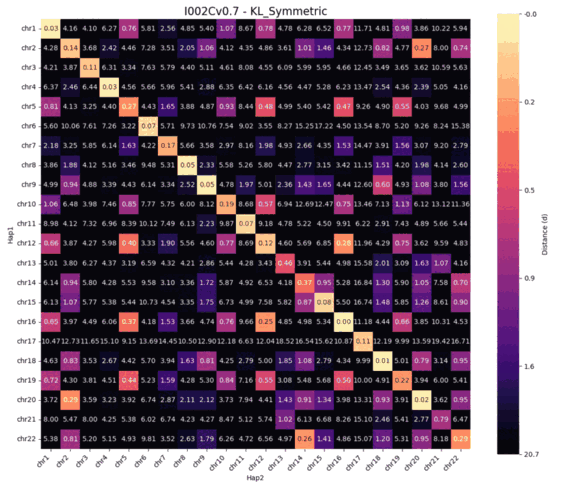}
  \caption{\textsc{i}\oldstylenums{002}v\oldstylenums{0.7} Symmetric \textsc{kl} }
\end{subfigure}
\caption{Distance heatmaps for \textsc{i}\oldstylenums{002}v\oldstylenums{0.7}.}
\end{figure}
\newpage

\begin{figure}[htbp]
\centering
\begin{subfigure}{.5\textwidth}
  \centering
  \includegraphics[width=\textwidth]{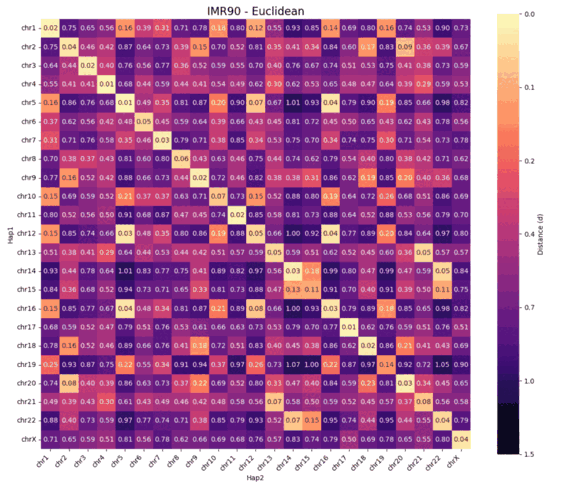}
  \caption{\textsc{imr}\oldstylenums{90} Euclidean}
\end{subfigure}

\begin{subfigure}{.5\textwidth}
  \centering
  \includegraphics[width=\textwidth]{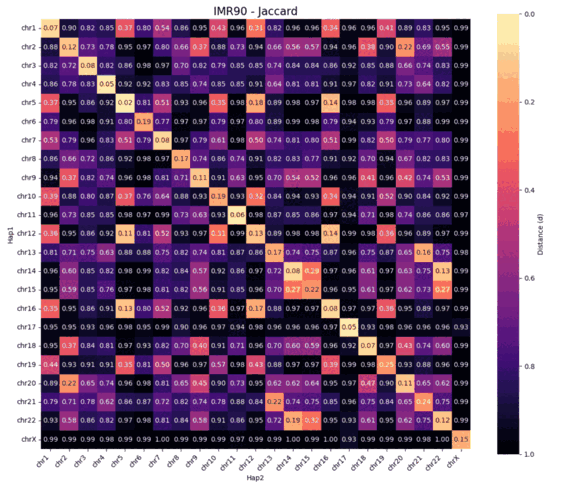}
  \caption{\textsc{imr}\oldstylenums{90} Jaccard}
\end{subfigure}

\begin{subfigure}{.5\textwidth}
  \centering
  \includegraphics[width=\textwidth]{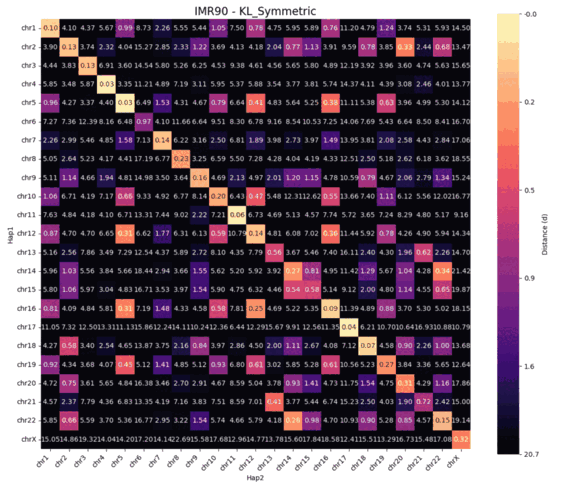}
  \caption{\textsc{imr}\oldstylenums{90} Symmetric \textsc{kl} }
\end{subfigure}
\caption{Distance heatmaps for the \textsc{imr}\oldstylenums{90} cell line.}
\end{figure}
\newpage

\begin{figure}[htbp]
\centering
\begin{subfigure}{.5\textwidth}
  \centering
  \includegraphics[width=\textwidth]{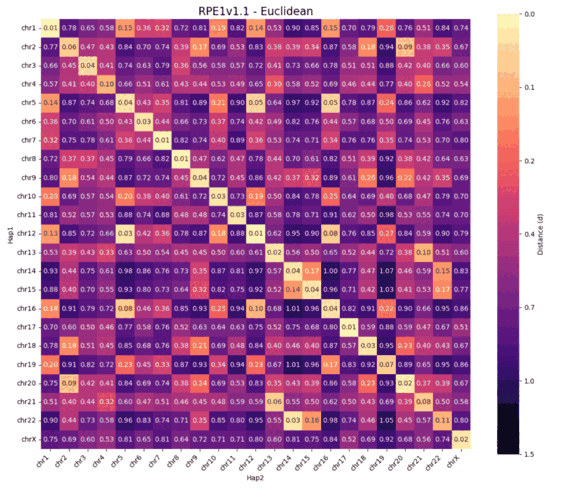}
  \caption{\textsc{rpe}\oldstylenums{1}v\oldstylenums{1.1} Euclidean}
\end{subfigure}

\begin{subfigure}{.5\textwidth}
  \centering
  \includegraphics[width=\textwidth]{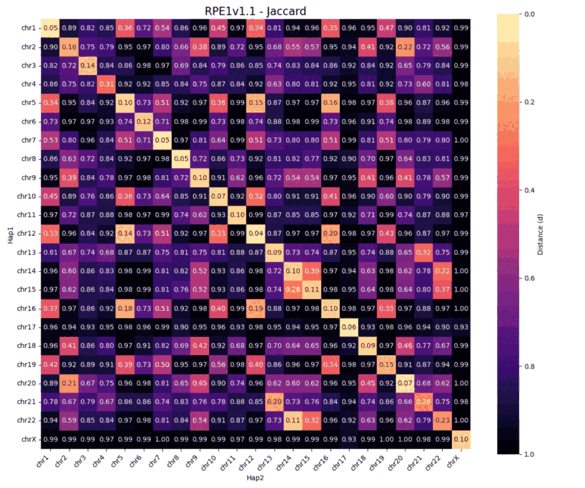}
  \caption{\textsc{rpe}\oldstylenums{1}v\oldstylenums{1.1} Jaccard}
\end{subfigure}

\begin{subfigure}{.5\textwidth}
  \centering
  \includegraphics[width=\textwidth]{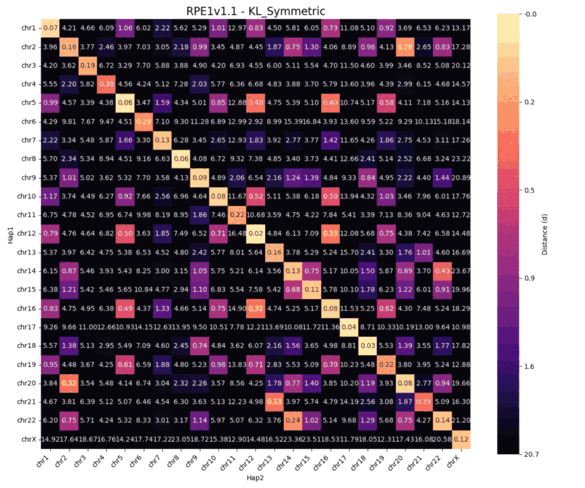}
  \caption{\textsc{rpe}\oldstylenums{1}v\oldstylenums{1.1} Symmetric \textsc{kl} }
\end{subfigure}
\caption{Distance heatmaps for the \textsc{rpe}\oldstylenums{1}v\oldstylenums{1.1} cell line.}
\end{figure}
\newpage

\begin{figure}[htbp]
\centering
\begin{subfigure}{.5\textwidth}
  \centering
  \includegraphics[width=\textwidth]{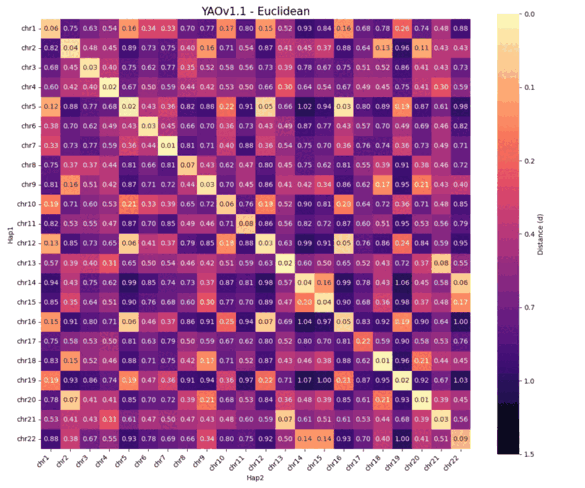}
  \caption{\textsc{yao}v\oldstylenums{1.1} Euclidean}
\end{subfigure}

\begin{subfigure}{.5\textwidth}
  \centering
  \includegraphics[width=\textwidth]{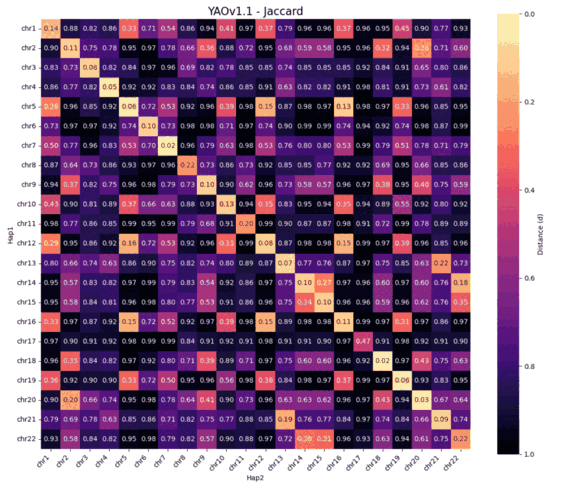}
  \caption{\textsc{yao}v\oldstylenums{1.1} Jaccard}
\end{subfigure}

\begin{subfigure}{.5\textwidth}
  \centering
  \includegraphics[width=\textwidth]{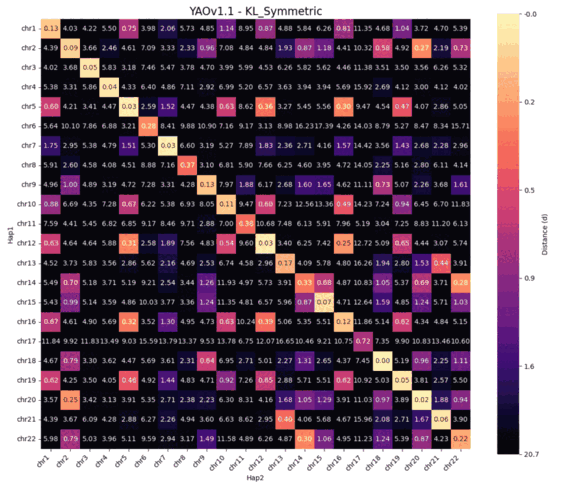}
  \caption{\textsc{yao}v\oldstylenums{1.1} Symmetric \textsc{kl} }
\end{subfigure}
\caption{Distance heatmaps for \textsc{yao}v\oldstylenums{1.1}.}
\end{figure}
\newpage

\begin{figure}[htbp]
\centering
\begin{subfigure}{.5\textwidth}
  \centering
  \includegraphics[width=\textwidth]{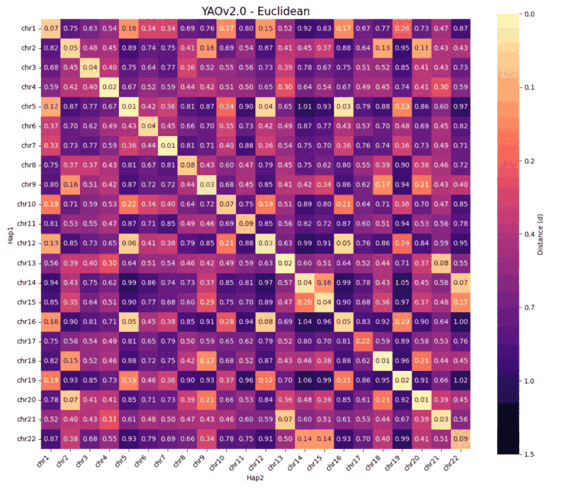}
  \caption{\textsc{yao}v\oldstylenums{2.0} Euclidean}
\end{subfigure} %
\begin{subfigure}{.5\textwidth}
  \centering
  \includegraphics[width=\textwidth]{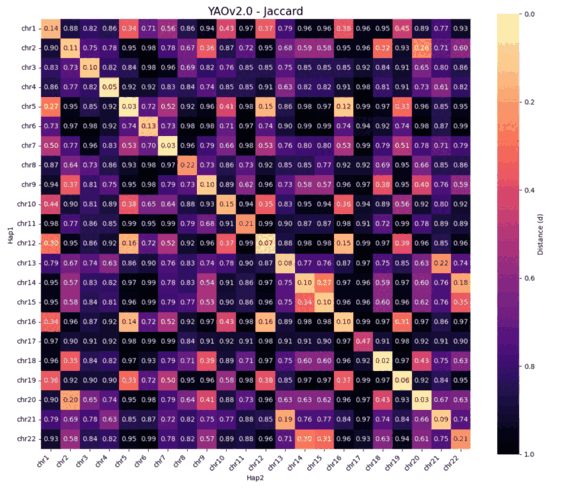}
  \caption{\textsc{yao}v\oldstylenums{2.0} Jaccard}
\end{subfigure}

\begin{subfigure}{.5\textwidth}
  \centering
  \includegraphics[width=\textwidth]{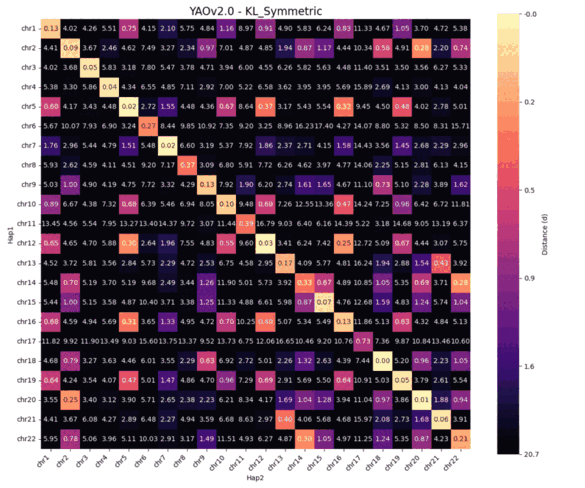}
  \caption{\textsc{yao}v\oldstylenums{2.0} Symmetric \textsc{kl} }
\end{subfigure}%
\begin{subfigure}{.5\textwidth}
  \centering
  \includegraphics[width=\textwidth]{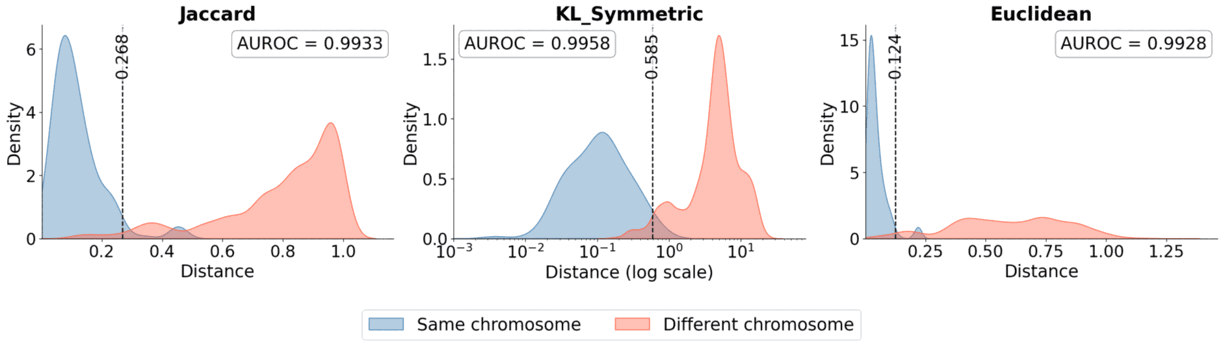}
  \caption{Separability analysis}
\end{subfigure}
\caption{Distance heatmaps for the \textsc{yao}v\oldstylenums{2.0} genome and overall metric separability analysis.}
\end{figure}

\end{document}